\documentclass[11pt, twoside]{scrbook}

\usepackage[utf8x]{inputenc}
\usepackage{graphicx}
\usepackage{amsfonts}
\usepackage{amsmath}
\usepackage{microtype}
\usepackage{arabtex}
\usepackage{xspace}
\usepackage{makeidx}
\usepackage{booktabs}
\usepackage{xspace}
\usepackage{float}
\usepackage{stmaryrd}
\usepackage{psfrag,subfigure,epsfig}
\usepackage{slashed}
\usepackage[page,header,title,titletoc]{appendix}
\usepackage{multicol}
\usepackage{fancyhdr}
\usepackage{mathtools}
\usepackage{makeidx}
\usepackage[numbers,square,comma, compress]{natbib}
\usepackage{hyperref}

\newcommand{\N}{\mathcal{N}}
\newcommand{\mc}{\mathcal}
\newcommand{\mb}{\mathbb}

\newcommand{\im}{\imath}
\newcommand{\jm}{\jmath}
\newcommand{\ST}{space-time\xspace}
\newcommand{\WS}{worldsheet\xspace}
\newcommand{\CG}{$\textrm{CG}_0$\xspace}
\newcommand{\CY}{CY$_3$\xspace}

\csname @addtoreset\endcsname{equation}{section}

\SetSymbolFont{stmry}{bold}{U}{stmry}{m}{n}

\newcommand{\overbar}[1]{\mkern 1.5mu\overline{\mkern-1.5mu#1\mkern-1.5mu}\mkern 1.5mu}

\newcommand{\prefrontmatter}{
   \KOMAoptions{twoside = false}
   \pagenumbering{roman}                                
   \thispagestyle{empty}                               
   \vspace*{\fill}
   \begin{center}
      \Huge{\bf{Topological Amplitudes and the String Effective Action}}\\*[2cm]
      \huge{\bf{Ahmad Zein Assi}}\\*[2.5cm]
      \includegraphics[width=3.2cm,height=3.2cm]{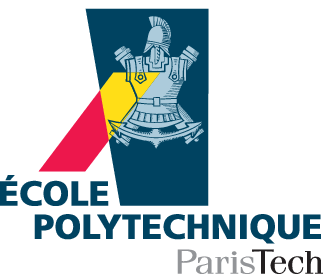}\hspace{1.5cm}\includegraphics[width=3.2cm,height=3.2cm]{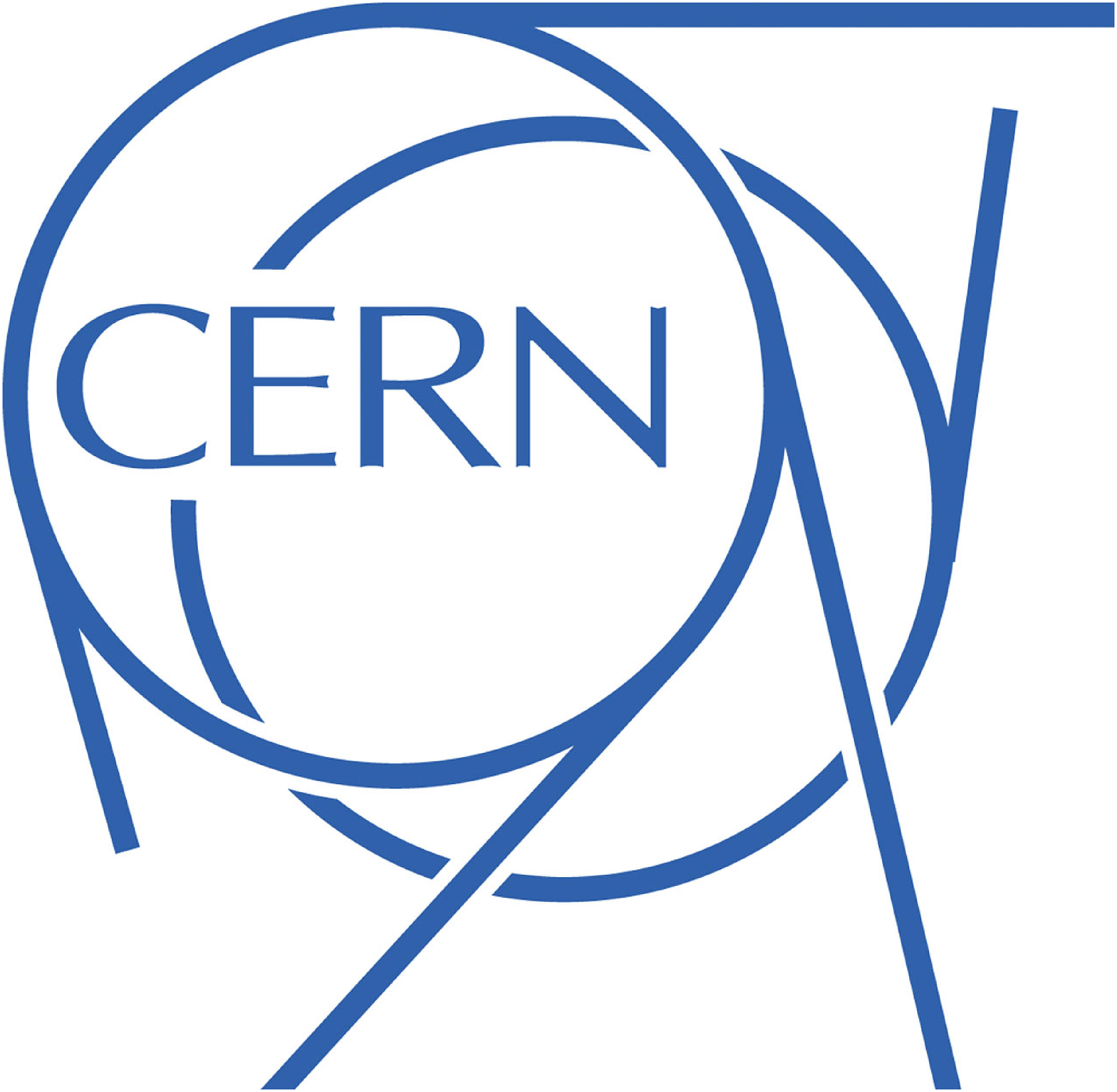}\\*[1cm]
      \Large{Ph.D. Dissertation}\\[0.25cm]
      \Large{Palaiseau, 2013}
   \end{center}
   \vspace*{\fill}
   \clearpage\mbox{}\thispagestyle{empty}\clearpage                                           
   \KOMAoptions{twoside = true}
}

    \newenvironment{newdedication}
        {\null\vspace{\stretch{1}}\begin{flushright}\textsc}
        {\par\end{flushright}\vspace{\stretch{2}}\null}

    \newenvironment{newdedicationar}
        {\null\vspace{\stretch{1}}\begin{flushright}\begin{arabtex}}
        {\par\end{arabtex}\end{flushright}\vspace{\stretch{2}}\null}

        \newenvironment{newquotation}
        {\null\vspace{\stretch{1}}}
        {\par\vspace{\stretch{2}}\null}

\makeindex

\begin{document}

\prefrontmatter
\newpage
\thispagestyle{empty}
\noindent \large{\bf Amplitudes Topologiques et l'Action Effective de la Th\'eorie des Cordes}\\[1cm]
\noindent\large{Th\`ese de doctorat pr\'epar\'ee par}\\[0.25cm]

\large{\bf Ahmad Zein Assi ${}^{*}$}\\[0.25cm]

\noindent\large en vue d'obtenir le grade de\\[0.25cm]

\textbf{\large Docteur De L'\'Ecole Polytechnique}\\[0.5cm]
\noindent{Sp\'ecialit\'e : Physique Th\'eorique\\[0.3cm]}
\hrule\vspace{0.7cm}

\noindent{Soutenue le 11 D\'ecembre 2013 devant la commission d'examen compos\'ee de}\\

\begin{multicols}{2}
 
\noindent Ignatios Antoniadis \\
Emilian Dudas\\
Albrecht Klemm\\
Jose Francisco Morales Morera\\
Kumar Shiv Narain\\
Nikita Nekrasov\\
Boris Pioline\\
Directeur de th\`ese\\
Pr\'esident du jury\\
Examinateur\\
Examinateur\\
Examinateur\\
Rapporteur\\
Rapporteur

\end{multicols}

\vspace{0.5cm}
\hrule
\vspace{0.5cm}
  
\vspace*{\fill}  
\begin{multicols}{2}
{\noindent\small ${}^*$ Centre de Physique Th\'eorique - UMR 7644}\\
{\small Ecole Polytechnique}\\
{\small Bat. 6, RDC, 91128 Palaiseau Cedex, France}\\
{\small Phone +33 (0)1 69 33 42 01\\ Fax +33 (0)1 69 33 49 49}\\
{\small www.cpht.polytechnique.fr}\\
{\noindent\small D\'epartement de Physique - Division Th\'eorie}\\
{\small CERN}\\
{\small CH-1211 Gen\`eve 23, Suisse}\\
{\small Phone +41 (0)22 767 42 22\\ Fax +41 (0)22 767 38 50}\\
{\small wwwth.cern.ch}

\vspace*{2.5cm}
\end{multicols}
\clearpage

\begin{newdedicationar}
\novocalize
\setverb
 {\Large 'il_A 'ahly Al'a\setarab\nospace`izzA'}
\end{newdedicationar}
\thispagestyle{empty}
\mbox{}\clearpage

\frontmatter
\setcounter{page}{5}
    \begin{newdedication}
    {To my loved ones}
    \end{newdedication}
\thispagestyle{empty}
\clearpage\mbox{}\thispagestyle{empty}\clearpage

\tableofcontents


\chapter*{Remerciements}
\addcontentsline{toc}{chapter}{\protect\numberline{}Remerciements}
Cette th\`ese n'aurait jamais pu voir le jour sans l'implication de toutes les personnes \`a qui je d\'edie tous mes remerciements ci-bas. Leurs soutien, expertise et encadrement durant mes trois ann\'ees d'\'etudes doctorales ont \'et\'e essentiels au bon d\'eroulement de mon travail.

Tout d'abord, je remercie mon directeur de th\`ese, Ignatios Antoniadis, pour m'avoir initi\'e aux divers aspects de la recherche en physique th\'eorique. Nos multiples discussions ont forg\'e mon intuition et mes connaissances en th\'eorie des cordes. Son soutien continu m'a permis d'avancer et de garder ma motivation, et ce pendant les p\'eriodes les plus difficiles m\^eme. Je lui suis extr\^emement reconnaissant.

Je voudrais \'egalement remercier Kumar Narain, mon collaborateur, pour sa disponibilit\'e, son accueil chaleureux \`a l'ICTP, Trieste et nos discussions fructueuses. Son regard critique et ses r\'eflexions profondes m'ont inspir\'e tout au long de mon doctorat. Ce fut un honneur de pouvoir travailler \`a ses c\^ot\'es et je lui serai toujours reconnaissant.

Mes sinc\`eres remerciements vont aussi \`a Emilian Dudas, Albrecht Klemm, Francisco Morales, Nikita Nekrasov et Boris Pioline pour avoir accept\'e de faire partie de mon jury de th\`ese.

Par ailleurs, je suis tr\`es reconnaissant \`a la Division Th\'eorique du CERN qui m'a accueilli durant mon doctorat. La qualit\'e exceptionnelle de ses chercheurs et leur disponibilit\'e m'ont permis de diversifier mes connaissances. Je remercie tout particuli\`erement Neil Lambert et Sir John Ellis pour les discussions stimulantes que j'ai pu mener avec eux. De plus, tout au long de mon s\'ejour, j'ai b\'en\'efici\'e du soutien administratif, informatique et moral exceptionnel de Michelle Connor, Nanie Perrin, Jeanne Rostant, Elena Gianolio et Myriam Soltani. Je les remercie du fond du c\oe ur.

Au sein du CPhT, je voudrais remercier Florence Auger, Fadila Debbou, Malika Lang et Jeannine Thomas pour leur disponibilit\'e \`a chaque fois que j'ai eu besoin de les solliciter.

Tout au long de ma th\`ese, j'ai pu b\'en\'efier d'un \'echange constant avec Ioannis Florakis, mon ami et collaborateur. Nos discussions interminables m'ont permis d'apprendre et d'avancer dans mes projets. Son intuition et sa pers\'ev\'erance ont \'et\'e une constante source d'inspiration pour moi. Je le remercie infiniment. Je voudrais \'egalement remercier Stefan Hohenegger, mon collaborateur, pour son regard exp\'eriment\'e et son soutien constant.

Je tiens \'egalement \`a remercier Dieter L\"ust, Stefanos Katmadas, Piotr Tourkine et Daniel Arean pour m'avoir permis de pr\'esenter mon travail \`a l'Institut de Max Planck, Munich, au Centre de Physique Th\'eorique de l'Ecole Polytechnique, Palaiseau, \`a l'Institut de Physique Th\'eorique du CEA, Saclay et \`a l'ICTP, Trieste respectivement. De plus, je remercie Stefan Stieberg et les organisateurs de l'atelier `Beyond the Standard Model' au Physikzentrum, Bad Honnef ainsi que George Zoupanos et les organisateurs de l'institut d'\'et\'e de Corfu 2013 pour m'avoir invit\'e \`a pr\'esenter mon travail lors de ces deux rencontres tr\`es stimulantes.

Tout au long de mon s\'ejour au CERN, j'ai eu le plaisir de c\^otoyer des jeunes physiciens talentueux. Je remercie Emanuele, Andrei, Sung, Ehsan, Fabian, Imtak, Sandeepan, Kwangwoo, Matthijs, Miki, David, Rob, Marius, Navid, Christine et Jean-Claude pour leur amiti\'e et leur sympathie. Nos diverses discussions m'ont permis d'\'elargir mes connaissances et nos sorties ont \'et\'e le meilleur moyen de se changer les id\'ees lorsque cela a \'et\'e n\'ecessaire.

Aujourd'hui, je ne peux nier le r\^ole pr\'epond\'erant que le Master à l'Ecole Normale Sup\'erieure a jou\'e dans ma formation, et ce d'autant plus que mon contact, mon amiti\'e avec mes camarades de classe ont été formidables. Je remercie Adrien, Axel, Benjamin, Fabien, Filippo, Flora, J\'er\'emie, Julius, Konstantina, Lorenzo, Marta, Maxime, M\'elody, Nicolas, $\Pi$-loop, Romain, Swann, Victor, et surtout Gogu, Lay-D, Federico, Piotr et Gautier pour cette année qui restera inoubliable. De plus, j'adresse ma plus grande reconnaissance \`a Mouche sans qui, toute ma vie durant, aurais-je \'et\'e innocent, forminable.

Dans la vie, il y a la physique et le reste, tout aussi important. Certaines amiti\'es et autres recontres ont ponctué mes dernières années et garderont dans ma personnalit\'e une marque ind\'el\'ebile. Kostas, Nikos, Remie, Evi, Thanasis ainsi que tous les vlaxos de la Terre et de l'Univers, vous m'avez prouv\'e que, crise ou pas crise, votre c\oe ur est grand. Je remercie aussi Daria, Anastasia, Jos\'e, Dino et l'ami de Dino qui ont été, pour moi et tant d'autres, une grande source d'inspiration. Pour Nabil, Fran\c{c}ois, Annis, Adrien, Petra, Mael, Maxence et Mehdi aussi avec j'ai partagé plus qu'un appartement. Dans ce contexte assaillant, c'est gr\^ace à Jean, le plus grand chef de tous les temps, que je suis parvenu \`a ma\^itriser les aspects les plus subtils de la th\'eorie des cordes.

Enfin, je remercie ma famille, mes parents, mes fr\`eres et mes s\oe urs. Leur pr\'esence a toujours \'et\'e le plus grand tr\'esor pour moi et leur soutien tout au long de mon doctorat a \'et\'e primordial dans l'accomplissement de mon travail. Jamais n'arriverai-je \`a trouver les mots pour leur exprimer, \`a leur juste valeur, mon respect et ma reconnaissance.

\thispagestyle{fancy}
\lhead{\textsl{Remerciements}}
\rhead{}

\chapter*{R\'esum\'e}
\addcontentsline{toc}{chapter}{\protect\numberline{}R\'esum\'e}
Cette th\`ese est d\'edi\'ee \`a l'\'etude d'une classe de couplages dans l'action effective de la th\'eorie des cordes qui se trouvent au croisement entre la th\'eorie des cordes topologique et les th\'eories de jauge supersym\'etriques. Ces couplages g\'en\'eralisent un ensemble de couplages gravitationnels qui calculent la fonction de partition de la th\'eorie des cordes topologique. Dans la limite de th\'eorie des champs, ces derniers reproduisent la fonction de partition de la th\'eorie de jauge dans le fond $\Omega$ lorsque l'un des param\`etres de ce dernier, $\epsilon_+$, est \'egal \`a z\'ero. Cela sugg\`ere naturellement l'existence d'une g\'en\'eralisation d\'enomm\'ee la corde topologique raffin\'ee. Les couplages \'etudi\'es dans ce manuscrit sont caract\'eris\'es par un multiplet vectoriel suppl\'ementaire et sont calcul\'es, en th\'eorie des cordes, aux niveaux perturbatif et non-perturbatif. De plus, leur limite de th\'eorie des champs donne la fonction de partition de la th\'eorie des champs dans un fond $\Omega$ g\'en\'eral. Ainsi, ces couplages ouvrent de nouvelles perspectives pour la d\'efinition, au niveau de la surface d'univers, de la th\'eorie des cordes topologiques raffin\'ee.\\

\noindent \textbf{Mots-Clefs :} Amplitudes de Th\'eorie des Cordes, Couplages Gravitationnels, Th\'eorie des Cordes Topologique, Fond Om\'ega, Th\'eories de Jauge Supersym\'etriques, Calcul d'Instantons, Corde Topologique Raffin\'ee, Dualit\'es.

\clearpage
\thispagestyle{plain}

\chapter*{Abstract}
\addcontentsline{toc}{chapter}{\protect\numberline{}Abstract}
In this thesis, we study a class of higher derivative couplings in the string effective action arising at the junction of topological string theory and supersymmetric gauge theories in the $\Omega$-background. They generalise a series of gravitational couplings involving gravitons and graviphotons, which reproduces the topological string theory partition function. The latter reduces, in the field theory limit, to the partition function of the gauge theory in the $\Omega$-background when one if its parameters, say $\epsilon_+$, is set to zero. This suggests the existence of a one-parameter extension called the refined topological string. The couplings considered in this work involve an additional vector multiplet and are evaluated, perturbatively and non-perturbatively, at the string level. In the field theory limit, they correctly reproduce the partition function of the gauge theory in a general $\Omega$-background. Hence, these couplings provide new perspectives toward a \WS definition of the refined topological string.\\

\noindent \textbf{Keywords:} String Amplitudes, Gravitational Couplings, Topological String Theory, Omega-Background, Supersymmetric Gauge Theories, Instanton Calculus, Refined Topological String, Dualities.

\clearpage
\thispagestyle{plain}

\chapter*{Summary}
\addcontentsline{toc}{chapter}{\protect\numberline{}Summary}
\label{ch:intro}
Since its early days, string theory has constantly been the source of multiple developments in our understanding of quantum gravity. Yet, not only has it been applied to explore the ultra-violet properties of quantum field theories in the presence of gravity, its unified framework has also shed light on the non-perturbative structure of gauge theories. Indeed, the spectrum of string theory contains certain extended objects, like D-branes, which are non-perturbative in the string coupling. They can be used at weak coupling to probe non-perturbative regimes in the low energy field theory. However, at the string level, a full understanding of the non-perturbative sectors is lacking and one is led to consider sub-sectors of string theory in which one can control the weak and strong coupling regimes using, for instance, supersymmetry. In this context, topological string theory can be viewed as a toy model where both regimes are well understood.

The structure of topological string theory is that of a topological field theory which is a quantum field theory whose correlation functions of physical observables are closed differential forms on the moduli space of \ST metrics considered up to diffeomorphisms. More specifically, coupling a twisted (topological) two-dimensional $\N=2$ supersymmetric sigma-model to gravity leads to the topological string.

The interplay between string theory and the topological string has led to fascinating developments in mathematics as well as in physics. For example, from the mathematical point of view, the observables of topological string theory can usually be understood as topological invariants of the space on which the theory is defined. Furthermore, these quantities acquire a clear physical meaning in string theory since the topological string can be viewed as a `twisted' version of string theory. In particular, the topological string captures properties of string theory that only depend on its topological, BPS sector\footnote{A Bogomol'nyi-Prasad-Sommerfield (BPS) sector is a set of states called \emph{BPS} that are invariant under a non-trivial subalgebra of the full supersymmetry algebra.}. The good understanding of these objects has also been used to test various string dualities, to study micro-state counting of BPS black holes or even wall-crossing.

On the other hand, one of the uses of topological string theory is to `geometrically engineer' supersymmetric gauge theories in four dimensions \cite{Gopakumar:1998ii,Gopakumar:1998jq}. This can be seen by supplementing the four-dimensional \ST with an internal Calabi-Yau manifold and consider the topological string defined on the resulting space. In this picture, gauge theory instantons are nothing but tree-level \WS instantons, and higher genus instantons can be viewed as `gravitational' corrections which are encoded in terms of higher derivative couplings in the string effective action. Since the latter only depend on the topological sector of string theory due to the decoupling of the string oscillators, they can be calculated exactly at the string level.

The typical example is the $\mc N=2$, $SU(2)$ gauge theory studied by Seiberg and Witten \cite{Seiberg:1994rs,Seiberg:1994aj}. Due to supersymmetry, the low energy energy effective action depends only on a holomorphic function called the prepotential which was calculated by Seiberg and Witten to the full non-perturbative level. In fact, the perturbative expansion stops at one-loop order and the prepotential reads
\begin{equation}
 \mc F^{\textrm{SW}}(a,\Lambda)=\frac{i}{2\pi}\,a^2\ln\frac{a^2}{\Lambda^2}+\sum_{k\geq1}\mc F_k\left(\frac{\Lambda}{a}\right)^{4k}a^2\,,
\end{equation}
where $a$ is the complex scalar of the $\mc N=2$ vector multiplet, $k$ is the instanton number and $\Lambda$ is the dynamically generated scale of the theory. The Seiberg-Witten solution gives precisely an ansatz for the instanton coefficients $\mc F_k$. The string compactification giving rise to this gauge theory is a Type IIA on an elliptically fibered $K3$ over $\mb P^1$, in which one studies an $A_1$ singularity in $K3$ corresponding to a vanishing $\mb P^1$. This is referred to as the local $\mb P^1\times\mb P^1$ model, since only the local description of the singularity is sufficient to recover the $\N=2$ gauge theory in four dimensions. In particular, this compactification has been used to generalise the Seiberg-Witten prepotential in the presence of gravity \cite{Kachru:1995fv,Katz:1996fh}. Indeed, decoupling the gravity/string modes leads to an $\N=2$ $SU(2)$ gauge theory. More specifically, the genus $g$ partition function of the local $\mb P^1\times\mb P^1$ is denoted $F_g$ and it has been shown that $F_0$, in the field theory limit, agrees with the Seiberg-Witten solution, with the Coulomb branch parameter $a$ being identified with the K\"ahler structure of the internal manifold. For $g>1$, $F_g$ can be interpreted as gravitational corrections.

For this, recall that the topological string is a sub-sector of string theory since it is obtained from the latter by performing the \emph{topological twist} \cite{Witten:1988xj}. It turns out that $F_g$ computes the (moduli dependent) coupling coefficients of a class of higher derivative F-terms in the string effective action \cite{Antoniadis:1995zn} coupling, at genus $g$, two gravitons and $2g-2$ graviphotons of definite self-duality, say anti-self-dual:
\begin{equation}
 F_g\,R_{-}^2\,F_G^{2g-2}\,.\label{GravBackground}
\end{equation}
Here, $R$ denotes the anti-self-dual Riemann tensor and $F_G$ is the anti-self-dual part of the graviphoton field strength. Due to supersymmetry, these couplings receive contributions from BPS states only so that they are protected against any perturbative and non-perturbative corrections. They have been used, for instance, to test the dualities between Type II, Heterotic and Type I string theories.\thispagestyle{fancy}\lhead{\textsl{Summary}}\rhead{}\fancyfoot[L]{\thepage} From the gauge theory point of view, it is rather surprising that these gravitational corrections survive the point particle limit. Their interpretation from the low energy effective action point of view goes beyond the Seiberg-Witten theory and becomes clear when one introduces the $\Omega$-background \cite{Moore:1997dj,Lossev:1997bz}. In fact, one may wonder whether it is possible to calculate the instanton coefficients $\mc F_k$ by evaluating the path integral over the instanton moduli space. However, the latter is not well-defined because of the non-compactness of the moduli space, and the $\Omega$-background acts as an appropriate regulator. It can most easily be seen as arising from a dimensional reduction of an $\N=1$ gauge theory in six dimensions on a $T^2$ torus fibration over $\mb R^4$, such that going around the cycles of $T^2$ is accompanied by a rotation in $\mb R^4$ parametrised by two complex parameters, $\epsilon_{1,2}$, and an R-symmetry rotation proportional to $\epsilon_1+\epsilon_2$. The latter is crucial in order to preserve supersymmetry. The integral over the instanton moduli space becomes well-behaved and, for instance, the effective volume of \ST is equal to $1/\epsilon_1\epsilon_2$. Moreover, the prepotential of this `deformed' gauge theory can be evaluated and takes the form
\begin{equation}
 \frac{\mc F^{\textrm{Nek}}(a,\Lambda,\epsilon_1,\epsilon_2)}{\epsilon_1\epsilon_2}\,.
\end{equation}
In the limit $\epsilon_{1,2}\rightarrow0$, it reduces to $\mc F^{\textrm{SW}}(a,\Lambda)/\epsilon_1\epsilon_2$, and this can be seen as a non-trivial check of the Seiberg-Witten solution. Moreover, for $\epsilon_1=-\epsilon_2$, it is given by the field theory limit of the topological string partition function:
\begin{align}
{\sum_{g=0}^\infty g_s^{2g-2}\mathcal{F}_g\bigr|_{\text{field theory}}}=\frac{\mc F^{\textrm{Nek}}(a,\Lambda,\epsilon_1,\epsilon_2)}{\epsilon_1\epsilon_2}\biggr|_{\epsilon_1=-\epsilon_2}\,,
\end{align}
and $\hbar\equiv\epsilon_1=-\epsilon_2$ is identified with the topological string coupling $g_s$. This limit is therefore referred to as the `topological string limit'. From the point of view of the string effective action, the $\Omega$-background can be identified, at the string level and for $\epsilon_1=-\epsilon_2$, with a constant anti-self-dual background of graviphotons as in \eqref{GravBackground}. From this perspective, one may wonder about the physical significance, in string theory, of a general $\Omega$-background. In other words, one would like to find, if any, a coupling in the string effective action in the form of \eqref{GravBackground} with an additional physical field $F$,
\thispagestyle{fancy}
\lhead{}
\rhead{\textsl{Summary}}
\fancyfoot[R]{\thepage}
\fancyfoot[L]{}
\begin{equation}
 F_{g,n}\,R^2\,F_G^{2g-2}\,F^{2n}\,,
\end{equation}
such that $F_{g,n}$ coincides, in the field theory limit, with the partition function of the $\Omega$-deformed gauge theory:
\begin{equation}
 \mc F^{\textrm{Nek}}(a,\Lambda,\epsilon_1,\epsilon_2)=\sum_{g,n}\epsilon_-^{2g}\epsilon_+^{2n}\,F_{g,n}\bigr|_{\text{field theory}}\,.
\end{equation}
Here, we have defined $\epsilon_\pm=\frac{\epsilon_1\pm\epsilon_2}{2}$. This task is quite tedious because of the non-trivial R-symmetry rotation in gauge theory. From the construction of the $\Omega$-background, $\epsilon_\pm$ parametrise rotations in the two independent planes of $\mb R^4$, respectively. By decomposing the Lorentz group accordingly as
\begin{equation}
 SU(2)_-\times SU(2)_+\,,
\end{equation}
this means that the graviphoton background couples only to $SU(2)_-$, consistently with its definite self-duality. Therefore, the additional physical field $F$ must couple to $SU(2)_+$ or, equivalently, be a self-dual field strength background.

Such a coupling would provide a physical realisation of the $\Omega$-background which, from the gauge theory point of view, can simply be regarded as a regulator of the instanton path integral. The natural question is whether the couplings $F_{g,n}$ are the partition functions of some topological string theory that one would call the \emph{refined topological string}. If that is the case, then the $F_{g,n}$ would be a \WS realisation of the latter. The aim of the present work is to achieve a first step in this direction by identifying, in string theory, a relevant self-dual field strength $F$ giving rise, in the field theory limit, to the $\Omega$-deformed gauge theory partition function. More specifically, we explore the connection of the topological string to the BPS-saturated couplings \eqref{GravBackground} and study a natural supersymmetric generalisation thereof. The additional insertion is then identified with a particular (universal) vector multiplet.

The plan of this work is the following. In Section \ref{ch:StringIntro}, we briefly review the quantisation of superstring theory. This sets the notation and provides with the basics needed in the subsequent parts of the manuscript. In Sections (\ref{ch:TFT}-\ref{ch:TopoAmp}), we present the main ingredients of topological string theory and discuss its known connection to higher derivative couplings in the string effective action. In Sections (\ref{Sect:Couplings}-\ref{Sec:TypeIamp}), based on \cite{Antoniadis:2013bja}, we identify the vector multiplet associated to $F$ and calculate, perturbatively, the corresponding coupling coefficients. In particular, it is shown that the field theory limit of the latter reproduces the perturbative part of the $\Omega$-deformed $\N=2$ gauge theory partition function introduced in Section \ref{ch:GaugeTheoryFromStringTheory}. Finally, in Section \ref{ch:ADHMinst}, we use a D-brane realisation of gauge theory instantons in order to calculate non-perturbative corrections to the generalised couplings and prove the precise matching with the non-perturbative part of the gauge theory partition function, as discussed in \cite{Antoniadis:2013mna}. The relation between these couplings and the refined topological string is discussed in Section \ref{Sec:RefTop}. Some complementary technical material used throughout the manuscript is gathered in a number of appendices.
\thispagestyle{fancy}
\lhead{\textsl{Summary}}
\rhead{}
\fancyfoot[L]{\thepage}
\fancyfoot[R]{}

\mainmatter

\part{Introduction}

\chapter{Elements of String Theory and Conformal Field Theory}
\label{ch:StringIntro}
The basic idea of string theory is to replace point particles by a one-dimensional object, or a string, embedded in a \emph{target space}. The world-line becomes a two-dimensional surface, the \emph{worldsheet}, describing the evolution of the string. Hence, the natural theory to study is a sigma-model in two dimensions describing the dynamics of a relativistic string.
Of course, the latter can have different shapes: it can be closed or open, oriented or unoriented. Many of the discussions below are valid for any type of strings, and we point out some of the peculiarities of each of them.
Similarly, a Feynman diagram describing an interaction of point particles is replaced by a surface. However, locally, the latter is simply the worldsheet of a string. In particular, there is no notion of vertex points and interactions seem to be purely geometric: an interaction term is solely fixed by the choice of the external string states, at each order in perturbation theory which is represented as a genus expansion or a sum over worldsheet topologies. Finally, in the limit where the string length is sent to zero, one expects to recover the usual properties of a quantum field theory.
In what follows, we briefly review the quantisation of the bosonic string, then we present some properties of superstring (perturbation) theory which is necessary in order to describe space-time fermions. Most of the material presented in this part of the manuscript can be found $\emph{e.g.}$ in \cite{Ibanez:2012zz,Kiritsis:2007zza,Polchinski:1998rq,Polchinski:1998rr}. For a more recent discussion of string perturbation theory, we refer the reader to \cite{Witten:2012bh}.

\section{The Bosonic String}

A free bosonic string propagating in a D-dimensional target space (or space-time) can be described by using a collection of bosonic fields $X^{\mu}(\xi^a)$, where $\mu=0\cdots D-1$, $a=1,2$ and $\xi^1=\sigma$, $\xi^2=\tau$ parametrise the worldsheet of the string. The latter take value in $[0,2\pi]\times\mb R$ for closed strings and $[0,\pi]\times\mb R$ for open strings. Physical quantities should not depend on a particular parametrisation of the worldsheet, so the sigma-model has to be $\xi$-reparametrisation invariant. The simplest choice is the \emph{Nambu-Goto} action that calculates the surface of the worldsheet:

\begin{equation}
 S_{\textrm{NG}}=-T\int d^2\xi\,\sqrt{-\det(g_{ab})}\,,
\end{equation}
where $T=\frac{1}{2\pi\alpha'}$ is the string tension and $g_{ab}=G_{\mu\nu}(X)\,\partial_{a}X^{\mu}\,\partial_{b}X^{\nu}$ is the pull-back of the \ST metric $G_{\mu\nu}$ on the \WS. The constant $\alpha'$ has the units of (\ST) length-squared and is called the \emph{Regge slope}. This non-linear action is difficult to quantise. Fortunately, by introducing an auxiliary field $h_{ab}$ having the meaning of a two-dimensional metric, one can linearise the Nambu-Goto action:

\begin{equation}\label{PolyakovAction}
S_{\textrm{P}}=-\frac{1}{4\pi\alpha'}\int d^2\xi\,\sqrt{-h}\,h^{ab}\,\partial_aX^{\mu}\,\partial_bX^{\nu}\,G_{\mu\nu}(X)\,.
\end{equation}
This is the Polyakov action. The \ST coordinates $X^{\mu}$ are scalars on the \WS coupled to two-dimensional gravity. The action \eqref{PolyakovAction} is manifestly invariant under (global) general coordinate transformations which reduce to Poincar\'e in the case where the \ST metric is flat, $G_{\mu\nu}(X)=\eta_{\mu\nu}$ (which we assume in the quantisation of the sigma-model below). Moreover, \eqref{PolyakovAction} also enjoys (local) diffeomorphism invariance. In fact, one can write a more general action respecting these symmetries:

\begin{align}
 S=-\frac{1}{4\pi\alpha'}\int d^2\xi\,\sqrt{-h}&\bigl(\,h^{ab}\,\partial_aX^{\mu}\,\partial_bX^{\nu}\,G_{\mu\nu}(X)\nonumber\\
                                               &+\varepsilon^{ab}\,\partial_aX^{\mu}\,\partial_bX^{\nu}\,B_{\mu\nu}(X)+\alpha'\phi(X)R^{(2)}\bigr)\,,\label{GeneralPolyakov}
\end{align}
where $\varepsilon$ is the Levi-Civita symbol, $B_{\mu\nu}$ is an anti-symmetric tensor (Kalb-Ramond field), $\phi$ is a scalar field identified with the dilaton and $R^{(2)}$ is the two-dimensional Ricci scalar. $G,\,B,\,\phi$ are usually referred to as the string background fields. Notice that for a constant dilaton background, the last term in \eqref{GeneralPolyakov} is somehow trivial and, in particular, does not introduce any classical dynamics. This is due to the fact that in two dimensions, any Einstein-Hilbert term satisfies Einstein's equation $R_{ab}-\frac{1}{2}h_{ab}R=0$ because $R_{abcd}\propto h_{ac}h_{bd}-h_{ad}h_{bc}$. On the other hand, the Polyakov action \eqref{PolyakovAction} enjoys an additional local symmetry, Weyl rescalings, which is very peculiar to two dimensional sigma-models and is fundamental for the theory to be well-defined. Its action is defined as an arbitrary rescaling of the \WS metric:
\begin{equation}
 h_{ab}\rightarrow e^{2\omega(\xi)}h_{ab}\,.
\end{equation}
Notice that Weyl invariance forbids any potential term $\int d^2\xi\,\sqrt{-h}\,V(X)$ or a cosmological constant. Moreover, the dilaton term in \eqref{GeneralPolyakov} breaks, in general this symmetry:
\begin{equation}\label{WeylTrans}
 \sqrt{-h}\,R^{(2)}\rightarrow\sqrt{-h}\,R^{(2)}-2\,\Box\,\omega(\xi)\,,
\end{equation}
with $\Box=\sqrt{-h}\,h^{ab}\,\nabla_{a}\nabla_{b}=\partial_a(\sqrt{-h}h^{ab}\,\nabla_{b})$. Hence, for a constant dilaton and a worldsheet without boundaries, the Ricci scalar term preserves, classically, Weyl invariance. For a surface with boundaries, an additional term involving the geodesic curvature $k$ of the boundary is necessary to restore this symmetry:
\begin{equation}
 \frac{1}{2\pi}\int_{\textrm{boundary}}ds\,\phi\,k\,.
\end{equation}
In general, for a constant dilaton, the Weyl-invariant $\phi$-term in the action gives the Euler characteristic $\chi$ of the worldsheet $\Sigma$ through the Gauss-Bonnet theorem:
\begin{equation}
 \frac{1}{4\pi}\int_{\Sigma}d^2\xi\,\sqrt{-h}\,\phi\,R^{(2)}+\frac{1}{2\pi}\int_{\partial\Sigma}ds\,\phi\,k=\chi(\Sigma)\phi\,.
\end{equation}

\subsection*{Gauge fixing}

As usual, we can use the local symmetries on the \WS to fix some of the degrees of freedom, for example, by using $\xi$-diffeormorphims we can bring the \WS metric to a conformally flat form:
\begin{equation}
 h_{ab}=e^{2\varrho(\xi)}\eta_{ab}\,.
\end{equation}
Moreover, using a Weyl transformation, the remaining component can be fixed and the metric becomes flat. Notice that this can always be done locally, which can be seen \emph{e.g.} from \eqref{WeylTrans}. The residual symmetry preserving the unit metric on the \WS is a subgroup of the (Diffeomorphism x Weyl) group and corresponds to conformal transformations. Therefore, the gauge-fixed two-dimensional string theory is a conformal field theory (CFT) which we now briefly present.

In two dimensions, the conformal group (CG) is infinite. It is generated by infinitesimal holomorphic and anti-holomorphic coordinate transformations:
\begin{align}
 z&\rightarrow z+\epsilon(z)\,,\\
 \bar z&\rightarrow \bar z+\bar\epsilon(\bar z)\,.
\end{align}
By writing the Laurent expansion of $\epsilon(z)$ and $\bar\epsilon(\bar z)$, we can define the generators $l_n=-z^{n+2}\partial$ and $\bar l_n=-\bar z^{n+2}\bar\partial$ that factorise the CG into a left- and a right-moving part. The generators $l_n$ satisfy the classical Virasoro or Witt algebra
\begin{equation}
 [l_n,l_m]=(m-n)l_{m+n}\,.
\end{equation}
Notice that $l_{-1},\,l_0,\,l_1$ together with their right-moving counterparts generate a globally defined subalgebra corresponding to $SL(2,\mathbb R)\times SL(2,\mathbb R)$ (or $SL(2,\mathbb C)/\mathbb Z_2$ for a Euclidean signature). We denote this subalgebra \CG. A conformal theory has a conserved, traceless energy-momentum (EM) tensor $T$:
\begin{align}
 \partial\,T_{\bar z\bar z}=0=\bar\partial\,T_{zz}\,\\
 T_{z\bar z}=0=T_{\bar z z}\,.\label{Traceless}
\end{align}
From the point of view of the CFT on the \WS, these relations should hold at the quantum level in order for the theory to be well-defined.

Let us pause for a moment and consider, in string theory, a cylinder describing the \WS of a propagating closed string, parametrised by the complex coordinate $w=\tau-i\sigma$. Under the conformal transformation $z=e^w$, the cylinder is mapped to the Riemann sphere $\mb C\cup \{\infty\}$. We mostly work in the $z$-coordinate system keeping in mind the map to the canonical picture. Note that in the $z$-frame, time flows radially and the origin represents the infinite past. Similarly, the simplest open string \WS, the strip, is mapped to the upper-half plane $\mb H$.

Back to the CG, the EM tensor can be expanded in Laurent modes as well,
\begin{equation}
 T_{zz}(z)\equiv T(z)=\sum_{n\in\mb Z} \frac{L_n}{z^{n+2}}\,,
\end{equation}
where $L_n=\frac{1}{2i\pi}\oint dz\, z^{n+1}\,T_{zz}(z)$ are the \emph{Virasoro generators}. Finally, an essential ingredient is the operator product expansion (OPE) which expresses, to arbitrary accuracy, a product of two operators coming close to each other as a sum of local operators:
\begin{equation}
 \mc O_i(z,\bar z)\mc O_j(w,\bar w)=\sum_{k} C^k_{ij}(z-w,\bar z-\bar w)\mc O_k (w,\bar w)\,.
\end{equation}
Conformal invariance puts stringent constraints on the form of the OPE. It is convenient to define a basis of local operators $\phi$ called \emph{primary fields} which are $(h,\bar h)$-tensors for the full CG, \emph{i.e.}
\begin{equation}
\phi(z,\bar z)\rightarrow\left(\frac{\partial z'}{\partial z}\right)^{-h}\left(\frac{\bar\partial\bar z'}{\bar\partial\bar z}\right)^{-\bar h}\phi(z,\bar z).
\end{equation}
$h,\bar h$ are called the conformal weights of $\phi$. One also defines $\Delta\equiv h+\bar h$ and $s\equiv h-\bar h$, the canonical dimension and spin of the operator. One can then show that the OPE of the EM tensor with a primary field (which can be used as a definition of a primary field) is
\begin{equation}
 T(z)\phi(w,\bar w)=\frac{h}{(z-w)^2}+\frac{\partial\phi(w,\bar w)}{z-w}+\cdots\,,
\end{equation}
where the dots denote regular terms. However, the EM tensor is \emph{not} a primary field but rather a \emph{quasi-primary} meaning that it transforms as a tensor only under the finite conformal subgroup \CG. The OPE of the EM tensor with itself is
\begin{equation}
 T(z)T(w)=\frac{c/2}{(z-w)^4}+\frac{2T(w)}{(z-w)^2}+\frac{\partial T(w)}{z-w}+\cdots\,.
\end{equation}
$c$ is called the central charge of the conformal algebra and appears in the \emph{quantum} Virasoro algebra for the modes $L_n$:

\begin{equation}
 [L_n,L_m]=(m-n)L_{m+n}+\frac{c}{12}m(m^2-1)\delta_{m+n}\,.
\end{equation}
The Green functions of primary fields are constrained by conformal invariance. For instance, one can completely determine (up to a constant) two- and three-point functions using \CG. Recall from \eqref{Traceless} that the EM tensor is, classically, conserved and traceless. However, this is in general not true at the quantum level due the to presence of the central charge. One can show that

\begin{equation}
 \langle T_{z\bar z}\rangle=\frac{c}{12}\,h_{z\bar z}\,R^{(2)}\,,
\end{equation}
and similarly for $T_{\bar z z}$. Hence, if the \WS is curved, the EM tensor is anomalous unless $c=\bar c=0$. This is the \emph{Weyl anomaly}. In order for the string theory to be well-defined, the latter must cancel.

Consider a general (free, exact) CFT. Using the conformal map to the cylinder, one can define a Hamiltonian density
\begin{equation}
H=T_{ww}+T_{\bar w\bar w}=\frac{1}{z^2}T_{zz}+c+\frac{1}{\bar z^2}T_{\bar z\bar z}+\bar c\,.
\end{equation}
By unitarity, $L_n=L_{-n}^{\dag}$. Moreover, the states of the theory are the irreducible representations of the Virasoro algebra. Firstly, construct the vacuum as a minimal-energy and \CG-invariant state. Then define the primary states of conformal weight $(h,\bar h)$, $L_0|\phi\rangle=h|\phi\rangle$. The Virasoro generators $L_{n>0}$ and $L_{n<0}$ act as annihilation and creation operators. Acting with the creation operators on the primary states generates the descendants of the primaries that form, all together, the \emph{Verma module}. Thus, the spectrum of the CFT is determined by its primaries. In fact, one can show that there is a one-to-one correspondence between the states and the operators of the CFT. In this sense, CFT plays a crucial role in string theory as we shall see below.

Let us apply the previous ideas by considering the gauge-fixed action \eqref{GeneralPolyakov} for a flat background. This is a CFT of D free scalar fields whose EM momentum tensor is $T(z)=-\frac{1}{\alpha'}\colon \partial X^{\mu}\,\partial X_{\mu}\colon$, where $\colon\colon$ is the normal product. By calculating the OPE of the EM tensor with itself, one finds that this CFT has central charges $c=\bar c=D$. Moreover, $X^{\mu}$ is not a primary field but $\partial X^{\mu}$ and $\bar\partial X^{\mu}$ are. They carry conformal weights $(1,0)$ and $(0,1)$ respectively. The `coherent' states $e^{ipX(z,\bar z)}$ are also primaries of weight $(\frac{\alpha' p^2}{4},\frac{\alpha' p^2}{4})$. Finally, for later reference, the two-point function is given by
\begin{equation}\label{ScalarTwoPoint}
 \langle X^{\mu}(z,\bar z)\,X^{\nu}(w,\bar w)\rangle=-\frac{\alpha'}{2}\eta^{\mu\nu}\log|z-w|^2\,.
\end{equation}

Another important class of CFTs is the bc-ghost system where b and c are anti-commuting fields whose action is
\begin{equation}\label{bcSystem}
 S_{\textrm{bc}}=\frac{1}{2\pi}\int d^2z\,b\bar\partial c\,.
\end{equation}
For b and c carrying weights $(\lambda,0)$ and $(1-\lambda,0)$, this defines a CFT with EM tensor
\begin{equation}
 T=\colon\partial b\,c\colon-\lambda\,\partial\colon b\,c\colon
\end{equation}
and central charges $c=-3(2\lambda-1)^2+1$ and $\bar c=0$. The b-c OPE is simply $b(z)c(w)\sim(z-w)^{-1}$. For $\lambda=2$, this system describes the Fadeev-Popov ghosts arising in the gauge-fixing of the Polyakov action as we describe below. For $\lambda=\frac{1}{2}$, it describes a theory of \WS fermions relevant for superstring theory.

We would now like to perform the path integral for the Polyakov action
\begin{equation}
 Z=\frac{1}{V}\int DX\,Dh\,e^{-S_{P}[X,h]}\,.
\end{equation}
First of all, we locally fix the metric to $\hat h_{ab}=\delta_{ab}$ (using Euclidean signature) so that we are left with a residual symmetry $\epsilon$:
\begin{equation}\label{GaugeTransf}
 h^{\epsilon}_{ab}(\xi')=e^{2\varrho(\xi)}\frac{\partial\xi^c}{\partial{\xi'}^a}\frac{\partial\xi^d}{\partial{\xi'}^b}\,\hat h_{cd}\,.
\end{equation}
In other words, we integrate only over physically inequivalent field configurations. This amounts to inserting the Fadeev-Popov gauge-invariant determinant $J[h]$, corresponding to the Jacobian of the coordinate transformation \eqref{GaugeTransf}, in the path integral:
\begin{equation}
 Z=\frac{1}{V}\int DX\,Dh\,D\epsilon\,\delta(h^{\epsilon}-h)\,J[h]\,e^{-S_{P}[X,h]}=\int DX\,J[\hat h]\,e^{-S_{P}[X,\hat h]}\,,
\end{equation}
where we have used the gauge invariance of the Jacobian and the action, and the integral over $\epsilon$ cancels the volume factor. The Fadeev-Popov determinant can be represented by a Grassmann integral of ghost fields b, c of conformal dimensions 2, -1 respectively:
\begin{equation}
 Z=\int DX\,\,e^{-S_{P}[X,\hat h]-S_{gh}[b,c,\hat h]}\,.
\end{equation}
The ghost action is $S_{\textrm{gh}}=\frac{1}{2\pi\alpha'}\int d^2z\,(b\bar\partial c+\bar b\partial\bar c)$. The total EM tensor is the sum of the EM tensor of the X and bc CFTs. The new total central charge is thus given by the sum of the central charges
\begin{equation}
 c_{\textrm{tot}}=c_X+c_{\textrm{gh}}=D-26\,.
\end{equation}
Hence, the vanishing of the total central charge, or the decoupling of the scale factor of the metric $\varrho(\xi)$ at the quantum level requires the target space dimension to be $D=26$. More precisely, one can show that under a Weyl transformation, the path integral transforms in the conformal gauge as
\begin{equation}
 Z\rightarrow \exp\left[\frac{c_{\textrm{tot}}}{24\pi}\int d^2\xi\,\sqrt{-h}(h^{ab}\partial_a\varrho\,\partial_b\varrho+R(h)\varrho)\right]Z	\,.
\end{equation}
Finally, let us simply mention that the condition for the gravitational anomalies to cancel, $c=\bar c$, is automatically satisfied.

\subsection*{Curving the background}

So far, we have dealt with a flat \ST background. However, one is interested in studying more general curved backgrounds \eqref{GeneralPolyakov}. Again, one has to ensure that the theory is consistent by requiring anomaly cancellation. Here, one is forced to work perturbatively in a derivative expansion in \ST, and this corresponds to considering the limit where $\frac{\alpha'}{R}\ll1$, with $R$ being the order of the \ST curvature. The statement of quantum conformal invariance boils down to the tracelessness of the EM tensor, and this can be expressed in terms of the $\beta$-functions of the background fields:
\begin{equation}
 {T^a}_{a}=-\frac{\beta^{\phi}}{12}R^{(2)}-\frac{1}{2\alpha'}(\beta^{G}_{\mu\nu}\,h^{ab}+\beta^{B}_{\mu\nu}\,\varepsilon^{ab})\partial_{a}X^{\mu}\,\partial_{b}X^{\nu}\,,
\end{equation}
with the $\beta$-functions calculated perturbatively as
\begin{align}
 \beta^{\phi}&=c_{\textrm{tot}}+\frac{3}{2}\alpha'\left(4(\nabla\phi)^2-4\Box\phi-R^{(D)}-\frac{1}{12}H_{\mu\nu\rho}H^{\mu\nu\rho}+\mc O(\alpha')\right)\,,\\
 \beta^{B}_{\mu\nu}&=-\alpha'\nabla^{\rho}(e^{-2\phi}H_{\mu\nu\rho})+\mc O(\alpha')\,,\\
 \beta^{G}_{\mu\nu}&=\alpha'\left(R_{\mu\nu}-\frac{1}{4}H_{\mu\rho\sigma}{H_{\nu}}^{\rho\sigma}+2D_{\mu}D_{\nu}\phi+\mc O(\alpha')\right)\,.
\end{align}
Here, $H=dB$ is the three-form flux associated to the Kalb-Ramond field. The vanishing of the $\beta$-functions can be viewed as a set of equations of motion for the background fields which can be used to write down a \ST effective action:
\begin{equation}
 S_{\textrm{eff}}=\frac{1}{2\kappa^2}\int d^D X\,\sqrt{G}\,e^{-2\phi}(R^{(D)}+4(\nabla\phi)^2-\frac{1}{2}H^2-\frac{2c_{\textrm{tot}}}{3\alpha'}	)+\mc O(\alpha')\,.
\end{equation}

\subsection*{BRST symmetry and physical states}

We have heretofore quantised the bosonic string using a particular gauge fixing. More generally, one can implement an arbitrary gauge-fixing $F^A$ by introducing new Lagrange multipliers in the path integral. This amounts to adding a new action $S_{\textrm{gf}}=-\int i\,B_{A}F^{A}$. The full action including the ghosts and the \WS fields enjoys a global symmetry, called the BRST symmetry:
\begin{align}
 &\delta_{\textrm{BRST}}(X^{\mu},h_{ab})=-i\,\eta\,c\,\delta_{\textrm{gauge}}(X^{\mu},h_{ab})\,,\\
 &\delta_{\textrm{BRST}}(c)=i\,\eta\,c\,\partial\,c\,,\\
 &\delta_{\textrm{BRST}}(b_A)=\eta\,B_A\,,\\
 &\delta_{\textrm{BRST}}(B_A)=0\,.
\end{align}
The corresponding Noether current is
\begin{equation}
 j_{\textrm{BRST}}=c\,T_{X}+\frac{1}{2}\,c\,T_{\textrm{gh}}+\frac{3}{2}\,\partial^2c\,.
\end{equation}

By making a variation of the gauge-fixing functional $F^A$, the path integral should remain invariant. This implies that a physical state must be BRST-invariant:
\begin{equation}
 Q_{\textrm{BRST}}|\mc O_{\textrm{phys}}\rangle=0\,,
\end{equation}
with $Q_{\textrm{BRST}}=\oint\frac{dz}{2i\pi}\,j_{\textrm{BRST}}$. In fact, there is an additional constraint coming from kinematics that one needs to impose on physical states, that is a mass-shell condition
\begin{equation}\label{MassShellB}
 b_0|\mc O_{\textrm{phys}}\rangle=0\,.
\end{equation}
Using the fact that $L_0=\{Q_{\textrm{BRST}},b_0\}$, this leads to $ L_0|\mc O_{\textrm{phys}}\rangle=0$. In the absence of conformal anomalies, the BRST charge $Q_{\textrm{BRST}}$ is nilpotent, $Q_{\textrm{BRST}}^2=0$. In particular, any state of the form $Q_{\textrm{BRST}}|\Phi\rangle$ is annihilated by the BRST-charge, but is orthogonal to any physical state. Therefore, the Hilbert space of physical states is in one-to-one correspondence with the $Q_{\textrm{BRST}}$-cohomology:
\begin{equation}
 \mc H_{\textrm{phys}}\simeq H_{Q_{\textrm{BRST}}}\,.
\end{equation}

\subsection*{Spectrum}

Let us go back to the free bosonic string described by \eqref{PolyakovAction} in a flat background. A primary field $\phi$ of weight h can be expanded as
\begin{equation}
 \phi(z)=\sum_{n\in\mb Z} z^{-n-h}\,\phi_n\,.
\end{equation}
Using the fact that $\partial X^{\mu}$ are weight 1 primaries, one can write down the mode expansion for the free scalars. Alternatively, we derive it canonically from their equation of motion supplemented with the appropriate boundary conditions. The solution can be decomposed into a left-moving free boson and a right-moving one, and the mode expansions read
\begin{align}
 X^{\mu}_{L}(z)&=\frac{x^{\mu}}{2}-i\frac{\alpha'p^{\mu}_{L}}{2}\ln(z)+i\sqrt{\frac{\alpha'}{2}}\sum_{n\neq0}\frac{\alpha^{\mu}_{n}}{n}\,z^{-n}\,,\label{XLModes}\\
 X^{\mu}_{R}(\bar z)&=\frac{x^{\mu}}{2}-i\frac{\alpha'p^{\mu}_{R}}{2}\ln(\bar z)+i\sqrt{\frac{\alpha'}{2}}\sum_{n\neq0}\frac{\tilde{\alpha}^{\mu}_{n}}{n}\,\bar z^{-n} \label{XRModes}\,.
\end{align}
Again, one can see from the zero-modes that $X^{\mu}$ is not a primary field. Reality of $X^{\mu}$ imposes the reality of the zero-modes $x^{\mu},\,p_L^{\mu}\,,\,p_R^{\mu}\,$, and the conditions $(\alpha^{\mu}_{n})^*=\alpha^{\mu}_{-n}$ and  $(\tilde{\alpha}^{\mu}_{n})^{*}=\tilde{\alpha}^{\mu}_{-n}$ on the oscillators. We first focus on the case of closed oriented strings for which $X^{\mu}$ are periodic, $X^{\mu}(\tau,\sigma+2\pi)=X^{\mu}(\tau,\sigma)$. In this case, $p_L^{\mu}=p_R^{\mu}$ and it represents the center-of-mass momentum of the string. In the quantum theory, the oscillators $\alpha^{\mu}_{-n}$ and $\tilde\alpha^{\mu}_{-n}$ for $n$ positive are creation operators and the spectrum of the theory is built by acting with them on the vacuum. The levels $N,\tilde N$ of a given state is defined by the number of oscillators (on the left, right) one acts with on the ground states, or, equivalently, the eigenvalues of the operators $\sum_{n\geq1}\alpha_{-n}\cdotp\alpha_{n}$ and $\sum_{n\geq1}\tilde\alpha_{-n}\cdotp\tilde\alpha_{n}$. Moreover, a given state in the theory must be gauge-invariant, and this imposes, in particular, the \emph{level matching} condition\footnote{The zero-point energies must also match.}
\begin{equation}\label{LevelMatching}
 N=\tilde N\,,
\end{equation}
which can be equivalently formulated as $L_0=\tilde L_0$.

The above mode expansions correspond to tree-level solutions (cylinder) and one can generalise them for higher genera (see subsequent sections for more details). Let us now consider the case of open strings where one needs to impose boundary conditions on their endpoints. Here, the left- and right-movers are no longer independent. There are two possibilities that one can combine:
\begin{enumerate}
 \item Neumann (N) boundary condition: $\partial_{\sigma}X^{\mu}|_{\textrm{boundary}}=0$, the endpoint is free.
 \item Dirichlet (D) boundary condition: $\partial_{\tau}X^{\mu}|_{\textrm{boundary}}=0$, the endpoint is fixed. This breaks Poincar\'e invariance.
\end{enumerate}
Depending on the choice for each endpoint, we can have N-N, D-D, N-D or D-N boundary conditions and we now list the mode expansions for open strings in each case:
\begin{enumerate}
 \item N-N $$X^{\mu}(z,\bar z)=x^{\mu}-i\alpha'p^{\mu}\ln|z|^2+i\sqrt{\frac{\alpha'}{2}}\sum_{n\neq0}\frac{\alpha^{\mu}_{n}}{n}(z^{-n}+\bar z^{-n})\,.$$
 \item D-D $$X^{\mu}(z,\bar z)=x^{\mu}-i\alpha'p^{\mu}\ln\frac{z}{\bar z}+i\sqrt{\frac{\alpha'}{2}}\sum_{n\neq0}\frac{\alpha^{\mu}_{n}}{n}(z^{-n}-\bar z^{-n})\,.$$
 \item D-N $$X^{\mu}(z,\bar z)=x^{\mu}+i\sqrt{\frac{\alpha'}{2}}\sum_{n\in\mb Z+{\frac{1}{2}}}\frac{\alpha^{\mu}_{n}}{n}(z^{-n}-\bar z^{-n})\,.$$
\end{enumerate}
In the case where one imposes Dirichlet boundary conditions, no momentum is allowed (the $p^{\mu}$-term only depends on $\sigma$). 

As mentioned above, a state in the Hilbert space can be thought of as a (vertex) operator of the underlying CFT. In the closed string case (we set $\alpha'=2$ for convenience), it is constrained by conformal invariance to be a primary field of weight $(1,1)$ integrated\footnote{In some specific cases, one uses unintegrated vertex operators by attaching a $c$ ($\bar c$) ghost to the primary operator. This happens, for instance, in tree-level amplitudes as we discuss in Section \ref{ch:GaugeTheoryFromStringTheory}. In fact, it turns out that in some cases, there is an equivalence between the formalism using integrated vertex operators and the one with unintegrated ones, see Section 2.5 of \cite{Witten:2012bh}.} over the \WS, $\int d^2z\,V(z,\bar z)$. In order to find the vertex operators for the lowest mass states of the theory, recall that the vacuum carrying momentum $p^{\mu}$ is annihilated by all positive-mode oscillators. The mass operator is 
\begin{equation}
 M^2=2\left(\sum_{n\neq0}\alpha_{-m}.\alpha_{m}-1\right)\,.
\end{equation}
Consequently, the ground state has mass $m^2=-2$ and is tachyonic. It is a problematic aspect of the bosonic string as it signals an instability in the theory. In fact, it is one of the motivations for superstring theory. The tachyon corresponds to the vertex operator $e^{i\,p\cdotp X}$, with the on-shell condition $p^2=2$. The first excited state is constructed by acting with one oscillator on the left and another one on the right, and it corresponds to massless states whose vertex operator is
\begin{equation}
 \partial X^{\mu}\,\bar\partial X^{\nu}\,e^{i\,p\cdotp X}\,.
\end{equation}
The latter can be decomposed into a symmetric traceless tensor $G_{\mu\nu}$ identified with the graviton, an antisymmetric tensor $B_{\mu\nu}$ and a scalar $\phi$ corresponding to the dilaton. Even though the initial sigma-model \eqref{PolyakovAction} did not include these fields, string theory generates them automatically.

In the open string case, one can make the same arguments showing that the ground state is again tachyonic and the first excited state is a massless gauge boson. One already sees that the closed string probes gravity while the open string carries the gauge degrees of freedom. The latter can be implemented by attaching an internal index to the string endpoints which, by definition, have trivial \WS dynamics. They are called \emph{Chan-Paton} (CP) labels and are natural degrees of freedom one introduces for such distinguished points in a quantum system. Any state in the theory carries two additional indices $i,j$ taking values from 1 to n. One can view the index $i$ as being in the fundamental representation of $U(n)$. If the string is oriented, then the index at the other endpoint, $j$, must transform in the anti-fundamental. Therefore, a given state at level $N$
\begin{equation}
 |N;p,ij\rangle
\end{equation}
transforms as an adjoint representation of $U(n)$. In fact, if we define the basis of states
\begin{equation}
 |N;p,a\rangle=\sum_{ij}\lambda^a_{ij}|N;p,ij\rangle\,,
\end{equation}
then the matrices $\lambda^a$, called the \emph{Chan-Paton matrices}\footnote{$a$ should be viewed as a colour index.}, are elements of $U(n)$. Hence, the oriented open string theory acquires a $U(n)$ gauge symmetry. For the unoriented string, one must impose the \emph{orientifold} parity $\Omega$ reversing the orientation of the string:
\begin{equation}
 \Omega:X^{\mu}(\tau,\sigma)\rightarrow X^{\mu}(\pi-\sigma,\tau)\,,
\end{equation}
which translates on the oscillators in the open string sector as
\begin{equation}
 \Omega\,\alpha_{n}^{\mu}\,\Omega^{-1}=(-1)^n\,\alpha_n^{\mu}\,.
\end{equation}
Including the Chan-Paton indices, the states transform as
\begin{equation}
 \Omega|N;p,ij\rangle=\omega|N;p,ji\rangle\,,
\end{equation}
with $\omega=\pm1$ depending on the level of the state or, equivalently, on its mass. Consequently, the CP matrices have to be either symmetric or antisymmetric in order for the corresponding state $|N;p,a\rangle$ to be invariant under $\Omega$. The CP matrices are then $Sp(n)$ or $SO(n)$ respectively\footnote{The notation $Sp(\frac{n}{2})$ is also used in the literature.}. For instance, the massless gauge bosons that are $\Omega$-odd are in the adjoint of $SO(n)$ so we would have an $SO(n)$ gauge theory.

When dealing with open string amplitudes, we have to take into account the CP matrices which are not dynamical fields. Hence, adjacent states should have their corresponding CP matrices contracted (the right endpoint index of one should correspond to the left endpoint index of the other) as the CP labels are not affected by the propagation of the string, and the full amplitude simply includes a trace of the CP matrices of the scattering states.

\subsection*{Amplitudes}

The simplest amplitudes one can consider are tree-level ones. They correspond to scattering of physical states or, equivalently, to CFT correlators of vertex operators. Naively, one would consider the correlation function of a number of vertex operators and integrate over their positions. However, this is divergent since we must only integrate over inequivalent classes with respect to \CG. In other words, \CG allows one to fix three of the vertex operators to arbitrary positions. This leads to (in the closed string case, for the open string it's very similar)
\begin{equation}
 S(V_1,\cdots,V_N)=\int\frac{|\prod_{i=1}^{N}dz_i|^2}{dV_{\textrm{CKG}}}\langle\prod_{i=1}^{N} V_i(z_i,\bar z_i)\rangle\,.
\end{equation}
The correlator is to be evaluated in the free CFT action considered above. For instance, if we fix the positions of the vertices $V_1$, $V_2$ and $V_N$, then
\begin{equation}\label{VolumeCKG}
 dV_{\textrm{CKG}}=\frac{dz_1\,dz_2\,dz_N}{(z_1-z_2)(z_1-z_N)(z_2-z_N)}\,.
\end{equation}

\begin{figure}[ht]
\centering
\includegraphics[width=0.8\textwidth]{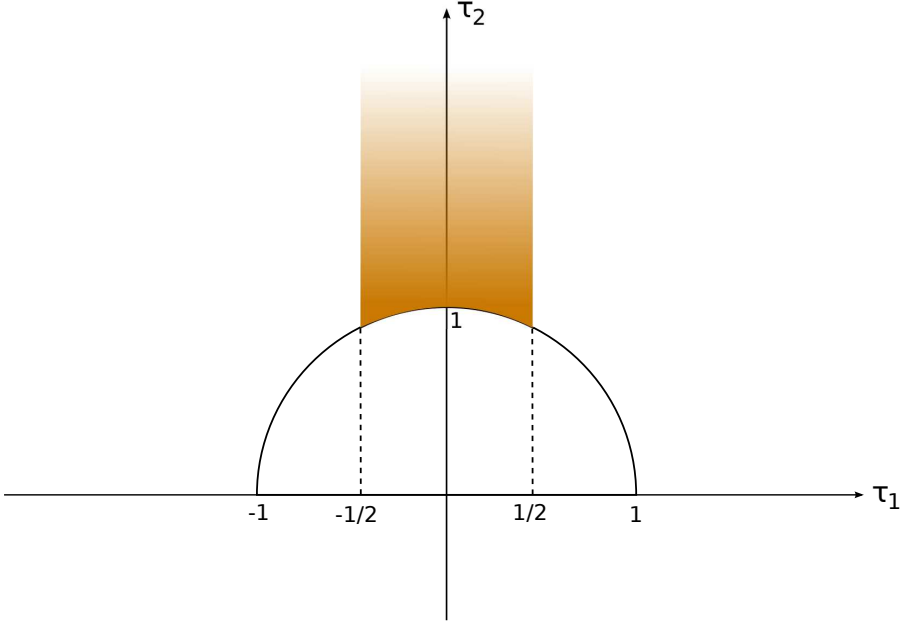}
\caption{A fundamental domain $\mc F$ of the two-torus.}
\label{fig:FundamentalDomain}
\end{figure}

Let us now turn to the more rich one-loop case and derive the partition function, or vacuum-vacuum amplitude, for the scalar and ghost CFTs. This is a quantity of interest not only as the simplest one-loop amplitude but also because it encodes the spectrum of the theory. Consider first the closed string case. The \WS is either a torus (oriented string) or a Klein bottle (unoriented string). In the path integral formalism, one must integrate over the space of metrics. On a genus $g$ Riemann surface, the metric can always be locally brought to a conformally flat form. However, this is not true globally due to the presence of inequivalent classes of metrics. Indeed, there exist $n$ conformal Killing vectors parametrising the transformations leaving the conformally flat metric invariant and $m$ moduli parametrising the conformally inequivalent metrics. Moreover, one can show that the conformal Killing vectors correspond to the $c$ zero modes whereas the moduli of the large gauge transformations are given by the $b$ zero modes. For a general bc-system as in \eqref{bcSystem}, the Riemann-Roch theorem states that
\begin{equation}
 N_0(b)-N_0(c)=(2\lambda-1)\chi(g)\,,
\end{equation}
where $\chi(g)$ is the Euler characteristic of the genus $g$ Riemann surface and $N_0$ counts the zero modes. In particular, for the case of interest $\lambda=2$, it yields $m-n=3\chi(g)$. Hence, the moduli space over which we integrate is the quotient of the space of metrics by the group of diffeomorphisms, Weyl transformations and large coordinate transformations that connect conformally inequivalent metrics. In the case of the two-torus, the latter is the modular group $SL(2,\mb Z)/\mb Z_2\equiv PSL(2,\mb Z)$ and the moduli space is the fundamental domain pictured in Fig. \ref{fig:FundamentalDomain}. It is parametrised by a complex parameter $\tau$, the complex structure of the torus. More generally, for a genus $g>1$ Riemann surface, there are $3g-3$ complex moduli (and no conformal Killing vectors).

\begin{figure}[ht]
\centering
\includegraphics[width=1\textwidth]{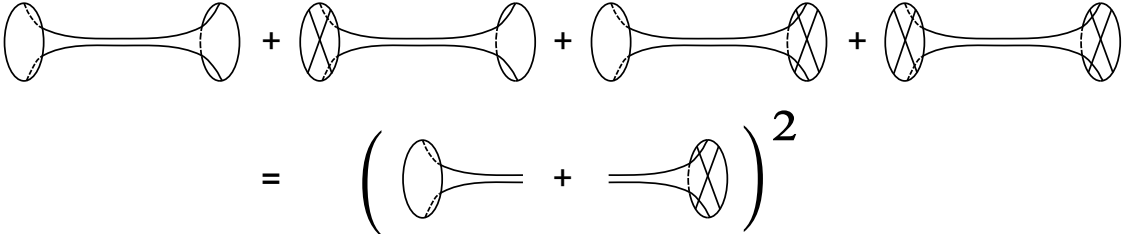}
\caption{Tadpole cancellation in Type I: the sum of the cylinder, M\"obius and Klein bottle worldsheets is the square of the sum of two tadpoles in the limit $t\rightarrow0$.}
\label{fig:TadpoleSum}
\end{figure}

The partition function can be canonically calculated by picturing the torus through its fundamental parallelogram: it is a cylinder of height $\tau_2$ whose boundary circles are identified after a twist by $\tau_1$. The identification produces a trace with the insertion of the generator of the translations for the cylinder given by the Hamiltonian density $H=L_0-\frac{c}{24}+\bar L_0-\frac{\bar c}{24}$. Hence, the partition function for the free scalars is (after performing a Wick rotation)
\begin{equation}
 Z_{T}^{X}=\textrm{Tr}[e^{-2\pi\tau_2(L_0-\frac{c}{24}+\bar L_0-\frac{\bar c}{24})+2i\pi\tau_1(L_0-\bar L_0)}]=\textrm{Tr}[q^{L_0-\frac{c}{24}}\,\bar q^{\bar L_0-\frac{\bar c}{24}}]\,,
\end{equation}
where $q=e^{2i\pi\tau}$ and $\tau=\tau_1+i\tau_2$. From this form, one can think of the partition function as a Fourier expansion, \emph{i.e.} a $q\bar q$-expansion whose exponents label the states in the spectrum with the \emph{integer} coefficients being the corresponding degeneracies. Using the fact that the generator $L_0=\frac{p^2}{2}+\sum_{n\geq0}\alpha_{-n}\cdotp\alpha_{n}$ and separating out the zero modes $p$, the partition function becomes
\begin{equation}
 Z_{T}^{X}=\frac{V_{D}}{(q\bar q)^{D/24}}\int \frac{d^{D}p}{(2\pi)^D}e^{-2\pi\tau_2\,p^2}\,\textrm{Tr'}\left[\prod_{i=1}^{D}\prod_{n\geq1}q^{\alpha_{-n}^i\,\alpha_{n}^i}\,\bar q^{\tilde\alpha_{-n}^i\,\tilde\alpha_{n}^i}\right]\,,
\end{equation}
where the prime ' denotes the fact that the trace excludes the zero modes. Performing the trace yields (after a Wick rotation back)
\begin{equation}\label{PartitionX}
 Z_{T}^{X}=\frac{iV_{D}}{(8\pi^2\tau_2)^{D/2}}\frac{1}{(\eta(\tau)\bar\eta(\bar\tau)^{D}}\,,
\end{equation}
$\eta(\tau)$ being the Dedekind eta function \eqref{DedekindEta}. Using the properties of the latter, it is easy to show that the partition function \eqref{PartitionX} is modular invariant, i.e. invariant under the group $PSL(2,\mb Z)$ of the torus, as expected from the consistency of the theory. In fact, the precise choice of the central charge $c$ is crucial for this to hold. In order to obtain the full vacuum amplitude, one should calculate the contribution of the ghost system (or, alternatively, choose a particular gauge-fixing in the previous calculation). Essentially, the ghosts removes the two transverse degrees of freedom so that they contribute
\begin{equation}\label{PartitionBC}
 Z_{T}^{\textrm{bc}}=|\eta(\tau)|^4
\end{equation}
to the torus partition function. Putting all the pieces together, the torus vacuum amplitude is the integral of the modular invariant partition function
\begin{equation}\label{PartitionT}
 Z_{T}=\frac{iV_{26}}{4(8\pi^2)^{13}}\int_{\mc F} \frac{d^2\tau}{\tau_2^2}\frac{1}{(\sqrt{\tau_2}\,\eta(\tau)\bar\eta(\bar\tau))^{24}}\,.
\end{equation}
Notice that the fundamental domain provides a natural regularisation for the field theory UV-divergences that would arise had we integrated the partition function around zero. One can do the same analysis for the Klein bottle. This is done by inserting the orientifold operator $\Omega$ in the partition function. The contributions are from those states whose left- and right-movers are in the same state. The fundamental domain of the Klein bottle is parametrised by one positive real number. Without going into the details, the result is
\begin{equation}\label{PartitionK}
 Z_{K}=\frac{iV_{26}}{4(8\pi^2)^{13}}\int_0^{\infty} \frac{dt}{t^2}\frac{1}{(\sqrt{t}\,\eta(2it))^{24}}\,.
\end{equation}
Contrary to the case of the torus, here the integral is divergent around $t=0$. In the open string case, one must implement the appropriate left-right identifications. The relevant worldsheets are the cylinder and the M\"obius strip. The results, very similar to \eqref{PartitionK}, are
\begin{align}
 Z_{C}&=\frac{iV_{26}\,n^2}{2(16\pi^2)^{13}}\int_0^{\infty} \frac{dt}{t^2}\frac{1}{(\sqrt{t}\,\eta(it))^{24}}\,,\label{PartitionC}\\
 Z_{M}&=\frac{\mp iV_{26}\,n}{4(16\pi^2)^{13}}\int_0^{\infty} \frac{dt}{t^2}\frac{1}{(\sqrt{t}\,\vartheta_{3}(0,2it)\,\eta(2it))^{12}}\,\label{PartitionM}\,.
\end{align}
Here, $n$ is the number of Chan-Paton degrees of freedom and the sign refers to the $SO(n)$ or $Sp(n)$ gauge groups, and $\vartheta_3$ is defined in \eqref{ThetaOtherNotation}. Once again, these quantities are divergent around $t=0$. For the theory to be well-defined (at least at one-loop), one can try to choose a gauge group such that the divergence in the sum of the vacuum amplitudes cancels out. Using the asymptotics of the partition functions (\ref{PartitionC}, \ref{PartitionK}, \ref{PartitionM}), it is easy to show that their behaviour is the same and the only difference is in the prefactor. The total divergence is dressed with a factor of (we reinstate the dimension D of \ST)
\begin{equation}\label{TadpoleCancellation}
 n^2-2n\,2^{D/2}+2^D=(n\mp2^{D/2})^2\,.
\end{equation}
Therefore, the bosonic open string theory is consistent only for an $SO(2^{13})$ gauge group. There is a useful diagrammatic way of representing \eqref{TadpoleCancellation}. Indeed, the limit $t\rightarrow0$ means that the string \WS develops a very long tube and the sum of the cylinder, M\"obius and Klein bottle contributions is represented in Fig. \ref{fig:TadpoleSum} below where a cross represents a \emph{cross-cap}, \emph{i.e.} an identification of the circle under the orientifold $\Omega$. The consistency of Type I is thus reinterpreted as a \emph{tadpole cancellation} or, equivalently, the absence of gauge anomalies in the low energy effective action\footnote{In fact, tadpole cancellation is stronger than the condition of absence of gauge anomalies.}.

The natural question is whether one can define an all-genus partition function
\begin{equation}
 Z=\sum_{\chi(g)}g_{s}^{-\chi(g)}Z_{g}\,,
\end{equation}
where $F_g$ is the genus $g$ partition function and $g_s$ is the vacuum expectation value (v.e.v) of the dilaton $\phi$ defining the string coupling constant. From the previous discussion, in order to have a sensible path integral, it is necessary to take into account the local and global \WS degrees of freedom that manifest themselves through the presence of zero-modes in the measure. For definiteness, let us focus on the closed string. On a genus $g$ Riemann surface $\Sigma_g$, there are $3g-3$ zero modes for the b-ghosts. Hence, one must insert $3g-3$ b-ghosts in the path integral, and these can be folded with the same number of Beltrami differentials parametrising deformations of the moduli space of genus $g$ Riemann surfaces $\mc M_g$ to form an $\mc M_g$-invariant measure. The genus $g$ partition function then reads
\begin{equation}\label{BosonicPartitiongloop}
 Z_g=\int_{\mc M_g}\langle\prod_{a=1}^{3g-3}b\cdotp\mu_a\prod_{a=1}^{3g-3}\bar b\cdotp\bar\mu_a\rangle\,,
\end{equation}
with $b\cdotp\mu_a=\int_{\Sigma_g}d^2z\,b_{zz}{(\mu_a)_{\bar z}}^{z}\,.$


\section{Superstring Theory}

The bosonic string theory, as it stands, has two main drawbacks. Firstly, its ground state is tachyonic so the theory is not viable or, at least, is not expanded around the right vacuum. Secondly, it lacks space-time fermions which are essential to describe matter. In fact, both problems are related as we now readily show.

\subsection*{Fermions in two dimensions}

Fermions on the \WS can be easily described by a bc-system as in \eqref{bcSystem} with spin $\lambda=\frac{1}{2}$. They can also carry a space-time index and the action is
\begin{equation}\label{FermionAction}
 S_{\textrm{F}}=\frac{T}{2}\int d^2\xi\,\bar\chi^{\mu}\,\gamma^{a}\partial_a\,\chi_{\mu}=-T\int d^2\xi\,(\psi^{\mu}\,\bar\partial\,\psi_{\mu}+\bar\psi^{\mu}\,\partial\,\bar\psi_{\mu})\,,
\end{equation}
where $\gamma^a$ are Pauli matrices and $\chi=\left(\begin{array}{ll}
                                                     \psi\\
						      \bar\psi
                                                    \end{array}
\right)$ is a Majorana-Weyl fermion. The equations of motion of the fermions $\psi$ and $\bar\psi$ imply that they are holomorphic and anti-holomorphic respectively. Viewed as a bc-system, \eqref{FermionAction} defines a CFT with EM tensor $T(z)=-\frac{1}{2}\psi^{\mu}\,\partial\,\psi_{\mu}$ and central charge $c=\frac{1}{2}$. On a closed string \WS, invariance of the action \eqref{FermionAction} imposes two possible boundary conditions for the fermions:
\begin{enumerate}
 \item Periodic or Ramond (R) boundary condition: $\psi^{\mu}(\tau,\sigma+2\pi)=\psi^{\mu}(\tau,\sigma)$\,,
 \item Anti-periodic or Neveu-Schwarz (NS) boundary condition: $\psi^{\mu}(\tau,\sigma+2\pi)=-\psi^{\mu}(\tau,\sigma)$\,.
\end{enumerate}
On the complex plane, this implies the following mode expansions
\begin{align}\label{PsiModes}
 \psi^{\mu}=\sum_{n\in\mb Z+\nu}\psi^{\mu}_n\,z^{-n-1/2}\,,\\
\end{align}
with $(\nu_{\textrm{R}},\nu_{\textrm{NS}})=(0,\tfrac{1}{2})$, and the modes satisfy the algebra
\begin{align}
 \{\psi_m^{\mu},\psi_n^{\nu}\}=\delta_{m+n,0}\,\eta^{\mu\nu}=\{\bar\psi_m^{\mu},\bar\psi_n^{\nu}\}\,.
\end{align}
Notice that the periodicities are exchanged when one goes from the cylinder to the complex plane so that a Ramond fermion becomes anti-periodic (when going around the origin of the complex plane) while an NS fermion becomes periodic. This stems from the fact that they carry conformal weight $1/2$. The states of the Hilbert space can be constructed as $|\psi\rangle=\lim_{z\rightarrow0}\psi(z)|0\rangle$. In the Ramond sector, this implies that $\psi_{n\geq0}$ act as annihilation operators since they lower the energy of the states:
\begin{equation}
 \left[L_0,\psi_n^{\mu}\right]=-n\psi_n^{\mu}\,,
\end{equation}
except for $\psi_0^{\mu}$ which is a zero mode. Therefore, the vacuum is degenerate and corresponds to a state $|\alpha\rangle$ with $2^{D/2}$ components generating a Clifford algebra in $D$ dimensions. This is precisely what is required to have space-time fermions.

In the NS sector, there are no zero modes because of the $1/2$-integer shift and $\psi^{\mu}_{n\geq1/2}$ act as annihilation operators. Combining left- and right movers, it is clear that \ST fermions arise from the $R\otimes NS$ and $NS\otimes R$ sectors whereas the $R\otimes R$ and $NS\otimes NS$ sectors can only generate \ST bosons.

\subsection*{RNS superstring}

Let us apply these ideas to construct the Ramond-Neveu-Schwarz (RNS) superstring. The starting point is a two-dimensional sigma-model with a flat target space, described by the supersymmetric version of Polyakov's action

\begin{align}
 S_{\textrm{SPol}}=-\frac{1}{4\pi\alpha'}\int d^2\xi\,\sqrt{-h}&(h^{ab}\partial_aX^{\mu}\,\partial_bX_{\mu}-i\bar\psi^{\mu}\,\gamma^{a}\partial_a\,\psi_{\mu}\nonumber\\
                                                               &+2\bar\chi_{a}\,\gamma^b\gamma^a\,\psi^{\mu}\,\partial_b\,X_\mu-\frac{i}{2}\bar\psi^{\mu}\psi_{\mu}\,\bar\chi_a\gamma^b\gamma^a\chi_b)\,,\label{SPolyakovAction}
\end{align}
where $\chi$ is the gravitino of the \WS gravitational multiplet. On top of the symmetries of \eqref{PolyakovAction} (\ST Lorentz, local reparametrisations and Weyl), this action has $\mc N=1$ supergravity and super-Weyl invariance
\begin{align}
 \delta_{\lambda}\chi_a&=i\lambda\,\gamma_a\,,\\
 \delta_{\lambda} X^{\mu}&=\delta_{\lambda}\psi^{\mu}=\delta_{\lambda} h=0\,.
\end{align}
Using these symmetries, one can locally bring the metric to a flat form and set $\chi_a=0$ so that the resulting action ($\alpha'=2$)
\begin{equation}
 S=\frac{1}{4\pi}\int d^2z\,(\partial X^{\mu}\,\bar\partial X_{\mu}+\psi^{\mu}\bar\partial\psi_{\mu}+\bar\psi^{\mu}\partial\bar\psi_{\mu})\,,
\end{equation}
represents a superconformal field theory (SCFT) generated by the EM tensor
\begin{equation}
 T(z)=-\frac{1}{2}(\partial X^{\mu}\,\partial X_{\mu}+\psi^{\mu}\partial\psi_{\mu})\,
\end{equation}
and its superpartner, the \WS supercurrent, which has conformal weight $3/2$
\begin{equation}
 T_{F}=i\psi^{\mu}\,\partial X_{\mu}\,.
\end{equation}
These conserved currents satisfy the following OPEs:
\begin{align}
 T(z)T(w)&=\frac{c/2}{(z-w)^2}+\frac{2T(w)}{(z-w)^2}+\frac{\partial T(w)}{z-w}\,,\\
 T(z)T_{F}(w)&=\frac{3/2}{(z-w)^2}+\frac{\partial T_{F}(w)}{z-w}\,,\\
 T_{F}(z)T_{F}(w)&=\frac{D}{(z-w)^3}+\frac{2T(w)}{z-w}\,.
\end{align}
Here, the central charge $c$ can be obtained by counting the contribution of $D$ free scalars and fermions $X^{\mu},\psi^{\mu}$ as $c=D+\frac{1}{2}D=\frac{3D}{2}$. Using the mode expansions of the \WS fields (\ref{XLModes}, \ref{XRModes}, \ref{PsiModes}), the EM tensor and the supercurrent can be expanded as
\begin{align}
 T(z)&=\sum_{n\in\mb Z}L_n\,z^{-n-2}\,,\\
 T_{F}(z)&=\sum_{r\in\mb Z+\nu}G_r\,z^{-n-3/2}\,.
\end{align}
The modes are defined as
\begin{align}
 L_m &=\frac{1}{2}\sum_{n\in\mb Z}\colon\alpha^{\mu}_{m-n}\,\alpha_{\mu\,n}\colon+\frac{1}{4}\sum_{r\in\mb Z+\nu}(2r-m)\colon\psi^{\mu}_{m-r}\,\psi_{\mu\,r}\colon+a_{\textrm{R,NS}}\,\delta_{m,0}\,,\\
 G_r&=\sum_{n\in\mb Z}\alpha^{\mu}_{n}\,\psi_{\mu\,r-n}\,,
\end{align}
where $a_{\textrm{R}}=1/16$ and $a_{\textrm{NS}}=0$ are the zero-point energies corresponding to the conformal weights of the ground state in the R and NS sectors.

In the Fadeev-Popov procedure, we implement the gauge-fixing by introducing the appropriate ghosts corresponding to the generators of the gauge symmetry. Here, on top of the b and c ghosts of the bosonic string, we introduce another couple of bosonic ghosts $\beta,\gamma$ which can be viewed as the superpartners of the b,c ghosts, and are themselves a (commuting) bc-system \eqref{bcSystem} with $\lambda=3/2$, so that $\beta,\gamma$ carry conformal weights $3/2,-1/2$. They introduce 11 units of central charge implying that the total central charge of the SCFT is $c_{\textrm{tot}}=\frac{3}{2}D-26+11=\frac{3}{2}D-15$. Consequently, the consistency of superstring theory requires the \ST dimension to be $D=10$. For future reference, let us write down the contribution of the ghost and superghost sectors to the EM tensor and the supercurrent:
\begin{align}
 T^{\textrm{gh}}&=-2\colon b\partial c\colon+\colon c\partial b\colon-\frac{3}{2}\colon\beta\partial\gamma\colon-\frac{1}{2}\colon\gamma\partial\beta\colon\,,\\
 T_{F}^{\textrm{gh}}&=-2\colon b\gamma\colon+\colon c\partial\beta\colon+\frac{3}{2}\colon\beta\partial c\colon\,.
\end{align}

As in the case of the bosonic string, a BRST symmetry survives the gauge fixing and its charge is given by
\begin{equation}\label{BRSTcharge}
 Q_{\textrm{BRST}}=\oint\frac{dz}{2i\pi}[cT+\gamma T_F+\frac{1}{2}(cT^{\textrm{gh}}+\gamma T_F^{\textrm{gh}})]\,.
\end{equation}
The BRST charge is again nilpotent when the total central charge of the SCFT vanishes, and physical states are classified by its cohomology. Moreover, one must impose a mass-shell condition  $L_0|\mc O_{\textrm{phys}}\rangle=0$ but also its fermionic counterpart in the R sector
\begin{equation}\label{MassShellF}
  G_0|\mc O_{\textrm{phys}}\rangle= \{Q_{\textrm{BRST}},\beta_0\}|\mc O_{\textrm{phys}}\rangle=0\,.
\end{equation}
Before closing this section, we give the mode expansions for the ghosts and superghosts
\begin{align}
 b(z)&=\sum_{n\in\mb Z}b_n\,z^{-n-2}\,,\label{ModeExpb}\\
 c(z)&=\sum_{n\in\mb Z}c_n\,z^{-n+1}\,,\label{ModeExpc}\\
 \beta(z)&=\sum_{n\in\mb Z+\nu}\beta_{n}\,z^{-n-3/2}\,,\label{ModeExpbeta}\\
 \gamma(z)&=\sum_{n\in\mb Z+\nu}\gamma_{n}\,z^{-n+1/2}\,,\label{ModeExpgamma}
\end{align}
with $\nu$ being $0$ in the R sector and $1/2$ in the NS sector.

\subsection*{Bosonisation}

One of the surprising yet very useful results of two-dimensional CFTs is the equivalence of field theories with different fields and actions, known as \emph{bosonisation}, which we now present. Consider a bc-system as in \eqref{bcSystem}. We would like to find a bosonic system equivalent to this fermionic one. A scalar field H has the OPE
\begin{equation}\label{HHOPE}
 H(z)H(w)\sim-\ln(z-w)\,.
\end{equation}
In order to recover the OPE of the bc-system, one can `exponentiate' \eqref{HHOPE}. Indeed, the coherent states $e^{\pm iH(z)}$ have the desired OPEs:
\begin{equation}
 e^{iH(z)}e^{-iH(w)}\sim\frac{1}{z}\,,
\end{equation}
the other ones being regular. One is thus led to assume the equivalence
\begin{align}
 b(z)&\cong e^{iH(z)}\,,\\
 c(z)&\cong e^{-iH(z)}\,.
\end{align}
However, a general bc-system contains a background charge $Q=2\lambda-1$, $\lambda$ being the conformal weight of the b-field, that a free scalar $H$ does not capture. This can be cured by shifting the EM tensor of the bosonic theory as
\begin{equation}
 T_{H}-i(\lambda-\frac{1}{2})\partial^2H\,.
\end{equation}
In other words, the equivalent bosonic CFT is that of a linear dilaton with $V=-i(\lambda-\frac{1}{2})$. Indeed, using the fact that $e^{iqH(z)}$ has conformal weight $q^2/2+iqV$, one can easily check that the conformal weights of b,c match precisely those of $e^{iH}$,$e^{-iH}$. Finally, the fermion or ghost number $j_{bc}=-bc$ is equivalent to $-i\partial H$. This applies in particular to the case of free fermions ($\lambda=1/2$), or the b,c ghost system ($\lambda=2$) for which the bosonic operators are denoted $e^{-\rho},e^{\,\rho}$. Note that for the superstring, one needs five free scalars $H_a$, $a=1,\ldots,5$ for the \WS fermions and one, $\rho$, for the b,c ghosts.

For the $\beta\gamma$ CFT, the bosonisation employs two decoupled CFTs, a bc-like CFT denoted $\eta\xi$ and a free scalar one $\varphi$. The equivalence is then formulated as
\begin{align}
 \beta(z)&\cong e^{-\varphi(z)}\partial\xi(z)\,,\label{BosBeta}\\
 \gamma(z)&\cong e^{\,\varphi(z)}\eta(z)\,.\label{BosGamma}
\end{align}
Once again, the conformal weights match since the operator $e^{q\varphi}$ has weight $-\frac{1}{2}q(q+2)$. The EM tensor of the $\beta\gamma$-system is identified with the sum of the $\varphi$ and $\eta\xi$ ones. Notice that the superghost number is given by $\partial\varphi$. In addition, one can further bosonise the $\eta\xi$ CFT in terms of a free scalar $\chi$ as $\eta\cong e^{-\chi}$ and $\xi\cong e^{\,\chi}$, hence obtaining
\begin{align}
 \beta(z)&\cong e^{-\varphi(z)+\chi(z)}\partial\chi(z)\,,\\
 \gamma(z)&\cong e^{\,\varphi(z)-\chi(z)}\,.
\end{align}

\subsection*{Spectrum, GSO projection}

The vacuum $|0\rangle$ of the SCFT corresponds to the identity operator. This is most easily seen by imposing regularity of $c(z)|0\rangle$ at the origin. Hence, $c_{n\geq2}|0\rangle=0$. However, $c_1$ is a lowering operator so that $|0\rangle$ is not the ground state of the theory. The latter is obtained by acting with $c_1$ as can be derived from the anti-commutation relations of the b,c ghosts. Therefore, the ground state for the ghost CFT is $c_1|0\rangle$ which can also be written as $c(0)|0\rangle$.

The situation is similar for the superghost CFT where, in the NS sector, $\gamma_{1/2}$ does not annihilate the identity operator. However, the commuting nature of the superghosts renders the problem more complicated because the ground state is degenerate. The operator $\alpha(z)$ one needs to insert should have a regular OPE with $\gamma(z)$ giving a simple zero (and a simple pole with $\beta(z)$). With the help of bosonisation, one can show that the operator $e^{-\varphi}$ has precisely these properties so that the ground state for the $\beta\gamma$ CFT is, in the NS sector, $|0\rangle_{\textrm{NS}}\equiv e^{-\varphi}|0\rangle$.

In the R sector, there are no branch cuts but the ground state is degenerate because of the fermionic zero modes. Consistency of the theory imposes having branch cuts in the OPE of the superghosts with the R ground state leading to $|0\rangle_{\textrm{R}}\equiv e^{-\frac{\varphi}{2}}\Theta_s|0\rangle$, $\Theta_s$ being the spin-field introducing a branch cut in the OPE with the \WS fermions. It can be bosonised in terms of the free scalars $H_a$:
\begin{equation}
 \Theta_s=e^{is^aH_a}\,,
\end{equation}
with $s^a=\pm1/2$. Let us give a more detailed, alternative derivation of this statement. Recall that in the R sector, the fermion zero modes $\psi_{0}^{\mu}$ generate a Clifford algebra $\{\gamma^{\mu},\gamma^{\nu}\}=-2\delta^{\mu\nu}$, with $\gamma^{\mu}\equiv i\sqrt{2}\,\psi^{\mu}_0$. Now define the complexified basis
\begin{align}
 \Gamma^{0}_{\pm}&=\frac{1}{2}(\gamma^1\pm\gamma^0)\,,\\
 \Gamma^{i}_{\pm}&=\frac{1}{2}(\gamma^i\pm i\gamma^{2i+1})\,,
\end{align}
with $i=1,\ldots,4$. $\Gamma_{\pm}$ play the role of creation and annihilation operators. The R ground state is then defined in terms of a \emph{spin field} $s$:
\begin{equation}
 |s\rangle=\prod_{a=0}^{4}(\Gamma_{+}^{a})^{s_a+\frac{1}{2}}|0\rangle\,.
\end{equation}
$|0\rangle$ is to be thought of as carrying a \emph{spin} $|\downarrow,\downarrow,\downarrow,\downarrow,\downarrow\rangle$ and each $\Gamma_{+}$ flips the corresponding spin. Recall that the $\mathbf{32}$ Dirac representation is reducible with respect to the chirality matrix
\begin{equation}
\Gamma_{11}=32\prod_{a=0}^{4}(\Gamma^a_+\Gamma^a_--\frac{1}{2})\,.
\end{equation}
Indeed, if one defines $|\alpha\rangle$ as being the state with even number of $s_a=1/2$ and $|\alpha'\rangle$ as carrying an odd number of them, then these two representations are irreducible and correspond to the $\mathbf{32}=\mathbf{16}\oplus\mathbf{16}'$ decomposition. In particular, $\Gamma_{11}|\alpha\rangle=|\alpha\rangle$ and $\Gamma_{11}|\alpha'\rangle=-|\alpha'\rangle$. The R ground state can thus be written as
\begin{equation}
 |0\rangle_{\textrm{R}}=|\alpha\rangle+|\alpha'\rangle\,.
\end{equation}
The operator $e^{-\frac{\varphi}{2}}\Theta_s$ is nothing but the vertex operator representing the R ground state.

In order to construct the spectrum of the superstring, we must include the right-movers and consider the four possible combinations of NS and R sectors. As we have seen previously only $R\otimes NS$ and $NS\otimes R$ generate \ST fermions, the other two sectors giving rise to \ST bosons.

First, consider the NS-NS sector. By taking into account the momentum degeneracy of the ground state, its vertex operator is (the c-ghost insertions are omitted)
\begin{equation}
 e^{-\varphi}e^{-\bar\varphi}\,e^{ip\cdotp X}\,.
\end{equation}
Using the mass-shell condition, one can show that this state is tachyonic ($m^2=-1$). The first excited state is obtained by acting with the oscillators $\psi^{\mu}_{-1/2}$ and corresponds to the vertex operator
\begin{equation}
 e^{-\varphi}\psi^{\mu}\,e^{-\bar\varphi}\bar\psi^{\nu}\,e^{ip\cdotp X}\,.
\end{equation}
It is a massless state generating a graviton, a B-field and a dilaton. Higher excitations are massive and we do not consider them here.

In the R-R sector, the ground state is a massless spinor (from the left and the right) and corresponds to the vertex operator

\begin{equation}
 e^{-\frac{\varphi}{2}}\Theta_s\,e^{-\frac{\bar\varphi}{2}}\bar\Theta_s\,e^{ip\cdotp X}\,.
\end{equation}

Finally, in the R-NS sector, the ground state is constrained by level matching \eqref{LevelMatching} to be
\begin{equation}
 e^{-\frac{\varphi}{2}}\Theta_s\,e^{-\bar\varphi}\bar\psi^{\mu}\,e^{ip\cdotp X}\,.
\end{equation}
This is a massless state decomposing into a spin-3/2 field, the gravitino, and a spin-1/2 field, the dilatino.

Notice that the ground states we obtain form a representation of $\mc N=1$ \ST supersymmetry which can be attributed \cite{Banks:1988yz} to the $\mc N=2$ SCFT on the \WS\footnote{The local $\mc N=1$ SCFT is in fact enhanced to a global $\mc N=2$ one.}. On the other hand, the presence of the tachyon is undesirable and it turns out that it is eliminated from the spectrum by the Gliozzi-Scherk-Olive (GSO) projection as we now briefly discuss.

The idea is to introduce a parity symmetry which selects only part of the spectrum that does not include the tachyon. This might seem \emph{ad hoc}, even though one can show that it is rooted in the consistency of the superstring\footnote{It is related to the conformal invariance of the \WS theory which shows up in particular at one-loop as the strong constraint of modular invariance of the partition function.}. Let us first focus on the NS sector in which the fermion number operator is
\begin{equation}
 F=\sum_{r\geq1/2}\psi^{\mu}_{-r}\,\psi_{r}^{i}\,.
\end{equation}
The fermion parity operator $(-)^{F}$ counts the number of fermionic oscillators of each state (modulo 2). Clearly, the NS ground state is even:
\begin{equation}
 (-)^F|0\rangle_{\textrm{NS}}=|0\rangle_{\textrm{NS}}\,.
\end{equation}
The natural projection is then to select only the states that are even under $G_{\textrm{NS}}\equiv(-)^{F+1}$. In the R sector, the ground state is massless and, naively, one is not forced to impose any projection. However, this would render the theory inconsistent as it would break one-loop modular invariance. Here, the relevant fermion parity is\footnote{The chirality matrix $\Gamma_{11}$ comes from the fermion zero modes present only in the R sector.} $G_{\textrm{R}}\equiv(-)^F\,\Gamma_{11}$ under which one of the irreducible spinor representations, denoted $S$ (for spinor), is even and the other one, $C$ (for conjugate spinor) is odd:
\begin{align}
 G_{\textrm{R}}|S\rangle&=|S\rangle\,,\\
 G_{\textrm{R}}|C\rangle&=-|C\rangle\,.
\end{align}
One has the freedom to choose either projection. This implies various projections when one includes the right-movers as well, leading to different consistent theories presented in the next subsections.

\subsection*{Partition Functions}

As in the bosonic string case, it is useful to calculate the contributions of the fermionic degrees of freedom of the SCFT to the one-loop partition function and we restrict our attention to the closed string case. Consider a free complex fermion $\psi$ on a two-torus whose action is given by
\begin{equation}
 S=-\frac{1}{2\pi}\int d^2z\,\psi\,\bar\partial\,\psi\,.
\end{equation}
$\psi$ can have two different boundary conditions on each cycle of the torus (periodic or anti-periodic) and we parametrise them by two $\mb Z_2$-valued integers a,b such that
\begin{align}
 \psi(\xi^1+2\pi,\xi^2)&=e^{i\pi(1-a)}\psi(\xi^1,\xi^2)\,,\label{aSpinStructure}\\
 \psi(\xi^1,\xi^2+2\pi)&=e^{i\pi(1-b)}\psi(\xi^1,\xi^2)\,.\label{bSpinStructure}
\end{align}
The choice of a couple $(a,b)$ is called a \emph{spin structure}. On a genus $g$ Riemann surface, there are $2g$ cycles so that the number of spin structures is $2^{2g}$. Mathematically, a fermion is a section of a line bundle $\eta$ such that $\eta\otimes\eta=K$, $K$ being the canonical line bundle, \emph{i.e.} the bundle of holomorphic one-forms on the Riemann surface\footnote{Roughly speaking, a fermion transforms as $dz^{1/2}$ so that it `squares' to a one-form.}. There are $2^{2g}$ ways of taking the `square-root', each one corresponding to a topologically inequivalent way of putting a spinor on the Riemann surface.

For a particular spin structure $(a,b)$, the partition function is
\begin{equation}\label{PsiPartition}
 Z_{\psi}[^a_b]=\textrm{Tr}[e^{i\pi b F}\,q^{H(a)}]\,,
\end{equation}
with the boundary condition $a$ shifting the zero-point energy in the Hamiltonian and $F$ being the fermion number. The canonical calculation of the trace is straightforward. However, we present here an alternative derivation using the path integral $\int d\psi\,e^{-S[\psi]}$. The integral is Gaussian and gives the determinant of the operator $\bar\partial$ which is an infinite product of its eigenvalues $m+\frac{a}{2}+(n+\frac{b}{2})\tau$, $\tau$ being the complex structure of the torus. Using $\zeta$-function regularisation as in Appendix \ref{sec:ZetaReg}, one obtains
\begin{equation}
 Z_{\psi}[^a_b]=\frac{\vartheta[^a_b](\tau,0)}{\eta(\tau)}\,.
\end{equation}
The general theta-function with characteristics is defined in \eqref{ThetaGenusOne}. Notice that for periodic boundary conditions $(a,b)=(0,0)$, the partition function is zero as expected due to the presence of zero-modes.

The $\beta\gamma$ CFT can be calculated in a similar fashion. From its bosonic nature, one expects the partition function to be proportional to $(Z_{\psi})^{-1}$. The relative phase is essentially fixed by the eigenvalue of the ground states under $G_{\textrm{NS}}$ or $G_{\textrm{R}}$ which depends not only on the boundary conditions $(a,b)$ but also on the choice of chirality for the R ground state which we parametrise by a $\mb Z_2$ integer $\mu$:
\begin{equation}
 Z_{\beta\gamma}[^a_b]=(-)^{b+\mu ab}\frac{\eta(\tau)}{\vartheta[^a_b](\tau,0)}\,.
\end{equation}
Notice that the partition function of the $\beta\gamma$ CFT cancels the one of a complex fermion. This is to be expected in the full superstring theory from the gauge-fixing procedure since the superghosts eliminate the two (unphysical) longitudinal degrees of freedom of the \WS fermions. Finally, the full partition function is obtained by including the right-movers (and also the partition functions (\ref{PartitionX}, \ref{PartitionBC})), implementing the GSO projection by inserting $\frac{1+G}{2}$ and summing over all spin structures.

\subsection*{Type II superstring}

The choice of boundary conditions for the fermions $\nu$ and GSO projection ($G$) for the left- and right-movers gives rise to 16 different sectors labeled by $((\nu, G),(\tilde\nu,\widetilde G))$, and only particular combinations of them lead to consistent superstring theories. First of all, notice that the sector $(\textrm{NS},-)$ which contains the tachyon, can only be paired with $(\widetilde{\textrm{NS}},-)$ due to level matching. Hence, there are only 10 possible sectors. In addition, we only focus on theories projecting out the tachyonic ground state. This is the Type II superstring which can be either chiral (IIB) or non-chiral (IIA) and these are the only inequivalent theories one can construct in this way. In what follows, we present the massless spectrum of each of them.

\subsubsection*{Type IIA}

The ground state of the NS-NS sector is a massless \ST vector from the left and the right, i.e. an $\mathbf{8}_v\otimes\mathbf{8}_v$ representation of the little group $SO(8)$ of $SO(1,9)$. It decomposes as
\begin{equation}
 \mathbf{8}_v\otimes\mathbf{8}_v=\mathbf{1}\oplus\mathbf{28}\oplus\mathbf{35}
\end{equation}
in terms of a dilaton (\textbf{1}), a two-form (\textbf{28}) and a symmetric traceless tensor (\textbf{35}). This is the string background as in the Polyakov action \eqref{GeneralPolyakov}. In the R-R sector, only the relative chirality matters and we choose the left GSO projection to be a spinor $S$ so that the right one is a conjugate spinor $\widetilde C$. This gives rise to a vector and a three-form as can be seen from the decomposition
\begin{equation}
 \mathbf{8}_s\otimes\mathbf{8}_c=\mathbf{8}_v\oplus\mathbf{56}_v\,.
\end{equation}
The fermionic states arise from the NS-R and R-NS sectors from which one obtains a dilatino (spin 1/2) and a gravitino (spin 3/2) according to the group-theoretic relation
\begin{equation}
 \mathbf{8}_v\otimes\mathbf{8}_{c,s}=\mathbf{8}_{s,c}\oplus\mathbf{56}_{s,c}\,.
\end{equation}
That is, \ST supports an $\mc N=2$ theory in a ten-dimensional \ST.

\subsubsection*{Type IIB}

The GSO projection is chosen such that the R ground state is a spinor. The massless spectrum is the same as in Type IIA apart from the states coming from the R-R sector (the theory is chiral). The latter gives rise to a scalar (or zero-form), a two-form and a self-dual four-form:
\begin{equation}
 \mathbf{8}_s\otimes\mathbf{8}_s=\mathbf{1}\oplus\mathbf{28}\oplus\mathbf{56}\,.
\end{equation}
Hence, $\mc N=2$ supergravity in ten dimensions arises as the massless or effective field theory of the Type IIB superstring.

\subsection*{Type I superstring}

\begin{figure}[H]
\centering
\includegraphics[width=0.8\textwidth]{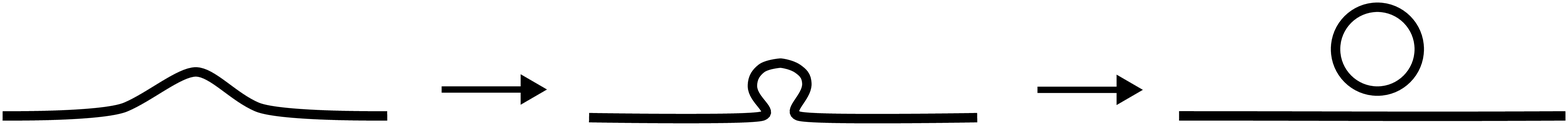}
\caption{Interaction between open and closed strings.}
\label{fig:OpenClosed}
\end{figure}

One can make the same analysis for the open string sector. In fact, an open superstring theory must also include closed strings because open and closed strings can interact, see Fig. \ref{fig:OpenClosed}. First of all, one must impose the orientifold projection $\Omega$ and then the appropriate GSO. In this case, the GSO has to be chiral as in Type IIB in order to be consistent with the orientifold projection (or else we break Poincar\'e invariance). In this sense, the theory one obtains is a Type IIB orientifold called the Type I superstring. The massless spectrum is thus very similar to the Type IIB and the $\Omega$-projection eliminates the B-field from the NS-NS sector and the zero- and four-forms from the R-R sector. The resulting effective field theory is an $\mc N=1$ supergravity.
In the previous section, we have seen that anomalies can appear as divergences in the partition function. In order for the theory to be well-defined, the latter must cancel and this occurs only when $n=32$, \emph{i.e.} when the Type I superstring supports an $SO(32)$ gauge theory, as can be seen from \eqref{TadpoleCancellation}. In fact, one can separate out two tadpoles, one coming from the RR sector and the other one from the NS-NS sector which are, however, equal because of supersymmetry. By Poincar\'e invariance, the RR tadpole corresponds to a term in the effective action sourced by a ten-form (in the NS-NS sector the source is given by the dilaton or the graviton):
\begin{equation}\label{RR10form}
 Q_{\textrm{tot}}\int d^{10}x\,C_{10}\,.
\end{equation}
In ten \ST dimensions, a ten-form is non-dynamical and appears in the action only through \eqref{RR10form}. Integrating it out yields the constraint
\begin{equation}
 Q_{\textrm{tot}}=0\,,
\end{equation}
precisely as computed by \eqref{TadpoleCancellation}.

\subsection*{Heterotic superstring}

There is a different hybrid superstring theory one can construct by using the fact that a closed string theory has independent left- and right-movers. The \emph{Heterotic} string is obtained by combining a SCFT on the left-moving side with a bosonic string on the right-moving one. The resulting \WS has $\mc N=(1,0)$ supersymmetry.

From the left-moving sector, the absence of conformal anomalies implies that the \ST dimension is 10. However, from the right-moving sector, the theory is inconsistent because the total central charge is non vanishing, $\tilde c_{\textrm{tot}}=\tilde c_{X}+\tilde c_{\textrm{gh}}=10-26=-16$. One is forced to introduce additional degrees of freedom which can either be bosonic or fermionic. In the first case, sixteen internal (compact) bosons are enough to cancel the total central charge and one can show that, due to modular invariance, they must live on a Euclidean even self-dual lattice. In the second case, one must add thirty-two free internal real fermions (i.e. without a \ST index) with the appropriate boundary conditions. In fact, both approaches are equivalent through bosonisation. Here, we focus on the fermionic approach.

The additional fermionic degrees of freedom are denoted $\tilde\lambda^A$, $A=1,\ldots,32$. In the right-moving sector, the CFT has an additional global symmetry associated to the new index $A$, which depends on the choice of boundary conditions. Notice that the vacuum structure is very different from the usual SCFT case because of the absence of superghosts. The simplest choice is to assign the same boundary conditions to all the fermions (NS or R). This implies that $\tilde\lambda^A$ transforms as an $SO(32)$ vector. The other possibility is to give NS boundary conditions to $n$ fermions and R to the $32-n$ remaining ones. In this case, one can show that the only consistent choice is $n=16$ so that the relevant internal group is $SO(16)\times SO(16)$. One way to see this is to calculate the zero-point energy
\begin{equation}
 \tilde a=\frac{8}{24}+\frac{n}{48}-\frac{32-n}{24}=\frac{n}{16}-1\,.
\end{equation}
Level-matching with the left-moving sector requires the mass-squared to be integer, $M^2=N-a=N-\tilde a$, and one sees that the only possible choice is $n=16$.

In the $SO(32)$ theory, the GSO projection in the supersymmetric (left-moving) sector is such that the R ground state is a spinor $|S\rangle$ and the NS ground state is a \ST vector
$\psi^{\mu}_{-1/2}|0,p\rangle$. On the bosonic (right-moving) side, one imposes either NS or R boundary conditions in all internal directions $A$. In the first case, the vacuum and the first excited states $\tilde\lambda^{A}_{-1/2}|0,p\rangle$ are tachyonic and only at level $\widetilde N=2$ one obtains massless states $\tilde\lambda^{A}_{-1/2}\,\tilde\lambda^{B}_{-1/2}|0,p\rangle$ and $\tilde\alpha^{\mu}_{-1}|0,p\rangle$. In the R sector, there are fermionic zero-modes $\tilde\lambda^{A}_0$ forming an $SO(32)$ Clifford algebra, and the ground state is a massive spinor $|A\rangle$. Modular invariance forces a GSO projection that projects the latter onto its positive chirality component $|A^+\rangle$.

The massless spectrum is then as follows. The states
\begin{align}
 \psi^{\mu}_{-1/2}|0,p\rangle\otimes\tilde\alpha^{\mu}_{-1}|0,p\rangle
\end{align}
give rise to the dilaton, the B-field and the graviton. Together with the gravitino and the dilatino in
\begin{align}
 |S\rangle\otimes\tilde\alpha^{\mu}_{-1}|0,p\rangle\,,
\end{align}
they form the $\mc N=1$ gravitational multiplet in ten dimensions. The gauge degrees of freedom arise from the states
\begin{align}
 \psi^{\mu}_{-1/2}|0,p\rangle\otimes\tilde\lambda^{A}_{-1/2}\,\tilde\lambda^{B}_{-1/2}|0,p\rangle\,,\label{VectorHet}\\
 |S\rangle\otimes\tilde\lambda^{A}_{-1/2}\,\tilde\lambda^{B}_{-1/2}|0,p\rangle\,.\label{SpinorHet}
\end{align}
\eqref{VectorHet} is a \ST SO(32) gauge boson which, together with the Majorana-Weyl spinor \eqref{SpinorHet}, form an $\mc N=1$ vector multiplet. Hence, the effective field theory of this Heterotic superstring is an $\mc N=1$ theory with gauge group $SO(32)$.

Let us turn to the $SO(16)\times SO(16)$ theory. The gravitational sector is unchanged and the massless gauge degrees of freedom arise from the NS-NS and R-NS/NS-R sectors (here, we are referring to the boundary conditions of the internal fermions in the right-moving sector only). The index $A$ is decomposed accordingly as $(a,\bar a)$ with $a=1,\ldots,16$ and $\bar a=17,\ldots,32$. In the NS-NS sector, the relevant states are
\begin{align}
 \tilde\lambda^{a}_{-1/2}\,\tilde\lambda^{b}_{-1/2}|0,p\rangle\,,\\
 \tilde\lambda^{\bar a}_{-1/2}\,\tilde\lambda^{\bar b}_{-1/2}|0,p\rangle\,,\\
 \tilde\lambda^{a}_{-1/2}\,\tilde\lambda^{\bar b}_{-1/2}|0,p\rangle\,.
\end{align}
Tensoring with $|S\rangle$ produces spinors $\chi^{ab},\chi^{\bar a\bar b}$ in the adjoint of each of the $SO(16)$, $(\textbf{120},\textbf{1})$ and $(\textbf{1},\textbf{120})$ respectively, and a mixed $(\textbf{16},\textbf{16})$ spinor $\chi^{a,\bar a}$. The tensor product with $\psi^{\mu}_{-1/2}|0,p\rangle$ gives the corresponding \ST gauge boson. However, the mixed states $(16,16)$ are eliminated by the GSO projection. From the R-NS sector, the relevant right-moving massless state is a spinor of the first $SO(16)$ and, together with the left-moving contribution, give a \ST spinor and a \ST vector $(\textbf{1},\textbf{128})$. Finally, the NS-R sector produces the same state with respect to the second $SO(16)$.

The Heterotic $SO(16)\times SO(16)$ has thus a gauge boson that transforms in the adjoint of the $\textbf{120}\oplus\textbf{128}=\textbf{248}$ representation which is the adjoint of the exceptional group $E_8$, and similarly for the other $SO(16)$. Consequently, \ST supports an $\mc N=1$, $E_8\times E_8$ gauge theory.

\section{Compactification}

We have so far described superstring theory with a target space having ten non-compact dimensions. Of course, one has to make contact with the low energy theory described by the Standard Model which is four-dimensional, non-supersymmetric and has been tested experimentally to a considerable accuracy. Moreover, the experimental bounds on possible extra dimensions are stringent (see \emph{e.g.} \cite{Aad:2012bsa}). Hence, one is led to consider \emph{compactification} of superstring theory, \emph{i.e.} making the six-dimensional internal space compact with its size being controlled by the Planck or the string scale. There are various ways of achieving this and we now point to some of them. For example, one can introduce an intersecting brane background breaking supersymmetry and implementing the SM gauge group at low energies. Another way is to turn on fluxes in the internal compact dimensions, leading to \emph{flux compactifications}. However, these two possibilities can only be studied effectively, at the level of supergravity and we do not discuss them further. Instead, we consider string compactifications in which the internal space needs only be consistent, in the sense that the internal (S)CFT has vanishing central charge. The most promising candidates are orbifolds and Calabi-Yau spaces.

In what follows, we first present toroidal compactifications as the simplest yet unrealistic example and then briefly review the orbifold and Calabi-Yau ones, showing how both are in fact related.

\subsection*{Toroidal compactification}

As a warm-up exercise, consider a closed string theory in which one of the target space directions, say 9, is a compact circle of radius $R$. Then the periodicity condition changes since the closed string can wind an arbitrary number of times around the compact circle. From the \WS point of view, these configurations correspond to topological solitons. For the compact boson, that is
\begin{equation}\label{windingBC}
 X^{9}(\tau,\sigma+2\pi)=X^{9}(\tau,\sigma)+2\pi nR\,.
\end{equation}
$n$ is called the winding number. The solution to the wave equation is still encoded in the mode expansions (\ref{XLModes}, \ref{XRModes}), the difference being only through the boundary conditions. The momentum of the string $\frac{p^9_L+p^9_R}{2}=\frac{m}{R}$ is quantised and the boundary condition \eqref{windingBC} gives $\frac{p^9_L-p^9_R}{2}=\frac{nR}{\alpha'}$ so that the expressions for the left- and right-moving momenta are
\begin{equation}
 p^9_{L,R}=\frac{m}{R}\pm\frac{nR}{\alpha'}\,.
\end{equation}
Notice that the solution acquires an extra symmetry
\begin{align}
 R\rightarrow \frac{\alpha'}{R}\,,\nonumber\\
 m\leftrightarrow n\,.\label{Tduality}
\end{align}
This perturbative symmetry is called \emph{T-duality}. Effectively, it turns $\partial_{\sigma}$ into $\partial_{\tau}$ so that Neumann and Dirichlet boundary conditions are exchanged. In particular, $X^9=X_L+X_R$ becomes $\tilde X^9=X_L-X_R$.

The level-matching condition \eqref{LevelMatching} now reads $N-\tilde N=mn$ and the mass operator is $M^2=(p^9_L)^2+(p^9_R)^2+\frac{2}{\alpha'}(N+\tilde N-2)$. The vacuum of the theory has additional labels given by the momentum and winding numbers, $|0,p,n,m\rangle$. The spectrum is essentially the same for $m=n=0$. For $n=0$ or $m=0$, one has additional scalar fields with masses
\begin{equation}
-\frac{4}{\alpha'}+\left(\frac{n}{R}\right)^2\,,\,-\frac{4}{\alpha'}+\left(\frac{mR}{\alpha'}\right)^2\,.
\end{equation}
For $n=\pm m=\pm1$, acting on the ground state with $\alpha_{-1}^{9}$, $\alpha_{-1}^{\mu\neq9}$ we obtain scalars and vectors with masses
\begin{equation}
 M^2=\frac{1}{R^2}+\frac{R^2}{\alpha'^2}-\frac{2}{\alpha'}=(\frac{1}{R}-\frac{R}{\alpha'})^2\,.
\end{equation}
Hence, at the self-dual radius $R_*=\sqrt{\alpha'}$, the above vectors become massless and we have an enhanced \ST gauge symmetry: the abelian $U(1)\times U(1)$ symmetry arising from the Kaluza-Klein reduction turns into a non-abelian $SU(2)\times SU(2)$. This is an inherently stringy regime. Notice that going away from the self-dual radius breaks the gauge symmetry as in the Higgs mechanism.

Moreover, the mass operator is invariant under T-duality \eqref{Tduality} so that the latter is a symmetry of the spectrum. In fact, one can show that it is preserved quantum mechanically through the OPE: T-duality is also a symmetry of the interactions, \emph{i.e.} valid to all orders in perturbation theory. Note that the self-dual radius $R_*$ is the smallest radius one can probe, and is indeed given by the string length. In the limit $R\gg R_*$, the winding states become very massive and, hence, decouple from the theory, whereas the momentum states (or Kaluza-Klein states) become very dense and one readily recovers the non-compact limit with a continuous momentum.

The partition function of the compact scalar can be written as in the non-compact case, the only difference being that the momentum integration is replaced by a sum over $p_{L,R}^9$:
\begin{align}
 Z_{X^9}(\tau,\bar\tau)&= \frac{1}{|\eta(\tau)|^2}\sum_{n,m\in\mb Z}q^{\frac{\alpha'(p^9_L)^2}{4}}\bar q^{\frac{\alpha'(p^9_R)^2}{4}}\nonumber\\
		       &= \frac{1}{|\eta(\tau)|^2}\sum_{n,m\in\mb Z}e^{-\pi\tau_2(\frac{\alpha'm^2}{R^2}+\frac{n^2R^2}{\alpha'})}e^{2i\pi\tau_1nm}\,.
\end{align}
This is the Hamiltonian representation of the partition function in which modular invariance is not manifest. In order to make it manifest, we go to the Lagrangian representation, which is the representation we would obtain had we calculated the partition function using the path integral. This is done by using the Poisson summation formula\footnote{It is often referred to as Poisson resummation.}~\eqref{PoissonResummation}:
\begin{equation}
 Z_{X^9}(\tau,\bar\tau)=\frac{2\pi R}{\sqrt{4\pi^2\alpha'\tau_2}}\frac{1}{|\eta(\tau)|^2}\sum_{n,\tilde m\in\mb Z}e^{-\frac{\pi R^2}{\alpha'\tau_2}|\tilde m-n\tau|^2}\,.
\end{equation}

The previous results can be generalised to the case where the internal space is a d-dimensional torus $T^{d}$ with the same philosophy. For this, consider the Kaluza-Klein ansatz for the metric element
\begin{equation}
 ds^2=g_{\mu\nu}dx^{\mu}dx^{\nu}+g_{ij}(dx^i+A_{\mu}^i\,dx^{\mu})(dx^j+A_{\mu}^j\,dx^{\mu})\,,
\end{equation}
with $A^i$ being commuting gauge fields parametrising Wilson lines along the cycles of $T^{d}$. The metric of the latter $g=e^{\dag}e$ is invariant under $SO(d,\mb R)$ transformations $e\rightarrow Me$, with $M\in SO(d,\mb R)$ and $e$ the vielbein. Using the Iwasawa decomposition
\begin{equation}
 GL(d,\mb R)=SO(d,\mb R)\times(\mb R^+)^d\times N_d\,,
\end{equation}
with $N_d$ being the unipotent group\footnote{It consists of upper triangular matrices with unit element on the diagonal.}, $e$ is accordingly written as $e=K\cdotp A\cdotp N$. Gauge-fixing the orthogonal matrix $K$ to the identity leaves $e$ as a product of a diagonal matrix with positive elements $A$ and a unipotent matrix. In fact, $A$ contains the radii of the torus whereas $N$ depends on the gauge fields. The moduli space of the toroidal compactification is thus parametrised by general linear transformations modded out by `gauge transformations' $SO(d,\mb R)$ together with the group of invariance $SL(d,\mb Z)$ of the torus. Recalling that $GL(\mb R)\cong \mb R^+\times SL(\mb R)$, the moduli space is
\begin{equation}
 \mb R^+\times[SO(d,\mb R)\backslash SL(d,\mb R)\slash SL(d,\mb Z)]\,,
\end{equation}
with the notation $\backslash,\slash$ for left, right coset. In fact, for the closed string background, the correct moduli space is
\begin{equation}
 \mb R^+\times[SO(d,\mb R)\times SO(d,\mb R)\backslash SO(d,d,\mb R)\slash SO(d,d,\mb Z)]\,,
\end{equation}

The partition function easily generalises to $d$ compact dimensions:
\begin{equation}\label{PartitionToroidalLattice}
 Z_{d}=\frac{1}{|\eta(\tau)|^{2d}}\sum_{p_{L,R}\in \Gamma}q^{\,p_L^2/2}\,\bar q^{\,p_R^2/2}\,,
\end{equation}
with $\Gamma$ some $2d$-dimensional lattice. Modular invariance of \eqref{PartitionToroidalLattice} requires the latter to be even self-dual\footnote{The dual lattice is the set of all points in $\mb R^{2d}$ such that the product in \eqref{LatticeProduct} with all points in $\Gamma$ is integer.} with $(d,d)$ signature, \emph{i.e.}
\begin{equation}
 \forall p=(p_L,p_R)\in \Gamma,\,p\circ p\in2\mb Z\,,
\end{equation}
with the lattice product $\circ$ being
\begin{equation}\label{LatticeProduct}
 p\circ p'\equiv p_L\cdotp p_L'-p_R\cdotp p_R'\,.
\end{equation}
The resulting lattice, denoted $\Gamma_{d,d}$ is the \emph{Narain} lattice. Notice that it is invariant under the action of $O(d,\mb R)\times O(d,\mb R)$ that rotates the left- and right-momenta separately as can be seen from \eqref{LatticeProduct}. However, the larger group $O(d,d,\mb R)$ produces inequivalent lattices. Moreover, the lattice is invariant under discrete orthogonal transformations $O(d,d,\mb Z)$ that map the lattice points to themselves. Therefore, the moduli space of Narain compactifications is
\begin{equation}\label{ModuliSpaceNarain}
 [O(d,\mb R)\times O(d,\mb R)]\backslash O(d,d,\mb R)\slash O(d,d,\mb Z)\,.
\end{equation}
This coset has dimension $d^2$ and $O(d,d,\mb Z)$ is the T-duality group. It contains, in particular, the inversion of radii of $T^d$ as in \eqref{Tduality}, but generates a richer symmetry group. The lattice momenta depend on the moduli arising from the compactification. In particular, for $d>1$, the B-field also gives rise to scalar fields and enter in the expression of the momenta
\begin{equation}
 p_{L,R}=\sqrt{\frac{\alpha'}{2}}(m_i+\frac{1}{\alpha'}(B_{ij}\pm g_{ij})n^j)\,.
\end{equation}
On the other hand, the number of moduli, \emph{i.e.} the number of components of $g_{ij}$ and $B_{ij}$ is $\frac{d(d+1)}{2}+\frac{d(d-1)}{2}=d^2$ and matches, as expected, the dimension of the coset space \eqref{ModuliSpaceNarain}. Finally, the mass and level-matching conditions are
\begin{align}
 M^2&=p_L\cdotp p_L+p_R\cdotp p_R+\frac{2}{\alpha'}(N+\tilde N-2)\,,\\
 p\circ p&=N-\tilde N\,,
\end{align}
and can be used to probe gauge symmetry enhancements as in the one-dimensional case.

\subsection*{T-duality and D-branes}

As an explicit example, consider the Type I superstring or, equivalently, a Type IIB orientifold compactified on a circle $S^1_R$ of radius $R$. One can show that T-duality turns a spinor $S$ into a conjugate spinor $C$ and flips the sign of the GSO projection. Hence, the T-dual theory is a Type IIA orientifold on $S^1_{\widetilde R}$, $\widetilde R$ being the dual radius. However, the dual orientifold $\widetilde\Omega$ is a composition of the \WS parity $\Omega$ and a $\mb Z_2$ action on the circle coordinate $x^9$ since, by T-duality, $\tilde X^9=X^9_L-X^9_R$ and $\Omega$ exchanges left- and right-movers. The $\mb Z_2$ action has two fixed points $x^9=0,\pi\widetilde R$ and any closed string state at a generic position $x^9$ should have an image with respect to the fixed point with opposite momentum (and orientation). From the \ST point of view, $x^9=0,\pi\widetilde R$ are hyperplanes where closed strings are of both orientations and play a special role in string compactifications. They have eight space dimensions and are called \emph{orientifold planes}, O8-planes for short.

In the open string sector, under T-duality, N and D boundary conditions are exchanged so that $X^9$ is now a DD direction. This implies that the string endpoints cannot move away from $x^9=0$ and this, again, defines a hyperplane called \emph{Dirichlet membrane}, D8-brane for short. In fact, there are as many D8-branes as CP labels ($N$). One might suspect that O-planes and D-branes are the same objects. This is however not the case. To see this, consider the original Type IIB orientifold and turn on $N=32$ non-trivial Wilson lines along the circle, $A_9^a=-\frac{\theta^a}{2\pi R}$, with $a=1,\ldots,N$. Because of the orientifold, one imposes $\theta^{a+16}=-\theta^a$ so that there are only 16 independent parameters. The coupling to the Wilson lines is given by
\begin{equation}
 -\frac{\theta^a-\theta^b}{2\pi R}\int d\tau\,\partial_{\tau}X^9\,,
\end{equation}
which implies that the masses and the center of mass momentum are shifted according to their CP labels. For an $ab$ open string, that is
\begin{equation}\label{ShiftedMomenta}
 p^9_{ab}=\frac{m}{R}+\frac{\theta^a-\theta^b}{2\pi R}\,.
\end{equation}
Notice that the presence of Wilson lines breaks the gauge symmetry to $U(1)^{16}$. Setting $M$ Wilson lines to equal values away from the fixed points of $\mb Z_2$ induces a $U(M)$ enhancement whereas setting them to a fixed point yields an $SO(2M)$ one. Intuitively, only strings attached to coincident branes can give rise to massless states and this can be seen from \eqref{ShiftedMomenta} (recall that the mass operator contains a $p^2$ term). Performing a T-duality exchanges momenta and windings and the latter are now shifted due to the presence of the Wilson lines. Alternatively, the boundary conditions for $X^9$ are modified and the open string endpoints are now fixed at $\theta^a\widetilde R$, \emph{i.e.} there are $N$ D8-branes sitting generically at different positions $x^9=\theta^a\widetilde R$.

By T-duality, the ten-form $C_{10}$ becomes a nine-form $C_9$. Clearly, a disc diagram is localised at a D8-brane whereas a cross-cap diagram, because of the orientifold, sits on an O8-plane. Therefore, a disc and a cross-cap tadpole as in Fig. \ref{fig:TadpoleSum} measure the charges of D8-branes and O8-planes respectively. This provides another interpretation of tadpole cancellation as Gauss's law, \emph{i.e.} the vanishing of the total charge of the RR form in the compact dimension. The calculation in \eqref{TadpoleCancellation} above shows that the O8- and D8-charges are related:
\begin{equation}
 Q_{O8}=-16\,Q_{D8}\,.
\end{equation}
Tadpole cancellation requires the presence of thirty-two D8-branes because of the two O8-planes.

This interpretation can be straightforwardly generalised to the case of $d\equiv9-p$ compact dimensions with a $\mb Z_2$ action\footnote{In Type IIA, $d\in2\mb Z+1$ whereas in Type IIB, $d\in2\mb Z$.}. The fixed points of the reflection symmetry lead to $2^d$ Op-planes and a number of Dp-branes for the open string DD boundary conditions in the compact directions. Under the corresponding RR $p$-form, their charges are related by
\begin{equation}
 Q_{Op}=-2^{p-4}Q_{Dp}\,.
\end{equation}
In this language, the Type I superstring in ten dimensions can be viewed as having a \ST filling O9-plane and sixteen D9-branes (plus their sixteen images) required by tadpole cancellation. Notice that, contrary to D-branes, orientifold planes are non-dynamical objects (they cannot fluctuate) and, therefore, only carry charges, in particular under the corresponding RR potential. The coupling of a Dp-brane to the RR $(p+1)$-form can naturally be realised as the minimal coupling
\begin{equation}
 \mu_{p}\int\,C_{p+1}\,,
\end{equation}
where the integral is over the Dp-brane world-volume and $\mu_p$ is the RR charge. In ten \ST dimensions, for $p=9$, there is no kinetic term since the corresponding field strength is trivial. Therefore, the field equations for $C_{10}$ yield $\mu_9=0$. In other words, the total RR charge must be zero in an anomaly-free theory. In fact, the presence of a RR charge implies that, at tree-level, a propagating closed string contributes $\mu_9^2/0$ where the zero in the denominator is due to the trivial dynamics of the ten-form (its propagator vanishes). This is precisely the divergence one obtains in the Type I one-loop calculation. Hence, tadpole cancellation can be equivalently regarded as:
\begin{enumerate}
 \item Gauss's law for $p$-forms,
 \item Consistency of the field equations for the RR potentials.
\end{enumerate}

In general, one might consistently study the simultaneous presence of branes of different dimensions. Each Dp-brane carries a RR charge giving rise to a tadpole that must be cancelled by objects carrying negative RR charge, for instance Op-planes. To illustrate this, consider a Type I theory compactified on $T^4/\mb{Z}_2$ (see next section for more details). The orientifold group is generated by the \WS parity $\Omega$ together with the $\mb Z_2$-action $g$. As in the previous sections, one can carry out the calculation of the one-loop divergences when one inserts $\frac{1+\Omega}{2}\frac{1+g}{2}$ in the partition function. The Klein bottle contribution (proportional to $\Omega$ in the closed string sector) gives rise to a divergence with the prefactor
\begin{equation}\label{KBbis}
 V_6(V_4+\frac{1}{V_4})\,,
\end{equation}
where $V_4$ is the volume of the internal $T^4$. The first term in \eqref{KBbis} is the same as the one found in \eqref{PartitionK} and necessitates the introduction of D9-branes. The second term, however, is proportional to the inverse volume of $T^4$ and this operation takes a D9- to a D5-brane. Therefore, this theory is consistent only with the presence of sixteen D5-branes \cite{Bianchi:1990tb,Pradisi:1988xd,Angelantonj:2002ct}, the second term in \eqref{KBbis} originating from sixteen O5-planes. This is the tadpole cancellation constraint for D5-branes. Notice that in the decompactification limit $V_4\rightarrow\infty$, the D5-brane contribution in \eqref{KBbis} is trivial and there is no constraint on the number of D5-branes\footnote{The generalisation to the general $p<9$ case is straightforward.}. This is consistent with the fact that, due to the non-compactness of the transverse space (identified with $\mb C^2/\mb Z_2$), the flux lines of the six-form can escape to infinity (Gauss's law applies for compact spaces only). Alternatively, the corresponding equations of motion can be solved in the presence of a non-trivial RR charge.

Finally, let us mention that D-branes are non-perturbative objects in string theory. By calculating the exchange of a closed string between two Dp-branes, one can show that the Dp-brane tension is
\begin{equation}\label{DbraneTension}
 T_p=\frac{1}{(2\pi)^p(\alpha')^{\frac{p+1}{2}}g_s}\,.
\end{equation}
Notice that the behaviour as $1/g_s$ differs from the usual one of solitons in field theory whose mass scales as $1/g^2$ (recall that in field theory instanton effects are of order $e^{-1/g^2}$), and this stems from the stringy nature of D-branes. On the other hand, RR gauge fields can be represented in terms of vertex operators which depend only on the field strength of the $(p+1)$-form potential \cite{Polchinski:1995mt}:
\begin{equation}
 V_{RR}=e^{-\frac{\varphi}{2}}\Theta_s\,\Gamma^{\mu_1\ldots\mu_{p+2}}\,e^{-\frac{\tilde\varphi}{2}}\tilde{\Theta}_s\,F_{\mu_1\ldots\mu_{p+2}}\,,
\end{equation}
$e^{-\frac{\varphi}{2}}\Theta_s$ being the vertex operator of the RR ground state. Hence, all perturbative states are neutral with respect to RR gauge fields. However, dualities in string theories exchanging NSNS and RR states suggest that there should be (non-perturbative) states carrying RR charges. These are precisely the D-branes presented above and they play an important role in probing non-perturbative aspects of string theory and gauge theory using the language of perturbation theory. In particular, in the weak coupling regime, $g_s\ll 1$, D-branes are heavy and can be studied semi-classically as \emph{rigid} planes. For instance, gauge theory instantons can be realised as D-branes bound states as discussed in Section \ref{ch:GaugeTheoryFromStringTheory}. In addition, one can show that they are BPS states\footnote{Orientifold planes are also BPS objects.} (having equal mass and charge) and, therefore, preserve a subset of supersymmetries. For example, in Type I, due to the left-right identifications (and DD boundary conditions), only a linear combination of the supercharges $Q_{L,R}$ is preserved:
\begin{equation}\label{SusyDbranes}
 Q=\epsilon_L\,Q_L+\epsilon_R\,Q_R\,,
\end{equation}
$\epsilon_{L,R}$ being arbitrary spinors in \ST. For a Dp-brane, the condition is
\begin{equation}
 \epsilon_L=(\prod_{i=0}^{p}\Gamma^{i})\epsilon_R\,.
\end{equation}
For instance, for Type IIB orientifolds in ten dimensions ($p=9$), this is given by the chirality matrix $\Gamma^{10}$ producing a preserved combination as can be seen from the orientifold action.

\subsection*{Orbifolds}

Toroidal compactifications correspond to models having maximal \ST supersymmetry, since all the gravitini of the spectrum are preserved. For example, in Type II one obtains an $\mc N=(4,4)$ theory whereas in Heterotic, the four-dimensional theory is $\N=4$ supersymmetric. In order to construct more realistic models in which the number of \ST (super)-symmetries is reduced, one must project out some of the states and one way to perform this is to gauge some discrete subgroup $H$ of the CFT symmetries. The resulting (compact) space can thus be viewed as an orbit space $\mc M_6/H$, called \emph{orbifold}\footnote{Mathematically, an orbifold is a more general topological space, generalising the notion of manifolds, which looks only locally like a quotient of a Euclidean space with the action of a finite group.}. It may have singular points but the necessary condition is, once again, the consistency of the internal CFT. The action of $H$ is usually taken to be left-right symmetric, even though it is possible to construct asymmetric orbifolds \cite{Narain:1986qm}, at least at specific points in the moduli space of the compactification.

To illustrate this, consider the case of an $S^1/\mb Z_2$ compactification in which the $\mb Z_2$-action is defined on the compact coordinate by
\begin{equation}\label{Z2action}
 X^9\rightarrow-X^9\,.
\end{equation}
Under the $\mb Z_2$-quotient, the circle becomes a line segment $[0,\pi R]$ with the boundary points $0,\pi R$ being fixed by the action \eqref{Z2action}. This is the simplest example of a $\mb Z_2$-orbifold. This theory contains a new class of states in the spectrum which obey anti-periodic boundary conditions:
\begin{equation}
 X^9(\tau,\sigma+2\pi)=-X^9(\tau,\sigma)\,.
\end{equation}
We refer to them as \emph{twisted states} and they belong to the so-called twisted sector of the spectrum. In the untwisted (periodic) sector, one must enforce \eqref{Z2action} by projecting onto invariant states. From the mode expansions (\ref{XLModes}, \ref{XRModes}), one sees that the orbifold action reverses the sign of the zero-modes and oscillators so that
\begin{equation}
 \mb Z_2: |N,\tilde N,p,n,m\rangle\rightarrow(-)^{N+\tilde N}|N,\tilde N,p,-n,-m\rangle\,.
\end{equation}
Hence, only states having an even number of oscillators are preserved. For instance, the Kaluza-Klein vectors are projected out. In the twisted sector, the mode expansion is different. As in the case of NS-fermions, it has a half-integer moding and no (momentum) zero-modes:
\begin{equation}
 X^9_t(z,\bar z)=i\sqrt{\frac{\alpha'}{2}}\sum_{n\in\mb Z+1/2}\frac{1}{n}(\alpha^9_n\,z^{-n}+\tilde\alpha^9_n\,\bar z^{-n})\,.
\end{equation}
In particular, the position can only have small fluctuations around the fixed points of the orbifold. By calculating the zero-point energy, one can show that the ground states of this sector, called \emph{twist fields}, are primary fields of dimension $1/16$.

The partition function of this orbifold is obtained by calculating the contributions of the twisted and untwisted sectors and, in each of them, projecting onto invariant states by inserting the operator $\frac{1+g}{2}$, with $g\in\mb Z_2$. If we label the sector by a $\mb Z_2$-index $h$, then the partition function is given by
\begin{equation}
 Z^{S^1/\mb Z_2}=\sum_{h\in\mb Z_2} \textrm{Tr}_h\left[\frac{1+g}{2}q^{L_0-1/24}\,\bar q^{\bar L_0-1/24}\right]\,.
\end{equation}
$h=0$ corresponds, conventionally, to the untwisted sector. There are four separate terms labeled by $g$ and $h$. For $h=g=0$, this is the same contribution as in the toroidal case \eqref{PartitionToroidalLattice}, with a $1/2$ factor due to the projection. The second term of the untwisted sector, $g=1$, corresponds to the insertions of $g$ in the trace over the Hilbert space which changes the signs of the oscillator sum due to \eqref{Z2action}. The result is
\begin{equation}
 \left|\frac{\eta(\tau)}{\vartheta[^1_0](\tau,0)}\right|\,.
\end{equation}
Finally, the twisted sector has a half-integer moding and its contributions can be obtained in a similar fashion. The total result is
\begin{align}
 Z^{S^1/\mb Z_2}_{X^9}&=\frac{1}{2}Z_1(R,\tau)+\left|\frac{\eta(\tau)}{\vartheta[^1_0](\tau,0)}\right|+\left|\frac{\eta(\tau)}{\vartheta[^0_1](\tau,0)}\right|+\left|\frac{\eta(\tau)}{\vartheta[^1_1](\tau,0)}\right|\,\nonumber\\
                &=\frac{1}{2}Z_1(R,\tau)+\sum_{(h,g)\neq(0,0)}\left|\frac{\eta(\tau)}{\vartheta[^{1+h}_{1+g}](\tau,0)}\right|\,.
\end{align}
Notice that only the `toroidal' part $Z_1$ of the partition function depends on the unique modulus $R$ of $S^1$ and is, by itself, modular invariant. This is in fact also true in more complicated compactifications. In addition, the other three terms are all together modular invariant as can be seen using the identities in Appendix \ref{appendix:ModularFunctionsGenusOne}. Indeed, under modular transformations, they get exchanged. In other words, the inclusion of the twisted sector is crucial for the consistency of the theory. From the path integral point of view, the partition function can be obtained by integrating over the compact boson with the following boundary conditions around the cycles of the internal $T^2$:
\begin{align}
 X^9(\xi^1+2\pi,\xi^2)=(-)^{1-h}X^9(\xi^1,\xi^2)\,,\\
 X^9(\xi^1,\xi^2+2\pi)=(-)^{1-g}X^9(\xi^1,\xi^2)\,.
\end{align}

In general, in order to preserve the superconformal invariance on the \WS, the internal fermions must also transform under the orbifold action. In the $\mb Z_2$ case, for a complex fermion $\psi$ with spin structure $[^a_b]$ (recall (\ref{aSpinStructure}, \ref{bSpinStructure})), that is
\begin{align}
 \psi(\xi^1+2\pi,\xi^2)=(-)^{1-a-h}\psi(\xi^1,\xi^2)\,,\\
 \psi(\xi^1,\xi^2+2\pi)=(-)^{1-b-g}\psi(\xi^1,\xi^2)\,.
\end{align}
Hence, the partition function \eqref{PsiPartition} is slightly modified:
\begin{equation}
 Z^{S^1/\mb Z_2}_{\psi}[^{a,h}_{b,g}]=\frac{\vartheta[^{a+h}_{b+g}](\tau,0)}{\eta(\tau)}\,.
\end{equation}

For a general orbifold, the logic is the same. The action of the discrete group $H$ generates $|H|$ twisted sector, where $|H|$ is the order of $H$. The projection onto $H$-invariant states is realised via the insertion of
\begin{equation}
 \frac{1}{|H|}\sum_{g\in H} g\,.
\end{equation}
Including twisted sectors ensures modular invariance and, in particular, the partition function is a sum over two parameters $h,g=0,\ldots,|H|$ labelling the untwisted/twisted sectors and the projection over invariant states. They can also be viewed as twists of the boundary conditions of the \WS fields. Of course, in a specific superstring theory, only particular orbifolds are consistent, \emph{c.f.} \cite{Walton:1987bu} for more details.

Before closing this section, we apply the procedure above to write the full partition function for Heterotic string theory compactified on the orbifold $T^4/\mb Z_2\times T^2$ with gauge group $E_8\times E_8$. In this case, the orbifold breaks one of the $E_8$ factors down to $E_7\times SU(2)$, and the result is
\begin{align}
	Z=&\frac{1}{\eta^{2}\bar\eta^{18}}\frac{1}{2}\sum\limits_{h,g=0,1}\Biggr[\frac{1}{2}\sum\limits_{a,b=0,1}(-)^{a+b}
\frac{\vartheta^2[^a_b]}{\eta^2}\frac{\vartheta[^{a+h}_{b+g}]}{\eta}\frac{\vartheta[^{a-h}_{b-g}]}{\eta}\Biggr]Z^{T^4/\mb Z_2}[^h_g]~Z_{2}(T,U)~\nonumber\\
          &\times Z_{E_8}~Z_{E_7\times SU(2)}~,\label{HetPartitionFunction}
\end{align}
where $T,U$ parametrise the moduli of the $T^2$ and depend on its radii. The partition function of the orbifold $T^4/\mb Z_2$ depends, as in the previous case, on the parameters $h,g$. In the $h=g=0$ sector, it depends on the $T^4$ moduli parametrised by $G,B$ (the $1/2$ averaging factor is already taken into account):
\begin{equation}
 Z^{T^4/\mb Z_2}[^0_0]=Z_4(G,B)\,,
\end{equation}
whereas in the sectors $(h,g)\neq (0,0)$ it is independent of the moduli:
\begin{equation}
 Z^{T^4/\mb Z_2}[^h_g]=\left|\frac{2\eta(\tau)}{\vartheta[^{1+h}_{1+g}](\tau,0)}\right|^4\,.
\end{equation}
The expressions of the gauge lattices $Z_{E_8}$ and $Z_{E_7\times SU(2)}$ are given in Section \ref{Sec:Hetamps}. Notice that in \eqref{HetPartitionFunction}, we have cancelled the ghost and superghost contributions. In the left-moving (supersymmetric) sector, every complex fermion contributes $\frac{\vartheta}{\eta}$, and the bosons in the \ST directions contribute $\frac{1}{\eta}$ each. Including the right-moving fields, one verifies that \eqref{HetPartitionFunction} contains all the degrees of freedom of the theory.

\subsection*{Calabi-Yau compactifications}

It is interesting to analyse the constraints arising from imposing $\N=1$ supersymmetry in the four-dimensional \ST. This is relevant in a realistic model since it is believed that supersymmetry might arise at low energies, \emph{e.g.} at the TeV scale. First of all, the target space must allow the existence of spinors. Secondly, the vacuum has to preserve supersymmetry. Let us consider the ($D=10$, $\N=1$) supergravity ($\alpha'\rightarrow0$) limit of (Heterotic) superstring theory. In order for the theory to preserve a fraction of supersymmetry, it is sufficient that the supersymmetry variations of the fermionic fields vanish. The reason for this is that if the theory is Lorentz invariant, then the vevs of vector and Fermi fields vanish, and the ones of the scalars can be at most constant:
\begin{equation}
 \langle A_{\mu}\rangle=\langle \psi^{\alpha}\rangle=\partial_{\mu}\langle\phi\rangle=0\,.
\end{equation}
Using the $\N=1$ supersymmetry transformations, this implies that a configuration preserves supersymmetry if $\delta\langle\psi^{\alpha}\rangle=0$, meaning that the supersymmetry variations of the fermionic fields vanish classically. For the gravitino, dilatino and gaugino these are
\begin{alignat}{3}
 &\delta_{\epsilon}\,\Psi_{M}\,&=&&&D_{M}\,\epsilon\,,\label{SusyGravitino}\\
 &\delta_{\epsilon}\,\lambda&=&\Bigr(&&-\frac{1}{2}\Gamma^{M}\partial_{M}\phi+\frac{1}{24}\Gamma^{MNP}H_{MNP}\Bigr)\epsilon\,,\label{SusyDilatino}\\
 &\delta_{\epsilon}\,\chi&=&&&\Gamma^{MN}F_{MN}\,\epsilon\,,\label{SusyGaugino}
\end{alignat}
where the spinor indices have not been displayed, $H$ is the three-form flux, $F$ is the gauge field strength and $D_M$ is the covariant derivative (with respect to the spin connection and $H$). In addition, assume that the target space factorises as
\begin{equation}
 M_4\times M_6\,,
\end{equation}
with $M_4$ being maximally symmetric, i.e. $R_{\mu\nu\rho\sigma}=\frac{R}{12}(g_{\mu\rho}g_{\nu\sigma}-g_{\mu\sigma}g_{\nu\rho})$. The ten-dimensional index is accordingly decomposed, $M={\mu,m}$. The possible non-vanishing bosonic fields are the dilaton $\phi$, $H_{mnp}$ and $F_{mn}$. Moreover, we choose the dilaton to be constant and set the \emph{torsion} $H_{mnp}$ to zero\footnote{This is not the only possibility, but is the one compatible with the ansatz for the cohomology constraint \eqref{CohomologyConstraint} in Heterotic, see below.}. In this case, the dilatino variation \eqref{SusyDilatino} vanishes and the covariant derivative $D_m$ only contains the spin connection. There are also additional constraints on the gauge field strength from \eqref{SusyGaugino}, but we do not consider them here for simplicity.  Hence, in order for supersymmetry to be preserved by the vacuum, \eqref{SusyGravitino} implies that $\epsilon$ is a covariantly constant (or Killing) spinor. The existence of such a spinor is very restrictive. Applying \eqref{SusyGravitino} twice yields
\begin{equation}\label{SpinorRiemann}
 [D_M,D_N]\epsilon\equiv R_{MNPQ}\Gamma^{PQ}\epsilon=0\,,
\end{equation}
with $R_{MNPQ}$ being the Riemann tensor viewed as a Lie algebra-valued two-form. It generates the holonomy group of the target space\footnote{In fact, if the space is not simply connected then it generates only the identity component of the holonomy group.}. The condition \eqref{SpinorRiemann} implies that $M_4$ is flat. However, the internal space $M_6$ need not be flat. Recall that the gravitino is a $\mathbf{16}$ spinor representation of the Lorentz (or holonomy) group $SO(1,9)\rightarrow SO(1,3)\times SO(6)$ and is decomposed as $(\mathbf{2},\mathbf{4})\oplus(\bar{\mathbf{2}},\bar{\mathbf{4}})$. Furthermore, the $\mathbf{4}$ of $SO(6)\cong SU(4)$ decomposes under $SU(3)$ as $\mathbf{4}=\mathbf{1}\oplus\mathbf{3}$. Notice that only the singlet part of the latter which is invariant under $SU(3)$ can give rise to a Killing spinor which can be group-theoretically written as $(\mathbf{2},\mathbf{1})\oplus(\bar{\mathbf{2}},\mathbf{1})$. Therefore, a covariantly constant spinor exists if and only if the holonomy group is contained in $SU(3)$. In fact, from \eqref{SpinorRiemann}, one can show that this is equivalent to Ricci-flatness, \emph{i.e.} $R_{mn}=0$, of the internal space. The covariantly constant spinor can be used to construct a K\"ahler form. A compact $2n$-dimensional K\"ahler manifold with holonomy in $SU(n)$ is called a \emph{Calabi-Yau} $n$-fold (CY$_n$).

In the previous discussion, we have restricted our attention to the lowest order contributions in $\alpha'$. In order for the full sigma-model to be well-defined, one has to study the fate of $\alpha'$ corrections in the supersymmetry transformations and the $\beta$-functions. In particular, the $\beta$-functions receive higher order corrections and it has been shown in \cite{Nemeschansky:1986yx} that, for a \CY, they can be made to vanish so that the underlying sigma-model is conformally invariant (to all orders in superstring perturbation theory) but its metric now differs from the Ricci-flat one\footnote{In some cases, the metric can be made Ricci-flat by a field redefinition.} (though the internal space remains K\"ahler). Moreover, the Ricci-flat condition gets modified. For instance, the lowest order correction is of the form $\alpha'^3 R^4$, $R$ being the Ricci curvature. This additional term, however, does not affect the cohomology class of the Ricci form (see also \cite{Grisaru:1986px,Grisaru:1986dk,Freeman:1986br,Lu:2003gp}).

We are interested in studying the massless spectrum of superstring theory compactified on a \CY which, in general, is composed of the graviton and a set of $p$-form potentials $C_p$ satisfying $\Delta_{10} C_p=0$ with $\Delta_{10}$ being the ten-dimensional Laplacian. The latter can be split as $\Delta_4+\Delta_6$ because the metric does. From the four-dimensional point of view, the massless spectrum is thus classified by the zero modes of $\Delta_6$ (a non-zero eigenvalue would give rise to a mass term). They are given by the cohomologies of the internal space whose dimensions define the \emph{Hodge numbers}\footnote{The dimension $h^{p,q}$ of the Dolbeault cohomology $H^{p,q}_{\bar\partial}$ is related to the dimension $b_p$ of the de Rham cohomology group $H^{p}$ through $b_p=\sum_{k=0}^p h^{k,p-k}$. $b_p$ are the \emph{Betti numbers}.} $h^{p,q}$. These topological numbers give an incomplete characterisation of the Calabi-Yau manifold since different CYs can have the same Hodge numbers. In addition, they are not all independent, and are usually displayed in the Hodge diamond. For a \CY ($p,q=0,\ldots,3$), that is

\begin{center}
\begin{tabular}[h]{ccccccc}
 & & & $h^{3,3}$ & & &\\
 & & $h^{3,2}$ & & $h^{2,3}$ & &\\
 & $h^{3,1}$ & & $h^{2,2}$ & & $h^{1,3}$ &\\
 $h^{3,0}$ & & $h^{2,1}$ & & $h^{1,2}$ & & $h^{0,3}$\,.\\
 & $h^{2,0}$ & & $h^{1,1}$ & & $h^{0,2}$ &\\
 & & $h^{1,0}$ & & $h^{0,1}$ & &\\
 & & & $h^{0,0}$ & & &
\end{tabular}
\end{center}

\noindent By complex conjugation, $h^{p,q}=h^{q,p}$ and Poincar\'e duality implies $h^{p,q}=h^{3-p,3-q}$. One can perform a thorough analysis of the cohomologies of the \CY showing that only $h^{1,1}$ and $h^{2,1}$ are undetermined:

\begin{center}
\begin{tabular}[h]{ccccccc}
 & & & 1 & & &\\
 & & 0 & & 0 & &\\
 & 0 & & $h^{1,1}$ & & 0 &\\
 1 & & $h^{2,1}$ & & $h^{2,1}$ & & 1\,.\\
 & 0 & & $h^{1,1}$ & & 0 &\\
 & & 0 & & 0 & &\\
 & & & 1 & & &
\end{tabular}
\end{center}

\noindent The Euler characteristic $\chi$ can be calculated as the alternating sum of the Betti numbers. In terms of the Hodge numbers, $\chi=2(h^{1,1}-h^{2,1})$. In addition, notice that a \CY compactification cannot support continuous isometries because $h^{1,0}=0$.

A CY can equivalently be defined as a compact K\"ahler manifold with vanishing first Chern class. By the Calabi-Yau theorem proven by Yau, this is equivalent to the definition above using Ricci flatness. More precisely, given a compact K\"ahler manifold, any representative of the first Chern class is the Ricci form of a unique K\"ahler metric.  Even though no explicit CY metric has ever been constructed in the compact case, many of its properties can be used to study the physics of the superstring target space. Moreover, in some cases, one can approximate the CY by an orbifold. This is referred to as the \emph{orbifold limit} and is discussed at the end of this section.

We now give some examples of CY manifolds. For $n=1$, the only compact CY is the two-torus $T^2$ (we do not consider non-compact CYs). In two complex dimensions, the only CYs are $T^4$ and $K3$ surfaces. The Hodge diamond of the latter is
\begin{center}
\begin{tabular}[h]{ccccccc}
 & & 1 & &\\
 & 0 & & 0 &\\
 1 & & 20 & & 1\,.\\
 & 0 & & 0 &\\
 & & 1 & &
\end{tabular}
\end{center}

\noindent In complex dimensions higher than two, many more examples are known but the complete classification is an open problem. In fact, it is not even known whether the number of \CY is finite. The holonomy group of a CY is a subgroup of $SU(n)$. If we restrict to spaces having $SU(n)$ holonomy only, then the torii having trivial holonomies are excluded, and this is the condition to have the minimal amount of preserved supersymmetry in four dimensions ($\N=1$ for Heterotic and $\N=(1,1)$ for Type II).

As in the simple toroidal case, CY compactifications give rise to a number of moduli that parametrise smooth deformations of a given CY or, equivalently, of its shape and size. In this sense, a Hodge diamond defines a continuous family of CYs, but we refer to it simply as CY keeping in mind the abuse of terminology. Consider first deforming the \CY metric without spoiling the Ricci flatness of the space. That is
\begin{equation}
 R_{MN}(g+\delta g)=0=R_{MN}(g)\,.
\end{equation}
The internal index is decomposed in terms of $SU(3)$ indices, $m=(i,\bar\im)$. Using the gauge fixing condition $D^{M}\delta g_{MN}=0$ in order to eliminate deformations parametrising coordinate transformations, the equation for the mixed (holomorphic-anti-holomorphic) indices reduces to
\begin{align}
 (\Delta \delta g)_{i\bar\jm}=0\,,\label{KaehlerDef}
\end{align}
meaning that $\delta g_{i\bar\jm}$ is a harmonic $(1,1)$-form parametrising changes in the K\"ahler structure. There are $h^{1,1}$ such forms called \emph{K\"ahler moduli} and they have to be chosen such that the metric is still positive definite. As for the other indices (holomorphic-holomorphic or anti-holomorphic-anti-holomorphic), the metric vanishes but one can still deform it by changing the complex structure (because a holomorphic change of coordinates would be irrelevant). These deformations are parametrised by the $h^{2,1}$ $(2,1)$-forms and are referred to as \emph{complex structure moduli}. Finally, deforming the B-field and its internal part $B_{i\bar\jm}$ yields $h^{1,1}$ moduli which, combined with the $h^{1,1}$ metric moduli form the \emph{complexified K\"ahler moduli} whose real part corresponds to $\delta B_{i\bar\jm}$ and imaginary part to $\delta g_{i\bar\jm}$. The moduli space of the CY compactifications is, by definition, parametrised by the scalar fields arising from the compactifications. Since K\"ahler and complex structure moduli do not mix (at least in the $\N=1$ case), the moduli space is factorised\footnote{It can be shown that the vector multiplet moduli space is special K\"ahler and the hypermultiplet one is quaternionic K\"ahler (but not K\"ahler).}:
\begin{equation}
 \mc M_{CY}=\mc M_{\mb C}\times\mc M_{K}\,.
\end{equation}

We now discuss the implications of \CY compactifications on the superstring spectrum. First, consider Type IIA in which the massless spectrum contains the graviton $G_{MN}$, the B-field $B_{MN}$, the dilaton $\phi$, a vector $C_M$ and a three-form potential $C_{MNP}$ (plus their fermionic partners). The reduction on a \CY using $SU(3)$ covariant indices $(i,\bar\im)$ yields
\begin{alignat}{3}
 &G_{MN} &\rightarrow&\, G_{\mu\nu}\,,\,G_{ij}\,,\, G_{i\bar\jm}\,,\\
 &B_{MN} &\rightarrow&\, B_{\mu\nu}\,,\,B_{i\bar\jm}\,,\\
 &C_M    &\rightarrow&\, C_{\mu}\,,\\
 &C_{MNP}\,&\rightarrow&\, C_{\mu\nu\rho}\,,\,C_{\mu i\bar\jm}\,,\,C_{ij\bar k}\,,\,C_{ijk}\,.
\end{alignat}
Notice that $C_M$ does not give rise to an internal field $C_i$ since it can be viewed as a $(1,0)$-form and we have seen that $h^{1,0}=0$. The scalar field $C_0$ dual to $B_{\mu\nu}$ and the dilaton combine into the axion-dilaton complex field $S=C_0+ie^{-\phi}$. Taking into account the gravitini and dilatini, we obtain a gravitational multiplet ($G_{\mu\nu},\,C_{\mu}$), $h^{1,1}$ vector multiplets ($C_{\mu i\bar\jm},\,G_{i\bar\jm},\,B_{i\bar\jm}$), $h^{2,1}$ hypermultiplets ($C_{ij\bar k},\,G_{ij}$) and the dilaton hypermultiplet ($\phi,\,B_{\mu\nu},\,C_{ijk}$). The latter is singled out because it corresponds to the string universal $S$ multiplet containing the dilaton. The vector field of the gravitational multiplet is the \emph{graviphoton}. All together, we have the gravitational multiplet, $h^{1,1}$ vector multiplets and $h^{2,1}+1$ hypermultiplets. The complex structure moduli belong to the hypermultiplets whereas the K\"ahler moduli are part of the vector multiplets. Therefore, the corresponding moduli spaces have real dimensions $4(h^{2,1}+1)$ and $2h^{1,1}$ respectively\footnote{The additional 2 in the former comes from the axion-dilaton.}.

In Type IIB, the massless spectrum contains in the R-R sector, instead, a zero-form $C$, a two-form $C_{MN}$ and a self-dual four-form $C_{MNPQ}$, and the reduction leads to
\begin{alignat}{3}
 &G_{MN}  &\rightarrow&\, G_{\mu\nu}\,,\,G_{ij}\,,\, G_{i\bar\jm}\,,\\
 &B_{MN}  &\rightarrow&\, B_{\mu\nu}\,,\,B_{i\bar\jm}\,,\\
 &C_{MN}  &\rightarrow&\, C_{\mu\nu}\,,\,C_{i\bar\jm}\,,\\
 &C_{MNPQ}\,&\rightarrow&\, C_{\mu\nu i\bar\jm}\,,\,C_{\mu ijk}\,,\,C_{\mu ij\bar k}\,,\,C_{ijkl}\,.
\end{alignat}
The resulting four-dimensional structure is made of the gravitational multiplet $(G_{\mu\nu},\,C_{\mu ijk})$, $h^{2,1}$ vector multiplets $(C_{\mu ij\bar k},\,G_{ij})$, $h^{1,1}$ hypermultiplets ($C_{\mu\nu i\bar\jm}$, $G_{i\bar\jm}$, $B_{i\bar\jm}$, $C_{i\bar\jm}$) and the universal dilaton multiplet. In total, there are $h^{2,1}$ vector multiplets and $h^{1,1}+1$ hypermultiplets. The complex structure moduli belong to the vector multiplets and the K\"ahler moduli to the hypermultiplets. Thus, the corresponding moduli spaces have dimensions $2h^{2,1}$ and $4(h^{1,1}+1)$ respectively.

Notice that the dilaton, which controls string perturbation theory, lies in a hypermultiplet in both Type II theories. This means that the hypermultiplet moduli space receives perturbative and non-perturbative corrections, whereas the vector multiplet moduli space is protected due to $\N=2$ supergravity (or the non-mixing between vector multiplets and hypermultiplets). We come back to this point in subsequent chapters.

Let us now turn to the Heterotic superstring and, for definiteness, we focus on the $E_8\times E_8$ case. The massless spectrum contains a three-form flux $H$ such that
\begin{equation}\label{CohomologyConstraint}
 dH=\textrm{tr}\,\mc R\wedge\mc R-\textrm{tr}F\wedge F\,,
\end{equation}
with $\mc R$ being the \ST curvature and $F$ the field strength of the gauge group connection. Since $dH$ is a closed four-form, both $\textrm{tr}\,\mc R\wedge\mc R$ and $\textrm{tr}F\wedge F$ should represent the same cohomology class. This is known as the \emph{cohomology constraint}. One way to achieve this is to embed the spin connection in the gauge group as follows. Recall that $\mc R$ is an SU(3) (holonomy) Lie algebra-valued two-form and $F$ is, instead, $E_8\times E_8$-valued. In order to satisfy the cohomology constraint, one can impose to have $F$ taking values in the internal space because $\textrm{tr}\,\mc R\wedge\mc R$ is non-trivial. Choose an $SU(3)$ subgroup of one of the $E_8$ factors and give the corresponding field strength the same background value as the curvature while setting the others (with respect to the rest of the gauge group) to zero. In other words, we identify the potential $A$ of $F$ with the spin connection (whose curvature is $\mc R$). Thus, one of the $E_8$'s is broken down to $E_6\times SU(3)$\footnote{For the $SU(2)$ holonomy case, \emph{e.g.} when the internal space is $K3\times T^2$, $E_8$ is broken down to $E_7\times SU(2)$, see \eqref{HetPartitionFunction}.} and the unbroken gauge group is $E_6\times E_8$. Under this breaking, the \textbf{248} adjoint representation of $E_8$ decomposes as
\begin{equation}
 \mathbf{248}=(\mathbf{78},\mathbf{1})\oplus(\mathbf{1},\mathbf{8})\oplus(\mathbf{27},\mathbf{3})\oplus(\overline{\mathbf{27}},\overline{\mathbf{3}})\,.
\end{equation}
On top of the gravitational multiplet and vector multiplets in the adjoint of $E_6\times E_8$, there are $h^{1,1}$ K\"ahler moduli and $h^{2,1}$ complex structure moduli all belonging to $\N=1$ chiral multiplets. One way to see this is to perform the reduction of the gauge potential
\begin{equation}
 A_{M}\rightarrow A_{\mu}\,,A_i\,,A_{\bar\im}\,,
\end{equation}
where $i$ is an $SU(3)$ holonomy index. The embedding of the spin connection into one of the $E_8$'s means that $A_i$ can carry an index with respect to the $(\overline{\mathbf{27}},\overline{\mathbf{3}})$, with $\mathbf{3}\in SU(3)$ of the gauge group. Hence, $A_i\equiv A_{i,\bar\alpha\bar\jm}$ can be viewed as a $(1,1)$-form taking values in the $\overline{\mathbf{27}}$ of $E_6$ and has $h^{1,1}$ zero modes corresponding to K\"ahler moduli. Similarly, taking $A_i$ to have an index in the $(\mathbf{27},\mathbf{3})$ gives rise to $h^{2,1}$ complex structure moduli. By supersymmetry, these moduli have fermions in the $\mathbf{27}$ that are chiral.

As mentioned above, it is not known how to explicitly construct a compact CY metric. However, the CY moduli space contains singularities at which the classical effective description breaks down due to the appearance of (non-perturbative) massless states. In particular, there exist points in the moduli space at which certain CYs can be described by orbifolds. Consider for instance a $T^4/\mb Z_2$ orbifold for which the $\mb Z_2$ action on the $T^4$ coordinates $z^i$ is
\begin{equation}
 z^i\rightarrow -z^i\,.
\end{equation}
This orbifold has sixteen singularities due to the sixteen fixed points of the $\mb Z_2$ action, given by
\begin{equation}
\left[0,\frac{1}{2},\frac{i}{2},\frac{1+i}{2}\right]\otimes\left[0,\frac{1}{2},\frac{i}{2},\frac{1+i}{2}\right]\,.
\end{equation}
In order to get a smooth space, one can \emph{blow up} the singularities by replacing each of them with a smooth space \cite{Gibbons:1979xn}. More precisely, one cuts away a ball around each singularity whose boundary is $S^3/\mb Z_2\cong\mb RP^3$ and replace it with an Eguchi-Hanson space \cite{Eguchi:1978xp}. The metric on the latter must be constructed in such a way that it matches across the sixteen boundaries. This can be done \emph{approximately} by using the flat metric on $T^4$ and the Eguchi-Hanson metric on the $\mb RP^3$'s. The Eguchi-Hanson metric contains an additional parameter, say $a$, such that when $a\rightarrow0$ one recovers the orbifold. One can show that the resulting smooth space has the topology of K3, but not its geometry. For this to be possible, one needs to smoothen up the metric everywhere and this is not known yet. On the other hand, K3 contains sixteen two-cycles \cite{Eguchi:1980jx}. By going to a point in the moduli space in which the size of these cycles goes to zero, one obtains a singular space having sixteen singularities, which is precisely $T^4/\mb Z_2$. Of course, one can take different limits of K3 corresponding to $T^4/G$, $G$ being a discrete subgroup of $SU(2)$, and the construction above can be generalised \cite{Walton:1987bu}. For $G=\mb Z_N$, $N$ can take the values 2, 3, 4 and 6.

In practice, one is often interested in studying properties of superstring theories that do not depend on the specific details of the internal manifold. In this case, one can take a singular limit of the CY in which it is easier to perform the analysis. For instance, it is in the orbifold limit described above that the partition function \eqref{HetPartitionFunction} is derived.

\section{Dualities}\label{sec:dualities}

So far we have described five different perturbative consistent string vacua in ten dimensions and some compactifications thereof. It turns out that they are not all independent. Instead, they seem to emerge as distinct vacua of the same underlying eleven-dimensional theory, usually referred to as \emph{M-theory}, which are connected through perturbative and non-perturbative string dualities. The precise description of M-theory is, however, unknown.

Beyond their conceptual interest, string dualities are of practical use since, in some cases, they connect weak and strong coupling regimes. For instance, results obtained in a weakly coupled theory can be reinterpreted in terms of the degrees of freedom of the dual, strongly coupled theory, if the latter exists. Of course, one must have proved in advance such a duality and this is a difficult task, since strongly coupled theories are hard to solve. Nevertheless, supersymmetry comes to our rescue as some quantities such as BPS states and couplings turn out to be protected against quantum corrections. In this sense, for lack of a full proof, the knowledge of BPS spectra is crucial to obtain strong arguments in favour of string dualities.

In fact, we have already encountered one such duality, namely T-duality, which is perturbative because it does not transform the string coupling. It relates, for example, Type IIA/B theories as a simple realisation of \emph{mirror symmetry} \cite{Hori:2003ic}. In what follows, we briefly present the dualities relating the aforementioned string vacua after compactification, together with some of their strong coupling limits.

Let us first focus on the ten dimensional theories. In Type IIB, the axion-dilaton field enjoys an $SL(2,\mb Z)$ symmetry\footnote{Together with T-duality in the compactified theory, they form the \emph{U-duality} group.} in the quantum theory, which includes S-duality (strong-weak symmetry as the inversion of the string coupling). In this sense, the full Type IIB theory (including its strong coupling regime) is $SL(2,\mb Z)$ self-dual. On the other hand, the perturbative Type IIA theory can be reached as an $S^1$ compactification of M-theory and the string coupling is given by the (small) $S^1$ radius, $R_{S^1}=l_s\,g_s$. If one considers, instead, a $\mb Z_2$ orbifold of the M-theory circle \cite{Horava:1995qa}, then the resulting theory is the $E_8\times E_8$ Heterotic string, with the strong/weak coupling regime being controlled by the size of the circle. Now T-duality relates the $E_8\times E_8$ Heterotic theory to the $SO(32)$ one\footnote{This can be seen upon compactification of both theories on a circle.} from which, by S-duality, one ends up in the Type I superstring \cite{Polchinski:1995df}. This web of dualities is a strong hint towards the uniqueness of string theory and offers a framework for exploring a number of its fundamental aspects.

When studying topological amplitudes in string theory, we make extensive use of four-dimensional string dualities relating Type II, Heterotic and Type I superstrings.

\subsection*{Type II/Heterotic duality}

Heterotic string theory compactified on $T^6$ is dual to Type IIA on $K3\times T^2$. Indeed, one can show that both compactifications share the same moduli spaces. In particular, the duality map exchanges the axion-dilaton $S$ and the K\"ahler structure $T$ of the $T^2$, and, for instance, the string coupling in Heterotic becomes the volume of $T^2$ in Type IIA. Hence, T-duality in Type II becomes S-duality in Heterotic.

In fact, this duality can be extended to the case where the Type II theory is compactified on a \CY which is a $K3$ fibration\footnote{A $K3$ fibration is a \CY which is locally a direct product of $K3$ and a two-dimensional surface.}. The dual theory is Heterotic on $K3\times T^2$, as it has been argued in many examples \cite{Kachru:1995wm}. As we discuss in Section \ref{ch:TopoAmp}, the study of BPS-saturated couplings provides a non-trivial check of this duality \cite{Antoniadis:1995zn}. An important feature in this context is that the dilaton, in Heterotic, belongs to a vector multiplet, whereas in Type II, it sits in a hypermultiplet. This leads to non-renormalisation theorems because of a non-mixing between hyper- and vector multiplets in $\N=2$ supergravity.

\subsection*{Heterotic/Type I duality}

The Heterotic and Type I superstrings, upon compactification on $K3\times T^2$, are dual. Even though the ten-dimensional duality is non-perturbative, in four dimensions there are weakly coupled regimes in both theories. In addition, the axion-dilatons are mapped to one another \cite{Antoniadis:1997gu,Antoniadis:1997nz}. On the other hand, the complex structure modulus U of $T^2$ is unchanged and the K\"ahler modulus T in Heterotic is mapped to another dilaton-like field denoted $S'$. Indeed, the presence of two dilatons in Type I is not surprising since the theory contains D9- and D5-branes whose coupling constants are given by the imaginary part of two different fields, namely $S$ and $S'$ respectively. This dictionary is summarised in the table below.

\begin{table}[ht]
 \begin{center}
  \begin{tabular}{cc}
  \toprule
  Heterotic &Type I\\\midrule
  S&S\\
  T&S'\\
  U&U\\\bottomrule
  \end{tabular}
 \end{center}
\caption{Mapping of the universal fields under Heterotic/Type I duality in four dimensions.}
\label{tb:HetTypeI}
\end{table}

On the Type I side, the four dimensional dilaton contains hyper- and vector multiplet components so that both moduli spaces receive quantum corrections. This is in contrast with the dual Heterotic theory where the dilaton belongs to a vector multiplet and the hypermultiplet moduli space is protected. We come back to these issues later.


\part{N=2 Topological String Theory and Gauge Theory: an Overview}
\chapter*{}
\vspace*{\fill}
In this section, we present a short review of $\mc N=2$ topological string theory (TST). First of all, we discuss general aspects of topological sigma models which are the building blocks for the definition of topological string theory once we couple the theory to two-dimensional gravity. Secondly, after defining the genus $g$ topological string partition function and its holomorphicity properties, we introduce a series of gravitational amplitudes in the `physical' string theory which turn out to reproduce the partition function of the `twisted' theory. In this sense, the topological string arises as a sub-sector of the superstring.

More specifically, we calculate the coupling, at genus $g$ in Type II, of two gravitons and $2g-2$ graviphotons and show that it reproduces the genus $g$ TST partition function. These gravitational couplings satisfy a non-renormalisation theorem due to their BPS nature. Moreover, it is shown that in the dual Heterotic theory they start receiving contributions at one-loop, opening up an interesting framework for explicit calculations.

On the other hand, the field theory limit of these gravitational couplings lead to corrections to the Seiberg-Witten prepotential encoded in the partition function of the $\N=2$ gauge theory in the $\Omega$-background. We briefly review the construction of the latter and show how the gauge instantons can be realised in string theory using particular D-brane bound states.
\vspace*{\fill}
\chapter{Topological Field Theories}
\label{ch:TFT}
Topological field theories (TFT) have been of extensive interest in the past few decades from the mathematical as well as the physical point of view. In fact, these theories have brought physicists and mathematicians to a common playground, though for different motivations.

From the mathematical point of view, a topological theory is important in order to count certain quantities called `topological invariants'\footnote{A topological invariant is a property of a topological space which is invariant under homeomorphisms.}. The word `topological' refers to an object that does not depend on the particular characteristics or \emph{moduli} of the space it lives in. In order to distinguish topologically inequivalent spaces, it is crucial to have a complete classification of all their topological invariants. In general, this is a tedious task but becomes simpler in the case of compact and connected spaces. In particular, for spaces of dimension lower than three, this is possible. For instance, in the two-dimensional case (the zero- and one-dimensional cases are rather trivial), there exists a complete classification: topologically inequivalent compact and connected surfaces have a different number of boundaries (b), handles (h) or crosscaps (c) (i.e. different orientability). An example of a topological invariant for two dimensional surfaces is the Euler characteristic $\chi=2-2h-b-c$. Conversely, the knowledge of the full set of topological invariants is important to decide whether two spaces are homeomorphic or not. In dimensions higher than two, this is an open question, and it is not known whether the set of topological invariants is even countable.

On the other hand, modern physics has been mainly based on the study of quantum field theory (QFT) which has rich yet very complex dynamics. In order to have a better understanding of the latter, one is led to consider more simplified models, and TFTs can be viewed as toy models for real, physical theories. In fact, TFTs are better to be thought of as subsectors of physical field theories so that one expects that they also calculate physical quantities that are inherited from the full theory, thus shedding light on complex aspects of the latter. This is in the core of the present work and is discussed in detail in the subsequent sections. 

In this chapter, we review the basic elements of topological field theories, in particular the ones underlying topological string theory (TST). We first present the main features of a TFT and then illustrate them with two classes of TFT: Chern-Simons theory in three dimensions and cohomological field theories (CohFT). The latter contain the salient features needed for the construction of TSTs. We only mention the main aspects and many of the technical details and mathematical proofs are omitted, as these can be found in the various references we point to.

\section{Generalities}

Consider a generic QFT defined on a certain background given by a metric, coupling constants, complex structures, etc. This theory has a set of \emph{observables} or physical operators $\mathcal O_i$ inserted at positions $x_i$, whose correlation functions are the quantities of interest:
$$\left<\mathcal{O}_1(x_1)\,\cdots\mathcal{O}_n(x_n)\right>\,,$$
and implicitly depend on the background of the theory. Physically speaking, this theory is said to be topological if all the above correlation functions of observables are independent of the choice of a background metric\footnote{In contrast with the case of TST, the metric is not integrated over.}. Of course, the background in which they are calculated may involve other quantities on which they could still depend. The theories we consider usually have general coordinate invariance. One can then show that in a topological theory, the physical correlation functions do not depend on the insertion points of the physical operators.

The class of TFTs is of course very special and one can split it into two categories \cite{Birmingham:1991ty}:
\begin{enumerate}
 \item Schwarz-type theories: defined without the use of a metric in the Lagrangian nor the observables,
 \item Witten-type theories: defined with a metric.
\end{enumerate}
In what follows, we present an example of each category.

\section{Chern-Simons Theory}

Perhaps the most direct way of obtaining a TFT is to define a theory without the use of a metric in its Lagrangian. Of course, this only guarantees that the theory is \emph{classically} topological and one has to ensure the absence of anomalies in order to preserve this property at the quantum level. Chern-Simons theory in three dimensions, on a generic manifold $\mathcal M$ and with gauge group $G$, is such an example. The Chern-Simons action is given by

\begin{equation}
 S=\frac{k}{4\pi}\int_{\mathcal M}\textrm{Tr}\left(A\wedge dA+\frac{2}{3}A\wedge A\wedge A\right)\,,
\end{equation}
where $k$ is the coupling constant\footnote{Gauge invariance constrains $k$ to be an integer.} and $A$ a gauge connection. In order to show that this theory is topological, one should verify that the topological nature is not spoiled by possible anomalies of the measure in the path integral. This is quite subtle and Witten \cite{Witten:1988hf} showed that there are no anomalies at the quantum level\footnote{More precisely, the partition function counts a topological invariant of $\mathcal M$ that also depends on a choice of trivialisation of the tangent bundle $T\mc M\oplus T\mc M$.}.

The observables of this theory are the Wilson loop operators $W_{\gamma}(A)$ defined as the trace of the holonomy of $A$ around a closed loop $\gamma\subset\mathcal M$:

\begin{equation}
 W_{\gamma}(A)=\textrm{Tr}\,P\, \exp\oint_{\gamma} A\,.
\end{equation}
These operators are gauge invariant. As mentioned above, the topological nature of the theory implies that the correlators of the latter are independent of their location in $\mathcal M$. In other words, one can continuously deform the closed loops without changing the correlation functions. Therefore, these topological properties count \emph{knot invariants}, and are given in terms of polynomials depending on the coupling constant $k$ and the gauge group. The simplest example is the Jones polynomial which arises in the case of $SU(2)$ gauge group.

\section{Cohomological Field Theories}
\label{sec:CohFT}
The TFTs of Witten type or CohFT are even richer in terms of topological invariants and are relevant for the study of TST. As its name suggests, a CohFT is based on the existence of a fermionic symmetry $\delta$ represented through the Noether procedure by a nilpotent operator $Q$:

$$Q^2=0.$$
We also assume that the vacuum is preserved by this symmetry. The latter acts on bosonic and fermionic fields as a commutator and anti-commutator with $Q$ respectively: 
\begin{align}
 \delta\mathcal O_B &=[Q,\mathcal O_B]\,,\nonumber\\
 \delta\mathcal O_F &=\{Q,\mathcal O_F\}\,,
\end{align}
even though we generically use the curly brackets $\{,\}$. $Q$ is to be thought of as the BRST-charge of ordinary QFT or the supercharge of the supersymmetry algebra\footnote{When one has more than one such charge, an appropriate linear combination has to be taken.}. The observables of the theory are defined as the $Q$-closed operators:

\begin{equation}
 \{Q,\mathcal O_{\textrm{phys}}\}=0\,.
\label{PhysCond}
\end{equation}
In fact, the physical states are classified by the cohomology of $\delta$ since any correlation function of physical operators $\mathcal O_i$ is invariant under

\begin{equation}
 \mathcal{O}_i\longrightarrow\mathcal{O}_i+\{Q,\Omega\}\,.
\end{equation}
This is due to the fact that the vacuum is preserved by $\delta$ and can be seen as follows:

\begin{align}
\langle0\lvert\cdots\{Q,\Omega\}\cdots\rvert0\rangle&=\langle0\lvert\cdots(Q\Omega-\Omega Q)\cdots\rvert0\rangle\nonumber\\
                                                    &=\langle0\lvert Q\cdots\Omega\rvert0\rangle-\langle0\lvert\cdots\Omega\cdots Q\rvert0\rangle\nonumber\\
                                                    &=0.
\end{align}
Here we have used the physical condition \eqref{PhysCond} to pass the charge over the physical operators (there are possible additional signs that can arise in the second step when $Q$ passes the physical operators, but this does not affect the final result). In a CohFT, given a local operator, one can construct a series of non-local ones using the so-called \emph{descent equations}. More precisely, starting from an observable $\mc{O}^{(0)}$ lying in the cohomology of $\delta$ and a series of $n$-forms $\mc O^{(n)}$ such that

\begin{equation}
\label{DescEq}
 d\mc O^{(n)}=\delta\mc O^{(n+1)}\,,
\end{equation}
for $n\in[\![1,m]\!]$, $m$ being the dimension of $\mc M$\footnote{$\mc M$ should be thought of as space-time when the TFT is described by a sigma-model, see next section.}, then one can define a set of non-local observables
\begin{equation}
 Y_{\gamma_n}=\int_{\gamma_n} \mc O^{(n)}\,,
\end{equation}
$\gamma_n$ being an element of the $n^{\textrm{th}}$ homology group of $\mc M$. It is easy to show that $Y_{\gamma_n}$ is indeed an observable of the topological theory:
\begin{equation}
 \delta Y_{\gamma_n}=\int_{\gamma_n}\delta\mc O^{(n)}=\int_{\gamma_n}d\mc O^{(n-1)}=\int_{\partial\gamma_n}\mc O^{(n-1)}=0\,.
\end{equation}
Therefore, provided we construct a solution for the descent equations \eqref{DescEq}, for every `scalar' physical observable there exists a number of non-local physical operators that are in one-to-one correspondence with the homology classes of $\mc M$. Finally, we require the energy-momentum tensor to be $Q$-exact:
\begin{equation}
 T=\frac{\delta S}{\delta h}=\{Q,G\}\,,
\label{EMcond}
\end{equation}
for some operator $G$, a background metric $h$ and an action $S$. Recall that the integrals of the energy-momentum tensor over space-like hypersurfaces are conserved charges that must commute with the (internal) symmetries of the theory. If the theory is local, one may impose the same requirement on the energy-momentum tensor itself. Therefore, the condition \eqref{EMcond} is simply a stronger version of this constraint.

In a CohFT, starting from \eqref{EMcond}, one can construct a solution to the descent equations using the momentum operator $P=\delta\,\cdotp H\equiv\{Q,H\}$, where $H=\int G$ is a space-like integral of the operator $G$ in \eqref{EMcond}. Indeed, the operator
\begin{equation}
 \mc O^{(1)}\equiv i\,H\cdotp\mc O^{(0)}
\end{equation}
is a solution to \eqref{DescEq} and all the others can be generated recursively. In particular, the non-local observables constructed from the top-form operators $Y_{\gamma_{m}}=t^a\int O^{m}_{a}$ are relevant deformations of the action that do not spoil the topological symmetry of the CohFT and play an important role in TST (see Section \ref{sec:holan}).

Using the conditions above, we can prove that the theory is topological. For this consider a generic correlation function of physical operators $\mathcal O_i$ and differentiate with respect to the metric $h_{\alpha\beta}$\footnote{The ordering of the operators in the path integral formalism is somewhat irrelevant because the correlation functions do not depend on the insertion points.}
\begin{align}
 \frac{\delta}{\delta h_{\alpha\beta}}\langle\mathcal{O}_1\,\cdots\mathcal{O}_n\rangle=\int D[\Phi]\frac{\delta}{\delta h_{\alpha\beta}}\left(\mathcal{O}_1\,\cdots\mathcal{O}_n \,e^{iS[\Phi]}\right)\,.
\end{align}
Assuming that the operators do not depend on the metric, the differentiation brings down a factor of the energy-momentum tensor in the correlator function. Using the conditions (\ref{EMcond}, \ref{PhysCond}), this correlator is vanishing\footnote{We neglect possible contributions from boundary terms that would spoil this property.}:
\begin{equation}
\frac{\delta}{\delta h_{\alpha\beta}}\langle\mathcal{O}_1\,\cdots\mathcal{O}_n\rangle\sim\langle\delta(\mathcal{O}_1\,\cdots\mathcal{O}_n\,G_{\alpha\beta})\rangle=0\,.
\end{equation}
Let us mention that in order to obtain \eqref{EMcond}, we may assume the stronger condition that the action is itself $Q$-exact. In this case, the theory is even simpler because the correlation functions are then independent of the coupling constant and, hence, can be be calculated in the classical limit\footnote{Recall that the coupling constant $g$ appears in the action as $\exp\{\frac{i}{g}S\}$.}.

Before closing this section, we discuss some specific properties arising in the two dimensional case. The CohFT now lives on an arbitrary Riemann surface (of genus $g$) $\Sigma_g$, and the correlation functions of physical operators are \emph{factorisable}:
\begin{equation}
 \label{Factorizable}
\big\langle\mc O_1\cdots\mc O_n\big\rangle_{\Sigma_g}=\big\langle \mc O_1\cdots\mc O_k\mc O_i\big\rangle_{\Sigma_{g_1}} \eta^{ij}\big\langle\mc O_j\mc O_{k+1}\cdots\mc O_n\big\rangle_{\Sigma_{g_2}}\,,
\end{equation}
with $g=g_1+g_2$. In general, a correlation function can be thought of as performing a path integral with a certain assignment of boundary conditions and insertions of the physical operators at various points of the Riemann surface. In the case where the boundary is a circle, a boundary condition for a state $\lvert\phi_i\rangle$ is given by a path integral over a hemisphere with the operator $\phi_i$ inserted in its `neck' (one can infinitely stretch the hemisphere to define an `asymptotic state' but in a topological theory this is irrelevant). In this picture, a path integral can be pictured by starting with a hemisphere and evolving it with possible insertions of other operators until the other hemisphere (or asymptotic state) is reached. This generalises straightforwardly to the case where the boundary is made of many circles giving rise to a path integral over higher genera Riemann surfaces. The factorisation property \eqref{Factorizable} can now be pictured by starting with a genus $g$ Riemann and a number of operator insertions. Then one uses topological invariance to deform the Riemann surface, developing a very long tube in which states labelled by $a,b$ can propagate. In the limit where the tube is of infinite length, the genus $g$ Riemann surface factorises into two lower genera ones with an additional operator insertion corresponding to the state propagating in the tube. In principal, one must insert all possible operators in the theory, but it is easy to show that only the physical ones can lead to a non vanishing contribution. Notice the insertion of the `metric' $\eta^{ij}$ which can be defined by applying \eqref{Factorizable} in the $n=2$ case:
\begin{equation}
 \big\langle\mc O_i\mc O_j\big\rangle_{\Sigma_g}=\big\langle\mc O_i\mc O_k\big\rangle_{\Sigma_{g_1}} \eta^{kl}\big\langle\mc O_l\mc O_{j}\big\rangle_{\Sigma_{g_2}}\,.
\end{equation}
Hence, $\eta^{ij}$ is the inverse of the metric given by the two point function\footnote{In fact, the metric $\eta_{ij}$ is given by the three point function  $\big\langle\mc O_i\mc O_j\,\mathbf{1}\big\rangle$ on the sphere. When the unit operator $\mathbf{1}$ belongs to the ring of observables, $\eta$ defines a complex metric over the latter. This is the case for CohFTs.}. An important consequence of the factorisation property is that all the correlation functions of the theory are essentially determined by the three-point functions ${C_{ij}}^k$ on the sphere. This can be shown by using \eqref{Factorizable} and stretching off a sphere with two of the operator insertions.

After this brief overview of TFTs, we now turn to topological sigma-models which are the building blocks for TST when one includes the coupling to gravity.

\chapter{Topological Sigma Models}
\label{ch:TSM}
\section[\texorpdfstring{$\N=(2,2)$ Supersymmetric Sigma Models}{N=(2,2) Supersymmetric Sigma Models}]{\texorpdfstring{$\boldsymbol{\N=(2,2)}$ Supersymmetric Sigma Models}{N=(2,2) Supersymmetric Sigma Models}}

We are now ready to explicitly construct a theory that has topological invariance. Since we are interested in studying topological sectors of string theory, we focus on two-dimensional (global) $\N=(2,2)$ sigma-models. As we discuss below, the natural target space is a K\"ahler manifold.

Consider a two-dimensional worldsheet given by $\mathbb R^2$ endowed with the flat Euclidean metric. A supersymmetry generator is a $\frac{1}{2}$-representation of the Lorentz group. In our case, the latter is $SO(2)$ and this representation is reducible. Irreducible representations are obtained by splitting the spinor into its two components having opposite Lorentz charges. Therefore, a theory with two fundamental supercharges has $\N=(2,2)$ supersymmetry and the irreducible supercharges are denoted $Q_+$, $Q_-$, $\bar Q_+$, $\bar Q_-$. Together with the usual Poincar\'e generators given by the translations ($H,P$) and rotations ($M$) Noether charges, and two internal $U(1)$ currents of axial and vector R-rotations $F_{A,V}$\footnote{They are defined in terms of the internal left- and right-moving internal currents, $F_{V,A}=F_L\pm F_R$.}, they for the $\N=(2,2)$ supersymmetry algebra (all the other relations are trivial):

\begin{align}
 &\{Q_{\pm},Q_{\pm}\}=\{\bar Q_{\pm},\bar Q_{\pm}\}=0\,,\\
 &\{Q_{\pm},\bar Q_{\pm}\}=H\pm P\,,\\
 &\{Q_{+},Q_{-}\}=\{\bar Q_{+},\bar Q_{-}\}=0\,,\label{Cc1}\\
 &\{Q_{-},\bar Q_{+}\}=\{Q_{+},\bar Q_{-}\}=0\,,\label{Cc2}\\
 &[M,Q_{\pm}]=\pm Q_{\pm}\,,\,[M,\bar Q_{\pm}]=\pm\bar Q_{\pm}\,,\label{Mrel}\\
 &[F_V,Q_{\pm}]=Q_{\pm}\,,\,[F_V,\bar Q_{\pm}]=-\bar Q_{\pm}\,,\\
 &[F_A,Q_{\pm}]=\pm Q_{\pm}\,,\,[F_A,\bar Q_{\pm}]=\mp\bar Q_{\pm}\,.
\end{align}
In fact, one can relax the relations (\ref{Cc1}, \ref{Cc2}) by introducing central charges:
\begin{align}
 \{Q_{+},Q_{-}\}&=\bar Z\,,\,\{\bar Q_{+},\bar Q_{-}\}=Z\,,\\
 \{Q_{-},\bar Q_{+}\}&=\tilde{Z}\,,\,\{Q_{+},\bar Q_{-}\}=\bar{\tilde{Z}}\,.
\end{align}
However, the central charges $Z$ and $\tilde Z$ must commute with all the operators of the theory. For generic central charges, the axial and vector currents are not conserved, unless $Z=0$ ($\tilde Z=0$) in which case $F_V$ ($F_A$) is conserved.

The supercharges of the theory generate supersymmetry transformations that are represented by an operator $\delta$, and depend on four parameters $\epsilon_{\alpha},\bar\epsilon_{\alpha}$, where $\alpha=+,-$:
\begin{equation}
 \delta\equiv\epsilon^{\alpha\beta}(\epsilon_{\alpha}Q_{\beta}+\bar\epsilon_{\alpha}\bar Q_{\beta})\,.
\end{equation}
In order to work in a theory that manifestly preserves supersymmetry, it is convenient to use the language of the $\N=(2,2)$ \emph{superspace} (in two dimensions) which is very similar to the $\N=1$ standard superspace (in four dimensions). We denote $(z,\bar z)$ the complexified (bosonic) coordinates on $\mathbb C\cong\mathbb R^2$ and supplement them with four fermionic coordinates $(\theta^+,\theta^-,\bar\theta^+,\bar\theta^-)$ that transform as spinors under Lorentz transformations with $+$ or $-$ chirality. This set of coordinates defines the superspace $\mathbb R^{2|4}$. In terms of the superspace coordinates, the supercharges can be represented as differential operators:

\begin{align}
 Q_{\alpha}&=\frac{\partial}{\partial\theta^{\alpha}}+i\bar\theta^{\alpha}\partial_{\alpha}\,,\nonumber\\
 \bar Q_{\alpha}&=-\frac{\partial}{\partial\bar\theta^{\alpha}}-i\theta^{\alpha}\partial_{\alpha}\,,
\end{align}
where $\partial_{\pm}$ are $z,\bar z$-derivatives. We define another set of differential operators that anti-commute with the supercharges:
\begin{align}
 D_{\alpha}&=\frac{\partial}{\partial\theta^{\alpha}}-i\bar\theta^{\alpha}\partial_{\alpha}\,,\nonumber\\
 \bar D_{\alpha}&=-\frac{\partial}{\partial\bar\theta^{\alpha}}+i\theta^{\alpha}\partial_{\alpha}\,.
\end{align}

A superfield is a function defined on the superspace, $F(z,\bar z,\theta^{\pm},\bar\theta^{\pm})$ which can be expanded in the fermionic coordinates, leading to (at most) sixteen bosonic components. It is instructive to write down the action of the R-rotations on a general superfield which might possess non-trivial R-charges $q_A$ and $q_V$:
\begin{align}
 e^{i\alpha F_V} F(z,\bar z,\theta^{\pm},\bar\theta^{\pm})&=e^{i\alpha q_V} F(z,\bar z,e^{-i\alpha}\theta^{\pm},e^{i\alpha}\bar\theta^{\pm})\,,\nonumber\\
 e^{i\alpha F_A} F(z,\bar z,\theta^{\pm},\bar\theta^{\pm})&=e^{i\alpha q_A} F(z,\bar z,e^{\mp i\alpha}\theta^{\pm},e^{\pm i\alpha}\bar\theta^{\pm})\,.\label{Rsymm}
\end{align}
In subsequent discussions, we mainly restrict our attention to \emph{chiral superfields} defined through the constraint
\begin{equation}
 \bar D_{\alpha}\Phi=0\,.
\end{equation}
They can therefore be expanded as\footnote{For convenience, we use the notation $z^{+}=z$ and $z^{-}=\bar z$.}
\begin{equation}
 \Phi(z^{\alpha},\theta^{\alpha},\bar\theta^{\alpha})=\phi(y^\alpha)+\theta^{\beta}\psi_{\beta}(y^\alpha)+\frac{1}{2}\epsilon_{\beta\gamma}\theta^{\beta}\theta^{\gamma}F(y^{\alpha})\,,
\end{equation}
with $y^{\alpha}=z^{\alpha}-i\theta^{\alpha}\bar\theta^{\alpha}$. The complex conjugate of a chiral superfield is called an \emph{anti-chiral superfield}. The supersymmetric transformation of a chiral superfield is again a chiral superfield and one can write the transformation on the components of the superfield as
\begin{align}
\label{SusyChiral}
 \delta\phi&=\epsilon_+\psi_--\epsilon_-\psi_+\,,\\
 \delta\psi_{\pm}&=-2i\bar\epsilon_{\mp}\partial_{z^{\pm}}\phi+\epsilon_{\pm}F\,,
\end{align}
and similarly for an anti-chiral superfield. For completeness, let us mention another class of superfields called \emph{twisted chiral superfields} $U$ defined by the condition
\begin{equation}
 \bar D_+ U=D_-U=0\,.
\end{equation}
With these definitions in mind, we now recall the construction of supersymmetric actions given by the so-called \emph{F-terms} and \emph{D-terms}. The latter are defined as functionals of superfields integrated over the full superspace:
\begin{equation}
 \int d^2 z d^4\theta K(F^I)\,.
\end{equation}
The integration measure $d^4\theta$ is simply $d\theta^+d\theta^-d\bar\theta^-d\bar\theta^+$. This term is invariant under the supersymmetry $\delta$. In the particular case where the function $K$ depends on chiral and anti-chiral superfields only, it is naturally interpreted as a K\"ahler potential. Indeed, performing the $\theta$-integrals and picking only the components depending on the scalar field $\phi^I$ of $\Phi^I$, one finds
\begin{equation}
 -\int d^2z\,g_{I\bar J}\,\eta^{\alpha\beta}\,\partial_{\alpha}\phi^I\partial_{\beta}\bar\phi^{J}\,,
\end{equation}
where $\eta$ is the Euclidean worldsheet metric and $g_{I\bar J}=\frac{\partial^2 K}{\partial\phi^I\partial\bar\phi^{J}}$. One recognises the action for a bosonic string (with gauge-fixed worldsheet metric) and a K\"ahler manifold target space.

On the other hand, an F-term is defined as an integral, over half of the superspace, of a function of chiral superfields only:
\begin{equation}
 \int d^2 zd^2\theta\left. W(\Phi^I)\right| _{\bar\theta^\alpha=0}\,,
\end{equation}
where $d^2\theta=d\theta^+d\theta^-$. Using the action of $\delta$ and the properties of chiral superfields, one can show that F-terms are also invariant. This class of supersymmetric terms turns out to be important in studying topological amplitudes in string theory.

So far we have only dealt with sigma-models on flat manifolds, even though our aim is to study string theory amplitudes defined in perturbation theory as a topological expansion over Riemann surfaces of arbitrary geometry. In other words, we would like to couple the supersymmetric sigma-model to gravity by using curved manifolds. In a bosonic theory, this is quite straightforward whereas in a supersymmetric theory, one must appropriately choose a spin structure in order to be able to put spinors on the worldsheet. However, the action would then break supersymmetry. One way to see this is to write down the supersymmetric variation of the action on a generic genus $g$ Riemann surface $\Sigma_g$ as
\begin{equation}
 \delta S=\int_{\Sigma_{g}}\epsilon^{\alpha\beta}\nabla_{\mu}(\epsilon_{[\alpha}G^{\mu}_{\beta]}-\bar\epsilon_{[\alpha}\bar G^{\mu}_{\beta]}).
\end{equation}
In order for this to vanish for a generic $\Sigma_g$, we must impose $\nabla_{\mu}\epsilon_{\alpha}=0=\nabla_{\mu}\bar\epsilon_{\alpha}$, meaning that $\epsilon_{\alpha}$ and $\bar\epsilon_{\alpha}$ are covariantly constant spinors. Therefore, the absence of such spinors on a curved Riemann surface implies that supersymmetry is broken. In the case of bosonic symmetries, this problem does not arise because an infinitesimal parameter of the latter is a number which can be viewed as a section of a trivial bundle $\mathbb C\times \Sigma$ over the worldsheet $\Sigma$ and so it can always be chosen to be constant. In this sense, in order to restore a fermionic symmetry, on may try to \emph{twist} the theory so that the supercharges become sections of a trivial bundle (instead of a spin bundle). Of course, this is at the cost of violating the spin-statistics theorem as we would obtain a scalar, Grassmann charge. In what follows, we show how this is performed.

\section{Topological Twist}\label{TopoTwist}

As we argued previously, in order to obtain a fermionic symmetry of the sigma-model on an arbitrary Riemann surface, one is led to twist the original theory by changing its Lorentz charges. This is done by redefining the $SO(2)\cong U(1)$ Lorentz generator $M$ using the axial or vector internal symmetries and there are two possible choices:
\begin{enumerate}
\label{TFTtwist}
 \item A-twist: $M_A\equiv M-F_V$,
 \item B-twist: $M_B\equiv M+F_A$.
\end{enumerate}
The resulting theories are said to be of type-A and type-B. Of course, this assumes that one of the internal currents is conserved. It turns out that all other possible twistings are equivalent to these two ones. Consequently, the twisted theory is obtained by gauging the new diagonal subgroup $\tilde U(1)_{A,B}$ of $U(1)\times U(1)_{V,A}$ (or, equivalently, by defining covariant derivatives with respect to this new Lorentz group) implying that the spin of the fields get modified. In order to see this, notice that the equations \eqref{Mrel} become
\begin{align}
 [M_A,Q_+]&=0\,,\nonumber\\
 [M_A,Q_-]&=-2Q_-\,,\nonumber\\
 [M_A,\bar Q_+]&=2Q_+\,,\nonumber\\
 [M_A,\bar Q_-]&=0\,,
\end{align}
for the type-A theory, and
\begin{align}
 [M_B,Q_{\pm}]&=\pm Q_{\pm}\,,\nonumber\\
 [M_B,\bar Q_{\pm}]&=0\,,
\end{align}
for the type-B one. From these relations, one can easily read off the new Lorentz charges of the supercharges, and these are listed in Table \ref{tb:supercharges} below. Similarly, using the transformation law \eqref{Rsymm}, one can deduce the new charges of the components of chiral and anti-chiral superfields which we list in Table \ref{tb:chiralcharges}.

\renewcommand{\arraystretch}{1.3}
\begin{table}[ht]
 \begin{center}
  \begin{tabular}{|c||c|c||c|c|}
  \hline
                &$\tilde U(1)_A$&Bundle&$\tilde U(1)_B$&Bundle\\\hline
   $Q_+$        &0&$\underline{\mathbb C}$&2&$K$\\\hline
   $Q_-$        &-2&$\overline{K}$&-2&$\overline{K}$\\\hline
   $\bar Q_+$   &2&$K$&0&$\underline{\mathbb C}$\\\hline
   $\bar Q_+$   &0&$\underline{\mathbb C}$&0&$\underline{\mathbb C}$\\\hline
  \end{tabular}
 \end{center}
\caption{Twisted Lorentz charges of the supercharges and the complex line bundles of which they are sections. Here, $\underline{\mathbb C}$ is the trivial bundle whereas $K$ is the canonical one.}
\label{tb:supercharges}
\end{table}

\begin{table}[ht]
 \begin{center}
  \begin{tabular}{|c||c|c|c||c||c|}
  \hline
                  &$U(1)$&$U(1)_V$&$U(1)_A$&$\tilde U(1)_A$&$\tilde U(1)_B$\\\hline
   $\phi$         &0&0&0&0&0\\\hline
   $\psi_+$       &-1&+1&+1&-2&0\\\hline
   $\psi_-$       &+1&+1&-1&0&0\\\hline
   $\bar\psi_+$   &-1&-1&-1&0&-2\\\hline
   $\bar\psi_-$   &+1&-1&+1&2&2\\\hline
  \end{tabular}
 \end{center}
\caption{Lorentz and internal charges of the components of chiral and anti-chiral superfields.}
\label{tb:chiralcharges}
\end{table}
Consequently, after the twisting, two supercharges are scalar operators and can serve as a definition of a fermionic symmetry on an arbitrary Riemann surface, whereas the other two can be combined into a vector-valued supercharge. The scalar supercharges define the \emph{topological charge}

\begin{align}
 \label{TopoCharge}
 \mc Q_A&=Q_++\bar Q_-\nonumber\,,\\
 \mc Q_B&=\bar Q_++\bar Q_-\,.
\end{align}
The vector supercharge is usually denoted $G^{A,B}_{\mu}$ and is defined in the type-A and type-B theories as
\begin{align}
 G_z^{A}&=\bar Q_+\,,\,G_{\bar z}^{A}=Q_-\,,\label{VectorSCA}\\
 G_z^{B}&= Q_+\,,\,G_{\bar z}^{B}=Q_-\,.\label{VectorSCB} 
\end{align}
Using the supersymmetry algebra, it is easy to show that the topological charge is nilpotent,
\begin{equation}
 \mc Q^2=0\,,
\end{equation}
and that the Hamiltonian and momentum operators are $\mc Q$-exact,
\begin{equation}
 \{\mc Q,G_{\mu}\}=P_{\mu}\,,
\end{equation}
where $P_0=H$, $P_{1}=P$, $G_0=\frac{1}{2}(G_z+G_{\bar z})$ and $G_1=\frac{1}{2}(G_z-G_{\bar z})$. Of course, this does not prove that the energy-momentum tensor of the theory is $\mc Q$-exact, but this turns out to be true in most of the cases of interest. Moreover, notice that the full Hilbert space of the theory is not altered by the topological twist but only the subset of physical states is affected.

\section{Type-A Model}

In the A-twisted sigma-models, the fermions $\psi^{I}_-$ and $\bar\psi^{I}_+$ become scalars and, for convenience, we relabel them as
\begin{align}
 \psi^{\bar I}_-=\chi^{\bar I}\,\nonumber\\
 \bar\psi^{I}_+=\chi^{ I}\,.
\end{align}
We also rename $\psi^{I}_+$ and $\bar\psi^I_-$ according to their new quantum numbers (they become one-forms):
\begin{align}
 \psi^{\bar I}_+=\rho^{\bar I}_{z}\,\nonumber\\
 \bar\psi^{I}_-=\rho^{I}_{\bar z}\,.
\end{align}

As in a CohFT, the set of physical operators are $\mc Q$-cohomology representatives and can be identified with differential forms on the complex manifold target space. Indeed, given a non-trivial $(p,q)$-form $\phi$ in $H^{p,q}(\mc M)$,
\begin{equation}
 \omega^{(p,q)}=\omega_{I_1\cdots I_p\,\bar J_1\cdots\bar J_q}(z)\,dz^{I_1}\wedge\cdots\wedge dz^{I_p}\wedge d\bar z^{J_1}\wedge\cdots\wedge d\bar z^{J_q}\,,
\end{equation}
the following operator belongs to the $\mc Q$-cohomology:
\begin{equation}
\label{OpA}
 \mc O_{\omega}=\omega_{I_1\cdots I_p\,\bar J_1\cdots\bar J_q}(\phi)\,\chi^{I_1}\cdots\chi^{I_p}\,\chi^{\bar J_1}\cdots\chi^{\bar J_q}\,.
\end{equation}
Notice that we do not use the $\rho$-fields in \eqref{OpA} since we would need to insert the worldsheet metric, thus spoiling the topological invariance. The correspondence is  achieved through the identification of $z^{I},dz^{I},d\bar z^{J}$ with $\phi^I,\chi^I,\chi^{\bar J}$ respectively. Indeed, since the action of the supercharges \eqref{SusyChiral} on the fields is changed,
\begin{align}
 \{\mc Q,\phi\}&=\chi\,\\
 \{\mc Q,\chi\}&=0\,,
\end{align}
the topological charge can only act non trivially on $\phi$,
\begin{equation}
\{\mc Q,\mc O_{\omega}\}= \partial_{\phi^{a}}(\omega_{I_1\cdots I_p\,\bar J_1\cdots\bar J_q}(\phi))\,\chi^{a}\,\chi^{I_1}\cdots\chi^{I_p}\,\chi^{\bar J_1}\cdots\chi^{\bar J_q}\,,
\end{equation}
with $a$ summed over two indices $I_{p+1},\bar J_{q+1}$. This implies that $\mc Q$ is identified with the de Rham differential operator $\partial+\bar\partial=d$ and the Hilbert space of physical operators is indeed isomorphic to $H_{dR}(\mc M)$. The operator $\mc O_{\phi}$ is said to be of degree $(p,q)$.

Consider the correlator of the topological sigma-model on an arbitrary Riemann surface $\Sigma_g$ of genus $g$
\begin{equation}
\label{ACor}
 \langle\mc O_{1}\cdots\mc O_{n}\rangle_{\Sigma_g}\,,
\end{equation}
which can be decomposed over a basis of homology classes $\beta$ of $\phi(\Sigma_g)$ in $H_{2}(\mc M,\mathbb Z)$. In order for this correlation function to be non-vanishing, one has to ensure that any possible anomalies coming from the global $U(1)$ symmetries are cancelled. In the case at hand, the axial $U(1)_A$ is a ghost number symmetry, \emph{i.e.} it is anomalous, and one is forced to cancel the corresponding ghost charge so that \eqref{ACor} gives a sensible contribution. In other words, one must soak up the fermionic zero-modes on $\Sigma_g$. For a fixed homology cycle and genus, this anomaly is given by the Riemann-Roch theorem:
\begin{equation}
 N_0(\chi)-N_0(\rho)|_{\beta,g}=2\,c_1(\mc M)\cdotp\beta+2m(1-g)\,,
\end{equation}
where $N_0$ counts the zero modes, $c_1(\mc M)$ is the first Chern class of $\mc M$ and $m$ its complex dimension. This means that under the action of an axial R-rotation $e^{i\alpha F_A}$, the path integral measure picks up a factor of $e^{i\alpha(N_0(\chi)-N_0(\rho))_{\beta,g}}$. Using the $U(1)_A$ charges of the $\chi$ fields (see Table \ref{tb:chiralcharges}), we deduce the following \emph{selection rule} for \eqref{ACor} to be non-vanishing:
\begin{equation}
 \label{ASelRuleA}
\sum_{i=1}^{n}(p_i+q_i)=2c_1(\mc M)\cdotp\beta+2m(1-g)\,.
\end{equation}
Similarly, $U(1)_V$ is not anomalous and the selection rule is simply
\begin{equation}
 \label{VSelRuleA}
\sum_{i=1}^{n}p_i=\sum_{i=1}^{n}q_i\,.
\end{equation}
This leads to very stringent constraints on the correlation functions of type-A topological sigma-models. In particular, if $c_1(\mc M)=0$, that is the target space is a Calabi-Yau manifold, all correlation functions vanish at genus $g>1$, whereas at genus one the partition function survives. This issue can be circumvented by coupling the sigma-model to gravity (see next chapter). At tree-level, the non-trivial correlation functions are the ones involving insertions of top-forms, and they count \emph{holomorphic maps} from the worldsheet to space-time (or worldsheet instantons) (see \emph{e.g.} \cite{Witten:1988xj}).

\section{Type-B Model}

In the case of a B-twist, the labelling of the fields is different \cite{Witten:1991zz}:
\begin{align}
 \eta^{\bar I}&=\psi^{\bar I}_++\psi^{\bar I}_-\,\nonumber\\
 \theta_{I}&=g_{I\bar J}(\psi^{\bar J}_+-\psi^{\bar J}_-)\,,\nonumber\\
 \rho^{I}_{z}&=2\bar\psi^{I}_+\,\nonumber\\
 \rho^{I}_{\bar z}&=2\bar\psi^{I}_-\,.
\end{align}
Here, the topological observables are defined, using elements of the Dolbeault cohomology $H^{p}_{\bar\partial}(\mc M,\wedge^{q}T\mc M)$
\begin{equation}
 \omega^{(p,q)}={\omega_{\bar I_1\cdots\bar I_p}}^{J_1\cdots J_q}(z,\bar z)\,d\bar z^{I_1}\wedge\cdots\wedge d\bar z^{I_p}\,\frac{\partial}{\partial z^{J_1}}\wedge\cdots\wedge\frac{\partial}{\partial z^{J_q}}\,,
\end{equation}
by
\begin{equation}
 \mc O_{\omega}={\omega_{\bar I_1\cdots\bar I_p}}^{J_1\cdots J_q}\,(\phi,\bar\phi)\,\eta^{\bar I_1}\cdots\eta^{\bar I_p}\,\theta_{J_1}\cdots\theta_{J_q}\,,
\end{equation}
with the identifications:
\begin{align}
\label{IdB}
 \eta&\longleftrightarrow d\bar z\,,\nonumber\\
 \theta&\longleftrightarrow \frac{\partial}{\partial z}\,.
\end{align}
As in the type-A model, one can show, by acting with the topological charge on $\mc O$, that the identification \eqref{IdB} provides an isomorphism between the Hilbert space of local topological observables and the $\bar\partial$-cohomology group. Moreover, the correlation functions have simpler selection rules because the target space must be a Calabi-Yau in order for the theory to be well-defined:

\begin{equation}
 \label{SelRuleB}
\sum p_i=\sum q_i=m(1-g)\,.
\end{equation}
In addition, the action is also $\mc Q$-exact and the semi-classical approximation is exact. However, the only worldsheet instantons are constant maps so that the path integral reduces to an integral over $\mc M$.

\chapter{Topological String Theory}
\label{ch:TST}
\section{Coupling to Topological Gravity}

As mentioned previously, the type-A and type-B topological sigma-models are trivial for $g>1$ and one is forced to introduce the degrees of freedom of an arbitrary Riemann surface rather than simply use a \emph{background} metric. In other words, in order to capture non-trivial information on an arbitrary Riemann surface in the twisted sigma-models, one should couple the latter to gravity, and perform the integral over the space of all possible metrics. Naively, one might think that this is trivial since the quantities we are calculating are metric independent. However, one should keep in mind that there exist metric configurations that are not connected by continuous deformations and that the volume of the corresponding space is infinite.

On the other hand, recall that when one includes the degrees of freedom of a metric in the sigma-model, the worldsheet theory becomes a (super-)conformal field theory (SCFT) whose symmetry group is very big. Therefore, by using these symmetries, one can reduce the metric path integral to a finite dimensional one over conformally inequivalent metric, \emph{i.e.} over complex structures. Obviously, one must ensure that conformal anomalies are absent and the guiding principle is \emph{string theory} since the structure of twisted sigma-models is extremely similar to the one of the bosonic string \cite{Dijkgraaf:1990qw,Witten:1992fb,Bershadsky:1993cx}. In particular, we would like to study $\N=1$ space-time compactifications in which case the worldsheet theory is an $\N=(2,2)$ SCFT \cite{Banks:1988yz}. The topological charge we have studied so far behaves very much like the BRST charge. Hence, the strategy to construct a TST is to twist the underlying $\N=(2,2)$ SCFT which is spanned by the energy momentum tensor ($T$), two supercharges ($G^{\pm}$) and an internal $U(1)$ current ($J$). Of course, these operators are supplemented with their right moving counterparts $(\tilde T,\tilde G^{\pm},\tilde J)$. The conformal weights and $U(1)$ charges are listed in Table \ref{tb:scft} below.

\begin{table}[ht]
 \begin{center}
  \begin{tabular}{ccc}
  \toprule
              &Conformal weight&$U(1)$ charge\\\midrule
  $T$         &2&0\\
  $G^{\pm}$   &$3/2$&$\pm1$\\
  $J$         &1&0\\\bottomrule
  \end{tabular}
 \end{center}
\caption{Conformal weights and $U(1)$ charges of the operators spanning the superconformal algebra.}
\label{tb:scft}
\end{table}

The (left-moving) superconformal algebra is given by the following OPEs, $c$ being the central charge of the algebra:

\begin{alignat}{3}
& T(z)T(w)&\sim&\,\frac{c/2}{(z-w)^4}+\frac{2T(w)}{(z-w)^2}+\frac{\partial_wT(w)}{z-w}\,,\label{TT}\\
& T(z)G^{\pm}(w)&\sim&\,\frac{\tfrac{3}{2} G^{\pm}(w)}{(z-w)^2}\pm\frac{\partial_wG^{\pm}(w)}{z-w}\,,\label{TG}\\
& T(z)J(w)&\sim&\,\frac{J(w)}{(z-w)^2}+\frac{\partial_wJ(w)}{z-w}\,,\label{TJ}\\
& G^{+}(z)G^{-}(z)\,&\sim&\,\frac{2c/3}{(z-w)^3}+\frac{2J(w)}{(z-w)^2}+\frac{\partial_w J(w)+2T(w)}{z-w}\,,\label{GG}\\
& J(z)G^{\pm}(w)&\sim&\,\pm\frac{G^{\pm}(w)}{z-w}\,,\label{JG}\\
& J(z)J(w) &\sim&\,\frac{c/3}{(z-w)^2}\,.\label{JJ}
\end{alignat}
The other OPEs are trivial. In particular, $G^{\pm}$ are nilpotent and can potentially serve as BRST operators had they had the correct conformal dimension. Now perform the topological twist by shifting the energy-momentum tensor using the $U(1)$ current $J$
\begin{equation}
 \label{twistSCFT}
T\longrightarrow T-\frac{1}{2}J\,,
\end{equation}
so that the new central charge vanishes, and \eqref{GG} becomes
\begin{align}
 G^{+}(z)G^{-}(z)&=\frac{2J(w)}{(z-w)^2}+\frac{2T(w)}{z-w}\,.\label{newGG}
\end{align}
Contrary to the case of the untwisted theory, the vanishing of the central charge implies that there is no constraint on the dimension of space-time and topological string theory is consistent on a target space of arbitrary dimension. However, Calabi-Yau threefolds play a very special role as we discuss below. Moreover, \eqref{twistSCFT} has the direct consequence of shifting the conformal weights of the operators in the theory with their $U(1)$ charges. The new weights are listed for the generators of the superconformal algebra in Table \ref{tb:tscft}.
\begin{table}[ht]
 \begin{center}
  \begin{tabular}{ccc}
  \toprule
               &Conformal weight&$U(1)$ charge\\\midrule
  $T$          &2&0\\
  $G^{+}$      &$1$&$+1$\\
  $G^{-}$      &$2$&$-1$\\
  $J$          &1&0\\\bottomrule
  \end{tabular}
 \end{center}
\caption{Conformal weights and $U(1)$ charges of the operators spanning the twisted superconformal algebra.}
\label{tb:tscft}
\end{table}

Consequently, $G^+$ can be identified, after the twist, with the BRST operator:
\begin{equation}
 \mc Q=\oint G^+\,.
\end{equation}
Using \eqref{newGG}, we find that the energy-momentum tensor is $\mc Q$-exact,
\begin{equation}
\label{Texact}
 \{\mc Q,G^{-}\}=T\,,
\end{equation}
which shows that this theory has all the ingredients of a topological theory. Moreover, the physical states of the theory are the \emph{chiral primaries} $\phi$ which have conformal weight $\frac{1}{2}$ and +1 charge. They satisfy $[G^+,\phi]=0$ and acquire dimension 0 after the twist. Therefore, they span the $\mc Q$-cohomology or the Hilbert space of physical states. The anti-chiral primaries, operators of conformal dimension $\frac{1}{2}$ and -1 charge before the twist, have dimension 1 after the twist and are unphysical. Hence, they should decouple from theory, and this is true only up to an anomaly (see Section \ref{sec:holan}). On the other hand, $G^-$ can be identified with the reparametrisation ghost $b$. Finally, the worldsheet theory also contains a right-moving SCFT that one has to twist, leading to the definition of two different topological string theories, namely, the A- and B-model:
\begin{align}
 \textrm{A-model: }\tilde T\longrightarrow\tilde T+\frac{1}{2}\tilde J\,,\\
 \textrm{B-model: }\tilde T\longrightarrow\tilde T-\frac{1}{2}\tilde J\,.
\end{align}
As the notation suggests, one can show that the relative sign between the A- and B-twist precisely corresponds to twisting the Lorentz generator with the axial or vector R-currents as in Section \ref{TFTtwist}. The identification of the right-moving reparametrisation ghost (and BRST charge) is also different:
\begin{align}
 \textrm{A-model: }\tilde b\longrightarrow\tilde G^{+}\,,\\
 \textrm{B-model: }\tilde b\longrightarrow\tilde G^{-}\,.
\end{align}

Since the central charge of TST is zero, the integration over conformally equivalent metrics is well-defined and one is only left with an integral over the moduli space $\mc M_g$ of conformally inequivalent metrics parametrised by $3(g-1)$ complex parameters on a genus $g>1$ Riemann surface\footnote{For g=0,1, this number is 0,1. In fact, $3(g-1)$ is the virtual dimension of $\mc M_g$ and is the actual dimension only for $g>1$.}. In order to understand the role of these complex parameters, recall that, in two dimensions, a conformal map is a holomorphic one so that an element of the tangent bundle over $\mc M_g$ describes an infinitesimal change in the complex structure:
\begin{equation}
 \delta_{\epsilon}dz^{\alpha}=\epsilon{(\mu_{a})_{\bar\alpha}}^{\alpha}\,dz^{\bar\alpha}\,,
\end{equation}
with $a=1,\ldots,3(g-1)$. $\mu_a$ is called a \emph{Beltrami differential}. Together with the reparametrisation ghosts, we can build a $(1,1)$ form $G^{-}_{\alpha\beta}{(\mu_{a})_{\bar\alpha}}^{\beta}dz^{\alpha}dz^{\bar\alpha}\equiv G^{-}(\mu_a)$ that we can integrate over the Riemann surface. This provides the $\mc M_g$-invariant\footnote{Strictly speaking, we integrate over the \emph{Deligne-Mumford} compactification of the moduli $\overline{\mc M}_g$ of the moduli space of genus $g$ Riemann surfaces.} measure necessary to construct the genus $g$ partition function of the A- and B-models ($m^a,\bar m^a$ form a basis of moduli):
\begin{equation}
 \label{Fg}
F_g^{A,B}=\int_{\mc M_g}\langle\prod_{a=1}^{3g-3}dm^{a}\,d\bar m^a\int_{\Sigma_g}G^{-}(\mu_a)\int_{\Sigma_g}\tilde G^{\pm}(\bar\mu_a)\rangle\,.
\end{equation}
Naively, one might think that $F_g^{B}$ vanishes because of the non-conservation of the $U(1)$ charge. However, recall that the $U(1)$ current $J$ is anomalous, \emph{c.f.} \eqref{JJ}, and provides, at genus $g$, a background charge of $\hat c(g-1)$, with $\hat c\equiv\tfrac{2}{3}c$. Therefore, for $g\neq1$, the genus $g$ topological string partition function is non-vanishing only for $\hat c=3$, \emph{i.e.} for a Calabi-Yau threefold. Apart from their phenomenological applications in string compactifications, this gives Calabi-Yau threefolds a very special role.


\section{Holomorphic Anomaly Equations}
\label{sec:holan}
\begin{figure}[ht]
\centering
\includegraphics[width=1\textwidth]{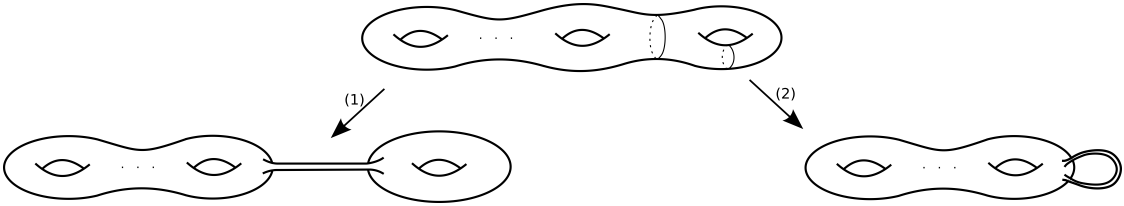}
\caption{The boundaries of a Riemann surface in the moduli space are obtained either through a dividing geodesic (1) or by pinching a handle (2).}
\label{fig:Pinching}
\end{figure}

In Section \ref{sec:CohFT}, we have seen that for every physical operator in the theory, it is easy to construct a set of non-local observables using the descent equations \eqref{DescEq}. In particular, the top-form operators one obtains can be considered as deformations of the theory that do not spoil the topological invariance. In our case, these operators are integrated two-form operators that can be constructed by starting from a `scalar' operator $\mc O^{(0)}$ and acting with the vector supercharge $G_{\alpha}$ (\ref{VectorSCA}, \ref{VectorSCB}):
\begin{align}
 \mc O^{(2)}_{\alpha\beta}=i\{G_{\alpha},\mc O^{(1)}_{\beta}\}=-\{G_{\alpha},[G_{\beta},\mc O^{(0)}]\}\,.
\end{align}
One can show (see \emph{e.g.} \cite{Vonk:2005yv}) that this physical operator is $\bar{\mc Q}$-exact. By unitarity, if we deform the theory with $t^a\int_{\Sigma_g}\mc O^{(2)}_a$, we must also include the complex conjugate operator
\begin{equation}
 \bar t^{a}\int_{\Sigma_g}\bar{\mc O}^{(2)}_a\,,
\end{equation}
with $(\bar t^{a})^{*}=t^a$. This operator is $\mc Q$-exact, \emph{i.e.} it corresponds to an unphysical state in the sense of CohFT:
\begin{equation}
 \bar{\mc O}^{(2)}=\oint G^{+}\oint\tilde G^{\pm}\cdotp\hat{\mc O}^{(2)}\,,
\end{equation}
for some chiral primary $\hat{\mc O}^{(2)}$, and the sign refers to the A- or B-model. Therefore, by topological invariance, any physical correlation function should only depend \emph{holomorphically} on $t$: the unphysical states should decouple, for example, from the topological string partition function. It turns out that this is true only up to an anomaly \cite{Bershadsky:1993ta,Bershadsky:1993cx}. Naively, differentiating the partition function with respect to $\bar t^{a}$ brings down a $\mc Q$-exact operator in the path integral so that by `integrating by parts', one obtains a vanishing result unless there are contributions from the boundaries of the moduli space $\mc M_g$. The latter can be obtained in two ways as illustrated in Fig. \ref{fig:Pinching}. They correspond to the limit of degeneration of one of the cycles of the Riemann surface. Stretching the resulting `tube' to infinity, the degenerate cycle is replaced by two punctures.

Concretely, for the B-model, we have
\begin{align}
 \frac{\partial F_g}{\partial\bar t^{a}}&=\int_{\mc M_g}[Dm]\left\langle\int d^2z\oint G^{+}\oint\tilde G^{+}\cdotp\hat{\mc O}^{(2)}(z)\prod_{a=1}^{3g-3}\int G^{-}(\mu_a)\int\tilde G^{-}(\bar\mu_a)\right\rangle\nonumber\\
                                        &=\int_{\mc M_g}[Dm]\left\langle\int \hat{\mc O}^{(2)}\sum_{c,d=1}^{3g-3}\int T(\mu_{c})\int\tilde T(\mu_d)\prod_{a\neq c,d}\int G^{-}(\mu_a)\int\tilde G^{-}(\bar\mu_a)\right\rangle\,,
\end{align}
where we deformed the contour integrals of $G^+$ and $\tilde G^+$ away and used the topological property \eqref{Texact} of the energy-momentum tensor.  Recall that the Beltrami differentials parametrise deformations of the worldsheet metric $h$ under a change in the moduli $m^a,\bar m^a$. Hence, $\frac{\partial S}{\partial m^a}=\frac{\partial S}{\partial h}\frac{\partial h}{\partial m^a}$ implies that
\begin{equation}
 T(\mu_a)=\frac{\partial S}{\partial m^a}\,.
\end{equation}
Consequently, the insertions of the energy-momentum tensor can be recast as derivatives with respect to the worldsheet moduli:
\begin{align}
 \frac{\partial F_g}{\partial\bar t^{a}}&=\int_{\mc M_g}[Dm]\sum_{c,d=1}^{3g-3}\frac{\partial^2}{\partial m^c\partial\bar m^d}\left\langle\int \hat{\mc O}^{(2)}\prod_{a\neq c,d}\int G^{-}(\mu_a)\int\tilde G^{-}(\bar\mu_a)\right\rangle\,,
\end{align}
and the integral reduces to one on the boundary of the moduli space which, as discussed previously, has two contributions. A thorough analysis of each contribution leads to the BCOV holomorphic anomaly equations \cite{Bershadsky:1993cx}:
\begin{equation}
 \label{BCOV}
\frac{\partial F_g}{\partial\bar t^{a}}=\frac{1}{2}{C_{\bar a}}^{bc}\left(D_bD_cF_{g-1}+\sum_{k=1}^{g-1}D_b F_k\, D_c F_{g-k}\right)\,.
\end{equation}
$D_a$ are K\"ahler covariant derivatives with respect to the chiral primaries and ${C_{\bar a}}^{bc}$ are defined using the tree-level three point function of chiral primaries $C_{abc}$, the K\"ahler metric\footnote{This metric is called the Zamolodchikov metric.} $G^{a\bar a}$ and its K\"ahler potential $K$:
\begin{equation}
 {C_{\bar a}}^{bc}=C_{\bar a\bar b\bar c}\,e^{2K}G^{b\bar b}\,G^{c\bar c}\,.
\end{equation}
These objects can be defined and studied using special geometry and we refer the reader to \cite{Bershadsky:1993cx} for more details.

\chapter{Topological Amplitudes and Higher Derivative Couplings}
\label{ch:TopoAmp}
\section{TST Partition Function from Type II Amplitudes}

Consider a generic \CY compactification of Type II string theory. The `twisted' or topological theory is obtained by performing the topological twist on the underlying SCFT as described above. A natural question is whether the resulting TST is related in some sense to the parent, `untwisted' theory. In other words, we would like to understand if the quantities calculated in TST can be reinterpreted in terms of physical ones in the original compactification.

Recall that the TST partition function is non-vanishing because of the $3(g-1)$ background charge acquired, after the twist, by the $U(1)$ current $J$. The latter can be bosonised in terms of a free field $H$,
\begin{equation}
J=i\sqrt{3}\,\partial H\,.
\end{equation}
In the original Type II theory, the background charge of $J$ can be obtained, at genus $g$, by coupling $H$ to the two-dimensional scalar curvature through the \WS sigma-model deformation
\begin{equation}
\delta S=\frac{i\sqrt{3}}{2}\int_{\Sigma_g} R^{(2)}\,H\,.
\end{equation}
Choosing a set of punctures $\{x_i, i=1,\ldots,2g-2\}$ on $\Sigma_g$, one can select a metric on the latter such that (recall that $\chi_g=-(2g-2)$)
\begin{equation}
R^{(2)}=-\sum_{i=1}^{2g-2} \delta^{(2)}(x-x_i)\,.
\end{equation}
Hence, the sigma-model deformation becomes
\begin{equation}
\delta S=-\frac{i\sqrt{3}}{2}\,\sum_{i=1}^{2g-2} H(x_i)\,,
\end{equation}
which amounts to the insertion of
\begin{equation}\label{UntwistedAmp}
\prod_{i=1}^{2g-2}e^{-\frac{i\sqrt{3}}{2}H}(x_i)
\end{equation}
in the partition function of the untwisted theory. The operator in \eqref{UntwistedAmp} is the internal part of the graviphoton vertex operator written in the $(-\tfrac{1}{2},-\tfrac{1}{2})$-picture, and this requires the insertion of $(g-1)$ PCOs in order to cancel the total background charge of the superghost. Together with the $(2g-2)$ PCOs necessary to soak up the $(\beta,\gamma)$ zero-modes at genus $g$, we have a total number of $3g-3$ PCOs to be inserted in \eqref{UntwistedAmp}, which necessarily contribute their $G^{-}$ (or $T_F$) part, as expected from the TST partition function. We are thus led to consider a genus $g$ scattering amplitude, in the untwisted theory, involving $2g-2$ graviphotons. In fact, the latter must be of definite self-duality which we choose to be anti-self-dual (see Appendix \ref{appendix:spinors}). In addition, one must insert two Riemann tensors, or else the amplitude would vanish by supersymmetry (or, equivalently, by the lack of zero-modes). This is also consistent with the supersymmetric term in the effective action calculated by this amplitude as presented in the next section.

We now prove this explicitly following \cite{Antoniadis:1993ze}. Namely, we calculate the genus $g$ scattering amplitude of two Riemann tensors and $2g-2$ graviphotons and show that it reproduces the TST partition function. For simplicity, we work in Type IIB and take the orbifold limit of the \CY, even though the result is valid for a generic \CY compactification. The orbifold acts on the complexified bosonic coordinates as
\begin{equation}
Z^a\rightarrow e^{2i\pi h_a}\,Z^a\,,
\end{equation}
with $h_3+h_4+h_5=0$ due to supersymmetry. Moreover, the complexified ten-dimensional \ST index, $M=1,\ldots,5$, is decomposed in terms of a \ST, $\ell=1,2$, and an internal one, $a=1,2,3$. The amplitude of interest is
\begin{equation}
\left<(V_{R_-})^2\,(V_{G_-})^{2g-2}\right>\,,
\end{equation}
where $R_-$ and $F^{G}_-$ are the anti-self-dual graviton and graviphoton. Their vertex operators are
\begin{alignat}{3}
&V_{R}(p)&=&\,h_{\mu\nu}(\partial X^{\mu}+i(p\cdotp\psi)\psi^{\mu})(\bar\partial X^{\nu}+i(p\cdotp\tilde\psi)\tilde\psi^{\nu})e^{ip\cdotp X}\,,\\
&V_{G}(p)&=&\,\epsilon_{\mu}p_{\nu}\,e^{-\frac{1}{2}(\varphi+\tilde\varphi)}\left(S^{\alpha}{(\sigma^{\mu\nu})_{\alpha}}^{\beta}\tilde S_{\beta}\,\Sigma\,\tilde\Sigma+S_{\dot\alpha}{(\sigma^{\mu\nu})^{\dot\alpha}}_{\dot\beta}\tilde S^{\dot\beta}\,\bar\Sigma\,\bar{\tilde\Sigma}\right)e^{ip\cdotp X}\,.
\end{alignat}
Here, because of anti-self-duality, we only use the first term of the graviphoton vertex operator, and the symmetric traceless polarisation of the graviton $h_{\mu\nu}$ is also chosen accordingly. Moreover, the polarisations satisfy the transversality conditions
\begin{equation}
\epsilon\cdotp p=0=p^{\mu}h_{\mu\nu}\,.
\end{equation}
Upon bosonisation of the \WS fermions,
\begin{align}
\psi^{M}=e^{\,i\phi_M}\,,
\end{align}
the (anti)-chiral \ST spin fields can be expressed as
\begin{align}
S_{1,2}&=e^{\pm\frac{i}{2}(\phi_1+\phi_2)}\,,\\
S^{\dot1,\dot2}&=e^{\pm\frac{i}{2}(\phi_1-\phi_2)}\,,
\end{align}
whereas the internal one is
\begin{equation}
\Sigma=e^{\,\frac{i}{2}(\phi_3+\phi_4+\phi_5)}\,.
\end{equation}
It is convenient to work in a particular kinematic configuration in which the anti-self-duality constraints are satisfied\footnote{Different kinematics can be obtained by Lorentz transformations.}, which we choose to be $p_1\neq0$ for the first graviton and $\bar p_1\neq 0$ for the second one, all the other momenta being set to zero. The contributions of the vertex operators are summarised in Table \ref{table:TopoAmpCharges}. Notice that in order to cancel the background charge of the superghost, the PCOs can only contribute $e^{\,\varphi}\,T_{F}$. All the other contributions can be deduced by charge conservation. For this, the positions $r_I$, $I=1,\ldots,3g-3$ are split into three different sets $r_{ai}$.

\begin{table}[H]
\begin{center}
\begin{tabular*}{\textwidth}{l @{\extracolsep{\fill}}llrrrrrr}\hline
Field & Pos.&Num.&$\phi_1$&$\phi_2$&$\phi_3$&$\phi_4$&$\phi_5$&$\varphi$\\[0.2cm]
\hline
$R_-$&$z$&1&$1$&$1$&$0$&$0$&$0$&$0$\\[0.2cm]
&$w$&1&$-1$&$-1$&$0$&$0$&$0$&$0$\\[0.2cm]
$G_-$&$x_i$&$g-1$&$+\frac{1}{2}$&$+\frac{1}{2}$&$+\frac{1}{2}$&$+\frac{1}{2}$&$+\frac{1}{2}$&$-\frac{1}{2}$\\[0.2cm]
&$y_i$&$g-1$&$-\frac{1}{2}$&$-\frac{1}{2}$&$+\frac{1}{2}$&$+\frac{1}{2}$&$+\frac{1}{2}$&$-\frac{1}{2}$\\[0.2cm]
PCO&$r_{3i}$&$g-1$&$0$&$0$&$-1$&$0$&$0$&$1$\\[0.2cm]
&$r_{4i}$&$g-1$&$0$&$0$&$0$&$-1$&$0$&$1$\\[0.2cm]
&$r_{5i}$&$g-1$&$0$&$0$&$0$&$0$&$-1$&$1$\\
\hline
\end{tabular*}
\end{center}
\caption{Contributions of the vertex operators to the amplitude.}
\label{table:TopoAmpCharges}
\end{table}

Factoring out the polarisations and the power of momenta corresponding to the effective coupling of interest $p_1^2\,\bar p_1^2\prod_{i}p_1^{(i)}\bar p_1^{(i)}$, the amplitude becomes\footnote{The bosonic part of the gravitons does not contribute, see discussion below. In addition, an integral over the positions of the physical vertex operators is understood.}

\begin{align}
\tilde F_g=&\,\int_{\mc M_g}\Bigl<\psi_1\,\psi_2\,\tilde\psi_1\,\tilde\psi_2(z,\bar z)\,\bar\psi_1\,\bar\psi_2\,\bar{\tilde\psi}_1\,\bar{\tilde\psi}_2(w,\bar w)\prod_{i}e^{-\frac{1}{2}(\varphi+\tilde\varphi)}\,S_1\,\tilde S_1\,\Sigma\,\tilde\Sigma(x_i,\bar x_i)\nonumber\\
&\times\prod_{i}e^{-\frac{1}{2}(\varphi+\tilde\varphi)}\,S_2\,\tilde S_2\,\Sigma\,\tilde\Sigma(y_i,\bar y_i)\prod_{I}e^{\,\varphi}\,T_F\,e^{\,\tilde\varphi}\,\tilde T_F(r_I,\bar r_I)\Bigr>\,,\label{TopoCoupling}
\end{align}
where the correlator is to be calculated at genus $g$ and a sum over the $2^{2g}$ spin structures is understood. Correlation functions on an arbitrary genus $g$ Riemann surface have been studied, for instance, in \cite{Verlinde:1986kw,Eguchi:1986ui,Dugan:1987qe}. Consider, an arbitrary spin-$\lambda$ bc-system \eqref{bcSystem} which has a background charge $Q=2\lambda-1$. At genus $g$, the Riemann-Roch theorem implies that the number of b zero-modes minus the number of c zero-modes is $I_{g,Q}=Q(g-1)$. Hence, a non-vanishing correlation function of b,c fields is such that the number of b-insertions minus the number of c-insertions is equal to $I_{g,Q}$. For a generic spin structure s, that is~\cite{Verlinde:1986kw,Eguchi:1986ui}
\begin{align}
&\Bigl<\prod_{i=1}^{N+I_{g,Q}}b(z_i)\prod_{k=1}^{N}c(w_k)\Bigr>_{g,s}=\nonumber\\
&\frac{\vartheta_s\left(\sum z_i-\sum w_k-Q\Delta\right)}{Z_1^{1/2}}\times\frac{\prod_{i<j}E^{\lambda(1-\lambda)}(z_i,z_j)\prod_{k<l}E^{\lambda(1-\lambda)}(w_k,w_l)}{\prod_{i,k}E^{\lambda(1-\lambda)}(z_i,w_k)}\frac{\prod_i\sigma^{Q}(z_i)}{\prod_k\sigma^Q(w_k)}\,,
\end{align}
where $\vartheta_s$ is the genus $g$ theta function, $E$ the prime form, $\sigma$ a $g/2$-differential, $Z_1$ the chiral determinant of the spin $(1,0)$ system, which can be thought of as the partition function of a chiral scalar field\footnote{In fact, the partition function of a chiral scalar field is equal to $Z_1^{-1/2}$.}, and $\Delta$ the Riemann class, \emph{c.f.} Appendix \ref{appendix:ModularFunctionsGeneral}. Using this identity, we calculate \eqref{TopoCoupling} by factorising the \ST, internal and superghost correlators.

In the \ST directions, one obtains
\begin{align}
&\prod_{\ell=1,2}\Bigl<e^{\,i\phi_{\ell}(z)}\prod_{i}e^{\,\frac{i}{2}\phi_{\ell}(x_i)}\,e^{-i\phi_{\ell}(z)}\prod_{i}e^{-\frac{i}{2}\phi_{\ell}(y_i)}\Bigr>=\nonumber\\
&\frac{\vartheta_s^2\left(\frac{1}{2}\sum_{i}(x_i-y_i)+z-w\right)}{Z_1}\times\frac{\prod_{i<j} E^{1/2}(x_i,x_j)E^{1/2}(y_i,y_j)\prod_{i}E(x_i,z)E(y_i,w)}{\prod_{i,j}E^{1/2}(x_i,y_j)E(z,w)^2\prod_{i}E(x_i,w)E(y_i,z)}\,,
\label{STcont}
\end{align}
whereas in the internal space, the contributions are
\begin{align}
&\prod_{a}\Bigl<\prod_{i}e^{\,\frac{i}{2}\phi_a(x_i)}\,e^{\,\frac{i}{2}\phi_a(y_i)}\,e^{-i\phi_a(r_{ai})}\Bigr>=\prod_{a}\frac{\vartheta_{s,h_a}(\sum_i\frac{1}{2}(x_i+y_i)-r_{ai})}{Z_1^{1/2}}\nonumber\\
&\times\prod_{a,i}\partial X_a(r_{ai})\times\frac{\prod_{i,j}E^{3/4}(x_i,y_j)\prod_{i<j}E^{3/4}(x_i,x_j)E^{3/4}(y_i,y_j)\prod_{a,i<j}E(r_{ai},r_{aj})}{\prod_{a}\prod_{i,j}E^{1/2}(x_i,r_{aj})E^{1/2}(y_i,r_{aj})}\,.
\label{Internalcont}
\end{align}
Finally, the superghost yield the correlation function
\begin{align}
&\Bigl<\prod_{i}e^{-\frac{1}{2}\varphi(x_i)}\,e^{-\frac{1}{2}\varphi(y_i)}\prod_a e^{\,\varphi(r_{ai})}\Bigr>=\frac{Z^{1/2}_1}{\vartheta_s\left(\frac{1}{2}\sum_{i}(x_i+y_i)-\sum_{I}r_I+2\Delta\right)}\nonumber\\
&\times\frac{\prod_{a,i,j}E^{1/2}(x_i,r_{aj})E^{1/2}(y_i,r_{aj})}{\prod_{i,j}E^{1/4}(x_i,y_j)\prod_{i<j}E^{1/4}(x_i,x_j)E^{1/4}(y_i,y_j)\prod_{I<J}E(r_I,r_J)}\times\frac{\prod_{i}\sigma(x_i)\sigma(y_i)}{\prod_{I}\sigma^2(r_I)}\,,
\label{Superghostcont}
\end{align}
and the integral over the zero-modes of the \ST bosons gives
\begin{equation}
\Bigl<\bigl|\prod_{\ell}\partial X^{\ell}_0\bigr|^2\Bigr>=\frac{1}{(\det\,\textrm{Im}\tau)^2}\,,
\end{equation}
$\tau$ being the period matrix of the Riemann surface. Joining all contributions, the effective coupling is
\begin{align}
\tilde F_g&=\int_{\mc M_g}\Bigl|\sum_{s}\frac{\vartheta_s^2\left(\frac{1}{2}\sum_{i}(x_i-y_i)+z-w\right)\prod_a\vartheta_{s,h_a}(\sum_i\frac{1}{2}(x_i+y_i)-r_{ai})}{Z_1^2\,\det(\textrm{Im}\,\tau)\,\vartheta_s\left(\frac{1}{2}\sum_{i}(x_i+y_i)-\sum_{I}r_I+2\Delta\right)}\nonumber\\
   &\times\frac{\prod_{i<j}E(x_i,x_j)E(y_i,y_j)\prod_i E(x_i,z)E(y_i,w)\prod_{a,i<j}E(r_{ai},r_{aj})}{E(z,w)^2\prod_i E(x_i,w)E(y_i,z)\prod_{I<J}E(r_I,r_J)}\nonumber\\
   &\times\frac{\prod_{i}\sigma(x_i)\sigma(y_i)}{\prod_{I}\sigma^2(r_I)}\prod_{a,i}\partial X_a(r_{ai})\epsilon_{a_1\ldots a_{3g-3}}\prod_{I}\int\mu_I(h_{a_I})\Bigr|^2\,.
\end{align}
The last term stems from the b-ghost zero-modes, with $\mu_I$ being the Beltrami differentials and $h_{a_I}$ the quadratic differentials, and serves as a measure over the moduli space $\mc M_g$. In order to be able to perform the sum over the spin structures using the Riemann summation identity \eqref{RiemannSummation}, it is convenient to fix the positions of the PCOs such that
\begin{equation}
-z+w+\sum_iy_i+2\Delta=\sum_I r_I\,,
\end{equation}
and the theta function in the denominator cancels out. In addition, the theta functions above are monodromy-invariant and one can readily perform the spin structure sum without any additional phase\footnote{Had we chosen the bosonic part of the graviton vertex operator, the result would have vanished after performing the spin-structure sum due to the Riemann vanishing theorem.}:
\begin{align}
\sum_{s}\Biggl[\vartheta_s(\frac{1}{2}\sum_{i}(x_i-y_i)+z-w)&\prod_{a=3}^{5}\vartheta_{s,h_a}(\sum_i\frac{1}{2}(x_i+y_i)-r_{ai})\Biggr]\nonumber\\
&=\vartheta(\sum_ix_i+z-w-\Delta)\prod_{a=3}^{5}\vartheta_{-h_a}(\sum_ir_{ai}-\Delta)\,.
\end{align}

Furthermore, the amplitude can be further simplified using the bosonisation identities of \cite{Verlinde:1986kw,AlvarezGaume:1987vm,Eguchi:1986ui}. More specifically, we identify a superghost system with background charge $Q=3$,
\begin{align}
\frac{Z^{1/2}_1}{\vartheta(-\sum r_I+3\Delta)\prod_{I<J}E(z_I,z_J)\prod_{I}\sigma^3(r_I)}&=Z_2^{-1}\,{\det}^{-1}\,h_{I}(z_J)\,,
\end{align}
and a bc-system with $Q=1$,
\begin{align}
\prod_{a=3}^{5}Z^{-1/2}\vartheta_{-h_a}(\sum r_{ai}-\Delta)\prod_{i<j}E(r_{ai},r_{aj})\prod_{i}\sigma(r_{ai})&=\prod_{a=3}^{5} Z_1\,\det\,\omega_{-h_a,i}(r_{aj})\,.
\end{align}
Similarly, the remaining parts of the amplitude can be recognised as the following chiral determinants:
\begin{align}
\frac{\vartheta(\sum x_i+z-w-\Delta)}{Z^{1/2}_1}\frac{\prod_{i<j}E(x_i,x_j)\prod_{i}E(x_i,z)\sigma(x_i)}{E(z,w)\prod_i E(x_i,w)}&=Z_1\,\det\,\omega_i(\hat x_j)\,,\\
\frac{\vartheta(\sum y_i-z+w-\Delta)}{Z^{1/2}_1}\frac{\prod_{i<j}E(y_i,y_j)\prod_{i}E(y_i,w)\sigma(y_i)}{E(z,w)\prod_i E(y_i,z)}&=Z_1\,\det\,\omega_i(\hat y_j)\,.
\end{align}
Here, we have defined for simplicity
\begin{align}
\hat x_i&\equiv x_i,z\,,\\
\hat y_i&\equiv y_i,w\,.
\end{align}
The total amplitude can now be compactly expressed in terms of the chiral determinants above:
\begin{align}
\tilde F_g=\int\Bigl|\frac{\det\,\omega_i(\hat x_j)\det\,\omega_i(\hat y_j)\,\prod_a\det(\partial X_a\,\omega_{-h_a,i}(r_{aj}))}{\det(\textrm{Im}\tau)\,\det\,h_I(r_J)}\,\epsilon_{a_1\ldots a_{3g-3}}\prod_{I}\int\mu_I\cdotp h_{a_I}\Bigr|^2\,.
\end{align}
Using the definition of the period matrix, it is easy to show that
\begin{equation}
\int_{\hat x}\det\,\omega_i(\hat x_j)=g!\,\det\,\textrm{Im}\tau\,.
\end{equation}
Moreover, summing over all the possible partitions of the positions $r_{ai}$ cancels the remaining determinants, which leads to
\begin{align}
\tilde F_g&=(g!)^2\int_{\mc M_g}\Bigl<\Bigl|\det\int_{\Sigma_g}\mu_I\cdotp h_{J}\Bigr|^2\Bigr>=(g!)^2\,F_g\,.
\end{align}
Hence, the coupling of two anti-self-dual gravitons and $2g-2$ anti-self-dual graviphotons in Type II computes the genus $g$ topological string partition function\footnote{In the last step, we have used the definition $G^{-}=\partial\,X^a\,\psi^a$ in the twisted theory and the bosonisation identities of \cite{Verlinde:1986kw} in \eqref{Fg}.}.

One can repeat the calculation above at higher genus by considering the same scattering amplitude (with $2g-2$ graviphotons) at genus $g'>g$. It is easy to show that it vanishes identically due to a lack of fermionic zero-modes. In fact, in turns out that this class of couplings is super-protected: it receives neither perturbative nor non-perturbative corrections, as discussed in the following section.

\section{Effective Field Theory}

Following Gopakumar and Vafa \cite{Gopakumar:1998ii,Gopakumar:1998jq}, the generating function of the A-model partition function on a \CY is obtained by integrating out all massive BPS states corresponding to D-branes wrapping two-cycles of the \CY in the background of a constant anti-self dual graviphoton field strength. Due to the 
anti-self duality, the latter only couples to the spin of the D-brane states along a particular $SU(2)$ of the four-dimensional Lorentz group. Specifically, in terms of the $\Omega$-supergravity background \cite{Moore:1997dj,Lossev:1997bz}, this means that the topological partition function only depends on one deformation parameter, $\epsilon_-$, that is identified with the topological string coupling. Thus, from the point of view of the  string effective action, we are naturally led to study $\mathcal{N}=2$ higher derivative F-terms including the anti-self-dual graviphoton field strength tensor.

Consider the following series of effective couplings in the standard four-dimensional superspace $\mathbb{R}^{4|8}\sim\{x^\mu,\theta^i_\alpha,\bar{\theta}_i^{\dot{\alpha}}\}$ \cite{Antoniadis:1993ze}:
\begin{align}
\mathcal{I}_g=&\int d^4x \int d^4\theta\, \mathcal{F}_g(X)\,(W_{\mu\nu}^{ij} W_{ij}^{\mu\nu})^g &&\text{for} && g\geq1\,,  \label{BpsStandard}
\end{align}
which is a $\frac{1}{2}$-BPS F-term since it is invariant under half of the supercharges. In addition, $W_{\mu\nu}^{ij}$ is the supergravity multiplet and we have introduced (anti-symmetrised) indices $i,j=1,2$ for the $SU(2)_R$ R-symmetry group. $W_{\mu\nu}^{ij}$ contains the graviphoton field-strength $F^G$, the field strength tensor $B_{\mu\nu}^i$ of an $SU(2)$ doublet of gravitini and the Riemann tensor:
\begin{align}
W_{\mu\nu}^{ij}=F^{G
,ij}_{(-),\mu\nu}+\theta^{[i} B_{(-),\mu\nu}^{j]}-(\theta^i\sigma^{\rho\tau}\theta^j) R_{(-),\mu\nu\rho\tau} + \cdots\,
\end{align}
The subscript $(-)$ denotes the anti-self-dual part of the corresponding field strength tensor. The coupling function $\mathcal{F}_g$ in (\ref{BpsStandard}) only depends on holomorphic vector multiplets, which contain a complex scalar $\phi$, an $SU(2)_R$ doublet of chiral spinors $\lambda_\alpha^i$ as well as an anti-self-dual field-strength tensor of a space-time vector $F^{\mu\nu}_{(-)}$:
\begin{align}
X^I=\phi^I+\theta^i\lambda_i^I+\tfrac{1}{2}F_{(-)\,\mu\nu}^I \epsilon_{ij}(\theta^i\sigma^{\mu\nu}\theta^j) +\cdots\,
\end{align}
We have added an additional label $I$ to indicate that there are several vector multiplets. In fact one of them, denoted $X^0$, is not physical but rather serves as a compensator of degrees of freedom in the formulation of ${\mc N}=2$ supergravity \cite{Cremmer:1984hj,de Wit:1984px}. The physical moduli are then the lowest components of the projective multiplets:
\begin{align}
\hat{X}^I:= \frac{X^I}{X^0}\,.
\end{align}
Upon explicitly performing the integral over the Grassmann variables, (\ref{BpsStandard}) induces a component term that was calculated in the previous section:
\begin{align}\label{BpsTerm}
\mathcal{I}_g=\int d^4x\, \mathcal{F}_g(\phi)\,R_{(-)\,\mu\nu\rho\tau}R_{(-)}^{\mu\nu\rho\tau}\,
\left[F^G_{(-)\,\lambda\sigma}F^{G\,\lambda\sigma}_{(-)}\right]^{g-1}+\cdots\,
\end{align}

Recall that the Weyl multiplet $W$ has conformal weight $1$. Since the integrand in \eqref{BpsStandard} must have weight $2$, the coupling $\mc F_g(X)$ is a homogeneous function of degree $2-2g$. Moreover, upon fixing the superconformal gauge, the compensator field $\phi^0$ can be expressed in terms of the K\"ahler potential $K$ and the string coupling:
\begin{equation}
\phi^0=g_s^{-1}\,e^{\,K/2}\,.
\end{equation}
Therefore, the coupling function can be written in terms of the physical moduli and, to the lowest order in the Grassmann variables,
\begin{equation}\label{DilatonDependence}
\mc F_g(\phi)=(\phi^0)^{2-2g}\mc F_g(\hat\phi)=g_s^{2(g-1)}\,e^{(1-g)K}\,\mc F_g(\hat\phi)\,.
\end{equation}
The K\"ahler potential depends only on vector multiplets and, in Type II, the dilaton belongs to a hypermultiplet. Moreover, due to the absence of mixing between vector and hypermultiplets in $\N=2$ supergravity, the K\"ahler potential cannot depend on $g_s$. Consequently\footnote{Recall that string loops are counted by the power of $g_s$.}, the couplings $F_g$ calculated in the previous section appear, as expected, at loop order $g$. In fact, they do not receive any further perturbative or non-perturbative corrections.

One way to see this is to analyse the dependence of the gravitational fields on $g_s$. The term \eqref{BpsTerm} in the effective action is written in the string frame. One can reach the Einstein frame by reabsorbing the dilaton dependence of the Einstein-Hilbert term:
\begin{equation}
\sqrt{\det\,G}\,R\,g_s^{-2}\rightarrow\sqrt{\det\,G}\,R\,,
\end{equation}
which is obtained by rescaling the metric as $G\rightarrow G\,g_s^{2}$. On the other hand, the graviphoton, as a RR field, contains a single power of $g_s$ such that $R^2\,(F^{G})^{2g-2}$ has a power of $g_s^{2g-2}$. Going to the Einstein frame, there are additional powers of the dilaton arising from the metric factors that contract the Riemann tensors and the graviphotons, leading to a power of $g_s^{2(2-2g)}$. All in all, including the loop factor of $g_s^{2g-2}$, the total dilaton dependence vanishes and the couplings $F_g$ are indeed super-protected.

\section{Heterotic Dual}

The gravitational couplings \eqref{BpsTerm} can be used to test string dualities. For instance, as discussed in Section \ref{sec:dualities}, Heterotic string theory on $K3\times T^2$ is dual to Type II on a $K3$ fibration. Hence, in the latter case, it is interesting to calculate the gravitational couplings above in the dual Heterotic theory. However, the non-renormalisation theorem above does not carry over through the duality map since the Heterotic dilaton belongs to a vector multiplet\footnote{Under Heterotic-Type II duality, the heterotic dilaton is mapped to the volume modulus of the $K3$ fibration base.}. In fact, the K\"ahler potential possesses a dilaton dependence of $\ln g_s^2$ so that, using \eqref{DilatonDependence}, $F_g$ already appears at genus one\footnote{This is true for $g>1$. For $g=0,1$, there are additional tree-level contributions.}. Consequently, in the weak coupling regime,
\begin{equation}
F_g^{\textrm{Het}}|_{\textrm{1-loop}}=F_{g}^{\textrm{Type II}}|_{\textrm{Im}\,S\rightarrow\infty}\,.
\end{equation}

One can check this explicitly by calculating the one-loop effective coupling of two gravitons and two anti-self-dual graviphotons \cite{Antoniadis:1995zn}. Notice that since this is a one-loop amplitude, upon evaluating the correlation function of the vertex operators, one can perform the integral over the fundamental domain of the torus, using, for example, the techniques developed in \cite{Angelantonj:2011br,Angelantonj:2012gw}, and obtain an explicit expression for the perturbative part of $F_g$. The derivation of the $F_g$'s in Heterotic is, however, not shown in this section since the more general setup of `refined couplings' is discussed in detail in Section \ref{Sec:Hetamps}.

In a word, we have seen the TST partition function is calculated, in Type II, by gravitational coupling which are super-protected due to $\N=2$ supergravity, and arise as higher derivative F-terms in the effective action. In addition, when the Type II compactification admits a Heterotic dual, these couplings start receiving contributions at one-loop. This renders an explicit evaluation possible at the perturbative level and gives a way to generalise them in a natural way as described in Section \ref{Sec:Hetamps}.

From the low energy field theory point of view, $F_0$ is nothing but a generalisation of the Seiberg-Witten prepotential in the presence of gravity. In addition, the terms $F_{g>0}$ are $R^2$ corrections to the latter. In terms of the underlying $\mc N=2$ gauge theory, they arise as a deformation of the Seiberg-Witten theory using the $\Omega$-background and which we briefly present in the following section.

\chapter{N=2 Gauge Theory from String Theory}
\label{ch:GaugeTheoryFromStringTheory}
As mentioned previously, a full-fledged non-perturbative definition of superstring theory is lacking. However, one can still probe some effects beyond perturbation theory using D-branes. The latter can be used to describe field theory instantons but also purely stringy, or exotic ones.

Recall that in field theory, instantons are finite action solutions of the equations of motion. We are interested in four-dimensional gauge theories in which they are given by gauge fields whose field strength is self-dual\footnote{Anti-instantons are solutions with opposite self-duality.}:
\begin{equation}
 F_{\mu\nu}=(\ast F)_{\mu\nu}\equiv\frac{1}{2}{\epsilon_{\mu\nu}}^{\rho\sigma}F_{\rho\sigma}\,.
\end{equation}
Consider the (Euclidean) Yang-Mills action with gauge group $G$ and a theta angle
\begin{equation}
 S_{YM}=\frac{1}{g^2}\int\textrm{Tr}(F\wedge\ast F)-\frac{i\theta}{8\pi^2}\int\textrm{Tr}(F\wedge F)\,.
\end{equation}
A finite action solution is a pure gauge because the field strength must vanish at infinity:
\begin{equation}
 A_{\mu}(x)\sim \bar{U}(x)\partial_{\mu}U(x)\,,
\end{equation}
with $x\in S^3$. Since $U$ defines a map from $S^3$ to $G$, non-trivial instanton configurations are classified by the third homotopy group of $G$. In most cases of interest, the latter is a simple group so that
\begin{equation}
 \pi_3(G)=\mb Z\,.
\end{equation}
Therefore, an instanton is characterised by a topological quantity, its \emph{winding number} $k\in\mb Z$ given by\footnote{For instantons, $k>0$, whereas for anti-instantons $k<0$.}
\begin{equation}
 k=\frac{1}{8\pi^2}\int_{\mb R^4} F\wedge F\,,
\end{equation}
and the corresponding (positive) action is
\begin{equation}\label{InstAction}
 S_{k}=\left(\frac{8\pi^2}{g^2}+i\theta\right)|k|\,.
\end{equation}
Notice that the factor $e^{-S_k}$ breaks the perturbative $U(1)$ symmetry of the theta angle to a discrete one.


\section{D-brane bound states}

String theory must also contain the same instanton configurations since its low energy limit is a gauge/field theory. Intuitively, they correspond, from the four-dimensional \ST point of view, to point-like objects that can be extended in some internal dimensions. The simplest such objects are (Euclidean) strings wrapping an internal two-cycle, but these are perturbative objects from the string theory point of view\footnote{They are, however, non-perturbative in $\alpha'$.}. In order to obtain stringy non-perturbative objects, we must consider, instead, D-branes wrapping some cycles in the internal space, and they are sometimes referred to as Euclidean branes (E-branes) or D-instantons. The action of this configuration is given by the Dp-brane couplings and, in the absence of background fields, that is
\begin{equation}
 T=\frac{\mu_p}{g_s}\int d^{p+1}\xi\sqrt{G}+i\mu_p\int C_{p+1}\,.
\end{equation}
The first term gives the volume of the cycle wrapped by the Dp-brane and $T$ can be regarded as a complexified modulus\footnote{In Type IIA/B, it is a complex structure/K\"ahler modulus.} generalising \eqref{InstAction}. Moreover, the second term breaks the continuous shift symmetry of the RR scalar $a_{p+1}\sim\int C_{p+1}$ to a discrete subgroup generated by $a_{p+1}\rightarrow a_{p+1}+2\pi$.

The gauge theory can be realised in terms of a stack of D-branes extending in the four-dimensional \ST and, possibly, in some internal dimensions. One can show that gauge theory instantons arise as Dp-D(p+4) bound states \cite{Witten:1995gx,Douglas:1996uz}. For more general brane configurations, one obtains exotic instantons. In fact, the D(p+4)-brane effective action contains the following term:
\begin{equation}
 \mu_{p+4}\int C_{p+1}\wedge \textrm{Tr}(F\wedge F)\,.
\end{equation}
Hence, an instanton configuration on a D(p+4)-brane carries a charge under $C_{p+1}$: it is a Dp-brane. Usually, one considers a stack of $N$ D(p+4)-branes in order to realise, for example, an $(S)O(N)$ gauge theory (or a subgroup thereof), together with $k$ Dp-branes to probe k-instantons. Of course, in a consistent string theory, tadpole cancellation restricts $N$ to a particular value and this is implicitly understood throughout the manuscript. Moreover, as shown in \cite{Witten:1995gx}, the $k$ Dp-brane, when collapsing to the same point, generate an $Sp(k)$ gauge group\footnote{For $U(N)$, $Sp(N)$ gauge theories, the instanton group is $U(k)$, $O(k)$ respectively.}.

In the open string sector, there are different kind of states depending on the location of the string endpoints. When the latter are on the D(p+4)-branes\footnote{We focus on the $\N=2$ case.}, the corresponding states are vector multiplets whose scalars $X_{p+4}$ parametrise the position of the branes in the transverse space. On the other hand, mixed states arising from open strings stretching between Dp- and D(p+4)-branes belong to hypermultiplets. If $X_{p}$ denotes the position of the Dp-branes in the transverse space to the D(p+4)-branes, then $X_{p+4}=X_p$ is a configuration where the D-instantons are `stuck' to the gauge branes: this is the \emph{Higgs branch} of the moduli space, and the instanton has a finite size parametrised by the hypermultiplets. On the other hand, $X_{p+4}\neq X_p$ is the \emph{Coulomb branch} and the instanton is a point-like configuration located away from the D(p+4)-branes.


\section{ADHM instantons}

In this section we briefly review the Atiyah-Drinfeld-Hitchin-Manin (ADHM) construction \cite{Atiyah:1978ri} of gauge theory instantons and show, quantitavely, how the instanton effective action arises, in string theory, from the degrees of freedom of a Dp-D(p+4) system. In addition, for later purposes, we also implement a non-trivial $\Omega$-background \cite{Nekrasov:2002qd} as a $U(1)^2$-deformation of the ADHM instantons. Our notation follows \cite{Atiyah:1978ri} (see \emph{e.g.} \cite{Dorey:2002ik} for a review). At the core of the (undeformed) ADHM construction is a specific ansatz for the gauge connection. Requiring this ansatz to be a solution of the Yang-Mills equations of motion gives rise to a number of constraints that can be encoded in an action principle. To obtain a deformation of this ADHM action, we implement a particular space-time $U(1)^2$ rotation, with parameters $\epsilon_{1,2}$ (the $\Omega$ background).\footnote{We also use the notation $\epsilon_{\pm}=\frac{\epsilon_1\pm\epsilon_2}{2}$.}

To be more specific, consider the following ansatz for the $SU(N)$ gauge connection \begin{align}\label{GaugeConnection}
&(A_{\mu})_{uv}=\bar U_{u}^a\partial_{\mu}U_{av}\,,&&\text{with} &&\bar U^a_uU_{a,v}=\delta_{uv}\,.
\end{align}
Here we have introduced the ADHM index $a=1,\ldots,2k+N$ and $u,v=1\ldots,N$, with $k$ being the instanton number.\footnote{In the following we mostly suppress the indices $u,v$ to keep the notation simple.} The ansatz \eqref{GaugeConnection} is a solution of the Yang-Mills equations, $DF=0$, with $D$ being the covariant derivative (with respect to the gauge connection), if the corresponding field-strength is self-dual, \emph{i.e.} $F=\ast F$. This can most easily seen by using the Bianchi identity as follows:
\begin{align}
 0=D\ast F=DF=0\,.
\end{align}

To find a solution for the matrix $U$ which has this property, we first notice that the operator ${P_{a}}^b\equiv U_a\,\bar U^b$ is a projector preserving $U$, \emph{i.e.} $P^2=P$ and $PU=U$. The field strength of $A_{\mu}$ can be written in terms of $P$ as
\begin{equation}\label{FieldStrength}
 F_{\mu\nu}=\partial_{[\mu}\bar U^a({\delta_{a}}^b-{P_a}^b)\partial_{\nu]}U_b\,.
\end{equation}
One is thus led to write the ansatz $1\!\!1-P=\Delta\,f\,\bar\Delta$, where the $[N+2k]\times[2k]$ matrix $\Delta$ is called the  \emph{ADHM matrix}, while $f$ is an arbitrary Hermitian matrix. If we assume that $\Delta$ is linear in the space-time coordinates,
\begin{equation}
 \Delta_{ai\dot\alpha}=\mathfrak{a}_{ai\dot\alpha}+\mathfrak{b}^{\alpha}_{ai}\,x_{\alpha\dot\alpha}\,,
\end{equation}
and that the matrix $f$ is `diagonalised' as
\begin{equation}
 {{f_{ij}}^{\dot\alpha}}_{\dot\beta}=f_{ij}{\delta^{\dot\alpha}}_{\dot\beta}\,,
\end{equation}
then the field strength \eqref{FieldStrength} is self-dual, being proportional to the matrix $\sigma_{\mu\nu}$. Here, the space-time coordinates $x_{\mu}$ are written with spinor indices, \emph{i.e.} $x_{\alpha\dot\alpha}=x_{\mu}(\sigma^{\mu})_{\alpha\dot\alpha}$ and we have introduced the  instanton index $i=1,\ldots,k$. Notice that $f$ can be written in terms of the ADHM matrix as
\begin{equation}\label{Constraint}
 \bar\Delta\Delta=f^{-1} 1\!\!1\,.
\end{equation}
In this way, the construction of the ADHM solution is reduced to finding a consistent set of matrices $\mathfrak{a}$ and $\mathfrak{b}$. The symmetries of the problem allow us to write \cite{Dorey:2002ik}
\begin{align}
&\mathfrak{a}=\left(\begin{array}{c}
w \\
a' \end{array} \right)\,,&&\mathfrak{b}=\left(\begin{array}{c}
0_{[N]\times[2k]} \\
1\!\!1_{[2k]\times[2k]} \end{array} \right) \,,\label{ADHMansatz}
\end{align}
where $w$ and $a'_{\alpha\dot\alpha}$ are $[N]\times[2k]$ and $[2k]\times[2k]$ matrices respectively. They are usually parametrised as follows:
\begin{align}
 w&=\left(\begin{array}{cc}
 J & I^{\dag} \end{array} \right)\,,\\
 a'&=\left(\begin{array}{cc}
B_1 & -B_2^{\dag} \\
B_2 & B_1^{\dag} \end{array} \right)\,.
\end{align}
Here $B_{1,2}$ are $[k]\times[k]$ and $J,I^{\dag}$ are $[N]\times[k]$. In terms of the symmetry group $U(k)$ of the instantons, $I$, $J$ transform in the fundamental, anti-fundamental representation, whereas $B_{1,2}$ transform in the adjoint. The condition \eqref{Constraint} now translates into the \emph{ADHM equations}
\begin{align}
 \mu_{\mb C}=&IJ+[B_1,B_2]=0\,,\nonumber\\
 \mu_{\mb R}=&II^{\dag}-J^{\dag}J+\sum_{l=1,2}[B_l,B_l^{\dag}]=0\,.\label{ADHMequations}
\end{align}

The equations \eqref{ADHMequations} are invariant under the action of $U(k)\times SU(n)$, where $SU(n)$ acts as a global gauge transformation and $U(k)$ reflects the residual symmetry of the ADHM data. In fact, it is important to quotient the space one obtains after implementing \eqref{ADHMequations} by this $U(k)$ in order to obtain the ADHM manifold $\mc M_k$ of dimension $4Nk$:
\begin{equation}
 \mc M_k=\frac{M}{U(k)}\,.
\end{equation}
In addition, one can also include an equivariant $U(1)^2$ action parametrising rotations in space-time with angles $\epsilon_{1,2}$. The total resulting action can be written on the ADHM data as
\begin{eqnarray}
 &B_l&\rightarrow e^{i\phi}\,B_l\,e^{-i\phi}e^{i\epsilon_l}\,,\\
 &I&\rightarrow e^{i\phi}\,I\,e^{-ia}e^{-i\epsilon_+}\,,\\
 &J&\rightarrow e^{ia}\,J\,e^{-i\phi}e^{-i\epsilon_+}\,,
\end{eqnarray}
where $a=\textrm{diag}(a_1,\ldots,a_N)$ is the $U(1)^{N-1}\subset SU(N)$ parameter of the gauge transformation $e^{ia}$, and, similarly, $e^{i\phi}\in U(k)$. Physically, $a$ corresponds to the vev of the gauge theory vector multiplet and $\phi$ is the adjoint complex scalar (Higgs field). In the context of $\N=2$ gauge theory, the previous bosonic moduli are supplemented with their fermionic superpartners appearing in the action below. The ADHM instanton action can finally be written in the form of a total BRST variation:
\begin{equation}\label{DeformedADHM}
 S=Q[\chi\cdotp\mathfrak{m}+\Psi\cdotp V(\bar\phi)]\,.
\end{equation}
Here $\chi$ acts as a Lagrange multiplier reinforcing (\ref{ADHMequations}) as formulated with the ansatz (\ref{ADHMansatz}), while $\phi$ can be interpreted as a gauge connection implementing an equivariant $U(1)^2$ action on the instantons. In addition, we have defined
\begin{equation}
 \mathfrak{m}\equiv\left(\begin{array}{c}
                          \mu_{\mb R}\\ \mu_{\mb C}
                         \end{array}
		    \right)\,.
\end{equation}
The action of $Q$ and the definition of the vector field $V$ are \cite{Nekrasov:2002qd}

\begin{center}
\begin{tabular}{ll}
$QB_l=\Psi_{B_l}\,,$ & $Q\Psi_{B_l}=[\phi,B_l]+\epsilon_lB_l\,,$\\[2ex]
$QI=\Psi_{I}\,,$ & $Q\Psi_{I}=\phi I-Ia-\epsilon_+I\,,$\\[2ex]
$QJ=\Psi_{J}\,,$ & $Q\Psi_{J}=-J\phi+aJ-\epsilon_+J\,,$\\[2ex]
$Q\chi_{\mb R}=H_{\mb R}\,,$ & $QH_{\mb R}=[\phi,\chi_{\mb R}]\,,$\\[2ex]
$Q\chi_{\mb C}=H_{\mb C}\,,$ & $QH_{\mb C}=[\phi,\chi_{\mb C}]+2\epsilon_{+}\,\chi_{\mb C}\,,$\\[2ex]
$Q\phi=0\,,$ & $Q\bar\phi=[\phi,\bar\phi]\,,$\\[2ex]
\multicolumn{2}{c}{$\Psi\cdotp V(\bar\phi)=\textrm{Tr}\left[\sum_{l=0,1}\Psi_{B_l}[\bar\phi,B_l^{\dag}]+ \bar\Psi_{B_l}[\bar\phi,B_l]-\bar\Psi_I\bar\phi I+I^{\dag}\bar\phi\Psi_{I}-J\bar\phi\bar\Psi_J+\Psi_J\bar\phi J^{\dag}\right]$.}
\end{tabular}
\end{center}
Here, $l=1,2$ and
\begin{equation}
 \mathfrak{H}\equiv\left(\begin{array}{c}
                          H_{\mb R}\\ H_{\mb C}
                         \end{array}
		    \right)
\end{equation}
implements the fermionic constraints as a `superpartner' of $\mathfrak{m}$. Moreover, the corresponding `scalar product' is
\begin{equation}
 \chi\cdotp\mathfrak{m}=\textrm{Tr}\left[\chi_{\mb R}\,\mu_{\mb R}+\frac{1}{2}(\chi_{\mb C}^{\dag}\,\mu_{\mb C}+\chi_{\mb C}\,\mu_{\mb C}^{\dag})\right]\,.
\end{equation}
Finally, defining $A=(B_l,I,J)$ and $G_k$ to be the instanton group, the path integral leading to Nekrasov's partition function is \cite{Nekrasov:2002qd}
\begin{equation}\label{NekPartK}
 Z^{\textrm{Nek}}_{k}(a,\epsilon_1,\epsilon_2)=\frac{1}{\textrm{Vol}(G_k)}\int\,D\phi\,D\bar\phi\,D\eta\,D\chi\,DH\,DA\,D\Psi\,e^{-Q\left(\chi\cdotp\mathfrak{m}+\Psi\cdotp V(\bar\phi)+\eta[\phi,\bar\phi]\right)}\,.
\end{equation}


\section{String theory realisation of ADHM instantons}

As discussed previously, a Dp-D(p+4) system is a string theory realisation of gauge theory instantons. Hence, one should be able to reproduce the aforementioned ADHM action by identifying the relevant massless fields arising from such a construction, at least in the absence of the $\Omega$-background ($\epsilon_{1,2}=0$)\footnote{This case is studied in Chapter \ref{ch:ADHMinst}.}. For definiteness, we focus on the D(-1)-D3 case \cite{Douglas:1996uz}, though it is easy to generalise it to any Dp-D(p+4) system. For instance, the D5/D9 system used in Chapter \ref{ch:ADHMinst} is related to the latter by T-duality.

Consider a system of $N$ D3-branes and $k$ D(-1)-branes in a Type IIB orientifold. In the simplest case of toroidal compactification on $T^6$, this leads to the maximally supersymmetric case of $\mc N=4$ $SU(N)$ gauge theory. For a $\mb Z_2$ orbifold as in Chapter \ref{ch:ADHMinst}, the amount of supersymmetry is reduces by half. In addition, we have already seen that the D-brane charge is usually constrained, at the quantum level, by tadpole cancellation, though for the purpose of classical dimensional reduction this is not relevant. With this remark in mind, we keep $N$ generic. The Euclidean Lorentz group is broken,
\begin{equation}\label{LorentzDecomp}
 SO(10)\rightarrow SO(4)\times SO(6)\,
\end{equation}
whose covering group is $\mathfrak{G}=SU(2)_{L}\times SU(2)_R\times SU(4)$. Here $SU(4)$ plays the role of the R-symmetry for the $\mc N=4$ theory. The ten-dimensional index $M\in\llbracket0,9\rrbracket$ is decomposed into longitudinal and transverse directions with respect to the \ST-filling D3s:
\begin{equation}
 M\rightarrow(\mu,a)\in\llbracket0,3\rrbracket\times\llbracket4,9\rrbracket\,.
\end{equation}
The string coordinates obey Neumann boundary conditions only along the longitudinal directions of the D3-branes. In particular, the boundary conditions for an open string can be NN, ND or DD depending on the location of its endpoints. The ten-dimensional spin field preserved by the GSO projection is decomposed by \eqref{LorentzDecomp} as
\begin{equation}
 S\rightarrow(S_{\alpha}S_{A},S^{\dot\alpha}S^{A})\,,
\end{equation}
where $\alpha,\dot\alpha$ denote chiral, anti-chiral spinors in four dimensions and the upper, lower $A$ index refers to fundamental, anti-fundamental representations of $SO(6)$, see Appendix \ref{appendix:spinors} for more details. The vertex operators for the supersymmetry currents are
\begin{align}
 j_{\alpha A}&=e^{-\frac{\varphi}{2}}S_{\alpha}S_{A}\,,\nonumber\\
 j^{\dot\alpha A}&=e^{-\frac{\varphi}{2}}S^{\dot\alpha}S^{A}\,,\label{SusyCurrents}
\end{align}
and similarly for the right-movers. The supercharges $Q$ are simply the integrals of these currents. Each set of D-branes preserves different supercharges according to \eqref{SusyDbranes}. In particular, due to the left-right identifications, these are combinations of the form
\begin{align}
 Q_{\alpha A}\pm \widetilde Q_{\alpha A}\,,&&Q^{\dot\alpha A}\pm\widetilde Q^{\dot\alpha A}\,.
\end{align}
For Dp-branes, the boundary conditions for the spin fields is of the form
\begin{equation}
 S=(\prod_{i=0}^{p}\Gamma^{i})\widetilde S
\end{equation}
at the boundary ($z=\bar z$). More precisely, for the D(-1)-branes, that is
\begin{align}
 S_{\alpha}S_{A}&=\widetilde S_{\alpha}\widetilde S_{A}\,,\\
 S^{\dot\alpha}S^{A}&=\widetilde S^{\dot\alpha}\widetilde S^{A}\,,
\end{align}
so that the preserved supercharges are $Q_{\alpha A}-\widetilde Q_{\alpha A}$ and $Q^{\dot\alpha A}-\widetilde Q^{\dot\alpha A}$. For the D3-branes, the boundary conditions are\footnote{In fact, one could choose the opposite sign with respect to the D(-1) boundary conditions, and this configuration would correspond to anti-instantons.}
\begin{align}
 S_{\alpha}S_{A}&=-\widetilde S_{\alpha}\widetilde S_{A}\,,\\
 S^{\dot\alpha}S^{A}&=\widetilde S^{\dot\alpha}\widetilde S^{A}\,.
\end{align}
Hence, the D3 boundary preserves $Q_{\alpha A}+\widetilde Q_{\alpha A}$ and $Q^{\dot\alpha A}-\widetilde Q^{\dot\alpha A}$. Notice that only the latter is preserved by both sets of branes. In other words, $\mc N=4$ supersymmetry of the gauge theory is broken down to $\mc N=2$ by the instantons.

Let us now discuss the massless spectrum obtained from the open string sector and which come from the reduction of the ten-dimensional $\mc N=1$ multiplets. The endpoints of the open string can be located on the D3- or the D(-1)-branes, and this leads to three different sectors denoted by 3-3, 3-(-1)/(-1)-3 and (-1)-(-1).
\begin{enumerate}
\item 3-3 sector\\
The massless excitations consist of a number of $\mathcal{N}=4$ vector multiplets, each of which containing a vector field $A^\mu$, six real scalars $\phi^a$, as well as two gaugini $(\Lambda^{\alpha A}, \Lambda_{\dot\alpha A})$ which transform in the $(\mathbf{2},\mathbf{1},\mathbf{4})\oplus(\mathbf{1},\mathbf{2},\mathbf{\bar 4})$ representation of $\mathfrak{G}$. The bosonic degrees of freedom stem from the NS sector, while the fermionic ones from the R sector. This sector taken separately realises an $\mc N=4$ super-Yang-Mills theory living on the four-dimensional \ST.
\item (-1)-(-1) sector\\
These states are moduli (i.e. non dynamical fields) from a string perspective, due to the instantonic nature of the corresponding D(-1)-branes. Indeed, the states in this sector cannot carry any momentum because of the Dirichlet boundary conditions in all directions. From the NS sector, we have ten bosonic moduli, which we write as a real vector $a^\mu$ and six scalars $\chi^a$. From the point of view of the gauge theory living on the world-volume of the D3-branes, $a^\mu$ corresponds to the position of gauge theory instantons. In the Ramond sector, there are sixteen fermionic moduli, which are conveniently written as $M^{\alpha A}$, $\lambda_{\dot\alpha A}$.
\item (-1)-3 and 3-(-1) sectors\\
Also this sector contains moduli from a string point of view. From the NS sector, the fermionic coordinates have integer-moded expansions whose zero-modes give rise to two Weyl spinors of $SO(4)$ and are usually called $(\omega_{\dot\alpha},\bar\omega_{\dot\alpha})$. Notice that these fields all have the same chirality, which in our case is anti-chiral, owing to the specific choice of boundary conditions above. The opposite choice in order to describe anti-instantons would have lead to chiral fields instead\footnote{Recall that instantons and anti-instantons have, by definition, opposite self-duality which translates into an opposite chirality of the corresponding field strengths when expressed in a spinor basis using sigma matrices.}. In fact this can be derived by imposing locality of the OPE of these fields with the conserved supersymmetry currents (see below). From a SYM point of view, these fields control the size of the instanton \cite{Dorey:2002ik}. In the R sector, fields are half-integer moded leading to two Weyl fermions $(\mu^A, \bar\mu^A)$ transforming in the fundamental representation of $SO(6)$.
\end{enumerate}
This field content is compiled in Table~\ref{Tab:Fields} below.
\begin{table}[ht]
\begin{center}
\begin{tabular}{|c||c||c|c|}\hline
&&&\\[-11pt]
\bf{Sector} & \bf{Field} & \bf{Statistic} & \bf{R / NS} \\\hline\hline
&&&\\[-11pt]
3-3 & $A^\mu$ & boson & NS \\
&&&\\[-11pt]\hline
&&&\\[-11pt]
 & $\Lambda^{\alpha A}$ & fermion & R \\
 &&&\\[-11pt]\hline
 &&&\\[-11pt]
 & $\Lambda_{\dot{\alpha} A}$ & fermion & R \\
 &&&\\[-11pt]\hline
 &&&\\[-11pt]
 & $\phi^a$ & boson & NS \\ &&&\\[-11pt]\hline\hline
 &&&\\[-11pt]
 (-1)-(-1) & $a^\mu$ & boson & NS \\
&&&\\[-11pt]\hline
&&&\\[-11pt]
 & $\chi^a$ & boson & NS \\
&&&\\[-11pt]\hline
&&&\\[-11pt]
& $M^{\alpha A}$ & fermion & R \\
&&&\\[-11pt]\hline
&&&\\[-11pt]
& $\lambda_{\dot{\alpha} A}$ & fermion & R \\
&&&\\[-11pt]\hline\hline
&&&\\[-11pt]
(-1)-3/3-(-1) & $\omega_{\dot{\alpha}},\bar\omega_{\dot{\alpha}}$ & boson & NS \\
&&&\\[-11pt]\hline
&&&\\[-11pt]
 & $\mu^A$ & fermion & R \\[2pt]\hline
\end{tabular}
\end{center}
\caption{Overview of the massless spectrum of the D(-1)/D3 system. The last two columns denote whether the field is bosonic or fermionic and the sector it stems from.}
\label{Tab:Fields}
\end{table}

The vertex operators for the fields in the 3-3 and (-1)-(-1) sectors can be obtained by dimensional reduction. In the former, these are
\begin{align}
 V_{A}(z)&= \frac{A_{\mu}(p)}{\sqrt{2}}\psi^{\mu}(z)\,e^{ip\cdotp X(z)}\,e^{-\varphi(z)}\,,&V_{\phi}(z)&= \frac{\phi_a(p)}{\sqrt{2}}\psi^a(z)\,e^{ip\cdotp X(z)}\,e^{-\varphi(z)}\,,\label{AVop}\\
 V_{\Lambda}(z)&= \Lambda^{\alpha A}S_{\alpha}(z)S_{A}(z)\,e^{ip\cdotp X(z)}\,e^{-\tfrac{1}{2}\varphi(z)}\,,&V_{\bar\Lambda}(z)&= \Lambda_{\dot\alpha A}S^{\dot\alpha}(z)S^{A}(z)\,e^{ip\cdotp X(z)}\,e^{-\tfrac{1}{2}\varphi(z)}\label{LambdaVop}\,,
\end{align}
whereas in the (-1)-(-1), we have
\begin{align}
 V_{a}(z)&= g_0\,a_{\mu}\psi^{\mu}(z)e^{-\varphi(z)}\,,&V_{\chi}(z)&= \frac{\chi_a}{\sqrt{2}}\psi^a(z)e^{-\varphi(z)}\,,\label{aVop}\\
 V_{M}(z)&= \frac{g_0}{\sqrt{2}}\,M^{\alpha A}S_{\alpha}(z)S_{A}(z)e^{-\tfrac{1}{2}\varphi(z)}\,,&V_{\lambda}(z)&= \lambda_{\dot\alpha A}\,S^{\dot\alpha}(z)S^{A}(z)\,e^{-\tfrac{1}{2}\varphi(z)}\,.\label{MlambdaVop}
\end{align}
Here, $g_0$ is the D(-1)-brane coupling constant. For a Dp-brane, it is defined as
\begin{equation}
 g_{p+1}^2=4\pi(2\pi\sqrt{\alpha'})^{p-3}\,g_s\,.
\end{equation}
Moreover, we have set $2\pi\alpha'=1$. In order to recover the usual dimensions, one must rescale the fields with $(2\pi\alpha')^{\frac{3-2\nu}{4}}$ and, as usual, $\nu_{\textrm{R,NS}}=0,\frac{1}{2}$. The vertex operators for the mixed sectors contain the so-called \emph{twist operators} that implement a change in the coordinates boundary conditions from D to N and vice-versa. These are bosonic fields denoted $\Delta,\bar\Delta$ and carry conformal dimension 1/4. The vertex operators are \cite{Green:2000ke}
\begin{align}
 V_{\omega}(z)=& \frac{g_0}{\sqrt{2}}\,\omega_{\dot\alpha}\Delta(z)S^{\dot\alpha}(z)e^{-\varphi(z)}\,,& V_{\bar\omega}(z)=& \frac{g_0}{\sqrt{2}}\,\bar\omega_{\dot\alpha}\bar\Delta(z)S^{\dot\alpha}(z)e^{-\varphi(z)}\,\label{OmegaVop}\\
 V_{\mu}(z)=& \frac{g_0}{\sqrt{2}}\,\mu^A\Delta(z)S_{A}(z)e^{-\frac{1}{2}\varphi(z)}\,,&V_{\bar\mu}(z)=& \frac{g_0}{\sqrt{2}}\,\bar\mu^{A}\bar\Delta(z)S_{A}(z)e^{-\frac{1}{2}\varphi(z)}\,.\label{muVop}
\end{align}

Consider the OPE of \eqref{OmegaVop} with the supersymmetry currents \eqref{SusyCurrents} which can be easily derived as follows:
\begin{align}
 j^{\dot\alpha A}(z)\,V_{\omega}(w)&\sim \frac{\omega^{\dot\alpha}}{z-w}\,S^A(w)\,\Delta(w)\,e^{-\frac{3}{2}\varphi(w)}\,,\label{LocalOPE}\\
 j_{\alpha A}(z)\,V_{\omega}(w)&\sim \frac{\omega_{\dot\alpha}{(\sigma_{\mu})^{\dot\alpha}}_{\alpha}}{\sqrt{2}(z-w)^{\frac{1}{2}}}\,S_A(w)\,\Delta(w)\,\psi^{\mu}(w)\,e^{-\frac{3}{2}\varphi(w)}\,.\label{NonLocalOPE}
\end{align}
Hence, the anti-chiral moduli $\omega,\bar\omega$ are compatible with the preserved supersymmetries in the D(-1)/D3 system since the OPE \eqref{LocalOPE} of their vertex operators with the corresponding supersymmetry currents are local. However, notice that the OPE \eqref{NonLocalOPE} is non-local because of the branch-cut, but this is not problematic because the corresponding supercharges are broken. On the other hand, the situation is reversed for chiral mixed moduli: the OPE of the latter with the preserved supersymmetry currents is non-local and, therefore, is inconsistent. This justifies the fact that the mixed bosonic moduli have a definite chirality stemming from the choice of boundary conditions for the spin fields.

We would like to describe the gauge theory interactions between the various states arising in the construction above, and compare with the action obtained in the ADHM construction. In the string theory language, these are tree-level open string interactions and are described by disc amplitudes \cite{Green:2000ke}. In addition, one must take the limit $\alpha'\rightarrow0$ in order to project out the string theory corrections. Let us first focus on 3-3 sector describing the perturbative gauge theory degrees of freedom. The disc diagrams have their boundary lying on the D3-branes only. Consider, for instance, the interaction term of two gaugini $\Lambda^{\alpha A}$ and a scalar field $\phi^a$, that is
\begin{align}
 \mathfrak{D}_{\Lambda\phi\Lambda}&=\langle\langle V_{\Lambda}\,V_{\phi}\,V_{\Lambda}\rangle\rangle\nonumber\\
 &\equiv N_4\int\frac{dz^1dz^2dz^3}{dV_{123}}\langle V_{\Lambda}(z_1)\,V_{\phi}(z_2)\,V_{\Lambda}(z_3)\rangle\,,\label{LambdaPhiLambda}
\end{align}
where $N_4$ is the normalisation for D3-branes disc amplitudes, $dV_{abc}$ is the volume of the conformal Killing group \eqref{VolumeCKG} and $\langle\cdots\rangle$ denotes a disc CFT correlator. For a Dp-brane, the normalisation factor $N_{p+1}$ is \cite{Billo:2002hm}
\begin{equation}
 N_{p+1}=\frac{1}{2\pi^2\alpha'^2}\frac{1}{C_{p+1}g_{p+1}^2}\,,
\end{equation}
with $C_{p+1}$ being the Casimir invariant of the fundamental representation $\mc F$ of the Dp-branes gauge group:
\begin{equation}
 \textrm{Tr}_{\mc F}(T^a\,T^b)=C_{p+1}\,\delta^{ab}\,.
\end{equation}
In our current case, the generators are normalised such that $C_0=1$ and $C_4=\frac{1}{2}$. At tree-level, the total superghost picture must be -2, and the vertex operators (\ref{AVop}, \ref{LambdaVop}) are already expressed in convenient pictures. Using the results listed in Appendix \ref{appendix:OPEs}, the correlator in \eqref{LambdaPhiLambda} can be evaluated, to the leading order in $\alpha'$, as
\begin{align}
 \langle V_{\Lambda}(z_1)\,V_{\phi}(z_2)\,V_{\Lambda}(z_3)\rangle&=\frac{1}{\sqrt{2}}\Lambda^{\alpha A}\,\phi_{a}\,\Lambda^{\beta B}\langle e^{-\frac{1}{2}\varphi(z_1)}\,e^{-\varphi(z_2)}\,e^{-\frac{1}{2}\varphi(z_3)}\rangle\,\nonumber\\
 &\times \langle S_{\alpha}(z_1)\,S_{\beta}(z_3)\rangle\langle S_{A}(z_1)\,\psi^a(z_2)\,S_{B}(z_3)\rangle\,\nonumber\\
 &=\frac{i}{2}\epsilon_{\alpha\beta}\,(\bar\Sigma^a)_{AB}\,\Lambda^{\alpha A}\,\phi_{a}\,\Lambda^{\beta B}\,(z_{12}\,z_{13}\,z_{23})^{-1}\,.
\end{align}
Including all the normalisation factors, this yields
\begin{equation}
  \mathfrak{D}_{\Lambda\phi\Lambda}=\frac{2i}{g_4^2}(\bar\Sigma^a)_{AB}\,\Lambda^{\alpha A}\,\phi_{a}\,{\Lambda_{\alpha}}^{B}\,.
\end{equation}
Finally, in order to obtain the full coupling, one must also calculate the inequivalently ordered term $\mathfrak{D}_{\Lambda\Lambda\phi}$ which gives through a similar calculation the same result with the opposite sign due to the different ordering of the fields:
\begin{equation}
  \mathfrak{D}_{\Lambda\phi\Lambda}=-\frac{2i}{g_4^2}(\bar\Sigma^a)_{AB}\,\Lambda^{\alpha A}\,{\Lambda_{\alpha}}^{B}\,\phi_{a}\,.
\end{equation}
Hence, the term in the effective action is
\begin{equation}
 \frac{2i}{g_{4}^2}(\bar\Sigma^a)_{AB}\,\Lambda^{\alpha A}[\phi_{a}\,,{\Lambda_{\alpha}}^{B}]\,.
\end{equation}
Computing all possible tree-level diagrams involving the vector multiplet components, taking the field theory limit and identifying $g_4$ with the Yang-Mills coupling $g_{\textrm{YM}}$, one recovers the tree-level super-Yang-Mills (SYM) effective Lagrangian:
\begin{align}
 \mc L_{\textrm{SYM}}=&\frac{1}{g_{\textrm{YM}}^2}\textrm{Tr}\left\{\frac{1}{2}F^2-2\bar\Lambda_{\dot\alpha A}\bar{\slashed{D}}^{\dot\alpha\beta}{\Lambda_{\beta}}^A+D_{\mu}\phi^a\,D^{\mu}\bar\phi_a-\frac{1}{2}[\phi_a,\phi_b]^2\right.\nonumber\\
&+\left.2i(\Sigma^a)^{AB}\bar\Lambda_{\dot\alpha A}[\phi_a,{{\bar{\Lambda}}^{\dot\alpha}}_{\,\,\,B}]+2i(\bar\Sigma^a)_{AB}\Lambda^{\alpha A}[\phi_a,{\Lambda_{\alpha}}^{B}]\right\}\,.\label{SYMaction}
\end{align}
Therefore, the stack of D3-branes supports, as expected, an $\N=4$ $SU(N)$ gauge theory on their world-volume. We now show that, in this picture, the D-instantons provide a good description of the corresponding gauge instantons.

\begin{figure}[h!t]
\begin{center}
\includegraphics[width=0.8\textwidth]{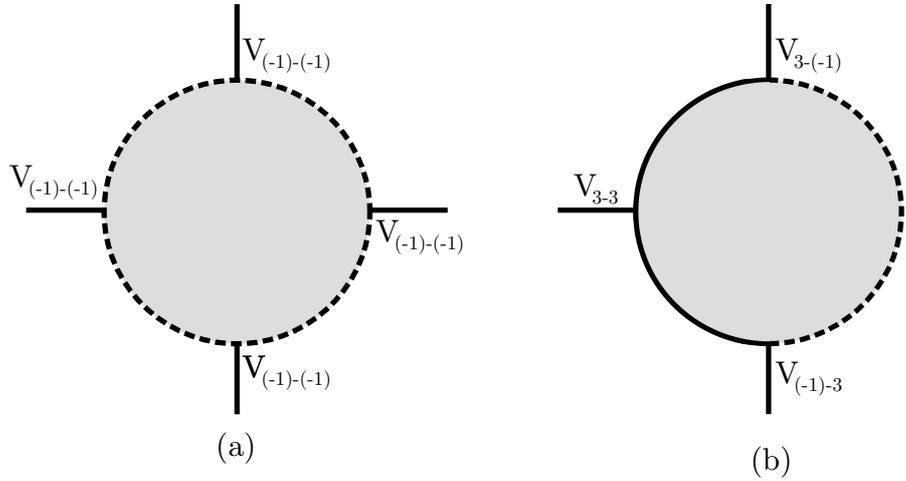}
\end{center}
\caption{Mixed and unmixed disc diagrams with boundary insertions only. Diagram (a) involves four boundary insertions from the (-1)-(-1) sector, whereas diagram (b) two insertions from the 3-(-1) and (-1)-3 sectors. While the whole boundary of diagram (a) lies on the D(-1)-branes, diagram (b) lies partly on the D3- and partly on the D(-1)-branes. Notice that the latter mixed boundary conditions are only relevant in the space-time directions. } 
\label{fig:mixedunmixed}
\end{figure}

The instanton effective action is obtained similarly by calculating all tree-level disc diagrams involving the moduli above, and these reduce, in the field theory limit, to three- and four-point amplitudes only. There are two types of disc diagrams as depicted in Fig.~\ref{fig:mixedunmixed}, mixed and unmixed ones. The latter are the ones in which the full boundary of the disc lies on the D(-1)-branes, whereas the former correspond to the case where the disc boundary is split between the D3- and the D(-1)-branes. This means that there are insertions of (-1)/3 and 3/(-1) moduli which are responsible, through the twist fields, for the jump of the boundary conditions from N to D and vice-versa. Consequently, mixed diagrams always contain an even number of such moduli.

Let us first focus on the cubic interactions in the (-1)-(-1) sector  which, by dimensional analysis, involve one NS and two R moduli. For instance, consider the following correlation function:
\begin{align}
 \mathfrak{D}_{M\chi M}&=\langle\langle V_{M}\,V_{\chi}\,V_{M}\rangle\rangle\nonumber\\
 &\equiv N_0\int\frac{dz^1dz^2dz^3}{dV_{123}}\langle V_{M}(z_1)\,V_{\chi}(z_2)\,V_{M}(z_3)\rangle\,.\label{MChiM}
\end{align}
This correlator has exactly the same structure as the one in \eqref{LambdaPhiLambda} and leads, similarly, to
\begin{align}
 \langle V_{M}(z_1)\,V_{\chi}(z_2)\,V_{M}(z_3)\rangle&=\frac{ig_0^2}{4}\epsilon_{\alpha\beta}\,(\bar\Sigma^a)_{AB}\,M^{\alpha A}\,\chi_{a}\,M^{\beta B}\,(z_{12}\,z_{13}\,z_{23})^{-1}\,,
\end{align}
so that the interaction term \eqref{MChiM} is
\begin{equation}
 \mathfrak{D}_{M\chi M}=\frac{i}{2}(\bar\Sigma^a)_{AB}\,M^{\alpha A}\,\chi_a\,{M_{\alpha}}^B\,.
\end{equation}
Including the other ordering of the polarisations (containing the CP degrees of freedom) yields the term
\begin{align}
 \frac{i}{2}(\bar\Sigma^a)_{AB}\,M^{\alpha A}[\chi_{a}\,,{M_{\alpha}}^{B}]\,.
\end{align}
All other three-point interactions can be derived similarly. Let us turn to the more involved case of four-point diagrams in which one of the positions of the vertex operators is integrated, and we focus on the amplitude
\begin{align}
 \mathfrak{D}_{aaaa}&=\langle\langle V_{a}\,V_{a}\,V_{a}\,V_{a}\rangle\rangle\nonumber\\
 &\equiv N_0\int\frac{dz^1dz^3dz^4}{dV_{134}}\langle V_{a}(z_1)\,V_{a}(z_2)\,V_{a}(z_3)\,V_{a}(z_4)\rangle\,,\label{aaaa}
\end{align}
where we choose the position $z_2$ to be the integrated one. Naively, the zero-picture vertex operator is $\partial X^{\mu}$. However, in order to regularise the worldsheet integrals, we need to turn on some momenta in the internal directions\footnote{See Section 2.5 of \cite{Witten:2012bh} for a recent discussion of integrated vertex operators in string theory.}. In principle, the Dirichlet boundary conditions due to the D(-1)-branes forbid the presence of momenta in any direction. Nevertheless, changing pictures for zero momentum vertex operators is quite subtle and one may formally define this procedure by allowing momenta in some directions and then taking the zero momentum limit \cite{Polyakov:1997rk}:
\begin{equation}
 V^{(0)}(p=0)=\lim_{p\rightarrow0} V_{PCO}\,V^{(-1)}(p)\,.
\end{equation}
Here, the superscript displays the picture of the vertex operator. The limit of vanishing momenta is then taken after evaluating the correlation functions and the worldsheet integrals. Alternatively, one may work in the dual D5/D9 system which is obtained by performing a T-duality along all the internal directions. In this case, the boundary conditions in the latter do not forbid any momenta\footnote{One must also take the decompactification limit of the $T^6$ in order for this to be possible.}. Here, it is important to notice that momenta are never allowed in the \ST directions (where the gauge theory lives). In general, the vertex operator at non-zero momentum is of the form
\begin{equation}
 (\partial X^{\mu}-i(p\cdotp \psi)\psi^{\mu})e^{ip\cdotp X}\,.
\end{equation}
One of the zero-picture vertices in \eqref{aaaa} is at fixed position, \emph{i.e.} it has conformal dimension zero, and this amounts to attaching to it a c-ghost. In fact, this naive step is incorrect since the resulting vertex operator would not lie in the BRST cohomology:
\begin{equation}
 Q_{\textrm{BRST}}\,\left[c\,(\partial X^{\mu}-i(p\cdotp \psi)\psi^{\mu})e^{ip\cdotp X}\right]\neq0\,.
\end{equation}
The correct physical zero-picture vertex operator of dimension zero is
\begin{equation}\label{PhysFixedVertexOperator}
 \left[c\,(\partial X^{\mu}-i(p\cdotp \psi)\psi^{\mu})-\gamma\,\psi^{\mu}\right]e^{ip\cdotp X}\,,
\end{equation}
as one can readily check by evaluating the action of the BRST charge on \eqref{PhysFixedVertexOperator} ($\gamma$ is the superconformal ghost). \eqref{aaaa} now becomes
\begin{equation}
 \mathfrak{D}_{aaaa}=N_0\lim_{\genfrac{}{}{0pt}{}{p_i\rightarrow0}{i=1,\ldots,4}}\int dz_2\langle \tilde V_{a}^{(0)}(z_1,p_1)\,V_{a}^{(0)}(z_2,p_2)\,\tilde V_{a}^{(-1)}(z_3,p_3)\,\tilde V_{a}^{(-1)}(z_4,p_4)\rangle\,,\label{Physaaaa}
\end{equation}
where the tilded vertex operators are the unintegrated ones:
\begin{alignat}{3}
 &\tilde V_{a}^{(0)}(z,p)&=&\,g_0\,a_{\mu}\left[c\,(\partial X^{\mu}-i(p_a\psi^a)\psi^{\mu})-\gamma\,\psi^{\mu}\right]e^{ip_aX^a}(z)\,,\label{Vtildea0}\\
 &\tilde V_{a}^{(-1)}(z,p)\,&=&\,g_0\,a_{\mu}\,c\,e^{-\varphi}\,\psi^{\mu}\,e^{ip_aX^a}(z)\,\label{Vtildea-1}\\
 &V_{a}^{(0)}(z,p)&=&\,g_0\,a_{\mu}(\partial X^{\mu}-i(p_a\psi^a)\psi^{\mu})e^{ip_a\, X^a}(z)\,.\label{Va0}
\end{alignat}
Consider the contribution of $\gamma\,\psi^{\mu}$ to \eqref{Physaaaa} for which only the fermionic part of \eqref{Va0} can give rise to a non-trivial correlation function between the fermions in the \ST directions. However, $\psi^a$ cannot contract with any other field, and this yields a vanishing contribution. Similarly, the terms $\partial X^{\mu}$, $(p_a\psi^a)\psi^{\mu}$ in \eqref{Vtildea0} can only contract with the corresponding ones in \eqref{Va0} resulting in the two following terms:

\begin{align}
 A_{1}&=\langle\partial X^{\mu}(z_1)\partial X^{\nu}(z_2)\rangle\langle\psi^{\rho}(z_3)\psi^{\lambda}(z_4)\rangle=-\frac{\delta^{\mu\nu}\delta^{\rho\lambda}}{(z_{12})^2 z_{34}}\,,\label{A1}\\
 A_{2}&=-p_{1a}\,p_{2b}\langle\psi^{a}(z_1)\psi^{b}(z_2)\rangle\langle\psi^{\mu}(z_1)\psi^{\nu}(z_2)\psi^{\rho}(z_3)\psi^{\lambda}(z_4)\rangle\nonumber\\
      &=-\frac{p_1\cdotp p_2}{z_12}\left[\frac{\delta^{\mu\nu}\delta^{\rho\lambda}}{z_{12} z_{34}}-\frac{\delta^{\mu\rho}\delta^{\nu\lambda}}{z_{13} z_{24}}+\frac{\delta^{\mu\lambda}\delta^{\nu\rho}}{z_{14} z_{23}}\right]\,.\label{A2}
\end{align}
Both $A_1$ and $A_2$ are multiplied by the contributions of the c-ghost, superghosts and the exponentials in the momenta: 
\begin{align}
 A_0&=\langle c(z_1)c(z_3)c(z_4)\rangle\langle e^{-\varphi(z_3)}\,e^{-\varphi(z_4)}\rangle\langle\prod_{i=1}^{4} e^{ip_{ia} X^a(z_i)}\rangle \nonumber\\
    &=z_{13}z_{14}\prod_{1\leq i<j\leq4} (z_{ij})^{p_i\cdotp p_j}\,.
\end{align}
Focusing on the ordering $1234$, the total correlation function is
\begin{equation}
 \langle \tilde V_{a}^{(0)}(z_1,p_1)\,V_{a}^{(0)}(z_2,p_2)\,\tilde V_{a}^{(-1)}(z_3,p_3)\,\tilde V_{a}^{(-1)}(z_4,p_4)\rangle=g_0^4\,\textrm{Tr}[a_{\mu}\,a_{\nu}\,a_{\rho}\,a_{\lambda}]A_0(A_1+A_2)\,.
\end{equation}
In this case, the range of integration of $z_2$ is $]z_1,z_3[$. Notice that only the product $p_1\cdotp p_2$ appears above and we can already set $p_3$ and $p_4$ to zero. The term proportional to $A_1$ yields
\begin{equation}
 -\frac{z_{14}}{z_{34}}\delta^{\mu\nu}\delta^{\rho\lambda}\,.
\end{equation}
This contribution cancels against the same one coming from the ordering $4321$ so that only the term proportional to $A_2$ can lead to a non-trivial result. This occurs when the integral over $z_2$ gives a pole in $p_1\cdotp p_2$. Clearly, the first term in \eqref{A2} cannot result in such a pole. Integrating the other two terms over $z_2$ and taking the limit of vanishing momenta, one obtains
\begin{equation}
 -\delta^{\mu\rho}\delta^{\nu\lambda}+\delta^{\mu\lambda}\delta^{\nu\rho}\,.
\end{equation}
Including the polarisations, the final result, after some simple manipulations, can be recast as
\begin{equation}
 \mathfrak{D}_{aaaa}=-\frac{g_0^2}{4}[a_{\mu},a_{\nu}]^2\,.
\end{equation}

Summarising the above results, the total instanton effective action in the (-1)-(-1) sector is
\begin{align}
 \mc S_{\textrm{ADHM}}^{(-1)}=&i\,\textrm{Tr}\left\{\lambda_{\dot\alpha A}[a^{\dot\alpha\beta},{M_{\beta}}^{A}]+\frac{1}{2}(\Sigma^a)^{AB}\,\lambda_{\dot\alpha A}[\chi_a,{\lambda^{\dot\alpha}}_{B}]+\frac{1}{2}(\bar\Sigma^a)_{AB}M^{\alpha A}[\chi_a,{M_{\alpha}}^{B}]\right\}\nonumber\\
&-\textrm{Tr}\left\{\frac{g_{0}^2}{4}[a_{\mu},a_{\nu}]^2+\frac{1}{2}[a_{\mu},\chi_a]^2+\frac{1}{4g_{0}^2}[\chi_a,\chi_b]^2\right\}\,,\label{UnmixedADHM}
\end{align}
where we have used the notation $a^{\dot\alpha\beta}\equiv a_{\mu}(\bar\sigma^{\mu})^{\dot\alpha\beta}$. Notice that for instanton number $k=1$, this action vanishes and, as we show below, only the effective action involving the moduli of the mixed sectors survives.

As explained above, the mixed disc diagrams contain an even number of twist fields. Thus, by dimensional analysis, the only mixed diagrams surviving the field theory limit are the ones involving two mixed moduli and either a (-1)-(-1) modulus or an NS field from the 3-3 sector. In the latter case, it corresponds to a vacuum expectation value (vev) for the scalars $\phi_a$ of the vector multiplet. Consider, for example, the following amplitude:
\begin{align}
 \mathfrak{D}_{\bar\mu\omega\lambda}&=\langle\langle V_{\bar\mu}\,V_{\omega}\,V_{\lambda}\rangle\rangle\nonumber\\
 &\equiv N_0\int\frac{dz^1dz^2dz^3}{dV_{123}}\langle V_{\bar\mu}(z_1)\,V_{\omega}(z_2)\,V_{\lambda}(z_3)\rangle\,.\label{BarmuOmegaLambda}
\end{align}
Notice that the normalisation is the same as in the unmixed sector. The correlator in \eqref{BarmuOmegaLambda} can be evaluated as before:
\begin{align}
 \langle V_{\bar\mu}(z_1)\,V_{\omega}(z_2)\,V_{\lambda}(z_3)\rangle&=-\frac{g_0^2}{2}\,\bar\mu^A\,\omega_{\dot\alpha}\,\lambda_{\dot\beta B}\langle\bar\Delta(z_1)\Delta(z_2)\rangle\langle S^{\dot\alpha}(z_2)S^{\dot\beta}(z_3)\rangle\langle S_A(z_1)S^B(z_3)\rangle\nonumber\\
&~\times\langle e^{-\frac{1}{2}\varphi(z_1)}e^{-\varphi(z_2)}e^{-\frac{1}{2}\varphi(z_3)}\rangle\nonumber\\
&=-i\frac{g_0^2}{2}\,\bar\mu^A\,\omega_{\dot\alpha}\,\lambda_{\dot\beta B}\,\epsilon^{\dot\alpha\dot\beta}\,{\delta^B}_A\,(z_{12}\,z_{13}\,z_{23})^{-1}\,.
\end{align}
Hence, this disc amplitude yields the following effective coupling:
\begin{equation}
 \mathfrak{D}_{\bar\mu\omega\lambda}=i\,\bar\mu^A\,\omega_{\dot\alpha}\,{\lambda^{\dot\alpha}}_B\,.
\end{equation}
Finally, calculating all the other mixed diagrams including the vev for the gauge multiplet, we obtain the instanton effective action for the mixed moduli:
\begin{align}
 \mc S^{\textrm m}_{\textrm{ADHM}}=i\,\textrm{Tr}\left\{(\bar\mu^A\,\omega_{\dot\alpha}+\bar\omega_{\dot\alpha}\,\mu^A){\lambda^{\dot\alpha}}_A+\frac{1}{2}(\bar\Sigma^a)_{AB}\,\bar\mu^A\,\mu^B\,\chi_a-D_c\,W^c+\frac{i}{2}\chi_a\,\bar\omega^{\dot\alpha}\,\omega_{\dot\alpha}\,\chi^a\right\}\,,\label{MixedADHM}
\end{align}
and the total action for the instanton moduli is
\begin{equation}
 \mc S_{\textrm{ADHM}}=\mc S_{\textrm{ADHM}}^{(-1)}+\mc S^{\textrm m}_{\textrm{ADHM}}\,.
\end{equation}
In \eqref{MixedADHM}, we have introduced $W^c\equiv\omega_{\dot\alpha}{(\tau^c)^{\dot\alpha}}_{\dot\beta}\,\bar\omega^{\dot\beta}$, together with a set of three auxiliary fields $D_c$ ($c=1,2,3$). In fact, integrating out the latter yields the ADHM equations \eqref{ADHMequations} as shown \emph{e.g.} in \cite{Billo:2002hm}.

In order to compare with the instanton effective action used in gauge theory, we need to define a clear mapping between the ADHM moduli and the ones arising from the brane construction. We merely perform this for the bosonic moduli since the rest follows by supersymmetry. Furthermore, we focus in this manuscript on $\mc N=2$ theories which can be obtained, in the string theory construction, by compactifying on a $T^2\times T^4/\mb Z_2$ orbifold. In this case, some of the states are projected out from the spectrum. In particular, only two of the bosonic moduli $\chi_a$ are preserved, namely $\chi_4,\chi_5$. The latter are complexified and readily identified with the ADHM moduli $\phi,\bar\phi$:
\begin{align}
 \phi=\frac{\chi_4-i\chi_5}{\sqrt{2}}\,,&&\bar\phi=\frac{\chi_4+i\chi_5}{\sqrt{2}}\,.
\end{align}
In addition, the moduli $a_{\mu}$ and $B_l$ are mapped to each other as follows:
\begin{align}
 B_l=(-)^l\,a_{2l-1}-i\,a_{2l-2}\,,&& \textrm{with } l=1,2\,,
\end{align}
with the inverse map being
\begin{align}
 a_{2l-2}=\,\frac{1}{2}(-)^l(B_l+B_l^{\dag})\,,&&a_{2l-1}=\,\frac{i}{2}(B_l-B_l^{\dag})\,.
\end{align}
We then plug these identifications in \eqref{UnmixedADHM} or \eqref{DeformedADHM}, for example
\begin{align}
 \textrm{Tr}\,\sum_{l=1,2}[\phi,B_l][\bar\phi,B_l^{\dag}]&=\textrm{Tr}\,\sum_{l=1,2}\left([\phi,a_{2l-1}][\bar\phi,a_{2l-1}]+[\phi,a_{2l-2}][\bar\phi,a_{2l-2}]\right)\nonumber\\
                                                                      &=\frac{1}{2}\,\textrm{Tr}\,[\chi_4-i\chi_5,a_{\mu}][\chi_4+i\chi_5,a^{\mu}]\nonumber\\
                                                                      &=\frac{1}{2}\,\textrm{Tr}\,[\chi_a,a_{\mu}][\chi^a,a^{\mu}]\,,
\end{align}
which appears in \eqref{UnmixedADHM}. Hence, to the leading order in the coupling constant, the terms of the effective action \eqref{UnmixedADHM} precisely match the ones coupling the moduli $B_l$ and their fermionic superpartners in \eqref{DeformedADHM}.

We now turn to the mixed sector in which the identifications are
\begin{alignat}{2}
 I&=\omega_{\dot1}\,,\,I^{\dag}&=\bar\omega_{\dot2}\,,\\
 J&=\bar\omega_{\dot1}\,,\,J^{\dag}&=\omega_{\dot2}\,,
\end{alignat}
which are used to map \eqref{MixedADHM} to the terms of \eqref{DeformedADHM} involving the moduli $I,J$ and their fermionic superpartners. As an illustration, recalling that spinor indices are raised and lowered using the Levi-Civita symbol, we obtain
\begin{align}
 \textrm{Tr}\left[\chi_a\,\bar\omega^{\dot\alpha}\,\omega_{\dot\alpha}\,\chi^a\right]&= \textrm{Tr}\left[\phi\,\bar\omega^{\dot\alpha}\,\omega_{\dot\alpha}\,\bar\phi+\bar\phi\,\bar\omega^{\dot\alpha}\,\omega_{\dot\alpha}\,\phi\right]\nonumber\\
 &=\textrm{Tr}\left[\phi\,I^{\dag}\,I\,\bar\phi+\phi\,J\,J^{\dag}\,\bar\phi+h.c.\right]\,.
\end{align}
Consequently, the Dp-D(p+4) brane setup is a consistent string theory description of the gauge theory instantons. Moreover, one can show that the BRST symmetry $Q$ appearing in \eqref{DeformedADHM} arises as a common supercharge preserved by the D-branes.

\section[\texorpdfstring{N=2 gauge theory in the $\Omega$-background}{N=2 gauge theory in the Omega-background}]{\texorpdfstring{N=2 gauge theory in the $\boldsymbol{\Omega}$-background}{N=2 gauge theory in the Omega-background}}

In this section, we briefly present the classical description of the gauge theory in the $\Omega$-background, whose instanton corrections were discussed above.
A standard way of obtaining the action for a pure $\mc N=2$ gauge theory in four dimensions is to start from an $\mc N=1$ theory in six dimensions whose action is
\begin{equation}\label{SixDimSYM}
 \mc S_{\textrm{YM}}^{6D,\,\mc N=1}=\frac{1}{g_6^2}\int\textrm{Tr}\left\{-\frac{1}{4}F_{MN}F^{MN}+\frac{i}{2}\bar\Psi_A\slashed{\nabla}\Psi^A\right\}+\frac{\theta}{8\pi^2}\int\textrm{Tr}\left(F\wedge F\right)\,,
\end{equation}
with $\slashed\nabla\equiv\gamma^{M}\nabla_M$, and dimensionally reduce it on $\mb R^{1,3}\times T^2$ with a flat metric
\begin{equation}\label{FlatMetric}
 ds_6^2=g_{\mu\nu}dx^\mu dx^\nu+dz d\bar z\,.
\end{equation}
Here, $z$ is a complex coordinate for $T^2$. Upon shrinking the cycles of the torus to zero, one recovers a four dimensional gauge theory with eight supercharges since the reduction on the torus is maximally supersymmetric. The four dimensional action is reached as follows. Consider the bosonic terms in \eqref{SixDimSYM} and reduce it on \eqref{FlatMetric} :
\begin{equation}
 -\frac{1}{4}F_{MN}F^{MN}=-\frac{1}{4}F_{\mu\nu}F^{\mu\nu}-\frac{1}{4}F_{\mu z}{F^\mu}_{\bar z}-\frac{1}{4}F_{z\bar z}F^{z\bar z}\,.
\end{equation}
The $z,\bar z$ components of the field strength define a complex scalar $\phi$ such that
\begin{eqnarray}
 F_{\mu z}=2\nabla_{\mu}\phi\,&\textrm{and}&\,F_{z\bar z}=\sqrt{2}[\phi,\bar\phi]\,.
\end{eqnarray}
One can perform the same analysis for the symplectic Majorana spinor $\Psi$ and this leads to the $\mc N=2$ Yang-Mills action in four dimensions
\begin{align}
 \mc S_{\textrm{YM}}^{4D,\,\mc N=2}&=\frac{1}{g_4^2}\int\textrm{Tr}\left\{-\frac{1}{4}F_{\mu\nu}F^{\mu\nu}-|\nabla_\mu\phi|^2-\frac{1}{2}[\phi,\bar\phi]^2\right.\nonumber\\
 &+\left.i\psi_A\slashed{\nabla}\bar\psi^A-\frac{i}{\sqrt{2}}\psi_A[\bar\phi,\psi^A]+\frac{i}{\sqrt{2}}\bar\psi^A[\phi,\psi_A]\right\}\nonumber\\
 &+\frac{\theta}{8\pi^2}\int\textrm{Tr}\left(F\wedge F\right)\,.
\end{align}
This is the unique renormalisable $\mc N=2$ action in four dimensions up to two derivatives. However, we are interested in studying the gauge theory deformed by the $\Omega$-background. For this, instead of the flat metric \eqref{FlatMetric}, we choose the torus to be fibered over \ST:
\begin{equation}
  ds_6^2=g_{\mu\nu}(dx^\mu+\Omega^\mu_{~\rho}\, x^\rho dz+{\bar\Omega^\mu}_{~\rho}\, x^\rho d\bar z )(dx^\nu+\Omega^\nu_{~\rho}\, x^\rho dz+{\bar{\Omega}^\nu}_{~\rho}\, x^\rho d\bar z)+dz d\bar z\,,
\end{equation}
such that going around the cycles of the torus is accompanied with a rotation in \ST parametrised by the matrices $\Omega$ and $\bar\Omega$:
\begin{eqnarray}
 \Omega^{\mu\nu}=\left(
 \begin{array}{cccc}
  0 & \epsilon_1 & 0 & 0\\
  -\epsilon_1 & 0 & 0 & 0\\
 0 & 0 & 0 & \epsilon_2 \\
 0 & 0 & -\epsilon_2 & 0 \\
 \end{array}
 \right)\,,\,&&\bar\Omega^{\mu\nu}=\left(
 \begin{array}{cccc}
  0 & \bar\epsilon_1 & 0 & 0\\
  -\bar\epsilon_1 & 0 & 0 & 0\\
 0 & 0 & 0 & \bar\epsilon_2 \\
 0 & 0 & -\bar\epsilon_2 & 0 \\
 \end{array}
 \right)\,.
\end{eqnarray}

Since the background is now curved, all supersymmetries are broken. Nevertheless, one can preserve a fraction of the latter by using the $SU(2)$ R-symmetry of the $\mc N=2$ algebra. This procedure is similar to the topological twist \cite{Nekrasov:2002qd} introduced in Section \ref{TopoTwist}. In addition, it is essential in order to define a nilpotent charge $Q$ as in \eqref{NekPartK} and calculate the corresponding partition function using localisation.

Similarly to the pure gauge theory case, the partition function of the $\N=2$ gauge theory in the $\Omega$-background factorises as
\begin{equation}\label{NekFull}
 Z^{\textrm{Nek}}=Z^{\textrm{Nek}}_{\textrm{Pert}}\times Z^{\textrm{Nek}}_{\textrm{NP}}\,,
\end{equation}
with $Z^{\textrm{Nek}}_{\textrm{Pert}}$ arising at one-loop only as in the Seiberg-Witten theory. Moreover, the non-perturbative part is a sum over the instanton sectors labelled by the instanton number $k$,
\begin{equation}
 Z^{\textrm{Nek}}_{\textrm{NP}}(\epsilon_1,\epsilon_2,a,q)=\sum_{k\geq0}q^k Z^{\textrm{Nek}}_k(\epsilon_1,\epsilon_2,a)\,,
\end{equation}
with $Z^{\textrm{Nek}}_k$ defined in \eqref{NekPartK}. Using localisation \cite{Lossev:1997bz,Moore:1997dj,Nekrasov:2002qd,Nekrasov:2003rj}, it was shown that the leading order expansion of the $\Omega$-deformed prepotential
\begin{equation}
 \mc F^{\textrm{Nek}}(\epsilon_1,\epsilon_2,a,q)\equiv\log Z^{\textrm{Nek}}
\end{equation}
matches the Seiberg-Witten prepotential:
\begin{equation}
 \mc F^{\textrm{Nek}}(\epsilon_1,\epsilon_2,a,q)=\frac{1}{\epsilon_1\epsilon_2}\mc F^{\textrm{SW}}(a,q)+\mc O(1)\,.
\end{equation}
In the case $\epsilon_{1}=-\epsilon_2$, the higher order corrections in $\epsilon_{1,2}$ can be interpreted as gravitational corrections and arise as the field theory limit of the topological string partition function
\begin{equation}
 {\sum_{g=0}^\infty g_s^{2g-2}\mathcal{F}_g|_{\text{field theory}}}=\mc F^{\text{Nek}}(\epsilon_+=0,\epsilon_-=g_s)\,,
\end{equation}
with the parameter $\epsilon_1=-\epsilon_2$ identified with the topological string coupling $g_s$. Hence, they encode the class of higher derivative couplings in the string effective action presented in Section \ref{ch:TopoAmp}. The case of a general $\Omega$-background is analysed in the following sections.

Finally, for future reference, the perturbative part of the Nekrasov partition function in the case of an $SU(2)$ gauge theory is

\begin{equation}\label{NekPertPart}
 \mathcal{F}^{\textrm{Nek}}_{\textrm{Pert}}\left(\epsilon_-,\epsilon_+,a\right)= -\frac{1}{2}\int_0^\infty\frac{dt}{t}~
\frac{\cos\left( 2\epsilon_+ t\right) }{\sin\left( \epsilon_--\epsilon_+\right)t ~\sin\left( \epsilon_-+
\epsilon_+\right)t} ~e^{-a t}\,,
\end{equation}
where $a$ is the vev of the scalar of the gauge multiplet and a proper regularisation of the Schwinger integral around zero is understood. It is important to mention that the partition function of the $\N=2$ gauge theory in the $\Omega$-background is holomorphic, \emph{i.e.} it does not depend on $\bar\Omega$ since $\bar\epsilon_{1,2}$ deformations are $Q$-exact \cite{Nekrasov:2002qd}. Hence, in the subsequent discussions, we only consider $\Omega$-deformations and set $\bar\Omega$ to zero.

\part{Refined Amplitudes as Generalized F-terms}
\label{ch:RefCouplings}
\chapter*{}
\vspace*{\fill}
In the last decade, our understanding of topological string theory has dramatically increased both from a physical and a mathematical point of view. A more recent development, inspired through the work on the partition function of supersymmetric gauge theories~\cite{Moore:1997dj,Lossev:1997bz,Nekrasov:2002qd,Nekrasov:2003rj}, is the realisation that an interesting one-parameter extension exists, known as the \emph{refined topological string}. Indeed, the field theory limit of the genus $g$ topological string partition function $\mathcal{F}_g^{\text{ft}}$ for a Calabi-Yau manifold $X$ is related to Nekrasov's partition function of a gauge theory on $\mathbb{R}^4\times S^1$ through \cite{Nekrasov:2002qd,Losev:2003py,Nekrasov:2003rj,Iqbal:2003ix,Iqbal:2003zz}:
\begin{align}
{\sum_{g=0}^\infty g_s^{2g-2}\mathcal{F}_g^{\text{ft}}}=\log Z^{\text{Nek}}(\epsilon_+=0,\epsilon_-=g_s)\,,
\end{align}
where $\epsilon_{\pm}$ are equivariant rotation parameters of $\mathbb{C}^2\sim \mathbb{R}^4$ (see Section \ref{ch:GaugeTheoryFromStringTheory}). Thus, the `unrefined' topological string only captures  one parameter, $\epsilon_-$, which is identified with the topological string coupling~$g_s$.  The refinement then consists in adding a deformation that also captures the second parameter, $\epsilon_+$. 

Most descriptions of the refinement do not follow along with the lines of the worldsheet approach towards the topological string (see \emph{e.g.} \cite{Witten:1988xj,Bershadsky:1993cx}). For instance, the refined A-model is defined via a lift to M-theory on $X\times S^1\times \text{TN}$, where the Taub-NUT space TN is twisted along $S^1$ to give rise to the two parameters $\epsilon_\pm$. The refined partition function is related to the BPS spectrum of M-theory on $X$ \cite{Gopakumar:1998ii,Gopakumar:1998jq,Hollowood:2003cv} and is equivalent to the BPS index of M2-branes wrapping 2-cycles of the Calabi-Yau manifold $X$ \cite{Dijkgraaf:2006um}. Explicitly it can be computed using a generalisation of the topological vertex formalism \cite{Awata:2005fa,Iqbal:2007ii}. Moreover, a non-perturbative definition of the refined topological string was recently proposed in \cite{Lockhart:2012vp}. However, what is still lacking is a convincing worldsheet description in terms of some twisted two-dimensional theory. There is a number of properties one would expect from such a description:
\begin{enumerate}
\item[\emph{(i)}] \emph{Unrefined limit}: Upon switching off the deformation, one expects to recover the worldsheet description of the `unrefined' topological string theory.
\item[\emph{(ii)}] \emph{(Exact) $\sigma$-model description}: We expect the refined topological string to be described by an exactly solvable $\sigma$-model. Strictly speaking, such a model is not guaranteed to exist, however, it is strongly desirable for purely practical purposes. 
\item[\emph{(iii)}] \emph{Field theory limit}: Near a point of enhanced gauge symmetry the worldsheet expression should \emph{precisely} reduce to the Nekrasov partition function of $\mathcal{N}=2$ gauge theories.
\end{enumerate}
To date, attempts to formulate a worldsheet description that possesses these properties have been inspired by the connection of the unrefined topological string to BPS-saturated amplitudes in string theory \cite{Antoniadis:1993ze,Antoniadis:1995zn,Antoniadis:1996qg}. Indeed, it has been proposed to consider perturbative string theory amplitudes as a definition of the worldsheet partition function of the refined topological string. Two different proposals have been brought forward so far \cite{Antoniadis:2010iq,Nakayama:2011be}. Both consider one-loop BPS-saturated amplitudes in Heterotic string theory compactified on $K3\times T^2$ (and their dual incarnations in Type II theory compactified on $K3$-fibered Calabi-Yau manifolds) of the form:
\begin{align}
&\mathcal{F}_{g,n}\sim\langle R_{(-)}^2 (F^G_{(-)})^{2g-2} V_{(+)}^{2n}\rangle\,,&&\text{with} &&g\geq 1\,,\label{SchematicAmplitude}
\end{align}
where $R$ stands for insertions of graviton vertices and $F^G$ for vertices of the graviphoton field strength tensor. For both fields the $(-)$ subscript indicates that only the anti-self-dual part of these tensors is used. To be precise, upon writing the four-dimensional Lorentz group as $SO(4)\sim SU(2)\times SU(2)$, these insertions are only sensitive to 
one of the $SU(2)$ Lorentz subgroups which, from the point of view of the $\Omega$-background, implies that they only couple to one of the deformation parameters, say $\epsilon_{-}$. In fact, in the absence of self-dual insertions $V_{(+)}$, \emph{i.e.} for $n=0$, the amplitude $\mathcal{F}_{g,0}$ in (\ref{SchematicAmplitude}) reduces to the class of amplitudes presented in Section \ref{ch:TopoAmp}. Thereby, property \emph{(i)} above is automatically manifest in all amplitudes of the form (\ref{SchematicAmplitude}). Coupling to the second deformation 
parameter (or sensitivity to the second $SU(2)$) is achieved through the additional insertions $V_{(+)}$. The main difference between the works \cite{Antoniadis:2010iq} and \cite{Nakayama:2011be} lies precisely in the choice of the $V_{(+)}$ insertions. \thispagestyle{fancy}\lhead{}\rhead{}\fancyfoot[L]{\thepage}\fancyfoot[R]{} In \cite{Antoniadis:2010iq}, based on the work \cite{Morales:1996bp}, it was proposed to use insertions of the self-dual field-strength of the vector partner of the Heterotic dilaton, whereas the authors of \cite{Nakayama:2011be} instead considered insertions of the field strengths of the vector partners of the K\"ahler and complex structure moduli of the internal $T^2$ as well as the $U(1)$ current of the superconformal algebra. Unfortunately, neither of these two proposals satisfies \emph{all} of the properties outlined above, with each of them only meeting two out of the three requirements. More specifically, while the amplitudes in \cite{Antoniadis:2010iq} fail to exactly reproduce the Nekrasov partition function -- the match is exact up to an $\epsilon_+$-dependent phase factor -- the ones in \cite{Nakayama:2011be} cannot be exactly evaluated at the string level due to higher order corrections in the $\sigma$-model. Conversely,  while the former can be computed exactly as string amplitudes, the latter reproduce the correct phase factor of the Nekrasov partition function in the field theory limit. 

In the following, we consider a  class of $N=2$ scattering amplitudes in Heterotic and Type I string theory compactified on $K3\times T^2$, involving the vector superpartner of the K\"ahler modulus $\bar T$ of the $T^2$ torus as the additional insertions $V_{(+)}$ introduced in (\ref{SchematicAmplitude}):
\begin{align}\label{AnsatzI}
&\mathcal{F}_{g,n}\sim\langle R_{(-)}^2 (F^G_{(-)})^{2g-2} (F^{\bar T}_{(+)})^{2n}\rangle\,,&&\text{with} &&g\geq 1\,,n\geq0\,.
\end{align}
We show that these amplitudes can be calculated \emph{exactly}\footnote{The term `exact' is used here  to stress that these particular  one-loop couplings  are evaluated exactly to all orders in $\alpha'$.} within string perturbation theory. Moreover, they precisely reproduce the expected gauge theory result of Nekrasov in the field theory limit around a point 
of enhanced gauge symmetry in the moduli space of the Heterotic compactification, where the torus Wilson lines take special values. We emphasise, however, that unlike \cite{Nakayama:2011be}, exact agreement with Nekrasov's partition function is achieved despite the fact that we do not turn on any R-symmetry current. In particular, we show that the additional vertices $V_{(+)}$ correspond to insertions of an $\mathcal{N}=2$ chiral superfield $\Upsilon$, defined as a chiral projection of the anti-chiral vector superfield $\bar T$.

After introducing a series of generalised F-terms defining a class of higher derivative couplings, we compute the latter at the one-loop level in a Heterotic theory compactified on $K3\times T^2$. In particular, we show that, in the field theory limit around an $SU(2)$ gauge group enhancement point, they reproduce the perturbative part of the Nekrasov partition function and the radius deformation of the Nekrasov-Okounkov formula \cite{Nekrasov:2003rj}, associated to the $\Omega$-background. Finally, we provide a further check of the universality of our ansatz by computing the couplings (\ref{AnsatzI}) at the one-loop level in the context of Type I superstring theory compactified on $K3\times T^2$ and reproduce the same results in the field theory limit. In fact, the Type I setup is a natural framework to study non-perturbative corrections to the refined couplings and this is discussed in detail in Section \ref{ch:ADHMinst}.
\thispagestyle{fancy}
\lhead{}
\rhead{}
\fancyfoot[R]{\thepage}
\fancyfoot[L]{}
\vspace*{\fill}

\clearpage
\thispagestyle{plain}

\chapter{Generalised Supersymmetric Effective Couplings}\label{Sect:Couplings}

In order to achieve a refinement corresponding to the second parameter $\epsilon_{+}$ of the $\Omega$-background (\emph{i.e.} a coupling to the spin of the second $SU(2)$ in the Gopakumar-Vafa picture), it is necessary to generalise (\ref{BpsStandard}) by including self-dual field strength tensors of vector multiplet fields. To this end, we introduce the following superfields which are defined as chiral projections of an arbitrary function $h(\hat{X}^I,(\hat{X}^I)^\dagger)$ of (anti-chiral) vector superfields:
\begin{align}
\Upsilon:=\Pi\frac{h(\hat{X}^I , (\hat{X}^I)^\dagger)}{ (X^0)^2} \,.
\end{align}
The projection operator $\Pi$ is defined in terms of the spinor derivatives of the $\mathcal{N}=2$ superconformal algebra:
\begin{align}
\Pi:= (\epsilon_{ij} \bar{D}^i \bar{\sigma}_{\mu\nu} \bar{D}^j)^2 \,,
\end{align}
such that we have the following action on the vector superfields:
\begin{align}
& \Pi\hat{X}^I =0 &&\text{and} && \Pi (\hat{X}^I)^\dagger = 96 \Box \hat{X}^I \,.
\end{align}
In terms of the $\Upsilon$ superfields, the following effective coupling was considered in \cite{Morales:1996bp}:
\begin{align}
\mathcal{I}_{g,n} = \int d^4x \int d^4\theta\,\tilde{\mathcal{F}}_{g,n}(X)\, (W^{ij}_{\mu\nu} W_{ij}^{\mu\nu})^g \Upsilon^n\,,\label{ProjectCoupling}
\end{align}
where $\tilde{\mathcal{F}}_{g,n}$ is a function of chiral vector multiplets. Once expressed in components, $\mathcal{I}_{g,n}$ contains particularly the terms
\begin{align}
\mathcal{I}_{g,n} = \int d^4x\,&\mathcal{F}_{g,n}(\varphi,\varphi^\dagger)\, \left[\left(R_{(-)\, \mu\nu\rho\tau} 
R_{(-)}^{\mu\nu\rho\tau}\right)\left(F^G_{(-)\,\lambda\sigma} F^{G\,\lambda\sigma}_{(-)}\right)+\left(B_{(-)\,\mu\nu}^{i\,\alpha} 
B_{(-)\,i\,\alpha}^{\mu\nu}\right)^2\right]\nonumber\\
&\times \left[F^G_{(-)\,\lambda\sigma} F^{G\,\lambda\sigma}_{(-)}\right]^{g-2}\,\left[F_{(+)\,\rho\sigma}F^{\rho\sigma}_{(+)}\right]^n 
+\ldots\label{EffectiveCoupling}
\end{align}
Here, we have explicitly displayed a term involving two Riemann tensors as well as the (supersymmetrically related) term with four gravitino field-strengths.\footnote{We implicitly assume $g\geq 2$, even though we expect our results to remain valid also for $g=1$.}  Concerning the precise nature of the vector field $F_{(+)}$ appearing in (\ref{EffectiveCoupling}), there are \emph{a priori} several different possibilities. As we have already mentioned in 
(\ref{AnsatzI}), we identify $F_{(+)}$ with the vector superpartner of the $\bar{T}$-modulus of the $T^2$ compactification.


\chapter{Heterotic Realisation of the Refinement}\label{Sec:Hetamps}

In this section we  compute the coupling (\ref{EffectiveCoupling}) in Heterotic string theory compactified on $K3\times T^2$ 
in the presence of a Wilson line. Since our one-loop Heterotic calculations only capture the perturbative part of the refined 
amplitudes, we keep in mind that a study of the dual Type II theory would eventually be required in order to probe non-perturbative 
effects. On the other hand, our results are exact to all orders in $\alpha'$, which we henceforth conveniently set to $\alpha'=1$.

As mentioned in the previous section, instead of directly computing  (\ref{AnsatzI}), we consider the amplitude 
obtained by replacing two Riemann tensors and two graviphotons with four gravitini insertions (for simplicity, we omit all indices)
\begin{align}\label{HetProposal}
\langle R_{(-)}^2(F^G_{(-)})^{2g-2} (F_{(+)}^{\bar{T}})^{2n}\rangle_{\text{1-loop}}^{\text{het}}\longrightarrow \langle B_{(-)}^4(F^G_{(-)})^{2N} 
(F_{(+)}^{\bar{T}})^{2M}\rangle_{\text{1-loop}}^{\text{het}}\,.
\end{align}
In the following, we first introduce our notation and setup of the relevant vertex operator insertions and proceed to 
evaluate the one-loop amplitude (\ref{HetProposal}), using an exact CFT realisation of $K3$ in terms of a $T^4/\mathbb{Z}_2$ 
orbifold. In order to make contact with gauge theory, we then expand  around a point of $SU(2)$ gauge symmetry 
enhancement, parametrised by Wilson lines wrapping the $T^2$. This should be contrasted with \cite{Antoniadis:2010iq}, 
where the amplitude is expanded around the $SU(2)$ enhancement point at $T=U$. In Section \ref{NekPart}, we show that 
our ansatz (\ref{HetProposal})  indeed reproduces the expected singularity structure, which is characterised by two BPS 
states becoming massless at the enhancement point (defined in \eqref{EnhancementPoint}), and then proceed to discuss 
radius deformations in Section \ref{NekrasovOkounkov}.


\section{Setup and Generating Functions}

In addition to the worldsheet coordinates $(\sigma,t)$, we introduce a ten-dimensional basis of complex bosonic coordinates 
$(Z^1,Z^2,X,Z^4,Z^5)$ for the target space\footnote{The reason for using a notation that singles out the $T^2$ super-coordinates 
$(X,\psi)$ lies in the fact that, for the special amplitudes we consider and with our chosen kinematics, $(X,\psi)$  turns 
out to contribute to the correlators only through their zero modes.}. Here $Z^{1,2},\, X\text{ and }Z^{4,5}$ parametrise the 
four-dimensional space-time, the torus $T^2$ and $K3$ of the $E_8\times E_8$ Heterotic string compactification, respectively. 
The (left-moving) superpartners of the coordinates mentioned above are denoted by $(\chi^1,\chi^2,\psi,\chi^4,\chi^5)$.  
 We can realise K3 as a $T^4/\mathbb{Z}_k$  orbifold  with $k=2,3,4,6$ and standard embedding, acting on K3  
coordinates as:
\begin{align}
	& (Z^4,\chi^4) ~\longrightarrow~ e^{2i\pi  g/k} (Z^4,\chi^4) ~,\\
	& (Z^5,\chi^5) ~\longrightarrow~ e^{-2i\pi g/k} (Z^5,\chi^5)~,
\end{align}
where $g\in\mathbb{Z}_k$. For simplicity, we  explicitly work with the $\mathbb{Z}_2$ realisation, even though our results are 
valid for general $\mathbb{Z}_k$ orbifold realisations  and are even expected to hold for generic K3 compactifications. It is 
convenient to bosonise the fermions in terms of free chiral bosons $\phi_i$ by writing
\begin{align}
&\psi=e^{i\phi_3}\,,&&\text{and} &&\chi^j=e^{i\phi_j}\,\hspace{0.5cm} \text{for} \hspace{0.5cm}j=1,2,4,5\,.
\end{align}
In a similar fashion,  the superghost is also bosonised via a free boson $\varphi$. 

We now present the vertex operators relevant to our amplitude.
It is important to separate these into self-dual and anti-self-dual parts with respect to the four-dimensional space-time. 
Indeed, anti-self-dual gauge fields carry $U(1)$ R-charge $+1$ and their charges with respect to the two SU(2) subgroups of 
the Lorentz group acting on the two planes are $(+1,+1)$. Similarly, the vertices for self-dual vector partners  carry $U(1)$ R-charge $+1$ and Lorentz charges $(+1,-1)$.  Using these conventions, the gravitino vertex operator in the $(-\frac{1}{2})$-picture is given by
\begin{align}
&V_{\psi^\pm}(\xi_{\mu\alpha},p)=\xi_{\mu\alpha}e^{-\varphi/2}S^\alpha e^{i\phi_3/2} \Sigma^\pm \,\bar{\partial}Z^\mu e^{ip\cdot Z}\,,
\end{align}
and is parametrised by a four-momentum $p$ and a polarisation tensor $\xi_{\mu\alpha}$.
Here $S^\alpha$ and $\Sigma^\pm$ are the space-time and internal spin fields respectively:
\begin{align}
	S^1=e^{i(\phi_1+\phi_2)/2} \qquad, \qquad S^2=e^{-i(\phi_1+\phi_2)/2} \qquad,\qquad \Sigma^\pm=e^{\pm i(\phi_4+\phi_5)/2} ~.
\end{align}
The vertex operators of the graviphotons and $\bar T$-vectors are respectively given by
\begin{align}\label{VectorsHet} 
V^G(p,\epsilon) &=\epsilon_{\mu}\left(\partial X-i(p\cdot\chi)\psi \right)\bar\partial Z^\mu e^{ip\cdot Z}~,\notag\\
V^{\bar T}(p,\epsilon) &=\epsilon_\mu\left(\partial Z^\mu-i(p\cdot \chi)\chi^\mu \right)\bar\partial X e^{ip\cdot  Z}~,
\end{align}
where $p$ is the four-momentum and $\epsilon_\mu$ the polarisation vector, satisfying $\epsilon\cdot p=0$.
As in \cite{Antoniadis:2010iq}, we choose a convenient kinematic configuration such that the amplitude can be written as
\begin{align}
\left\langle (V_{\psi^+}(x_1)\cdot V_{\psi^+}(x_2))\,(V_{\psi^-}(y_1)\cdot V_{\psi^-}(y_2))\,(V^{G}(\epsilon_1,p_2)
V^{G}(\epsilon_{\bar{1}},p_{\bar{2}}))^N\,(V^{\bar{T}}(\epsilon_1,p_{\bar{2}})V^{\bar{T}}(\epsilon_{\bar{1}},p_2))^M\right\rangle\,.\nonumber
\end{align}
We consider the case where $2m\leq 2M$ of the $V^{\bar{T}}$ vertex operators contribute the fermion-bilinear piece and the structure of the different vertices is conveniently summarised  in Table~\ref{HetVertex}. The bosonic part 
of the amplitude takes the form:
\begin{align}
\langle (Z^1\bar{\partial} Z^2)^{N+2}(\bar{Z}^1\bar{\partial} \bar{Z}^2)^{N+2} (Z^1\partial\bar{Z}^2)^{M-m}(\bar{Z}^1\partial Z^2)^{M-m} 
(\partial X)^{2N+2} (\bar{\partial} X)^{2M}\rangle\,.
\end{align}
This correlator can be computed with the help of the generating function
\begin{align}\label{BosGen}
G^\textrm{bos}(\epsilon_{-},\epsilon_{+})=\left< \exp\Biggr[-\epsilon_{-}\int{d^2 z~ \partial X(Z^1 \bar\partial Z^2 + 
\bar{Z}^2\bar\partial\bar{Z}^1)} -  \epsilon_{+}\int{d^2 z~(Z^1 \partial \bar{Z}^2 + Z^2\partial\bar{Z}^1)\bar\partial X} 
\Biggr] \right>~.
\end{align}


\begin{table}[H]
\begin{center}
\begin{tabular}{|c|c|c||c|c||c||c|c||c|}\hline
\textbf{Field} & \textbf{Pos.} & \textbf{Number} &
\parbox{0.5cm}{\vspace{0.2cm}$\phi_1$\vspace{0.2cm}}&
\parbox{0.5cm}{\vspace{0.2cm}$\phi_2$\vspace{0.2cm}} &
\parbox{0.5cm}{\vspace{0.2cm}$\phi_3$\vspace{0.2cm}} &
\parbox{0.5cm}{\vspace{0.2cm}$\phi_4$\vspace{0.2cm}} &
\parbox{0.5cm}{\vspace{0.2cm}$\phi_5$\vspace{0.2cm}} &
\parbox{1.5cm}{\vspace{0.2cm}\textbf{Bosonic}\vspace{0.2cm}} \\\hline\hline
gravitino & \parbox{0.35cm}{\vspace{0.2cm}$x_1$\vspace{0.2cm}}
& $1$ & \parbox{0.7cm}
{\vspace{0.2cm}$+\frac{1}{2}$\vspace{0.2cm}} & \parbox{0.7cm}
{\vspace{0.2cm}$+\frac{1}{2}$\vspace{0.2cm}} & \parbox{0.7cm}
{\vspace{0.2cm}$+\frac{1}{2}$\vspace{0.2cm}} & \parbox{0.7cm}
{\vspace{0.2cm}$+\frac{1}{2}$\vspace{0.2cm}}& \parbox{0.7cm}
{\vspace{0.2cm}$+\frac{1}{2}$\vspace{0.2cm}} & \parbox{1.15cm}
{\vspace{0.2cm}$Z^1\bar{\partial}Z^2$\vspace{0.2cm}} \\\hline
 & \parbox{0.35cm}{\vspace{0.2cm}$x_2$\vspace{0.2cm}}
& $1$ & \parbox{0.7cm}
{\vspace{0.2cm}$-\frac{1}{2}$\vspace{0.2cm}} & \parbox{0.7cm}
{\vspace{0.2cm}$-\frac{1}{2}$\vspace{0.2cm}} & \parbox{0.7cm}
{\vspace{0.2cm}$+\frac{1}{2}$\vspace{0.2cm}} & \parbox{0.7cm}
{\vspace{0.2cm}$+\frac{1}{2}$\vspace{0.2cm}}& \parbox{0.7cm}
{\vspace{0.2cm}$+\frac{1}{2}$\vspace{0.2cm}} & \parbox{1.15cm}
{\vspace{0.2cm}$\bar{Z}^1\bar{\partial}\bar{Z}^2$\vspace{0.2cm}} \\\hline
 & \parbox{0.35cm}{\vspace{0.2cm}$y_1$\vspace{0.2cm}}
& $1$ & \parbox{0.7cm}
{\vspace{0.2cm}$+\frac{1}{2}$\vspace{0.2cm}} & \parbox{0.7cm}
{\vspace{0.2cm}$+\frac{1}{2}$\vspace{0.2cm}} & \parbox{0.7cm}
{\vspace{0.2cm}$+\frac{1}{2}$\vspace{0.2cm}} & \parbox{0.7cm}
{\vspace{0.2cm}$-\frac{1}{2}$\vspace{0.2cm}}& \parbox{0.7cm}
{\vspace{0.2cm}$-\frac{1}{2}$\vspace{0.2cm}} & \parbox{1.15cm}
{\vspace{0.2cm}$Z^1\bar{\partial}Z^2$\vspace{0.2cm}} \\\hline
 & \parbox{0.35cm}{\vspace{0.2cm}$y_2$\vspace{0.2cm}}
& $1$ & \parbox{0.7cm}
{\vspace{0.2cm}$-\frac{1}{2}$\vspace{0.2cm}} & \parbox{0.7cm}
{\vspace{0.2cm}$-\frac{1}{2}$\vspace{0.2cm}} & \parbox{0.7cm}
{\vspace{0.2cm}$+\frac{1}{2}$\vspace{0.2cm}} & \parbox{0.7cm}
{\vspace{0.2cm}$-\frac{1}{2}$\vspace{0.2cm}}& \parbox{0.7cm}
{\vspace{0.2cm}$-\frac{1}{2}$\vspace{0.2cm}} & \parbox{1.15cm}
{\vspace{0.2cm}$\bar{Z}^1\bar{\partial}\bar{Z}^2$\vspace{0.2cm}} \\\hline\hline
$F^{G}$ & \parbox{0.2cm}{\vspace{0.2cm}$z$\vspace{0.2cm}}
& $N$ & \parbox{0.15cm}
{\vspace{0.2cm}$0$\vspace{0.2cm}} & \parbox{0.15cm}
{\vspace{0.2cm}$0$\vspace{0.2cm}} & \parbox{0.15cm}
{\vspace{0.2cm}$0$\vspace{0.2cm}} & \parbox{0.15cm}
{\vspace{0.2cm}$0$\vspace{0.2cm}}& \parbox{0.15cm}
{\vspace{0.2cm}$0$\vspace{0.2cm}} & \parbox{2.05cm}
{\vspace{0.2cm}$\partial X\,Z^1\bar{\partial} Z^2$\vspace{0.2cm}} \\\hline
& \parbox{0.2cm}{\vspace{0.2cm}$z'$\vspace{0.2cm}}
& $N$ & \parbox{0.15cm}
{\vspace{0.2cm}$0$\vspace{0.2cm}} & \parbox{0.15cm}
{\vspace{0.2cm}$0$\vspace{0.2cm}} & \parbox{0.15cm}
{\vspace{0.2cm}$0$\vspace{0.2cm}} & \parbox{0.15cm}
{\vspace{0.2cm}$0$\vspace{0.2cm}}& \parbox{0.15cm}
{\vspace{0.2cm}$0$\vspace{0.2cm}} & \parbox{2.05cm}
{\vspace{0.2cm}$\partial X\,\bar{Z}^1\bar{\partial} \bar{Z}^2$\vspace{0.2cm}} \\\hline\hline
$F^{\bar{T}}$ & \parbox{0.2cm}{\vspace{0.2cm}$u$\vspace{0.2cm}}
& $m$ & \parbox{0.6cm}
{\vspace{0.2cm}$+1$\vspace{0.2cm}} & \parbox{0.6cm}
{\vspace{0.2cm}$-1$\vspace{0.2cm}} & \parbox{0.15cm}
{\vspace{0.2cm}$0$\vspace{0.2cm}} & \parbox{0.15cm}
{\vspace{0.2cm}$0$\vspace{0.2cm}}& \parbox{0.15cm}
{\vspace{0.2cm}$0$\vspace{0.2cm}} & \parbox{0.7cm}
{\vspace{0.2cm}$\bar{\partial} X$\vspace{0.2cm}} \\\hline
& \parbox{0.2cm}{\vspace{0.2cm}$u'$\vspace{0.2cm}}
& $m$ & \parbox{0.6cm}
{\vspace{0.2cm}$-1$\vspace{0.2cm}} & \parbox{0.6cm}
{\vspace{0.2cm}$+1$\vspace{0.2cm}} & \parbox{0.15cm}
{\vspace{0.2cm}$0$\vspace{0.2cm}} & \parbox{0.15cm}
{\vspace{0.2cm}$0$\vspace{0.2cm}}& \parbox{0.15cm}
{\vspace{0.2cm}$0$\vspace{0.2cm}} & \parbox{0.7cm}
{\vspace{0.2cm}$\bar{\partial} X$\vspace{0.2cm}} \\\hline
 & \parbox{0.2cm}{\vspace{0.2cm}$t$\vspace{0.2cm}}
& $M-m$ & \parbox{0.15cm}
{\vspace{0.2cm}$0$\vspace{0.2cm}} & \parbox{0.15cm}
{\vspace{0.2cm}$0$\vspace{0.2cm}} & \parbox{0.15cm}
{\vspace{0.2cm}$0$\vspace{0.2cm}} & \parbox{0.15cm}
{\vspace{0.2cm}$0$\vspace{0.2cm}}& \parbox{0.15cm}
{\vspace{0.2cm}$0$\vspace{0.2cm}} & \parbox{2.05cm}
{\vspace{0.2cm}$\bar{\partial} X\,Z^1\partial \bar{Z}^2$\vspace{0.2cm}} \\\hline
& \parbox{0.2cm}{\vspace{0.2cm}$t'$\vspace{0.2cm}}
& $M-m$ & \parbox{0.15cm}
{\vspace{0.2cm}$0$\vspace{0.2cm}} & \parbox{0.15cm}
{\vspace{0.2cm}$0$\vspace{0.2cm}} & \parbox{0.15cm}
{\vspace{0.2cm}$0$\vspace{0.2cm}} & \parbox{0.15cm}
{\vspace{0.2cm}$0$\vspace{0.2cm}}& \parbox{0.15cm}
{\vspace{0.2cm}$0$\vspace{0.2cm}} & \parbox{2.05cm}
{\vspace{0.2cm}$\bar{\partial} X\,\bar{Z}^1\partial Z^2$\vspace{0.2cm}} \\\hline\hline
PCO & \parbox{0.2cm}{\vspace{0.2cm}$P$\vspace{0.2cm}}
& $2$
& \parbox{0.15cm}
{\vspace{0.2cm}$0$\vspace{0.2cm}} & \parbox{0.15cm}
{\vspace{0.2cm}$0$\vspace{0.2cm}} & \parbox{0.6cm}
{\vspace{0.2cm}$-1$\vspace{0.2cm}} & \parbox{0.15cm}
{\vspace{0.2cm}$0$\vspace{0.2cm}}& \parbox{0.15cm}
{\vspace{0.2cm}$0$\vspace{0.2cm}} & \parbox{0.7cm}
{\vspace{0.2cm}$\partial X$\vspace{0.2cm}} \\\hline
\end{tabular}
\end{center}
\caption{Overview of the vertex contributions for the Heterotic amplitude.}
\label{HetVertex}
\end{table}
Notice that since no $\bar X$ appears in the correlator, the $T^2$ currents $\partial X$ and $\bar\partial X$ only contribute 
zero-modes. On the other hand, it is straightforward to perform the fermionic contractions and the corresponding correlator is 
expressed in terms of prime forms, \emph{cf.} Appendix \ref{appendix:ModularFunctions}:
\begin{align}
G^{\text{ferm}}_{s,(m)}=&\frac{\theta_s\!\left(\frac{x_1-x_2+y_1-y_2}{2}+u-u'\right)\theta_s\!\left(\frac{x_1-x_2+y_1-y_2}{2}-u+u'
\right)\ E^2(u,u) E^2(u',u')}{E(x_1,y_2)E(x_2,y_1)E^2(u,u')}\nonumber\\
&\times \theta_{h,s}\!\left(\frac{x_1+x_2-y_1-y_2}{2}\right)\theta_{-h,s}\!\left(\frac{x_1+x_2-y_1-y_2}{2}\right)\,,
\end{align}
where we have already cancelled the contribution of the superghosts against the contribution of the torus fermions. 
Moreover, we use the shorthand
\begin{align}
&E(u,u):=\prod_{i<j}^{m_1}E(u_i,u_j)\,,&&E(u',u'):=\prod_{i<j}^{m_2}E(u'_i,u'_j)\,,&&E(u,u'):=\prod_{i=1}^{m_1}\prod_{j=1}^{m_2}E(u_i,u'_j)\,.
\end{align}
The sum over spin structures can now be performed using  the Riemann summation identity \eqref{RiemannSummation} and the result can be further 
recast as a product of correlators:
\begin{align}
G^{\text{ferm}}_{(m)}=&\frac{\theta_1\!\left(x_1-y_2\right)\theta_1\!\left(x_2-y_1\right)\theta_{h}\!\left(u-u'\right)\theta_{-h}\!
\left(u-u'\right)\ E^2(u,u) E^2(u'u')}{E(x_1,y_2)E(x_2,y_1)E^2(u,u')}\nonumber\\
=&\left\langle\chi^1(x_1)\bar{\chi}^1(y_2)\,\chi^2(x_2)\bar{\chi}^2(y_1)\right\rangle\,\left\langle\prod_{i=1}^m\chi^4\chi^5(u_i)\,
\bar{\chi}^4\bar{\chi}^5(u'_i)\right\rangle_h\,,\label{fermCorrelatorm}
\end{align}
with both correlators evaluated in the odd spin structure. The first correlator involving $\chi^{1,2},\bar\chi^{1,2}$ yields a 
factor of $\eta^4$, since all fermions simply soak up the space-time zero modes. On 
the other hand, the fermionic correlators associated to $K3$ can be evaluated through the generating function
\begin{align}\label{FermGen}
G^{\text{ferm}}\left[\begin{matrix}
                             h \\
			     g
                            \end{matrix}
  \right](\epsilon_+)=\left\langle e^{-\epsilon_+\int (\chi^4\chi^5-\bar{\chi}^4\bar{\chi}^5)\bar{\partial}X}\right\rangle_{h,g}~.
\end{align}
Summing  the full correlator over $h,g\in\mathbb{Z}_2$ gives the contribution of the orbifold sectors and enforces the orbifold projections, 
respectively. In what follows, the bosonic and fermionic correlators (\ref{BosGen}) and (\ref{FermGen}) are calculated 
by directly evaluating the corresponding path integrals.


\section{Evaluation of the Couplings}

We are now ready to evaluate the generating functions (\ref{BosGen}) and (\ref{FermGen})  using a worldsheet path integral approach. 
In the case of the bosonic space-time directions, the worldsheet action receives a deformation of the form:
\begin{align}
	S_\textrm{def}^\textrm{bos} = \tilde\epsilon_{-}\int{d^2 z\left(\,Z^1 \bar\partial Z^2 + \bar{Z}^2\bar\partial\bar{Z}^1\,\right)} 
+  \check{\epsilon}_{+}\int{d^2 z\left(\,Z^1 \partial \bar{Z}^2 + Z^2\partial\bar{Z}^1\,\right)} ~,
\end{align}
where we have absorbed the zero-mode contribution of the $T^2$  currents into the deformation parameters
\begin{align}
	 \tilde\epsilon_{\pm} \equiv \langle \partial X\rangle\,\epsilon_{\pm} = \lambda_i (M+\bar\tau N)^i\, \epsilon_{\pm} \qquad,\qquad 
\check\epsilon_{\pm} \equiv \langle \bar\partial X\rangle\,\epsilon_{\pm} = \bar\lambda_i (M+\tau N)^i \,\epsilon_{\pm}~.
\end{align}
Here, $\lambda=(1,\bar{U})/(U-\bar{U})$ is the appropriate moduli-dependent vector picking the direction associated to $X$. One needs to 
keep in mind that in the path integral derivation, the $T^2$-lattice originally appears in its Lagrangian representation, with winding 
numbers $M^i, N^i\in\mathbb{Z}$. Upon Poisson resummation, $\lambda_i (M+\bar\tau N)^i$ and $\bar\lambda_i(M+\tau N)^i$ are effectively 
replaced by $\tau_2 P_L/\sqrt{(T-\bar{T})(U-\bar{U})-\tfrac{1}{2}(\vec{Y}-\vec{\bar{Y}})^2}$ and $\tau_2 P_R/\sqrt{(T-\bar{T})(U-\bar{U})-
\tfrac{1}{2}(\vec{Y}-\vec{\bar{Y}})^2}$, respectively, with $P_L$ and $P_R$ being the lattice momenta of the Heterotic $K3\times T^2$ 
compactification:
\begin{align}
 P_L&=\frac{m^2-U m^1-\vec Y\cdot\vec Q+T n_1+(TU-\frac{1}{2}\vec{Y}^2)n_2}{\sqrt{(T-\bar T)(U-\bar U)-\frac{1}{2}(\vec{Y}-\vec{\bar{Y}})^2}}\,,\\
 P_R&=\frac{m^2-U m^1-\vec Y\cdot\vec Q+(\bar T-\frac{1}{2}(\vec Y-\vec{\bar Y})\cdot\vec Y_1) n_1+((\bar T-\frac{1}{2}(\vec{Y}-\vec{\bar Y})\cdot \vec Y_1)U-\frac{1}{2}\vec{Y}^2)n_2}{\sqrt{(T-\bar T)(U-\bar U)-\frac{1}{2}(\vec{Y}-\vec{\bar{Y}})^2}}\,,
\end{align}
and $\vec p=\vec{Q}+\vec Y_in^i$ in the $E_8$ directions. Here, $m_i$ are the momenta, $n^i$ the windings and $Q^a$ the $U(1)$ Cartan charge vectors of $E_8$. The index $i=1,2$ parametrises the two $T^2$-directions, while $a=1,\ldots, 8$ runs over the Cartan subalgebra of $E_8$. In addition, $\vec{Y}\equiv \vec{Y}_2-U\vec{Y}_1$ is the complexified Wilson line. The above observation is important, in order to properly check modular invariance at each stage of the calculation. Hence, under 
$\tau\rightarrow -\frac{1}{\tau}$, the effective deformation parameters transform as
\begin{align}
 \tilde\epsilon_{\pm} \rightarrow \frac{\tilde\epsilon_{\pm}}{\bar\tau}  \qquad,\qquad \check\epsilon_{\pm} \rightarrow 
\frac{\check\epsilon_{\pm}}{\tau} ~.
\end{align}
The path integral over the bosonic modes $Z^1,\bar{Z}^1,Z^2,\bar{Z}^2$ can be straightforwardly performed and the resulting 
generating function can be conveniently factorised into an (almost) anti-holomorphic and a non-holomorphic piece:
\begin{align}\label{Factorization}
	G^\textrm{bos}(\epsilon_{-},\epsilon_{+})=&\, G_{\textrm{ahol}}(\epsilon_{-},\epsilon_{+})\times G_{\textrm{non-hol}}(\epsilon_{-},\epsilon_{+})~,
\end{align}
where the explicit expressions for the functional determinants $G_{\textrm{ahol}}$ and $G_{\textrm{non-hol}}$ are given in 
Appendix \ref{FourierExp}.
Using standard $\zeta$-function regularisation techniques \cite{Antoniadis:1995zn,Antoniadis:2010iq} as explained in Appendix \ref{sec:ZetaReg}, the~almost anti-holomorphic factor is simply given by
\begin{align}\label{Amodel}
	G_{\textrm{ahol}}(\epsilon_{-},\epsilon_{+})
&= \frac{(2\pi )^2(\epsilon_{-}^2-\epsilon_{+}^2)\,\bar\eta(\bar\tau) ^6}{\bar\theta_1(\tilde\epsilon_{-}-\tilde\epsilon_{+};
\bar\tau)\,\bar\theta_1(\tilde\epsilon_{-}+\tilde\epsilon_{+};\bar\tau)}~e^{-\frac{\pi}{\tau_2}(\tilde\epsilon_{-}^2+\tilde\epsilon_{+}^2)}~,
\end{align}
Moreover, as shown in Appendix \ref{FourierExp}, the non-holomorphic factor $G_{\textrm{non-hol}}$  of \eqref{Factorization} 
also admits a well-defined regularisation and, in fact, becomes trivial in the $\tau_2\rightarrow\infty$ limit at a 
point\footnote{Note that in the next section we expand around a Wilson line enhancement point, where 
$P_L=P_R\rightarrow 0$.} where $P_L = P_R$ :
\begin{align}\label{remainder}
	G_{\textrm{non-hol}}(\epsilon_{-},\epsilon_{+}) ~~\overset{\tau_2\rightarrow\infty}{\longrightarrow}~~1~.
\end{align}
We can now treat the fermionic generating function \eqref{FermGen} in a similar fashion by directly performing the 
path integral and using $\zeta$-function regularisation:
\begin{equation}
 G^\textrm{ferm}[^h_g](\check\epsilon_+)=\frac{\theta[^{1+h}_{1+g}](\check\epsilon_+ ;\tau)\theta[^{1-h}_{1-g}](\check\epsilon_+ ;\tau)}
{\eta^2}~ e^{\frac{\pi}{\tau_2}\check\epsilon_+^2}~.
\end{equation}
The full amplitude can then be written by including also the internal and gauge degrees of freedom:
\begin{align}\label{FullAmplitudeHet}
 \mathcal{F}\textrm(\epsilon_{-},\epsilon_{+})&=\sum_{g,n\geq0}\epsilon_-^{2g}\epsilon_+^{2n}\,\mathcal{F}_{g,n}\nonumber\\
  &=\int_{\mathcal{F}}\frac{d^2\tau}{\tau_2}\,G^\textrm{bos}(\epsilon_{-},\epsilon_{+})\frac{1}{\eta^4\bar\eta^{24}}\frac{1}{2}\sum_{h,g=0}^1 
G^\textrm{ferm}[^h_g](\check\epsilon_+)Z[^h_g]~\Gamma_{(2,2+8)}(T,U,Y)\,,
\end{align}
where 
\begin{align}\label{Zblock}
	Z[^h_g] = \Gamma_{K3}[^h_g]~\frac{1}{2}\sum\limits_{k,\ell=0,1}\bar\theta^6[^k_\ell]\bar\theta[^{k+h}_{\ell+g}]
\bar\theta[^{k-h}_{\ell-g}]~,
\end{align}
is the orbifold block of the K3-lattice together with the partition function of $E_7\times SU(2)$, as a result 
of the breaking of one of the $E_8$-group factors by the $\mathbb{Z}_2$-orbifold action. 
Furthermore, the K3-lattice is given explicitly by
\begin{align}\label{K3lattice}
	\Gamma_{K3}[^h_g] ~=~ \Biggr\{ \begin{array}{l l}
							\Gamma_{(4,4)}(G,B) & ,~(h,g)=(0,0) \\
							\left|\frac{2\eta^3}{\theta[^{1+h}_{1+g}]}\right|^4 & ,~(h,g)\neq(0,0) \\
							       \end{array} ~.
\end{align}
Notice that we have combined the $T^2$- and $E_8$- lattices\footnote{Conventionally, we do not include Dedekind $\eta$-function factors corresponding to oscillator contributions in the definition of the lattices.} into $\Gamma_{(2,2+8)}$, as this is convenient for incorporating non-trivial Wilson lines. The overall holomorphic Dedekind $\eta^{-4}$  factor in  (\ref{FullAmplitudeHet}) is the result of  a factor $\eta^{-4}$ arising from the bosons in the space-time directions, a factor $\eta^{-2}$ from the $T^2$ bosons, a factor $\eta^{-4}$ from the K3 bosons, a factor of $\eta^4$ from the correlator of  the fermions in the space-time direction (in the odd spin structure) and, finally, a contribution of $\eta^2$ by the bosonic $bc\,\,\textrm{-}$~ghost system. The superghost cancels the relevant $\eta$-contribution of the $T^2$ fermions. Expanding the various functional determinants in $\epsilon_{\pm}$ and extracting the power $\epsilon_-^{2g}\epsilon_+^{2n}$ yields
\begin{align}
 \mc F_{g,n}=\int_{\mc F}\frac{d^2\tau}{\tau_2}\,G_{g,n}(\tau,\bar\tau)\sum_{m_i,n^i,Q^a\in\mb Z}\left(\frac{\tau_2 P_L}{\xi}\right)^{2g-2}\left(\frac{\tau_2 P_R}{\xi}\right)^{2n}\,q^{|P_L|^2}\bar q^{|P_R|^2+\frac{1}{2}\vec{p}^{\,2}}\,.
\end{align}
Here, $G_{g,n}$ is a modular series of weights $(2n,2g-2)$. The latter can be expressed by using \eqref{FullRegBos} and the lattices in \eqref{FullAmplitudeHet}.

As a check, notice that upon taking the limit $\epsilon_+=0$, the non-holomorphic generating function trivialises, 
$G_{\textrm{non-hol}}(\epsilon_{-},0)=1$, the fermionic correlator $G^{\textrm{ferm}}$ cancels against the twisted $K3$ lattice 
and one readily recovers the result of \cite{Antoniadis:1995zn}.


\section{Field Theory Limit and the Nekrasov Partition Function}\label{NekPart}

In order to make contact with $\mathcal{N}=2$ gauge theory, we now turn to the field theory limit of the Heterotic 
amplitude (\ref{HetProposal}). We first recall that Nekrasov's partition function \eqref{NekPartK} was derived 
by starting from an $\mathcal{N}=1$ theory in six dimensions and compactifying it on a two-torus fibered over space-time 
with the $\Omega$-twist. In particular, the latter is accompanied by an R-symmetry rotation which is necessary in order to preserve a fraction of supersymmetry. In the limit where the volume of the two-torus goes to zero, one reaches a four-dimensional 
$\mathcal{N}=2$  gauge theory in the $\Omega$-background. In this  section, we start by considering the four-dimensional 
field theory limit of our amplitude at a point of enhanced gauge symmetry, where the contribution of the BPS states 
becoming massless  dominates, and we recover Nekrasov's partition function. Then, in Section \ref{NekrasovOkounkov}, 
we provide a higher dimensional extension of the latter, by keeping track of the contribution of the full tower of 
Kaluza-Klein states, thus obtaining a $\beta$-deformation thereof.

We now focus on the contribution of the full amplitude (\ref{FullAmplitudeHet}) in the field theory limit $\tau_2\rightarrow\infty$ and expand it around an $SU(2)$ Wilson line enhancement point $Y\rightarrow Y^\star$:
\begin{align}\label{EnhancementPoint}
	Y_{1}^{a\star}=Y_{2}^{a\star}=(\tfrac{1}{2},\tfrac{1}{2},y^3,\ldots,y^8) ~~,~~ (m_i,n^i)^\star =0 ~~,~~ {Q}^{a\star}=\pm( 1,- 1,0,\ldots,0)  ~,
\end{align} 
at which both left- and right- moving momenta vanish:
\begin{align}
	P_L=P_R\equiv P = \frac{a_2-U a_1}{\sqrt{(T-\bar{T})(U-\bar{U})-\tfrac{1}{2}(\vec{Y}-\vec{\bar{Y}})^2}} ~\longrightarrow~ 0~.
\end{align}
Here we have used the shorthand notation $a_i \equiv \vec{Y}_i\cdot \vec{Q}$.
It is easy to see that only the untwisted sector is relevant for the enhancement, so that it is sufficient to focus on $h=0$. 
Furthermore, since $ Z[^0_g] = 1+\mathcal{O}(e^{-2\pi\tau_2})$ we can effectively replace $Z[^0_g]\rightarrow 1$ in 
(\ref{FullAmplitudeHet}).  Using the behaviour of Jacobi theta functions in the large-$\tau_2$ limit \eqref{CIdentity}, we extract 
the $q$-expansion of the $\mathbb{Z}_2$-projected fermionic K3 correlator $G^{\textrm{ferm}}$ :
\begin{align}\label{cosine}
	\frac{1}{2}\sum\limits_{g=0,1} \theta[^{~\,1~}_{1+g}](\check\epsilon_{+};\tau) ~\theta[^{~\,1~}_{1-g}](\check\epsilon_{+};\tau) 
= -2\cos(2\pi\check\epsilon_{+})q^{1/4} + \mathcal{O}(q^{5/4})~,
\end{align}
where $q=e^{2\pi i\tau}$.
We  now take the $\tau_2\rightarrow\infty$ limit of the bosonic correlator:
\begin{align}
	G^\textrm{bos}(\epsilon_{-},\epsilon_{+})~~\overset{\tau_2\rightarrow\infty}{\longrightarrow}~~ 
\frac{\pi^2(\tilde\epsilon_{-}^2-\tilde\epsilon_{+}^2)}{\sin(\tilde\epsilon_{-}-\tilde\epsilon_{+})\sin(\tilde\epsilon_{-}+
\tilde\epsilon_{+})} + \mathcal{O}(e^{-2\pi \tau_2})~.
\end{align}
Adding all pieces together and, taking into account the remaining $\eta^{-6}$ factor,  the field theory limit of  
(\ref{FullAmplitudeHet}) at the Wilson-line enhancement point ($P_{L}= P_R=P\sim 0$) is:
\begin{align}\label{result}
\mathcal{F}\left(\epsilon_-,\epsilon_+\right)&~\sim~ (\epsilon_{-}^2-\epsilon_{+}^2)\int_0^\infty\frac{dt}{t}~
\frac{-2\cos\left( 2\epsilon_+ t\right) }{\sin\left( \epsilon_--\epsilon_+\right)t ~\sin\left( \epsilon_-+
\epsilon_+\right)t} ~e^{-\mu t}~,
\end{align}
after an appropriate rescaling by the BPS mass parameter:
\begin{align}
\mu\sim\sqrt{(T-\bar{T})(U-\bar{U})-\tfrac{1}{2}(\vec{Y}-\vec{\bar{Y}})^2}\,\bar{P}=a_2-\bar{U}a_1~,
\end{align}
 in order to exhibit the singularity behaviour of the amplitude.
The leading singularity for the $\mathcal{F}_{g,n}$-term, which is given by the coefficient of $\epsilon_-^{2g}\epsilon_+^{2n}$ 
in the expansion of (\ref{result}), is parametrised by $\mu^{2-2g-2n}$. Hence, the Heterotic amplitude (\ref{HetProposal}) 
around the $SU(2)$ enhancement point  (\ref{EnhancementPoint}) reproduces precisely the perturbative part of Nekrasov's 
partition function for an $SU(2)$ gauge theory without flavours, given in (A.7) of \cite{Nekrasov:2003rj}.

Notice that, similarly to \cite{Antoniadis:2010iq},  (\ref{result}) is still anti-holomorphic in the relevant modulus, 
which is here identified with the complexified Wilson line $Y$, even though our vertices for the graviphoton and $\bar{T}$ 
field strengths involve both $\partial X$ and $\bar\partial X$ and, hence, contribute both $P_L$ and $P_R$ to the correlation 
functions. This is to be expected, since at the Wilson line enhancement point, $P_L=P_R=P$. In addition, the invariance 
under $\epsilon_{\pm}\rightarrow -\epsilon_{\pm}$ is a consequence of the fact that $\epsilon_{-}$ and $\epsilon_{+}$ couple 
to anti-self-dual and self-dual field strengths and Lorentz invariance of the string effective action requires the presence 
of even numbers of self-dual and anti-self-dual tensors. On the other hand, contrary to \cite{Antoniadis:2010iq}, the 
generating function (\ref{result}) is not symmetric under the exchange $\epsilon_{-}\leftrightarrow\epsilon_{+}$, due to 
the presence of the $\epsilon_{+}$-dependent phase. This asymmetry can be traced back to the fact that our setup for the 
vertex operators involving graviphotons and $\bar{T}$-vectors breaks the exchange symmetry between the two Lorentz $SU(2)$'s.


\section{Radius Deformations and the Nekrasov-Okounkov Formula}\label{NekrasovOkounkov}

Let us now compare our amplitude with the partition function of a five-dimensional gauge theory with eight supercharges, compactified on a 
circle of radius $\beta$ with an $\Omega$-twist in the four non-compact dimensions, which is derived  in Section 7 of 
\cite{Nekrasov:2003rj}. To exhibit the connection, we first decouple the winding modes by taking  the $T^2$-volume to be 
sufficiently larger than the string scale, $T_2=\textrm{Vol}(T^2)\gg 1$. In this case, the Kaluza-Klein spectrum is dense 
and we have to retain the sum over the momentum modes. However, it is interesting to first consider the case where the 
modulus $U$ of the two-torus is held fixed and obtain a deformed version of Nekrasov's (four-dimensional) partition function \eqref{NekPertPart}:
\begin{align}\label{NOstart}
	\mathcal{F}\sim\int\frac{d\tau_2}{\tau_2}\sum\limits_{m_i\in\mathbb{Z} }\frac{-\epsilon_1 \epsilon_2~e^{-2\pi\tau_2|P|^2}}
{\sin(\pi\epsilon_1\tau_2 P/\xi)\sin(\pi\epsilon_2\tau_2 P/\xi)}~e^{-i\pi(\epsilon_1+\epsilon_2)\tau_2 P/\xi} + (\epsilon_i \rightarrow -\epsilon_i)~,
\end{align}
where $\xi \equiv 2i\sqrt{T_2 U_2-\tfrac{1}{2}(\text{Im}\vec Y)^2}$ and
\begin{align}
	P = \frac{1}{\xi}\Bigr( m_2+a_2-U(m_1+a_1)\Bigr)~.
\end{align}
Note that the second exponential of the cosine (\ref{cosine}) has been taken care of in (\ref{NOstart}) by symmetrising 
with respect to $\epsilon_i \rightarrow - \epsilon_i$. Expanding in the $\epsilon_i$-parameters, Poisson resumming the 
momenta $m_i$ and performing the $\tau_2$-integral, the volume dependence $T_2$ drops out and the result can be expressed as
\begin{align}\nonumber 
	\frac{\mathcal{F}}{\epsilon_1\epsilon_2}&\sim\sum\limits_{\genfrac{}{}{0pt}{}{g_1,g_2\geq 0}{g_1+g_2=0(\textrm{mod}\,2)}}\frac{B_{g_1}B_{g_2}}
{g_1!g_2!}\epsilon_1^{g_1-1}\epsilon_2^{g_2-1}\Bigr(\frac{i\pi }{U_2}\Bigr)^{g_1+g_2-2}\\ \nonumber
	& \qquad\qquad \qquad \times{\sum\limits_{m_i}}'e^{2\pi i(a\cdot m)}\left( m_1+U m_2\right)^{g_1+g_2-2}\frac{U_2}{|m_1+Um_2|^2} 
\\ \label{NO6d}
	=& \tfrac{1}{2}{\sum\limits_{m_i}}'\frac{U_2}{|m_1+Um_2|^2}\frac{e^{2\pi i(a_1 m_1+a_2 m_2)}}{\Bigr(e^{i\pi \epsilon_1(m_1+Um_2)/U_2}-1
\Bigr)\Bigr(e^{i\pi\epsilon_2(m_1+Um_2)/U_2}-1\Bigr)}+(\epsilon_i\rightarrow -\epsilon_i)~.
\end{align}
Notice that  $\mathcal{F}/(\epsilon_1\epsilon_2)$  is invariant under the T-duality transformation $U\rightarrow -1/U$ and 
$Y\rightarrow Y/U$, provided one also assigns an appropriate  transformation to the $\epsilon$ parameters,  
$\epsilon_i\rightarrow \epsilon_i/\bar{U}$. Hence,  (\ref{NO6d}) is a $U$-deformation of the Nekrasov partition function \eqref{NekPertPart}, describing a compactification of a six-dimensional theory on $T^2$.

In order to recover the result of \cite{Nekrasov:2003rj} as arising from a circle compactification of a five-dimensional theory, 
we choose a rectangular torus $T=iR_1 R_2$, $U=iR_2/R_1$ and send one of the radii to zero\footnote{Since we have 
already taken the limit $\alpha'\rightarrow 0$, we are implicitly assuming $\sqrt{\alpha'}\ll R_2 \ll R_1$.}, 
$R_2\rightarrow 0$, $a_2/R_2\rightarrow 0$. In this limit, the sum over $m_2$ can be approximated by an integral 
and one easily recovers the partition function\footnote{Here we are only concerned with the cut-off independent finite part.}
\begin{align}
\gamma_{\epsilon_1,\epsilon_2}(x|\beta) = \sum\limits_{n=1}^\infty \frac{1}{n}\frac{ e^{-\beta x}}{(e^{\beta n\epsilon_1}-1)(e^{\beta n \epsilon_2}-1)}~
\end{align}
 appearing in (A.12) of \cite{Nekrasov:2003rj}, arising from the compactification of a five-dimensional theory on a circle of 
circumference $\beta=2\pi R_1$, with the identifications $(i/R_2)\epsilon_i\rightarrow \epsilon_i$ and $x= -i a_1/R_1$.


\chapter{Type I Refined Amplitudes}\label{Sec:TypeIamp}
In this section, we  calculate the coupling (\ref{EffectiveCoupling}) at the one-loop level in Type~I string theory 
compactified on $K3\times T^2$. We first outline our conventions (which essentially follow 
\cite{Gava:1996hr}) and introduce the vertex operators for all relevant fields. In Section~\ref{TypeI:PathIntegral} 
we then evaluate a particular amplitude involving insertions $V_{(+)}$ of vector superpartners of the $T^2$ torus $\bar T$-moduli.

As before, we realise K3 as a $T^4/\mathbb{Z}_2$ orientifold, admitting both D9- and D5-branes. The 
starting point, in the absence of Wilson lines along the $T^2$, is  the $U(16)\times U(16)$ BSGP model \cite{Bianchi:1990tb}, 
obtained by setting all D5-branes to one of the $T^4/\mathbb{Z}_2$ fixed points.  The first $U(16)$ factor, associated to 
the D9-branes, can be further broken down to $U(1)\times U(1)\times U(14)$ by turning on appropriate Wilson lines for the D9-brane charges:
\begin{align}
	Y = \left(\begin{array}{c c c}
		a \sigma_3 & 0 & 0 \\ 
		0 & b \sigma_3 & 0 \\
		0 & 0 & c \sigma_3\otimes\textbf{1}_{14}
	       \end{array}\right)~,
\end{align}
where $\sigma_3$ is the Pauli matrix. We can now continuously vary the Wilson line to a point $a\rightarrow b \neq c$  
where a $U(1)$ gauge symmetry is enhanced to $SU(2)\subset U(2)$. Similarly to the Heterotic calculation of Section 
\ref{Sec:Hetamps}, we are interested in studying the field theory limit of the amplitude (\ref{EffectiveCoupling}) around this $SU(2)$  enhancement 
point\footnote{Of course, one may consider more general constructions and expand around different enhancement points, 
as discussed above eq.(\ref{FullFormula}). We refer to \cite{Pradisi:1988xd}, \cite{Bianchi:1990tb},\cite{Angelantonj:2002ct} 
for further details on the construction of consistent orientifold models.}. There, the BPS states becoming massless belong 
to vector multiplets only and, hence, the dominant contribution arises from the 9-9 sector of the annulus amplitude.

\section{Vertex Operators}\label{TypeI:Setup}
Following the discussion of the previous paragraph, we restrict our attention to the 9-9 sector of the annulus diagram. 
We represent the cylinder as a torus  acted upon by the $\mathbb{Z}_2$
involution 
\begin{align}
\Omega:\,(\sigma,t)\mapsto(-\sigma,t)\,.\label{WorldSheetInvolution}
\end{align}
A point on the worldsheet is then parametrised by $z=\sigma + \tau t$, with the worldsheet modulus $\tau=i\tau_2$ 
being purely imaginary. The $\mathbb{Z}_2$ image of
$z$ is accordingly given by $\hat{z}=-\sigma +\tau t$. By choosing this coordinate system we have fixed the analytic transformations of $z$ up to rigid translations and, hence, the formulae we obtain are not manifestly invariant under analytic transformations.

We employ the same notation  for the worldsheet super-coordinates as in Section \ref{Sec:Hetamps}. Using the `doubled picture' of a toroidal worldsheet, the right-moving superpartners are denoted by a tilde $(\tilde\chi^1,\tilde\chi^2,\tilde\psi,\tilde\chi^4,\tilde\chi^5)$. They correspond to the images of the worldsheet fermions $(\psi,\chi^i)$ under $\Omega$ and, in a similar fashion, we bosonise the superghost via a free boson $\varphi$, its mirror being $\tilde{\varphi}$. 

We are now ready to discuss the worldsheet emission vertex operators of physical fields in the  $\mathcal{N}=2$ Type I compactification. In particular, we  focus only on those states that are relevant for  later explicit computations, namely gravitini ($V^{\text{grav}}$), graviphotons ($V^G$) as well as the vector partners of the dilaton ($V^{\bar S}$), the complex structure modulus of $T^2$ ($V^U$) and the D5-gauge coupling ($V^{\bar S'}$) respectively, see Table \ref{tb:HetTypeI}. Using similar  conventions as in the Heterotic case, the anti-self-dual vertex operators for the graviphoton and the vector partner of the $U$-modulus take the form:
\begin{align}
	V^U(p,\epsilon)=V(a=+1;p,\epsilon) \qquad ,\qquad V^G(p,\epsilon)=V(a=-1;p,\epsilon)\,,\label{TypeIVertexGT}
\end{align}
where $V(a;p,\epsilon)$ is given by:
\begin{align}
V(a;p,\epsilon)=\epsilon_\mu &\bigg[\left(\partial X+i(p\cdot \chi) \psi\right) \left(\bar{\partial} Z^{\mu}+i(p\cdot\tilde{\chi}) 
\tilde{\chi}^\mu\right)\nonumber\\
+&a\, e^{-\tfrac{1}{2}(\varphi+\tilde{\varphi})} p_\nu S^\alpha {(\sigma^{\mu\nu})_\alpha}^\beta \tilde{S}_\beta\,e^{\frac{i}{2}(\phi_3+\tilde{\phi}_3)}\, 
\Sigma^+ \tilde{\Sigma}^-\bigg] \,e^{ip\cdot Z}+[\rm{left} \leftrightarrow \rm{right} ]\,.\label{TypeIVertexGTtype}
\end{align}
They are parametrised by a momentum vector $p_\mu$ and a polarisation vector $\epsilon_\mu$ satisfying the transversality condition $\epsilon\cdot p=0$. Moreover, we have introduced the space-time spin fields, for which we choose the explicit representation:
\begin{align}
&S^\alpha{(\sigma^{12})_\alpha}^\beta\tilde{S}_\beta=e^{\frac{i}{2}(\phi_1+\phi_2)}\times e^{\frac{i}{2}(\tilde{\phi}_1+\tilde{\phi}_2)}\,,\nonumber\\
&S^\alpha{(\sigma^{\bar{1}\bar{2}})_\alpha}^\beta\tilde{S}_\beta=e^{-\frac{i}{2}(\phi_1+\phi_2)}\times e^{-\frac{i}{2}(\tilde{\phi}_1+\tilde{\phi}_2)}\,,
\end{align}
and, similarly, for the spin fields of the internal $K3$:
\begin{align}
	\Sigma^\pm=e^{\pm\frac{i}{2}(\phi_4+\phi_5)} \qquad,\qquad \hat{\Sigma}^\pm=e^{\pm\frac{i}{2}({\phi}_4-{\phi}_5)}\,.
\end{align}
The two terms in the square bracket of (\ref{TypeIVertexGTtype}) come with different powers of the superghosts $e^{\varphi+\tilde{\varphi}}$ and correspond to the NS and R contributions, respectively. Notice that the difference between $V^G$ and $V^U$ lies in the relative sign between these two contributions, labeled by the parameter $a=\pm1$. \footnote{Note that this convention is compatible with space-time supersymmetry (or Heterotic/Type~I duality).}

Similarly, the vertices for self-dual vector partners of $\bar{S}$ and $\bar{S}'$ are
\begin{align}
&V^{\bar{S}'}(p,\epsilon)=\overbar{V}(b=+1;p,\epsilon) \qquad,\qquad V^{\bar{S}}(p,\epsilon)=\overbar{V}(b=-1;p,\epsilon)\,,\label{TypeIVertexUS}
\end{align}
where we have introduced
\begin{align}
\overbar{V}(b;p,\epsilon)=\epsilon_\mu &\bigg[\left(\partial X+i(p\cdot \chi) \psi\right) \left(\bar{\partial} 
Z^{\mu}+i(p\cdot\tilde{\chi}) \tilde{\chi}^\mu\right)\nonumber\\
+&b\, e^{-\tfrac{1}{2}(\varphi+\tilde{\varphi})} p_\nu S_{\dot{\alpha}} {(\bar{\sigma}^{\mu\nu})^{\dot{\alpha}}}_{\dot{\beta}} 
\tilde{S}^{\dot{\beta}}\,e^{\frac{i}{2}(\phi_3+\tilde{\phi}_3)}\, \hat{\Sigma}^+ \hat{\tilde{\Sigma}}^-\bigg] \,e^{ip\cdot Z}+
[\rm{left} \leftrightarrow \rm{right} ]\,,\label{TypeIVertexUStype}
\end{align}
with the following convention for the space-time spin fields:
\begin{align}
&S_{\dot{\alpha}}{(\bar{\sigma}^{1\bar{2}})^{\dot{\alpha}}}_{\dot{\beta}}\tilde{S}^{\dot{\beta}}=e^{\frac{i}{2}(\phi_1-\phi_2)}\times 
e^{\frac{i}{2}(\tilde{\phi}_1-\tilde{\phi}_2)}\,,\nonumber\\
&S_{\dot{\alpha}}{(\bar{\sigma}^{\bar{1}2})^{\dot{\alpha}}}_{\dot{\beta}}\tilde{S}^{\dot{\beta}}=e^{-\frac{i}{2}(\phi_1-\phi_2)}\times
e^{-\frac{i}{2}(\tilde{\phi}_1-\tilde{\phi}_2)}\,.
\end{align}
Once again, the relative sign between the NS and R sectors distinguishes between the two fields. To make this distinction more visible in explicit calculations, we denoted this relative sign through a parameter $b=\pm 1$, where $b=1$ corresponds to $F_{\bar{S}'}$ and $b=-1$ corresponds to $F_{\bar{S}}$.

At a technical level, fixing the relative signs between different spin structures turns out to be a non-trivial problem, even at the one-loop level since the absence of modular invariance does not allow one to fix all signs unambiguously. We circumvent this problem by inserting at least one fermion vertex operator into our amplitude. In this case monodromy invariance of the final answer  allows us to fix all relative signs. Hence, as in Section \ref{Sec:Hetamps}, instead of two gravitons --- as written schematically in (\ref{SchematicAmplitude}) --- we use four gravitini. As discussed in Section~\ref{Sect:Couplings} this is possible since both of these fields are part of the supergravity multiplet and the two terms 
(\ref{EffectiveCoupling}) in the  string effective action are related by supersymmetry. The vertex operator for the gravitino can be written as:
\begin{align}
&V_{\pm}^{\text{grav}}(\xi_{\mu\alpha},p)=\xi_{\mu\alpha}e^{-\varphi/2}S^\alpha e^{i\phi_3/2} \Sigma^\pm \,\left[\bar{\partial}
Z^\mu+i(p\cdot\tilde{\chi})\tilde{\chi}^\mu\right] e^{ip\cdot Z}+[\rm{left} \leftrightarrow \rm{right} ]\,,
\end{align}
which is parametrised by the four-momentum $p^\mu$ and the polarisation tensor $\xi_{\mu\alpha}$.

\section{Amplitude and Spin-Structure Sum}
We are now ready to compute the effective coupling (\ref{EffectiveCoupling}). To simplify the computation, we choose a particular kinematic configuration for all external fields. Specifically, we consider a setting of the form ($N=g-2$):
\begin{align}
\mathcal{F}_{M,N}=\bigg\langle &V^{\text{grav}}_{+}(\xi_{21},p_1)\,V^{\text{grav}}_{-}(\xi_{\bar{2}1},p_1)\,V^{\text{grav}}_{+}
(\xi_{22},p_{\bar{1}})\,V^{\text{grav}}_{-}(\xi_{\bar{2}2},p_{\bar{1}})\,
\nonumber\\
&\times \left[V^{G}(\epsilon_2,p_1)V^{G}(\epsilon_{\bar{2}},p_{\bar{1}})\right]^N\,\left[V^{\bar{S}',\bar{S}}(\epsilon_{\bar{2}},p_1)
V^{\bar{S}',\bar{S}}(\epsilon_{2},p_{\bar{1}})\right]^M\bigg\rangle\,,\label{AmplitudesDefinition}
\end{align}
where, for the moment, we consider inserting vector partners of either $\bar{S}$ or $\bar{S}'$. A major difficulty in computing this amplitude lies in the fact that all vertices contribute in all possible ways, some of them providing the R-R part while the rest the NS-NS part and, out of the NS-NS part, some contribute the bosonic Lorentz current and the others the fermionic one. To see this, let us consider a typical term with $(n_1,n_2,m_1,m_2)$ numbers of fermionic Lorentz currents at positions $(z,z',u,u')$ with kinematics $(\epsilon_2,p_1)$, $(\epsilon_{\bar{2}},p_{\bar{1}})$, $(\epsilon_{\bar{2}},p_1)$, $(\epsilon_{2},p_{\bar{1}})$, respectively, and $(n_3,n_4,m_3,m_4)$ R-R
vertices at positions $(w,w',v,v')$ with kinematics $(\epsilon_2,p_1)$, $(\epsilon_{\bar{2}},p_{\bar{1}})$, $(\epsilon_{\bar{2}},p_1)$, $(\epsilon_{2},p_{\bar{1}})$ respectively. The positions have indices, \emph{e.g.} $z_i$ where $i=1,...,n_1$ etc., but we suppress these in the
following to simplify the notation. Concerning the gravitini, we have to consider two different possibilities:
\begin{itemize}
\item[(i)] the gravitini only contribute bosonic Lorentz currents\,,
\item [(ii)] the gravitini also contribute fermionic currents\,,\footnote{We only discuss in detail the case where all 
of them contribute the fermionic currents.}
\end{itemize}
which we discuss in parallel and denote their worldsheet positions by $(x_1,x_2,y_1,y_2)$. For convenience, we have compiled an overview of the vertex operators in Tables~\ref{Tab:TypeIvertex} and \ref{Tab:TypeIvertexA}, respectively. In both cases Lorentz charge conservation implies
\begin{align}
	n_1-n_2+n_3-n_4=0 \qquad,\qquad m_1-m_2+m_3-m_4=0~.\label{ConstraintsSum}
\end{align}
As mentioned above, we have used the trick of doubling the cylinder and the right-moving part of the vertex at the image point is indicated through hatted variables.\footnote{Note for example that an R-R vertex of the type $S_L\times S'_R(w)$ is the same as $S_L(w)\times S'_L(\hat{w})$. Since our vertices are symmetrised between left and right sectors this amounts to integrating the worldsheet coordinates over the entire doubled cylinder (ie. $\sigma \in [-1/2,1/2]$).}  To balance the ghost charges, we also insert $m_{\text{PCO}}=(n_3+n_4+m_3+m_4+2)$ picture-changing operators (PCO) at some positions $P$. Moreover, we note that the total (\emph{i.e.} left plus right) $U(1)$ charge in the $T^2$ fermion sector $(\psi, \bar{\psi})$ can only be cancelled if all PCOs contribute the supercurrent of $T^2$,
\begin{align}
V_{\text{PCO}}=e^{\varphi} \partial X \bar{\psi} +e^{\tilde{\varphi}} \bar{\partial} X \tilde{\bar{\psi}}+\ldots\,,
\end{align}
as indicated in Tables~\ref{Tab:TypeIvertex} and \ref{Tab:TypeIvertexA}. Since in the vertices of the physical states, as well as in the PCOs,  only the holomorphic torus coordinate $X$ (but not $\bar{X}$) appears, the latter only contributes momentum zero modes:
\begin{equation}\label{P3}
P_3 =  \frac{\tau}{\sqrt{(T-\bar{T})(U-\bar{U})-\tfrac{1}{2}(\vec{Y}-\vec{\bar{Y}})^2}} \Bigr(\, m_2-Um_1+\vec{Y}\cdot\vec{Q} \,\Bigr)~.
\end{equation}
Here, $\vec{Y}=\vec{Y}_2-U\vec{Y}_1$ is the (complexified) Wilson line vector associated to the D9-brane gauge group along the two directions of  $T^2$ and $\vec{Q}$ is the associated charge vector of the open string states. Since we do not turn on a Ramond-Ramond $B$ field on the $T^2$, the modulus $T$ is purely imaginary, $T=i \text{Vol}(T^2)$.


\begin{table}[H]\begin{center}
\noindent\makebox[\textwidth]{%
\begin{tabular}{|c|c|c||c|c||c||c|c||c|c||c||c|c||c|}\hline
\textbf{Field} & \textbf{Pos.} & \textbf{$\#$} &
\parbox{0.5cm}{\vspace{0.2cm}$\phi_1$\vspace{0.2cm}}&
\parbox{0.5cm}{\vspace{0.2cm}$\phi_2$\vspace{0.2cm}} &
\parbox{0.5cm}{\vspace{0.2cm}$\phi_3$\vspace{0.2cm}} &
\parbox{0.5cm}{\vspace{0.2cm}$\phi_4$\vspace{0.2cm}} &
\parbox{0.5cm}{\vspace{0.2cm}$\phi_5$\vspace{0.2cm}} & \parbox{0.5cm}{\vspace{0.2cm}$\tilde{\phi}_1$\vspace{0.2cm}}&
\parbox{0.5cm}{\vspace{0.2cm}$\tilde{\phi}_2$\vspace{0.2cm}} &
\parbox{0.5cm}{\vspace{0.2cm}$\tilde{\phi}_3$\vspace{0.2cm}} &
\parbox{0.5cm}{\vspace{0.2cm}$\tilde{\phi}_4$\vspace{0.2cm}} &
\parbox{0.5cm}{\vspace{0.2cm}$\tilde{\phi}_5$\vspace{0.2cm}} &
\parbox{1.5cm}{\vspace{0.2cm}\textbf{Bosonic}\vspace{0.2cm}} \\\hline\hline
grav & \parbox{0.35cm}{\vspace{0.2cm}$x_1$\vspace{0.2cm}}
& $1$ & \parbox{0.7cm}
{\vspace{0.2cm}$+\frac{1}{2}$\vspace{0.2cm}} & \parbox{0.7cm}
{\vspace{0.2cm}$+\frac{1}{2}$\vspace{0.2cm}} & \parbox{0.7cm}
{\vspace{0.2cm}$+\frac{1}{2}$\vspace{0.2cm}} & \parbox{0.7cm}
{\vspace{0.2cm}$+\frac{1}{2}$\vspace{0.2cm}}& \parbox{0.7cm}
{\vspace{0.2cm}$+\frac{1}{2}$\vspace{0.2cm}} 
&  \parbox{0.15cm}{\vspace{0.2cm}$0$\vspace{0.2cm}} & \parbox{0.15cm}{\vspace{0.2cm}$0$\vspace{0.2cm}} & \parbox{0.15cm}
{\vspace{0.2cm}$0$\vspace{0.2cm}} & \parbox{0.15cm}{\vspace{0.2cm}$0$\vspace{0.2cm}}& \parbox{0.15cm}{\vspace{0.2cm}$0$\vspace{0.2cm}}
& \parbox{1.15cm}
{\vspace{0.2cm}$Z^1\bar{\partial}Z^2$\vspace{0.2cm}} \\\hline
 & \parbox{0.35cm}{\vspace{0.2cm}$x_2$\vspace{0.2cm}}
& $1$ & \parbox{0.7cm}
{\vspace{0.2cm}$+\frac{1}{2}$\vspace{0.2cm}} & \parbox{0.7cm}
{\vspace{0.2cm}$+\frac{1}{2}$\vspace{0.2cm}} & \parbox{0.7cm}
{\vspace{0.2cm}$+\frac{1}{2}$\vspace{0.2cm}} & \parbox{0.7cm}
{\vspace{0.2cm}$-\frac{1}{2}$\vspace{0.2cm}}& \parbox{0.7cm}
{\vspace{0.2cm}$-\frac{1}{2}$\vspace{0.2cm}} & 
\parbox{0.15cm}{\vspace{0.2cm}$0$\vspace{0.2cm}} & \parbox{0.15cm}{\vspace{0.2cm}$0$\vspace{0.2cm}} & \parbox{0.15cm}
{\vspace{0.2cm}$0$\vspace{0.2cm}} & \parbox{0.15cm}{\vspace{0.2cm}$0$\vspace{0.2cm}}& \parbox{0.15cm}{\vspace{0.2cm}$0$\vspace{0.2cm}} &
\parbox{1.15cm}
{\vspace{0.2cm}$Z^1\bar{\partial}Z^2$\vspace{0.2cm}} \\\hline
 & \parbox{0.35cm}{\vspace{0.2cm}$y_1$\vspace{0.2cm}}
& $1$ & \parbox{0.7cm}
{\vspace{0.2cm}$-\frac{1}{2}$\vspace{0.2cm}} & \parbox{0.7cm}
{\vspace{0.2cm}$-\frac{1}{2}$\vspace{0.2cm}} & \parbox{0.7cm}
{\vspace{0.2cm}$+\frac{1}{2}$\vspace{0.2cm}} & \parbox{0.7cm}
{\vspace{0.2cm}$+\frac{1}{2}$\vspace{0.2cm}}& \parbox{0.7cm}
{\vspace{0.2cm}$+\frac{1}{2}$\vspace{0.2cm}} &
\parbox{0.15cm}{\vspace{0.2cm}$0$\vspace{0.2cm}} & \parbox{0.15cm}{\vspace{0.2cm}$0$\vspace{0.2cm}} & \parbox{0.15cm}
{\vspace{0.2cm}$0$\vspace{0.2cm}} & \parbox{0.15cm}{\vspace{0.2cm}$0$\vspace{0.2cm}}& \parbox{0.15cm}{\vspace{0.2cm}$0$\vspace{0.2cm}}&
 \parbox{1.15cm}
{\vspace{0.2cm}$\bar{Z}^1\bar{\partial}\bar{Z}^2$\vspace{0.2cm}} \\\hline
 & \parbox{0.35cm}{\vspace{0.2cm}$y_2$\vspace{0.2cm}}
& $1$ & \parbox{0.7cm}
{\vspace{0.2cm}$-\frac{1}{2}$\vspace{0.2cm}} & \parbox{0.7cm}
{\vspace{0.2cm}$-\frac{1}{2}$\vspace{0.2cm}} & \parbox{0.7cm}
{\vspace{0.2cm}$+\frac{1}{2}$\vspace{0.2cm}} & \parbox{0.7cm}
{\vspace{0.2cm}$-\frac{1}{2}$\vspace{0.2cm}}& \parbox{0.7cm}
{\vspace{0.2cm}$-\frac{1}{2}$\vspace{0.2cm}} 
&  \parbox{0.15cm}{\vspace{0.2cm}$0$\vspace{0.2cm}} & \parbox{0.15cm}{\vspace{0.2cm}$0$\vspace{0.2cm}} & \parbox{0.15cm}
{\vspace{0.2cm}$0$\vspace{0.2cm}} & \parbox{0.15cm}{\vspace{0.2cm}$0$\vspace{0.2cm}}& \parbox{0.15cm}{\vspace{0.2cm}$0$\vspace{0.2cm}}
& \parbox{1.15cm}
{\vspace{0.2cm}$\bar{Z}^1\bar{\partial}\bar{Z}^2$\vspace{0.2cm}} \\\hline\hline
\parbox{0.8cm}{\vspace{0.2cm}$F^{G,U}$\vspace{0.2cm}} & \parbox{0.2cm}{\vspace{0.2cm}$z$\vspace{0.2cm}}
& $n_1$ & \parbox{0.6cm}
{\vspace{0.2cm}$+1$\vspace{0.2cm}} & \parbox{0.6cm}
{\vspace{0.2cm}$+1$\vspace{0.2cm}} & \parbox{0.15cm}
{\vspace{0.2cm}$0$\vspace{0.2cm}} & \parbox{0.15cm}
{\vspace{0.2cm}$0$\vspace{0.2cm}}& \parbox{0.15cm}
{\vspace{0.2cm}$0$\vspace{0.2cm}}
&  \parbox{0.15cm}{\vspace{0.2cm}$0$\vspace{0.2cm}} & \parbox{0.15cm}{\vspace{0.2cm}$0$\vspace{0.2cm}} & \parbox{0.15cm}
{\vspace{0.2cm}$0$\vspace{0.2cm}} & \parbox{0.15cm}{\vspace{0.2cm}$0$\vspace{0.2cm}}& \parbox{0.15cm}{\vspace{0.2cm}$0$
\vspace{0.2cm}} & \parbox{0.7cm}
{\vspace{0.2cm}$\partial X$\vspace{0.2cm}} \\\hline
& \parbox{0.2cm}{\vspace{0.2cm}$z'$\vspace{0.2cm}}
& $n_2$ & \parbox{0.6cm}
{\vspace{0.2cm}$-1$\vspace{0.2cm}} & \parbox{0.6cm}
{\vspace{0.2cm}$-1$\vspace{0.2cm}} & \parbox{0.15cm}
{\vspace{0.2cm}$0$\vspace{0.2cm}} & \parbox{0.15cm}
{\vspace{0.2cm}$0$\vspace{0.2cm}}& \parbox{0.15cm}
{\vspace{0.2cm}$0$\vspace{0.2cm}}
&  \parbox{0.15cm}{\vspace{0.2cm}$0$\vspace{0.2cm}} & \parbox{0.15cm}{\vspace{0.2cm}$0$\vspace{0.2cm}} & \parbox{0.15cm}
{\vspace{0.2cm}$0$\vspace{0.2cm}} & \parbox{0.15cm}{\vspace{0.2cm}$0$\vspace{0.2cm}}& \parbox{0.15cm}{\vspace{0.2cm}$0$
\vspace{0.2cm}} & \parbox{0.7cm}
{\vspace{0.2cm}$\partial X$\vspace{0.2cm}} \\\hline
 & \parbox{0.2cm}{\vspace{0.2cm}$w$\vspace{0.2cm}}
& $n_3$ & \parbox{0.7cm}
{\vspace{0.2cm}$+\frac{1}{2}$\vspace{0.2cm}} & \parbox{0.7cm}
{\vspace{0.2cm}$+\frac{1}{2}$\vspace{0.2cm}} & \parbox{0.7cm}
{\vspace{0.2cm}$+\frac{1}{2}$\vspace{0.2cm}} & \parbox{0.7cm}
{\vspace{0.2cm}$+\frac{1}{2}$\vspace{0.2cm}}& \parbox{0.7cm}
{\vspace{0.2cm}$+\frac{1}{2}$\vspace{0.2cm}}
& \parbox{0.7cm}
{\vspace{0.2cm}$+\frac{1}{2}$\vspace{0.2cm}} & \parbox{0.7cm}
{\vspace{0.2cm}$+\frac{1}{2}$\vspace{0.2cm}} & \parbox{0.7cm}
{\vspace{0.2cm}$+\frac{1}{2}$\vspace{0.2cm}} & \parbox{0.7cm}
{\vspace{0.2cm}$-\frac{1}{2}$\vspace{0.2cm}}& \parbox{0.7cm}
{\vspace{0.2cm}$-\frac{1}{2}$\vspace{0.2cm}}
 & \parbox{0.7cm}
{\vspace{0.2cm}$\partial X$\vspace{0.2cm}} \\\hline
 & \parbox{0.2cm}{\vspace{0.2cm}$w'$\vspace{0.2cm}}
& $n_4$ & \parbox{0.7cm}
{\vspace{0.2cm}$-\frac{1}{2}$\vspace{0.2cm}} & \parbox{0.7cm}
{\vspace{0.2cm}$-\frac{1}{2}$\vspace{0.2cm}} & \parbox{0.7cm}
{\vspace{0.2cm}$+\frac{1}{2}$\vspace{0.2cm}} & \parbox{0.7cm}
{\vspace{0.2cm}$+\frac{1}{2}$\vspace{0.2cm}}& \parbox{0.7cm}
{\vspace{0.2cm}$+\frac{1}{2}$\vspace{0.2cm}}
& \parbox{0.7cm}
{\vspace{0.2cm}$-\frac{1}{2}$\vspace{0.2cm}} & \parbox{0.7cm}
{\vspace{0.2cm}$-\frac{1}{2}$\vspace{0.2cm}} & \parbox{0.7cm}
{\vspace{0.2cm}$+\frac{1}{2}$\vspace{0.2cm}} & \parbox{0.7cm}
{\vspace{0.2cm}$-\frac{1}{2}$\vspace{0.2cm}}& \parbox{0.7cm}
{\vspace{0.2cm}$-\frac{1}{2}$\vspace{0.2cm}}
 & \parbox{0.7cm}
{\vspace{0.2cm}$\partial X$\vspace{0.2cm}} \\\hline\hline
\parbox{0.8cm}{\vspace{0.2cm}$F^{\bar{S},\bar{S}'}$\vspace{0.2cm}} & \parbox{0.2cm}{\vspace{0.2cm}$u$\vspace{0.2cm}}
& $m_1$ & \parbox{0.6cm}
{\vspace{0.2cm}$+1$\vspace{0.2cm}} & \parbox{0.6cm}
{\vspace{0.2cm}$-1$\vspace{0.2cm}} & \parbox{0.15cm}
{\vspace{0.2cm}$0$\vspace{0.2cm}} & \parbox{0.15cm}
{\vspace{0.2cm}$0$\vspace{0.2cm}}& \parbox{0.15cm}
{\vspace{0.2cm}$0$\vspace{0.2cm}}
&  \parbox{0.15cm}{\vspace{0.2cm}$0$\vspace{0.2cm}} & \parbox{0.15cm}{\vspace{0.2cm}$0$\vspace{0.2cm}} & \parbox{0.15cm}
{\vspace{0.2cm}$0$\vspace{0.2cm}} & \parbox{0.15cm}{\vspace{0.2cm}$0$\vspace{0.2cm}}& \parbox{0.15cm}{\vspace{0.2cm}$0$\vspace{0.2cm}}
 & \parbox{0.7cm}
{\vspace{0.2cm}$\partial X$\vspace{0.2cm}} \\\hline
& \parbox{0.2cm}{\vspace{0.2cm}$u'$\vspace{0.2cm}}
& $m_2$ & \parbox{0.6cm}
{\vspace{0.2cm}$-1$\vspace{0.2cm}} & \parbox{0.6cm}
{\vspace{0.2cm}$+1$\vspace{0.2cm}} & \parbox{0.15cm}
{\vspace{0.2cm}$0$\vspace{0.2cm}} & \parbox{0.15cm}
{\vspace{0.2cm}$0$\vspace{0.2cm}}& \parbox{0.15cm}
{\vspace{0.2cm}$0$\vspace{0.2cm}}
&  \parbox{0.15cm}{\vspace{0.2cm}$0$\vspace{0.2cm}} & \parbox{0.15cm}{\vspace{0.2cm}$0$\vspace{0.2cm}} & \parbox{0.15cm}
{\vspace{0.2cm}$0$\vspace{0.2cm}} & \parbox{0.15cm}{\vspace{0.2cm}$0$\vspace{0.2cm}}& \parbox{0.15cm}{\vspace{0.2cm}$0$\vspace{0.2cm}}
 & \parbox{0.7cm}
{\vspace{0.2cm}$\partial X$\vspace{0.2cm}} \\\hline 
& \parbox{0.2cm}{\vspace{0.2cm}$v$\vspace{0.2cm}}
& $m_3$ & \parbox{0.7cm}
{\vspace{0.2cm}$+\frac{1}{2}$\vspace{0.2cm}} & \parbox{0.7cm}
{\vspace{0.2cm}$-\frac{1}{2}$\vspace{0.2cm}} & \parbox{0.7cm}
{\vspace{0.2cm}$+\frac{1}{2}$\vspace{0.2cm}} & \parbox{0.7cm}
{\vspace{0.2cm}$+\frac{1}{2}$\vspace{0.2cm}}& \parbox{0.7cm}
{\vspace{0.2cm}$-\frac{1}{2}$\vspace{0.2cm}}
& \parbox{0.7cm}
{\vspace{0.2cm}$+\frac{1}{2}$\vspace{0.2cm}} & \parbox{0.7cm}
{\vspace{0.2cm}$-\frac{1}{2}$\vspace{0.2cm}} & \parbox{0.7cm}
{\vspace{0.2cm}$+\frac{1}{2}$\vspace{0.2cm}} & \parbox{0.7cm}
{\vspace{0.2cm}$-\frac{1}{2}$\vspace{0.2cm}}& \parbox{0.7cm}
{\vspace{0.2cm}$+\frac{1}{2}$\vspace{0.2cm}}
 & \parbox{0.7cm}
{\vspace{0.2cm}$\partial X$\vspace{0.2cm}} \\\hline
 & \parbox{0.2cm}{\vspace{0.2cm}$v'$\vspace{0.2cm}}
& $m_4$ & \parbox{0.7cm}
{\vspace{0.2cm}$-\frac{1}{2}$\vspace{0.2cm}} & \parbox{0.7cm}
{\vspace{0.2cm}$+\frac{1}{2}$\vspace{0.2cm}} & \parbox{0.7cm}
{\vspace{0.2cm}$+\frac{1}{2}$\vspace{0.2cm}} & \parbox{0.7cm}
{\vspace{0.2cm}$+\frac{1}{2}$\vspace{0.2cm}}& \parbox{0.7cm}
{\vspace{0.2cm}$-\frac{1}{2}$\vspace{0.2cm}}
& \parbox{0.7cm}
{\vspace{0.2cm}$-\frac{1}{2}$\vspace{0.2cm}} & \parbox{0.7cm}
{\vspace{0.2cm}$+\frac{1}{2}$\vspace{0.2cm}} & \parbox{0.7cm}
{\vspace{0.2cm}$+\frac{1}{2}$\vspace{0.2cm}} & \parbox{0.7cm}
{\vspace{0.2cm}$-\frac{1}{2}$\vspace{0.2cm}}& \parbox{0.7cm}
{\vspace{0.2cm}$+\frac{1}{2}$\vspace{0.2cm}}
 & \parbox{0.7cm}
{\vspace{0.2cm}$\partial X$\vspace{0.2cm}} \\\hline\hline
PCO & \parbox{0.2cm}{\vspace{0.2cm}$P$\vspace{0.2cm}}
& $m_{\text{PCO}}$
& \parbox{0.15cm}
{\vspace{0.2cm}$0$\vspace{0.2cm}} & \parbox{0.15cm}
{\vspace{0.2cm}$0$\vspace{0.2cm}} & \parbox{0.6cm}
{\vspace{0.2cm}$-1$\vspace{0.2cm}} & \parbox{0.15cm}
{\vspace{0.2cm}$0$\vspace{0.2cm}}& \parbox{0.15cm}
{\vspace{0.2cm}$0$\vspace{0.2cm}}
& \parbox{0.15cm}
{\vspace{0.2cm}$0$\vspace{0.2cm}} & \parbox{0.15cm}
{\vspace{0.2cm}$0$\vspace{0.2cm}} & \parbox{0.15cm}
{\vspace{0.2cm}$0$\vspace{0.2cm}} & \parbox{0.15cm}
{\vspace{0.2cm}$0$\vspace{0.2cm}}& \parbox{0.15cm}
{\vspace{0.2cm}$0$\vspace{0.2cm}}
 & \parbox{0.7cm}
{\vspace{0.2cm}$\partial X$\vspace{0.2cm}} \\\hline
\end{tabular}}
\end{center}
\caption{Overview of the vertex contributions for the Type I amplitude in case (i), \emph{i.e.} the gravitini only contribute 
bosonic Lorentz currents.}
\label{Tab:TypeIvertex}
\end{table}


Having fixed the precise setup of vertex operators, we now proceed to compute all possible contractions. Since this is a rather technical and tedious task, we only point out the salient features. First of all, one can check that the spin-structure dependent part of the $T^2$-contribution of $(\psi,\tilde{\psi})$ precisely cancels that of the superghosts. Therefore, the positions of the picture-changing operators $P$ drop out of the expression, as expected from physical consistency, and the contribution of the fermions takes the form:
\begin{align}
G_{(i)}[&s]=
\theta_s\left(\tfrac{1}{2}(x_1+x_2-y_1-y_2+w-w'+\hat{w}-\hat{w}'+v-v'+\hat{v}-\hat{v}')
+z-z'+u-u'\right)\nonumber\\*
&\times \theta_s\left(\tfrac{1}{2}(x_1+x_2-y_1-y_2+w-w'+\hat{w}-\hat{w}'-v+v'-\hat{v}+\hat{v}') +z-z'-u+u'\right)\nonumber\\*
&\times\theta_{h,s}\left(\tfrac{1}{2}(x_1-x_2+y_1-y_2+w+w'-\hat{w}-\hat{w}'+v+v'-\hat{v}-\hat{v}')\right)\nonumber\\*
&\times\theta_{-h,s}\left(\tfrac{1}{2}(x_1-x_2+y_1-y_2+w+w'-\hat{w}-\hat{w}'-v-v'+\hat{v}+\hat{v}')\right)\nonumber\\*
&\times\mathbb{B}_{(i)}(x_1,x_2,y_1,y_2,u,u',v,v',w,w',z,z',\hat{w},\hat{w}',\hat{v},\hat{v}')\,,\label{SpinStructureAi}\\ \nonumber~\\
G_{(ii)}[&s]=\theta_s\big(\tfrac{1}{2}(x_1+x_2-y_1-y_2+w-w'+\hat{w}-\hat{w}'+v-v'+\hat{v}-\hat{v}')
+z-z'\nonumber\\*
&\hspace{1.5cm}+u-u'+\hat{x}_1+\hat{x}_2-\hat{y}_1-\hat{y_2}\big)\nonumber\\*
&\times \theta_s\big(\tfrac{1}{2}(x_1+x_2-y_1-y_2+w-w'+\hat{w}-\hat{w}'-v+v'-\hat{v}+\hat{v}') +z-z'\nonumber\\*
&\hspace{1.5cm}-u+u'+\hat{x}_1+\hat{x}_2-\hat{y}_1-\hat{y_2}\big)\nonumber\\*
&\times\theta_{h,s}\left(\tfrac{1}{2}(x_1-x_2+y_1-y_2+w+w'-\hat{w}-\hat{w}'+v+v'-\hat{v}-\hat{v}')\right)\nonumber\\*
&\times\theta_{-h,s}\left(\tfrac{1}{2}(x_1-x_2+y_1-y_2+w+w'-\hat{w}-\hat{w}'-v-v'+\hat{v}+\hat{v}')\right)\nonumber\\*
&\times\mathbb{B}_{(ii)}(x_1,x_2,y_1,y_2,u,u',v,v',w,w',z,z',\hat{w},\hat{w}',\hat{v},\hat{v}')\,,\label{SpinStructureExpressions}
\end{align}
\noindent where $\mathbb{B}$ is independent of the spin structures and is essentially a quotient of prime forms,  depending on the various worldsheet positions. In order to keep the discussion simple, we  refrain from displaying their explicit expression.

Summing over all different spin structures and using various bosonisation identities (\emph{cf}. \cite{Verlinde:1986kw}) the result becomes:
\begin{align}
G_{(i)}=&\langle\chi^1(x_1)\, \chi^2(x_2)\,\bar{\chi}^2(y_1)\, \bar{\chi}^1(y_2)
\chi^1\chi^2(z)\, \bar{\chi}^1\bar{\chi}^2(z')\, \chi^1
\tilde{\chi}^2(w)\, \bar{\chi}^1
\tilde{\bar{\chi}}^2(w')\nonumber\\
&\times \chi^4\chi^5(u)\, \bar{\chi}^4\bar{\chi}^5(u')\, \chi^4
\tilde{\chi}^5(v)\, \bar{\chi}^4
\tilde{\bar{\chi}}^5(v') \rangle_{\text{odd}}\,,\label{SpinResult1}
\end{align}
\begin{align}
G_{(ii)}=&\langle\chi^1\tilde{\chi}^1\tilde{\chi}^2(x_1)\, \chi^2\tilde{\chi}^1\tilde{\chi}^2(x_2)\, \bar{\chi}^2
\tilde{\bar{\chi}}^1\tilde{\bar{\chi}}^2(y_1)\,\bar{\chi}^1\tilde{\bar{\chi}}^1\tilde{\bar{\chi}}^2(y_2)\,
\chi^1\chi^2(z)\, \bar{\chi}^1\bar{\chi}^2(z')\, \nonumber\\
&\times\chi^1
\tilde{\chi}^2(w)\, \bar{\chi}^1
\tilde{\bar{\chi}}^2(w')\, \chi^4\chi^5(u)\, \bar{\chi}^4\bar{\chi}^5(u')\, \chi^4
\tilde{\chi}^5(v)\, \bar{\chi}^4
\tilde{\bar{\chi}}^5(v') \rangle_{\text{odd}}\,,\label{SpinResult2}
\end{align}
which is to be evaluated in the odd-spin structure. Some more details on how to perform this sum can be found in Appendix \ref{App:ThetaFunctions}.
\begin{table}[H]\begin{center}
\noindent\makebox[\textwidth]{%
\begin{tabular}{|c|c|c||c|c||c||c|c||c|c||c||c|c||c|}\hline
\textbf{Field} & \textbf{Pos.} & \textbf{$\#$} &
\parbox{0.5cm}{\vspace{0.2cm}$\phi_1$\vspace{0.2cm}}&
\parbox{0.5cm}{\vspace{0.2cm}$\phi_2$\vspace{0.2cm}} &
\parbox{0.5cm}{\vspace{0.2cm}$\phi_3$\vspace{0.2cm}} &
\parbox{0.5cm}{\vspace{0.2cm}$\phi_4$\vspace{0.2cm}} &
\parbox{0.5cm}{\vspace{0.2cm}$\phi_5$\vspace{0.2cm}} & \parbox{0.5cm}{\vspace{0.2cm}$\tilde{\phi}_1$\vspace{0.2cm}}&
\parbox{0.5cm}{\vspace{0.2cm}$\tilde{\phi}_2$\vspace{0.2cm}} &
\parbox{0.5cm}{\vspace{0.2cm}$\tilde{\phi}_3$\vspace{0.2cm}} &
\parbox{0.5cm}{\vspace{0.2cm}$\tilde{\phi}_4$\vspace{0.2cm}} &
\parbox{0.5cm}{\vspace{0.2cm}$\tilde{\phi}_5$\vspace{0.2cm}} &
\parbox{1.5cm}{\vspace{0.2cm}\textbf{Bosonic}\vspace{0.2cm}} \\\hline\hline
grav & \parbox{0.35cm}{\vspace{0.2cm}$x_1$\vspace{0.2cm}}
& $1$ & \parbox{0.7cm}
{\vspace{0.2cm}$+\frac{1}{2}$\vspace{0.2cm}} & \parbox{0.7cm}
{\vspace{0.2cm}$+\frac{1}{2}$\vspace{0.2cm}} & \parbox{0.7cm}
{\vspace{0.2cm}$+\frac{1}{2}$\vspace{0.2cm}} & \parbox{0.7cm}
{\vspace{0.2cm}$+\frac{1}{2}$\vspace{0.2cm}}& \parbox{0.7cm}
{\vspace{0.2cm}$+\frac{1}{2}$\vspace{0.2cm}} 
&  \parbox{0.6cm}
{\vspace{0.2cm}$+1$\vspace{0.2cm}} & \parbox{0.6cm}
{\vspace{0.2cm}$+1$\vspace{0.2cm}} & \parbox{0.15cm}{\vspace{0.2cm}$0$\vspace{0.2cm}} & \parbox{0.15cm}
{\vspace{0.2cm}$0$\vspace{0.2cm}}& \parbox{0.15cm}{\vspace{0.2cm}$0$\vspace{0.2cm}}
& --- \\\hline
 & \parbox{0.35cm}{\vspace{0.2cm}$x_2$\vspace{0.2cm}}
& $1$ & \parbox{0.7cm}
{\vspace{0.2cm}$+\frac{1}{2}$\vspace{0.2cm}} & \parbox{0.7cm}
{\vspace{0.2cm}$+\frac{1}{2}$\vspace{0.2cm}} & \parbox{0.7cm}
{\vspace{0.2cm}$+\frac{1}{2}$\vspace{0.2cm}} & \parbox{0.7cm}
{\vspace{0.2cm}$-\frac{1}{2}$\vspace{0.2cm}}& \parbox{0.7cm}
{\vspace{0.2cm}$-\frac{1}{2}$\vspace{0.2cm}} & 
\parbox{0.6cm}
{\vspace{0.2cm}$+1$\vspace{0.2cm}} & \parbox{0.6cm}
{\vspace{0.2cm}$+1$\vspace{0.2cm}} & \parbox{0.15cm}{\vspace{0.2cm}$0$\vspace{0.2cm}} & \parbox{0.15cm}
{\vspace{0.2cm}$0$\vspace{0.2cm}}& \parbox{0.15cm}{\vspace{0.2cm}$0$\vspace{0.2cm}} &
--- \\\hline
 & \parbox{0.35cm}{\vspace{0.2cm}$y_1$\vspace{0.2cm}}
& $1$ & \parbox{0.7cm}
{\vspace{0.2cm}$-\frac{1}{2}$\vspace{0.2cm}} & \parbox{0.7cm}
{\vspace{0.2cm}$-\frac{1}{2}$\vspace{0.2cm}} & \parbox{0.7cm}
{\vspace{0.2cm}$+\frac{1}{2}$\vspace{0.2cm}} & \parbox{0.7cm}
{\vspace{0.2cm}$+\frac{1}{2}$\vspace{0.2cm}}& \parbox{0.7cm}
{\vspace{0.2cm}$+\frac{1}{2}$\vspace{0.2cm}} &
\parbox{0.6cm}
{\vspace{0.2cm}$-1$\vspace{0.2cm}} & \parbox{0.6cm}
{\vspace{0.2cm}$-1$\vspace{0.2cm}} & \parbox{0.15cm}{\vspace{0.2cm}$0$\vspace{0.2cm}} & \parbox{0.15cm}
{\vspace{0.2cm}$0$\vspace{0.2cm}}& \parbox{0.15cm}{\vspace{0.2cm}$0$\vspace{0.2cm}}&
 --- \\\hline
 & \parbox{0.35cm}{\vspace{0.2cm}$y_2$\vspace{0.2cm}}
& $1$ & \parbox{0.7cm}
{\vspace{0.2cm}$-\frac{1}{2}$\vspace{0.2cm}} & \parbox{0.7cm}
{\vspace{0.2cm}$-\frac{1}{2}$\vspace{0.2cm}} & \parbox{0.7cm}
{\vspace{0.2cm}$+\frac{1}{2}$\vspace{0.2cm}} & \parbox{0.7cm}
{\vspace{0.2cm}$-\frac{1}{2}$\vspace{0.2cm}}& \parbox{0.7cm}
{\vspace{0.2cm}$-\frac{1}{2}$\vspace{0.2cm}} 
&  \parbox{0.6cm}
{\vspace{0.2cm}$-1$\vspace{0.2cm}} & \parbox{0.6cm}
{\vspace{0.2cm}$-1$\vspace{0.2cm}} & \parbox{0.15cm}{\vspace{0.2cm}$0$\vspace{0.2cm}} & \parbox{0.15cm}
{\vspace{0.2cm}$0$\vspace{0.2cm}}& \parbox{0.15cm}{\vspace{0.2cm}$0$\vspace{0.2cm}}
& --- \\\hline\hline
\parbox{0.8cm}{\vspace{0.2cm}$F^{G,U}$\vspace{0.2cm}} & \parbox{0.2cm}{\vspace{0.2cm}$z$\vspace{0.2cm}}
& $n_1$ & \parbox{0.6cm}
{\vspace{0.2cm}$+1$\vspace{0.2cm}} & \parbox{0.6cm}
{\vspace{0.2cm}$+1$\vspace{0.2cm}} & \parbox{0.15cm}
{\vspace{0.2cm}$0$\vspace{0.2cm}} & \parbox{0.15cm}
{\vspace{0.2cm}$0$\vspace{0.2cm}}& \parbox{0.15cm}
{\vspace{0.2cm}$0$\vspace{0.2cm}}
&  \parbox{0.15cm}{\vspace{0.2cm}$0$\vspace{0.2cm}} & \parbox{0.15cm}{\vspace{0.2cm}$0$\vspace{0.2cm}} & \parbox{0.15cm}
{\vspace{0.2cm}$0$\vspace{0.2cm}} & \parbox{0.15cm}{\vspace{0.2cm}$0$\vspace{0.2cm}}& \parbox{0.15cm}{\vspace{0.2cm}$0$\vspace{0.2cm}} & \parbox{0.7cm}
{\vspace{0.2cm}$\partial X$\vspace{0.2cm}} \\\hline
& \parbox{0.2cm}{\vspace{0.2cm}$z'$\vspace{0.2cm}}
& $n_2$ & \parbox{0.6cm}
{\vspace{0.2cm}$-1$\vspace{0.2cm}} & \parbox{0.6cm}
{\vspace{0.2cm}$-1$\vspace{0.2cm}} & \parbox{0.15cm}
{\vspace{0.2cm}$0$\vspace{0.2cm}} & \parbox{0.15cm}
{\vspace{0.2cm}$0$\vspace{0.2cm}}& \parbox{0.15cm}
{\vspace{0.2cm}$0$\vspace{0.2cm}}
&  \parbox{0.15cm}{\vspace{0.2cm}$0$\vspace{0.2cm}} & \parbox{0.15cm}{\vspace{0.2cm}$0$\vspace{0.2cm}} & \parbox{0.15cm}
{\vspace{0.2cm}$0$\vspace{0.2cm}} & \parbox{0.15cm}{\vspace{0.2cm}$0$\vspace{0.2cm}}& \parbox{0.15cm}{\vspace{0.2cm}$0$\vspace{0.2cm}} & \parbox{0.7cm}
{\vspace{0.2cm}$\partial X$\vspace{0.2cm}} \\\hline
 & \parbox{0.2cm}{\vspace{0.2cm}$w$\vspace{0.2cm}}
& $n_3$ & \parbox{0.7cm}
{\vspace{0.2cm}$+\frac{1}{2}$\vspace{0.2cm}} & \parbox{0.7cm}
{\vspace{0.2cm}$+\frac{1}{2}$\vspace{0.2cm}} & \parbox{0.7cm}
{\vspace{0.2cm}$+\frac{1}{2}$\vspace{0.2cm}} & \parbox{0.7cm}
{\vspace{0.2cm}$+\frac{1}{2}$\vspace{0.2cm}}& \parbox{0.7cm}
{\vspace{0.2cm}$+\frac{1}{2}$\vspace{0.2cm}}
& \parbox{0.7cm}
{\vspace{0.2cm}$+\frac{1}{2}$\vspace{0.2cm}} & \parbox{0.7cm}
{\vspace{0.2cm}$+\frac{1}{2}$\vspace{0.2cm}} & \parbox{0.7cm}
{\vspace{0.2cm}$+\frac{1}{2}$\vspace{0.2cm}} & \parbox{0.7cm}
{\vspace{0.2cm}$-\frac{1}{2}$\vspace{0.2cm}}& \parbox{0.7cm}
{\vspace{0.2cm}$-\frac{1}{2}$\vspace{0.2cm}}
 & \parbox{0.7cm}
{\vspace{0.2cm}$\partial X$\vspace{0.2cm}} \\\hline
 & \parbox{0.2cm}{\vspace{0.2cm}$w'$\vspace{0.2cm}}
& $n_4$ & \parbox{0.7cm}
{\vspace{0.2cm}$-\frac{1}{2}$\vspace{0.2cm}} & \parbox{0.7cm}
{\vspace{0.2cm}$-\frac{1}{2}$\vspace{0.2cm}} & \parbox{0.7cm}
{\vspace{0.2cm}$+\frac{1}{2}$\vspace{0.2cm}} & \parbox{0.7cm}
{\vspace{0.2cm}$+\frac{1}{2}$\vspace{0.2cm}}& \parbox{0.7cm}
{\vspace{0.2cm}$+\frac{1}{2}$\vspace{0.2cm}}
& \parbox{0.7cm}
{\vspace{0.2cm}$-\frac{1}{2}$\vspace{0.2cm}} & \parbox{0.7cm}
{\vspace{0.2cm}$-\frac{1}{2}$\vspace{0.2cm}} & \parbox{0.7cm}
{\vspace{0.2cm}$+\frac{1}{2}$\vspace{0.2cm}} & \parbox{0.7cm}
{\vspace{0.2cm}$-\frac{1}{2}$\vspace{0.2cm}}& \parbox{0.7cm}
{\vspace{0.2cm}$-\frac{1}{2}$\vspace{0.2cm}}
 & \parbox{0.7cm}
{\vspace{0.2cm}$\partial X$\vspace{0.2cm}} \\\hline\hline
\parbox{0.8cm}{\vspace{0.2cm}$F^{\bar{S},\bar{S}'}$\vspace{0.2cm}} & \parbox{0.2cm}{\vspace{0.2cm}$u$\vspace{0.2cm}}
& $m_1$ & \parbox{0.6cm}
{\vspace{0.2cm}$+1$\vspace{0.2cm}} & \parbox{0.6cm}
{\vspace{0.2cm}$-1$\vspace{0.2cm}} & \parbox{0.15cm}
{\vspace{0.2cm}$0$\vspace{0.2cm}} & \parbox{0.15cm}
{\vspace{0.2cm}$0$\vspace{0.2cm}}& \parbox{0.15cm}
{\vspace{0.2cm}$0$\vspace{0.2cm}}
&  \parbox{0.15cm}{\vspace{0.2cm}$0$\vspace{0.2cm}} & \parbox{0.15cm}{\vspace{0.2cm}$0$\vspace{0.2cm}} & \parbox{0.15cm}
{\vspace{0.2cm}$0$\vspace{0.2cm}} & \parbox{0.15cm}{\vspace{0.2cm}$0$\vspace{0.2cm}}& \parbox{0.15cm}{\vspace{0.2cm}$0$\vspace{0.2cm}}
 & \parbox{0.7cm}
{\vspace{0.2cm}$\partial X$\vspace{0.2cm}} \\\hline
& \parbox{0.2cm}{\vspace{0.2cm}$u'$\vspace{0.2cm}}
& $m_2$ & \parbox{0.6cm}
{\vspace{0.2cm}$-1$\vspace{0.2cm}} & \parbox{0.6cm}
{\vspace{0.2cm}$+1$\vspace{0.2cm}} & \parbox{0.15cm}
{\vspace{0.2cm}$0$\vspace{0.2cm}} & \parbox{0.15cm}
{\vspace{0.2cm}$0$\vspace{0.2cm}}& \parbox{0.15cm}
{\vspace{0.2cm}$0$\vspace{0.2cm}}
&  \parbox{0.15cm}{\vspace{0.2cm}$0$\vspace{0.2cm}} & \parbox{0.15cm}{\vspace{0.2cm}$0$\vspace{0.2cm}} & \parbox{0.15cm}
{\vspace{0.2cm}$0$\vspace{0.2cm}} & \parbox{0.15cm}{\vspace{0.2cm}$0$\vspace{0.2cm}}& \parbox{0.15cm}{\vspace{0.2cm}$0$\vspace{0.2cm}}
 & \parbox{0.7cm}
{\vspace{0.2cm}$\partial X$\vspace{0.2cm}} \\\hline 
& \parbox{0.2cm}{\vspace{0.2cm}$v$\vspace{0.2cm}}
& $m_3$ & \parbox{0.7cm}
{\vspace{0.2cm}$+\frac{1}{2}$\vspace{0.2cm}} & \parbox{0.7cm}
{\vspace{0.2cm}$-\frac{1}{2}$\vspace{0.2cm}} & \parbox{0.7cm}
{\vspace{0.2cm}$+\frac{1}{2}$\vspace{0.2cm}} & \parbox{0.7cm}
{\vspace{0.2cm}$+\frac{1}{2}$\vspace{0.2cm}}& \parbox{0.7cm}
{\vspace{0.2cm}$-\frac{1}{2}$\vspace{0.2cm}}
& \parbox{0.7cm}
{\vspace{0.2cm}$+\frac{1}{2}$\vspace{0.2cm}} & \parbox{0.7cm}
{\vspace{0.2cm}$-\frac{1}{2}$\vspace{0.2cm}} & \parbox{0.7cm}
{\vspace{0.2cm}$+\frac{1}{2}$\vspace{0.2cm}} & \parbox{0.7cm}
{\vspace{0.2cm}$-\frac{1}{2}$\vspace{0.2cm}}& \parbox{0.7cm}
{\vspace{0.2cm}$+\frac{1}{2}$\vspace{0.2cm}}
 & \parbox{0.7cm}
{\vspace{0.2cm}$\partial X$\vspace{0.2cm}} \\\hline
 & \parbox{0.2cm}{\vspace{0.2cm}$v'$\vspace{0.2cm}}
& $m_4$ & \parbox{0.7cm}
{\vspace{0.2cm}$-\frac{1}{2}$\vspace{0.2cm}} & \parbox{0.7cm}
{\vspace{0.2cm}$+\frac{1}{2}$\vspace{0.2cm}} & \parbox{0.7cm}
{\vspace{0.2cm}$+\frac{1}{2}$\vspace{0.2cm}} & \parbox{0.7cm}
{\vspace{0.2cm}$+\frac{1}{2}$\vspace{0.2cm}}& \parbox{0.7cm}
{\vspace{0.2cm}$-\frac{1}{2}$\vspace{0.2cm}}
& \parbox{0.7cm}
{\vspace{0.2cm}$-\frac{1}{2}$\vspace{0.2cm}} & \parbox{0.7cm}
{\vspace{0.2cm}$+\frac{1}{2}$\vspace{0.2cm}} & \parbox{0.7cm}
{\vspace{0.2cm}$+\frac{1}{2}$\vspace{0.2cm}} & \parbox{0.7cm}
{\vspace{0.2cm}$-\frac{1}{2}$\vspace{0.2cm}}& \parbox{0.7cm}
{\vspace{0.2cm}$+\frac{1}{2}$\vspace{0.2cm}}
 & \parbox{0.7cm}
{\vspace{0.2cm}$\partial X$\vspace{0.2cm}} \\\hline\hline
PCO & \parbox{0.2cm}{\vspace{0.2cm}$P$\vspace{0.2cm}}
& $m_{\text{PCO}}$
& \parbox{0.15cm}
{\vspace{0.2cm}$0$\vspace{0.2cm}} & \parbox{0.15cm}
{\vspace{0.2cm}$0$\vspace{0.2cm}} & \parbox{0.6cm}
{\vspace{0.2cm}$-1$\vspace{0.2cm}} & \parbox{0.15cm}
{\vspace{0.2cm}$0$\vspace{0.2cm}}& \parbox{0.15cm}
{\vspace{0.2cm}$0$\vspace{0.2cm}}
& \parbox{0.15cm}
{\vspace{0.2cm}$0$\vspace{0.2cm}} & \parbox{0.15cm}
{\vspace{0.2cm}$0$\vspace{0.2cm}} & \parbox{0.15cm}
{\vspace{0.2cm}$0$\vspace{0.2cm}} & \parbox{0.15cm}
{\vspace{0.2cm}$0$\vspace{0.2cm}}& \parbox{0.15cm}
{\vspace{0.2cm}$0$\vspace{0.2cm}}
 & \parbox{0.7cm}
{\vspace{0.2cm}$\partial X$\vspace{0.2cm}} \\\hline
\end{tabular}}
\end{center}
\caption{Overview of the vertex contributions for the Type I amplitude in case (ii), \emph{i.e.} the gravitini contribute fermionic currents.}
\label{Tab:TypeIvertexA}
\end{table}
\noindent Thus, summarising the above computation, after putting together all the combinations the result is equivalent to computing the correlation function in the odd spin structure with the following identification of operators:
\begin{eqnarray}
V^{G,U}(\epsilon_2,p_1) &\rightarrow& P_3[ J^B_{++}+ (\chi^1 + a \tilde{\chi}^1)  (\chi^2 + a
\tilde{\chi}^2)]\,, \nonumber\\
V^{G,U}(\epsilon_{\bar{2}},p_{\bar{1}})& \rightarrow& P_3 [J^B_{--}+(\bar{\chi}^1 + a\bar{ \tilde{\chi}}^1)  (\bar{\chi}^2 + a
\bar{\tilde{\chi}}^2)]\,, \nonumber\\
V^{\bar{S}',\bar{S}}(\epsilon_{\bar{2}},p_1) &\rightarrow& P_3[J^B_{+-}+ (\chi^4 + b \tilde{\chi}^4)  (\chi^5 + b
\tilde{\chi}^5)]\,, \nonumber\\
V^{\bar{S}',\bar{S}}(\epsilon_2,p_{\bar{1}}) &\rightarrow&P_3[ J^B_{-+}+ (\bar{\chi}^4 + b \bar{\tilde{\chi}}^4)  (\bar{\chi}^5 + b
\bar{\tilde{\chi}}^5) ]\,.\label{RepVector}
\end{eqnarray}
We remind that $a=\pm 1$ and $b=\pm 1$ correspond to the two relative signs in $V^{G,U}$ and $V^{\bar{S}',\bar{S}}$, respectively. $J^B$ are the total (\emph{i.e.} left- plus right- moving) bosonic Lorentz currents\footnote{For $a=-1$, corresponding to graviphoton insertions, we see that the combinations that enter in the first two lines in (\ref{CurrentsRed}) are $\chi^1~-~\tilde{\chi}^1$ and $\chi^2-~\tilde{\chi}^2$. These combinations cannot soak the fermion zero modes in the odd spin structure, since for the zero modes one has $\chi =\tilde{\chi}$. This is consistent with the fact that the graviphoton is the lowest component of the Weyl multiplet. On the other hand for $a=+1$ the vertices $V^U$ represent a higher component of the vector multiplet.} with appropriate charges:
\begin{align}
J_{++}^{B}=&Z^1\bar\partial Z^2+(\text{left}\leftrightarrow\text{right}) \qquad,\qquad J_{--}^{B}=\bar Z^1\bar\partial\bar 
Z^2+(\text{left}\leftrightarrow\text{right})~,\nonumber\\ 
J_{+-}^{B}=&Z^1\bar\partial\bar Z^2+(\text{left}\leftrightarrow\text{right}) \qquad,\qquad J_{-+}^{B}=\bar Z^1\bar\partial 
Z^2+(\text{left}\leftrightarrow\text{right})~,\label{CurrentsRed}
\end{align}
and $P_3$ is the complex $T^2$-momentum, defined in (\ref{P3}). For convenience, we introduce
\begin{align}
J^{\rm{total}}_{++}= J^B_{++}+ (\chi^1 - \tilde{\chi}^1)  (\chi^2 -
\tilde{\chi}^2) \qquad,\qquad  J^{\rm{total}}_{--}= J^B_{--}+(\bar{\chi}^1 -\bar{ \tilde{\chi}}^1)  (\bar{\chi}^2 -
\bar{\tilde{\chi}}^2)\,.
\end{align}
Similarly, the gravitini vertices can be recast in a convenient form. Indeed, as we can see from (\ref{SpinResult1}) and (\ref{SpinResult2}), the vertices are replaced by:
\begin{eqnarray}
V^{\text{grav}}_{+}(\xi_{21},p_1;x_1)&\rightarrow &\chi^1(J^B_{++,R}+ \tilde{\chi}^1 \tilde{\chi}^2)+
\tilde{\chi}^1(J^B_{++,L}+\chi^1 \chi^2)\nonumber\\
&=&\chi^1 J^B_{++,R} + \tilde{\chi}^1
J^B_{++,L} - \chi^1 \tilde{\chi}^1 (\chi^2 - \tilde{\chi}^2)\,,\nonumber\\
V^{\text{grav}}_{-}(\xi_{\bar{2}1},p_1;x_2)&\rightarrow &\chi^2(J^B_{++,R}+ \tilde{\chi}^1 \tilde{\chi}^2)+
\tilde{\chi}^2(J^B_{++,L}+\chi^1 \chi^2)\nonumber\\
&=&\chi^2 J^B_{++,R} + \tilde{\chi}^2
J^B_{++,L} + \chi^2 \tilde{\chi}^2 (\chi^1 - \tilde{\chi}^1)\,,\nonumber\\
V^{\text{grav}}_{+}(\xi_{22},p_{\bar{1}};y_1)&\rightarrow &\bar{\chi}^2(J^B_{--,R}+ \tilde{\bar{\chi}}^1 \tilde{\bar{\chi}}^2)+
\tilde{\bar{\chi}}^2(J^B_{--,L}+\bar{\chi}^1 \bar{\chi}^2)\nonumber\\
&=&\bar{\chi}^2 J^B_{--,R} + \tilde{\bar{\chi}}^2
J^B_{--,L} + \bar{\chi}^2 \tilde{\bar{\chi}}^2 (\bar{\chi}^1 - \tilde{\bar{\chi}}^1)\,,\nonumber\\
V^{\text{grav}}_{-}(\xi_{\bar{2}2},p_{\bar{1}};y_2)&\rightarrow &\bar{\chi}^1(J^B_{--,R}+ \tilde{\bar{\chi}}^1 \tilde{\bar{\chi}}^2)+
\tilde{\bar{\chi}}^1(J^B_{--,L}+\bar{\chi}^1 \bar{\chi}^2)\nonumber\\
&=&\bar{\chi}^1 J^B_{--,R} + \tilde{\bar{\chi}}^1
J^B_{--,L} - \bar{\chi}^1 \tilde{\bar{\chi}}^1 (\bar{\chi}^2 - \tilde{\bar{\chi}}^2)\,,
\end{eqnarray}
where the subscripts $L$ and $R$ in $J^B$ denote the left- and right- moving parts of the bosonic Lorentz current. Notice that the zero modes of $(\chi_1,\chi_2,\bar{\chi}_1,\bar{\chi}_2)$ can only be soaked up by the operators at $(x_1,x_2,y_2,y_1)$ respectively. We  denote this by putting a superscript zero $(\chi^i)^0$ as follows:
\begin{align}
&V^{\text{grav}}_{+}(\xi_{21},p_1;x_1)\rightarrow (\chi^1)^0 J^{\rm{total}}_{++} \qquad,\qquad V^{\text{grav}}_{-}(\xi_{\bar{2}1},p_1;x_2)
\rightarrow (\chi^2)^0 J^{\rm{total}}_{++}\,,\nonumber\\
&V^{\text{grav}}_{+}(\xi_{22},p_{\bar{1}};y_1)\rightarrow (\bar{\chi}^2)^0 J^{\rm{total}}_{--}\qquad,\qquad V^{\text{grav}}_{-}
(\xi_{\bar{2}2},p_{\bar{1}};y_2)\rightarrow (\bar{\chi}^1)^0 J^{\rm{total}}_{--}~.\label{RepGravitini}
\end{align}
Now using the replacement rules (\ref{RepVector}) and (\ref{RepGravitini}) we can write the following generating function for the correlation functions introduced in eq.~(\ref{AmplitudesDefinition}):
\begin{align}
{\mathcal{F}}(\epsilon_-,\epsilon_+)&=\int_{{\cal{M}}_{\rm{cylinder}}}\frac{d\tau} {\tau}\Bigr<\sum_{P_3} \frac{\tau^2\,e^{-i\pi \frac{|P_3|^2}{\tau}}}{P_3^2} 
[ e^{P_3 S_I} - 1- \frac{P_3^2}{\tau^2}
\epsilon_-^2 ~ J^{\rm{total}}_{++}~J^{\rm{total}}_{--}]\Bigr>'~,
\label{Seff}
\end{align}
where the prime $\langle~\rangle'$ denotes the soaking of the space-time fermionic zero modes. Moreover, the $T^2$ correlators as well as the ghosts have disappeared as their non-zero mode determinants cancel each other. Hence, only the zero mode part of $T^2$ appears  in the  lattice sum above. The action deformation $S_I$ is given by:
\begin{equation}
S_I= \frac{\epsilon_{-}}{\tau} \int ( J^{\rm{total}}_{++}+J^{\rm{total}}_{--})
+\frac{\epsilon_+}{\tau}  \int(J^B_{+-}+J^B_{-+}+J_b^{K3})\,,\label{DeformationTypeI}
\end{equation}
where the integral is over the worldsheet cylinder and
\begin{equation}
J_b^{K3}=  (\chi^4 + b \tilde{\chi}^4)  (\chi^5 + b
\tilde{\chi}^5)+ (\bar{\chi}^4 + b
\bar{\tilde{\chi}}^4)(\bar{\chi}^5 + b \bar{\tilde{\chi}}^5) \,.
\end{equation}
Before proceeding with the actual computation of the generating function, let us make a few remarks. First of all, the operators in $S_I$ do not have a well-defined conformal dimension but are to be computed in a specific worldsheet coordinate system where conformal transformations are completely fixed, modulo rigid translations. Secondly, the right-hand side of (\ref{Seff}) starts at order $P_3^2$. This is to be expected since for $N=M=0$ the correlation function behaves as $P_3^2$ due to the two picture-changing operators needed to balance the ghost charges of the four gravitini vertices. Finally, only even powers of $\epsilon_-$ and $\epsilon_+$ survive in (\ref{Seff}) as a result of the structure of the non-zero mode correlators, \emph{i.e.} $\chi^1$ has a non-zero correlator only with $\bar{\chi}^1$ and similarly for the rest.


\section{Path Integral Evaluation of The Amplitudes}\label{TypeI:PathIntegral}

In this section, we explicitly evaluate the generating function (\ref{Seff}) using a worldsheet path integral approach. The path integrals can be performed exactly, since every term in $S_I$ in (\ref{DeformationTypeI}) is quadratic in the field variables. There are three major contributions, namely the bosonic and fermionic space-time parts as well as the contribution of the $K3$ fermions. In what follows, we separately deal with all three. We begin with the contribution of the space-time bosons:
\begin{align}\nonumber
\left\langle e^{P_3 S_I}\right\rangle_{\text{bos}}=\left\langle \exp\Biggr[\frac{\hat\epsilon_-}{\tau}\int d^2\sigma
\left(Z^1(\bar\partial-\partial) Z^2+ \bar Z^1(\bar\partial-\partial)\bar Z^2\right)\right. \\
	\left.+\frac{\hat\epsilon_+}{\tau}\int d^2\sigma \left( Z^1(\bar\partial-\partial)\bar Z^2+\bar 
Z^1(\bar\partial-\partial) Z^2\right)   \Biggr]\right\rangle ~,\label{TypeIBosSpaceTime}
\end{align}
where we defined $\hat\epsilon_{\pm}\equiv \epsilon_{\pm}P_3/\sqrt{(T-\bar{T})(U-\bar{U})-\tfrac{1}{2}(\vec{Y}-
\vec{\bar{Y}})^2}$. Plugging in the appropriate mode expansions:
\begin{align}\label{ModesZ}
	& Z^i = \sum\limits_{n,m} Z^i_{n,m}\cos(2\pi n\sigma)e^{2\pi i m t}\,,&& \bar Z^i = \sum\limits_{n,m} 
\bar{Z}^i_{n,m}\cos(2\pi n\sigma)e^{2\pi i m t}\,,
\end{align}
with $i=1,2$, corresponding to NN boundary conditions $\left.\partial_1 Z\right|_{\sigma_1=0,\frac{1}{2}}=0$ and carefully performing the path integral over the modes, we can express the space-time bosonic correlator in the form:
\begin{align}
	\left\langle e^{P_3 S_I}\right\rangle_{\text{bos}}= \Bigr[H_1\left(\tfrac{\hat\epsilon_{-}-\hat\epsilon_{+}}{2};0;
\tfrac{\tau}{2}\right)H_1\left(\tfrac{\hat\epsilon_{-}+\hat\epsilon_{+}}{2};0;\tfrac{\tau}{2}\right)\Bigr]^{-1}~
\frac{\pi^2(\hat\epsilon_{-}-\hat\epsilon_{+})(\hat\epsilon_{-}+\hat\epsilon_{+})}{\sin\pi(\hat\epsilon_{-}-\hat\epsilon_{+})\,
\sin\pi(\hat\epsilon_{-}+\hat\epsilon_{+})}~,
\end{align}
where the function $H_s(z;\frac{g}{2};\tau)$ is defined as:
\begin{align}\label{Hfunction}
	H_s(z;\tfrac{g}{2};\tau) \equiv
	\frac{\theta_1(z+\frac{g}{2};\tau)}{2 \eta^3(\tau) \sin\pi(z+\frac{g}{2})} 
 \prod\limits_{\genfrac{}{}{0pt}{}{m\in\mathbb{Z}}{n>0}} \left(1-\frac{z^2}{|m+\frac{g}{2}+z-n\tau|^{2s}}\right)~,
\end{align}
and is normalised such that $H_s(0;0;\tau)=1$. In Appendix \ref{TypeIDetReg}, it is shown that, in the full correlator, the functions $H_s(z;\frac{g}{2};\tau)$ trivialise in the limit $\tau_2\rightarrow\infty$, so that the contribution surviving in the field theory limit comes precisely from the integration of the $n=0$ modes\footnote{The fact that the $n=0$ mode in (\ref{ModesZ}) corresponds to the field theory limit is natural from a physical point of view, since it is precisely associated to the vibrations of the open string stretched between the two boundaries of the annulus.}:
\begin{align}
	\left\langle e^{P_3 S_I}\right\rangle_{\text{bos}} ~~\overset{\tau_2\rightarrow \infty}{\longrightarrow}~~ 
\frac{\pi^2(\hat\epsilon_{-}-\hat\epsilon_{+})(\hat\epsilon_{-}+\hat\epsilon_{+})}{\sin\pi(\hat\epsilon_{-}-\hat\epsilon_{+})\,
\sin\pi(\hat\epsilon_{-}+\hat\epsilon_{+})}~.
\end{align}
Let us now compute the correlators of space-time fermions $\chi^{1,2},\bar\chi^{1,2}$, generated by
\begin{align}\label{SpacetimeFermDef}
\left\langle e^{P_3 S_I}\right\rangle^{\textrm{s-t}}_{\textrm{ferm}} = \left\langle \exp\Biggr[\frac{\hat\epsilon_{-}}{\tau}\int 
d^2\sigma \left[ (\chi^1-\tilde\chi^1)(\chi^2-\tilde\chi^2)+(\bar\chi^1-\tilde{\bar{\chi}}^1)(\bar\chi^2-\tilde{\bar{\chi}}^2)
\right] \Biggr]\right\rangle~.
\end{align}
The relevant mode expansions are those for complex fermions in the Ramond sector with NN boundary conditions:
\begin{align}
&\chi^i = \sum\limits_{n,m}\chi^i_{n,m}\, e^{2\pi i(n\sigma+m t)}~,  &\tilde\chi^i = \sum\limits_{n,m}\chi^i_{n,m}\, 
e^{2\pi i(-n\sigma+m t)}\,,\\
&\bar\chi^i = \sum\limits_{n,m}\bar\chi^i_{n,m}\, e^{2\pi i (n\sigma+m t)}~,  &\tilde{\bar{\chi}}^i = 
\sum\limits_{n,m}\bar\chi^i_{n,m}\, e^{2\pi i(-n\sigma+m t)}\,.
\end{align}
Notice that the $n=0$ modes cancel out in the deformation (\ref{SpacetimeFermDef}) and, hence, their contribution is $\epsilon_{-}$-independent. Path integration over the $n\neq 0$ modes, on the other hand, yields a non-trivial contribution so that the correlator of the space-time fermions can be compactly written as:
\begin{align}
	\left\langle e^{P_3 S_I}\right\rangle^{\textrm{s-t}}_{\textrm{ferm}} = \Bigr[H_1(\tfrac{\hat\epsilon_{-}}{2};0;
\tfrac{\tau}{2})\Bigr]^2 ~~\overset{\tau_2\rightarrow \infty}{\longrightarrow}~~1~.
\end{align}
Hence, the net effect of the absence of $\epsilon_{-}$-dependent $n=0$ mode contributions in the deformed action is to render the space-time fermionic correlator trivial in the field theory limit.

Finally, we turn to the contribution of the worldsheet fermions in the $K3$ directions $\chi^{4,5},\bar\chi^{4,5}$ which 
are sensitive to the sign parameter $b=\pm 1$:
\begin{align}\label{K3FermDef}
	 \left\langle e^{P_3 S_I}\right\rangle^{K3,b}_{\textrm{ferm}}
	  = \left\langle \exp\Biggr[\frac{\hat\epsilon_{+}}{\tau}\int d^2\sigma \left[ (\chi^4+b\tilde\chi^4)
(\chi^5+b\tilde\chi^5)+(\bar\chi^4+b\tilde{\bar{\chi}}^4)(\bar\chi^5+b\tilde{\bar{\chi}}^5)\right] \Biggr]\right\rangle~.
\end{align}
Using similar mode expansions as previously for the fermions in the $K3$ direction, the path integral can be readily computed and the result cast in the following form:
\begin{align}
	\bigr\langle & e^{P_3 S_I}\bigr\rangle^{K3,b}_{\textrm{ferm}} = -4\sin^2(\tfrac{\pi g}{2}) H_1(\tfrac{\hat\epsilon_{+}}{2};
\tfrac{g}{2};\tfrac{\tau}{2})H_1(\tfrac{\hat\epsilon_{+}}{2};-\tfrac{g}{2};\tfrac{\tau}{2})\Bigr(\cos^2\pi\hat\epsilon_{+}-
\cot^2(\tfrac{\pi g}{2})\sin^2\pi\hat\epsilon_{+}\Bigr)^{(1+b)/2}~.\label{PathFermK3TypeI}
\end{align}
Here $g\in\mathbb{Z}_2$ is the orbifold projection parameter that twists the $K3$ fermions. When $\tau_2\rightarrow\infty$, the function $H_s\rightarrow 1$ and therefore 
\begin{align}
 \left\langle e^{P_3 S_I}\right\rangle^{K3,b=-1}_{\textrm{ferm}}~~\overset{\tau_2\rightarrow \infty}{\longrightarrow}~~ 1\,.
\end{align}
This is consistent with the fact that --- as for the correlators involving the fermions in the space-time directions $\chi^{1,2},\bar\chi^{1,2}$ --- setting $b=-1$ results in a cancellation of the $n=0$ modes in (\ref{K3FermDef}) which leads to a trivial field theory limit. The case $b=+1$, however, is much more interesting, since the $n=0$ modes give rise to a non-trivial $g$-dependent contribution that survives in the field theory limit. Indeed, from (\ref{PathFermK3TypeI}) we find
\begin{align}
	 \bigr\langle & e^{P_3 S_I}\bigr\rangle^{K3,b=+1}_{\textrm{ferm}}  ~~\overset{\tau_2\rightarrow \infty}{\longrightarrow}~~ -
4\Bigr(\sin^2(\tfrac{\pi g}{2}) \cos^2\pi\hat\epsilon_{+}-\cos^2(\tfrac{\pi g}{2})\sin^2\pi\hat\epsilon_{+}\Bigr)~.
\end{align}
Putting all the pieces together, the full correlator becomes:
\begin{align}\nonumber
	  & \mathcal{A}[^0_g] = (-4\sin^2\tfrac{\pi g}{2})\Bigr(\cos^2\pi\hat\epsilon_{+}-\cot^2(\tfrac{\pi g}{2})
\sin^2\pi\hat\epsilon_{+}\Bigr)^{(1+b)/2} Z_{K3}[^0_g] \\ \label{FullCorrelatorTypeI}
	&\times \frac{\pi^2(\hat\epsilon_{-}-\hat\epsilon_{+})(\hat\epsilon_{-}+\hat\epsilon_{+})}
{\sin\pi(\hat\epsilon_{-}-\hat\epsilon_{+})\,\sin\pi(\hat\epsilon_{-}+\hat\epsilon_{+})} 
\frac{\Bigr[H_1(\tfrac{\hat\epsilon_{-}}{2};0;\tfrac{\tau}{2})\Bigr]^2 H_1(\tfrac{\hat\epsilon_{+}}{2};
\tfrac{g}{2};\tfrac{\tau}{2})~H_1(\tfrac{\hat\epsilon_{+}}{2};-\tfrac{g}{2};\tfrac{\tau}{2})}
{H_1(\tfrac{\hat\epsilon_{-}-\hat\epsilon_{+}}{2};0;\tfrac{\tau}{2})~H_1(\tfrac{\hat\epsilon_{-}+\hat\epsilon_{+}}{2};0;\tfrac{\tau}{2})}~, 
\end{align}
where $Z_{K3}[^0_g]$ is the bosonic $K3$ lattice partition function, with $Z_{K3}[^0_1]= 4\eta^6(\tfrac{\tau}{2})/\theta_2^2(\tfrac{\tau}{2})$. Since its $q$-expansion begins with a constant term, $Z_{K3}[^0_g]=1+\mathcal{O}(q)$, and since we are interested in extracting the field theory limit around a point of enhanced gauge symmetry, the $Z_{K3}$ lattice does not play any substantial role in our subsequent analysis and, henceforth, we omit it.

The correlator (\ref{FullCorrelatorTypeI}) should now be weighted by appropriate Chan-Paton factors, together with the $T^2$-lattice accordingly Poisson-resummed to its Hamiltonian representation and with its momentum quantum numbers properly shifted by the Wilson line insertions, $a_i$. An overall factor of $1/4$ is also required from the insertion of the orientifold projections into the traces. Furthermore, this should be supplemented by the 5-5 and 9-5 correlators of the annulus and the 9-9 and 5-5 correlators of the M\"obius diagram. However, as argued above, only the 9-9 sector of the annulus diagram is relevant for the field theory limit in the vicinity of the $SU(2)$ enhancement point we consider, where the only extra massless states belong to vector multiplets. It is then straightforward to show that the net contribution of the extra massless vectors is:
\begin{align}\label{VectorContribution}
	n_V ~\frac{\mathcal{A}[^0_0]+\mathcal{A}[^0_1]}{2}~e^{-\pi\tau_2\mathcal{M}^2_V}~,
\end{align}
where $n_V$ is the number of extra vectors becoming massless at the enhancement point and $\mathcal{M}^2_V$ is their (physical) BPS mass squared.

Before extracting the field theory limit, it is useful to consider the case $\epsilon_{+}=0$  in (\ref{VectorContribution}). Indeed, independently of the choice of sign $b$, the non-zero mode $n\neq 0$ contributions of the fermionic and bosonic determinants cancel each other and one obtains:
\begin{align}
	 \int \frac{d\tau_2}{\tau_2}\left[\frac{\epsilon_{-}\bar\mu\tau_2}{\sin(\epsilon_{-}\bar\mu\tau_2)}\right]^2 
e^{-|\mu|^2\tau_2}=\sum\limits_{g=1}^\infty\frac{(2g-1)}{2g}B_{2g}\epsilon_{-}^{2g}\mu^{-2g}~,\label{EppLim0}
\end{align}
where $B_{2g}$ are the Bernoulli numbers and
\begin{align}
	\mu \sim a_2-U a_1~,
\end{align}
is the BPS mass parameter of the extra massless charged states. Indeed, (\ref{EppLim0}) agrees with the singularity structure of the higher derivative $F$-terms $F_g W^{2g}$ near a conifold singularity, which were computed in a similar setup in \cite{Gava:1996hr} by considering the solitonic state becoming massless as an open string stretched between intersecting D5-branes. Notice, however, that in our setup, the singularity arises at a Wilson line enhancement point, $\mu\rightarrow 0$.

Now we resume our analysis of the refined case $\epsilon_{+}\neq 0$. First recall that the case $b=-1$  corresponds to a scattering of vector partners of $\bar{S}$-moduli so that one expects to reproduce the results of  \cite{Antoniadis:2010iq}, where the corresponding amplitude involving $\bar{S}$-vectors was computed in a Heterotic setup. Indeed, it is easy to show that in the field theory limit (\ref{VectorContribution}) reduces to
\begin{align}\label{SvectorsTypeIresult}
	\frac{\mathcal{A}[^0_0]+\mathcal{A}[^0_1]}{2} \Biggr|_{b=-1} ~~\overset{\tau_2\rightarrow \infty}{\longrightarrow}~~2~ 
\frac{\pi \,(\hat\epsilon_{-}-\hat\epsilon_{+})}{\sin\pi(\hat\epsilon_{-}-\hat\epsilon_{+})}~\frac{\pi \,(\hat\epsilon_{-}+
\hat\epsilon_{+})}{\sin\pi(\hat\epsilon_{-}+\hat\epsilon_{+})}~,
\end{align}
in perfect agreement with \cite{Antoniadis:2010iq}. Turning to the more interesting case $b=+1$, corresponding to scattering vector partners of $\bar{S}'$-moduli, the non-trivial $n=0$ mode contributions play an important role. Extracting the field theory limit around an $SU(2)$ enhancement point yields:
\begin{align}\label{PhaseTypeI}
	\frac{\mathcal{A}[^0_0]+\mathcal{A}[^0_1]}{2}\Biggr|_{b=+1} ~~\overset{\tau_2\rightarrow \infty}{\longrightarrow}~~-
2\cos(2\pi\epsilon_{+})~\frac{\pi\,(\hat\epsilon_{-}-\hat\epsilon_{+})}{\sin\pi(\hat\epsilon_{-}-\hat\epsilon_{+})}~\frac{\pi\,
(\hat\epsilon_{-}+\hat\epsilon_{+})}{\sin\pi(\hat\epsilon_{-}+\hat\epsilon_{+})}~.
\end{align}
After the appropriate rescaling, the field theory limit of our Type I amplitude around the $SU(2)$ Wilson-line enhancement point permits one to extract the leading singularity for $\mathcal{F}_{N,M}$ as the $\epsilon_{-}^{2N}\epsilon_{-}^{2M}$ term in the expansion of the generating function:
\begin{equation}
 \mathcal{F}\left(\epsilon_-,\epsilon_+\right) \sim n_V(\epsilon_{-}^2-\epsilon_{+}^2) \int_0^\infty\frac{dt}{t}~\frac{-
2\cos\left( 2\epsilon_{+}t\right) }{\sin\left( \epsilon_{-}-\epsilon_{+}\right)t ~\sin\left( \epsilon_{-}+\epsilon_{+}\right)t} ~
e^{-\mu t}\label{GenFct}~.
\end{equation}
This precisely reproduces the perturbative part of the free energy of the pure $\mathcal{N}=2$, $SU(2)$ Yang-Mills theory in the $\Omega$-background \eqref{NekPertPart}. Notice that both (\ref{SvectorsTypeIresult}) and (\ref{PhaseTypeI}) are symmetric with respect to $\epsilon_{\pm}\rightarrow -\epsilon_{\pm}$. Unlike (\ref{SvectorsTypeIresult}), however, the generating function (\ref{PhaseTypeI}) is no longer symmetric with respect to the exchange $\epsilon_{-}\leftrightarrow\epsilon_{+}$. This asymmetry can be traced back to the different choice of vertices $a=-1$, $b=+1$, selecting the graviphotons and $\bar{S}'$-vectors, respectively.

Finally, let us mention that the above discussion generalises in a straightforward fashion when expansions around more general enhancement points are considered. In particular, if there are $n_V, n_H$ extra massless vector multiplets and hypermultiplets, respectively, the dominant contribution in the field theory limit becomes:
\begin{equation}\label{FullFormula}
\mathcal{F}\sim (\epsilon_{-}^2-\epsilon_{+}^2) \int_0^\infty\frac{dt}{t}~\frac{2\bigr(n_H-n_V\cos\left( 2\epsilon_{+}t\right)\bigr) }
{\sin\left( \epsilon_{-}-\epsilon_{+}\right)t ~\sin\left( \epsilon_{-}+\epsilon_{+}\right)t} ~e^{-\mu t}~,
\end{equation}
in accordance with  the results of \cite{Gopakumar:1998jq,Iqbal:2007ii}. It is worth noting that the relative coefficient between hyper- and vector multiplets agrees with the fact that in the unrefined limit $\epsilon_+ =0$, in the $\mathcal{N}=4$ theory, where $n_H = n_V$, the amplitude must vanish.  

Before ending this section, we give an alternative, more physical, derivation of the contributions $-2n_V\cos(2\epsilon_{+}t)$ and $2n_H$ of 
vectors and hypers  in the numerator of (\ref{FullFormula}) using the operator formalism.\footnote{Beyond the field theory limit, the operator formalism becomes rather complicated and it is actually simpler to use the path-integral approach as described above.} We first discuss the case where the end points of the open string are lying on two D9-branes or two D5-branes and  restrict our attention to the field theory limit, hence keeping only the constant modes of the K3 fermions $\chi^{4}_0,\chi^{5}_0,\bar\chi^{4}_0,\bar\chi^{5}_0$ with respect to the $\sigma$-direction. For zero modes, there is no difference between left- and right- movers ($\chi_0=\tilde\chi_0$) and, thus, only for $b=+1$ does the deformation (\ref{K3FermDef}) survive. Neglecting the oscillator part of the deformed Hamiltonian 
\begin{equation}
H=\epsilon_{+}(\chi^4_0
\chi^5_0+\textrm{c.c.})+\textrm{osc.}~,
\label{hamiltonian}
\end{equation}
we are led to evaluate
\begin{align}
	{\textrm{Tr}\,}_{\mathcal{H}}\,(-)^F\,e^{-2\pi \tau_2 H}~,
\end{align}
over the finite-dimensional Hilbert space $\mathcal{H}$ of the periodic K3-fermion zero modes (corresponding to the odd spin structure in the doubled annulus picture), which satisfy the standard anti-commutation relations. One may pick the vacuum $|0\rangle$ to be annihilated by $\chi^4_0$ and $\chi^5_0$. The Hilbert space $\mathcal{H}$ is  spanned by exactly four states, which can be chosen as follows:
\renewcommand{\arraystretch}{1.5}
\begin{align}
	\begin{array}{l  || c }
		 \textrm{~~~~~State} & ~~\mathbb{Z}_n \textrm{-action}~~\\ \hline
		|0\rangle &  1   \\
		|1\rangle= \bar\chi^4_0|0\rangle  & e^{-2\pi i/n} \\
		|2\rangle= \bar\chi^5_0|0\rangle & e^{2\pi i/n} \\
		|3\rangle= \bar\chi^4_0 \bar\chi^5_0 |0\rangle & 1\\
	\end{array}
\end{align}
\renewcommand{\arraystretch}{1.0}where the second column displays their transformation under the $\mathbb{Z}_n$-orbifold action. It is then easy to see that $\mathcal{H}$ can be decomposed into subspaces according to their $\mathbb{Z}_n$-action:
\begin{align}
	\mathcal{H}=\mathcal{H}_0\oplus\mathcal{H}_{-}\oplus\mathcal{H}_{+}~,
\end{align}
where $\mathcal{H}_0$ is the $\mathbb{Z}_n$-invariant subspace spanned by $|0\rangle$ and $|3\rangle$, whereas one-dimensional subspaces $\mathcal{H}_{-}$ and $\mathcal{H}_{+}$ are spanned by vectors $|1\rangle$ and $|2\rangle$, respectively. $\mathcal{N}=2$ vector multiplets are built by fermionic oscillators invariant under $\mathbb{Z}_n$ and lie in $\mathcal{H}_0$, whereas hypermultiplets, whose oscillators transform with $e^{\pm 2\pi i/n}$ under $\mathbb{Z}_n$,  belong to $\mathcal{H}_{-}\oplus\mathcal{H}_{+}$. While the Hamiltonian (\ref{hamiltonian})
annihilates states  $|1\rangle$ and $|2\rangle$, it mixes the two states  of $\mathcal{H}_0$, namely  $|0\rangle$ and $|3\rangle$. Diagonalising the Hamiltonian in each subspace and taking the trace, immediately yields the contributions of vectors and hypers appearing in the numerator of (\ref{FullFormula}). The relative minus sign in the latter comes from the fact that we are evaluating the trace with the $(-1)^F$~insertion, whose eigenvalues are $+1$ on $\mathcal{H}_0$ and $-1$ on $\mathcal{H}_{\pm}$. 

For the case where the two end points of the open string lie on D9- and D5-branes respectively, the massless states are hypermultiplets. In this case there is a half integer shift in the moding of the worldsheet bosonic and fermionic fields along the $K3$ directions. This implies that in the Ramond sector, the massless space-time fermions are singlets under the $SO(4)$ tangent group of $K3$ and, therefore, the  Hamiltonian obtained from the deformation (\ref{K3FermDef}) (which now involves half-integer mode oscillators) annihilates the ground state. Consequently, we see that the contribution of hypers appears without an $\epsilon_+$-dependent phase.

We would like to emphasise that in this computation, we have not inserted any R-symmetry currents, yet the result correctly reproduces the
Nekrasov-Okounkov partition function. This may seem surprising, since one usually attributes  different phase factors for hypers and
vectors to the fact that they transform as different $SU(2)_R$-representations (while gauginos are doublets, hyperinos are singlets). Even though in our amplitudes all vertices are $SU(2)_R$ neutral, after the spin structure sum one effectively finds an $SU(2)_R$ current in the Hamiltonian $H$ (\ref{hamiltonian}). This is also the case in the Heterotic computation as can be seen from (\ref{FermGen}).


\chapter{Deformed ADHM Instantons and the Topological String}
\label{ch:ADHMinst}
As discussed in Section \ref{ch:GaugeTheoryFromStringTheory}, gauge theory instantons have a natural description as bound states of D-branes in string theory \cite{Witten:1995gx,Douglas:1996uz}. For instance, four-dimensional super-Yang-Mills theory can be realised on a stack of D3-branes in Type IIB string theory, with additional D($-1$)-branes playing the role of instantons. Their corrections to the Yang-Mills action are captured by string disc diagrams with boundary field insertions. In~\cite{Billo:2006jm}, the effect of a non-trivial constant string background in this setup was considered, by including additional bulk vertices in the tree-level amplitudes. It was shown that the insertion of anti-self-dual graviphoton field strength tensors, in the point particle limit, correctly reproduces the ADHM action on an $\Omega$-background with one of its deformation parameters switched off (\emph{e.g.} $\epsilon_+=0$). Using localisation techniques~\cite{Moore:1997dj,Lossev:1997bz,Nekrasov:2002qd,Nekrasov:2003rj,Losev:2003py,Iqbal:2003ix,Iqbal:2003zz} this allows one to compute the non-perturbative Nekrasov partition function $Z^{\textrm{Nek}}(\epsilon_{+}=0,\epsilon_{-})$.

Obtaining the partition function for a gauge theory on a generic $\Omega$-background (with $\epsilon_+\neq 0$) from string theory remains an interesting question. Phrased differently, one would like to find a  modification of the anti-self-dual graviphoton background, considered in \cite{Billo:2006jm}, giving rise to the fully deformed ADHM action in the point-particle limit of the appropriate disc diagrams. A hint for answering this question comes from the series of higher derivative one-loop couplings in the effective action of the Heterotic string compactified on $K3\times T^2$, considered in Section \ref{Sec:Hetamps}. In the field theory limit, these one-loop amplitudes precisely reproduce the perturbative contribution of the $\mc N=2$ gauge theory partition function on the full $\Omega$-background, \emph{i.e.} with $\epsilon_{-},\epsilon_+\neq 0$. We therefore expect that, including $F^{\bar{T}}_{(+)}$ also as a background field in the instanton computation described above, allows us to extract the fully deformed ADHM action from string theory. In what follows, we show that this is indeed the case.

We work in Type I string theory compactified on $K3\times T^2$ and consider D9-branes together with D5-instantons wrapping $K3\times T^2$. As discussed in Section~\ref{sec:dualities}, this setting is dual to Heterotic string theory on $K3\times T^2$ and the corresponding background is given by anti-self-dual graviphotons and self-dual field strength tensors of the vector partner of $\bar{S}'$, which we refer to as $\bar{S}'$-vectors in the remainder of this section.  We compute all tree-level diagrams with boundary insertions in this background which, in the field theory limit, correctly reproduce the fully $\Omega$-deformed version of the ADHM action, which was used to compute Nekrasov's partition function \cite{Nekrasov:2002qd,Nekrasov:2003rj}. 

This result can also be interpreted as computing gauge theory instanton corrections to the higher derivative couplings discussed in Section \ref{Sec:TypeIamp}. The fact that we reproduce precisely the full Nekrasov partition function can also be seen as further evidence for the proposal that these couplings furnish a worldsheet description of the refined topological string. Indeed, our results show that the background introduced in \cite{Antoniadis:2013bja} can be understood as a physical realisation of the $\Omega$-background in string theory.

We would like to mention that various RR backgrounds were discussed in \cite{Ito:2010vx,Ito:2011cr} where the $\Omega$-deformed ADHM action was recovered using the language of D-instantons. However, contrary to our present work, the instanton calculation is performed without NS-NS field strengths. In addition, the interpretation in terms of a string effective coupling was lacking.

In what follows, we calculate the tree-level (disc) diagrams with bulk insertions of $\bar S'$-vectors. Contrary to \cite{Ito:2010vx,Ito:2011cr,Green:2000ke,Billo:2002hm,Billo:2006jm}, we do not use auxiliary fields since correlation functions involving the latter are in general not well-defined\footnote{Auxiliary fields are not BRST-closed objects.}. In addition, we show that, in a particular factorisation limit corresponding to a specific choice of ghost pictures, the OPE of two physical operators can be effectively interpreted in terms of an auxiliary field.


\section{String Theory Setup}\label{Sect:InstantonsSetup}
Gauge instantons in a theory living on a configuration\footnote{Before considering the effect of D5 instantons, one may start from a consistent string vacuum, \emph{e.g.} the model discussed in \cite{Bianchi:1990tb} with gauge group $U(16)\times U(16)$, where the total number of D5 branes (wrapping the space-time and $T^2$ directions) and D9 branes is fixed to be 16 by tadpole cancellation. The stack of 16 D9 branes may be further higgsed to $N<16$ by turning on appropriate Wilson lines along $T^2$, so that one may keep $N$ generic for the purposes of our analysis.} of $N$ D9-branes are the D5-instantons  wrapping  $K3 \times T^2$. For 
concreteness, we summarise the brane setup in the following table.
\begin{center}
\parbox{14.5cm}{\begin{center}\begin{tabular}{|c||c||c|c|c|c||c|c||c|c|c|c|}\hline
&&&&&&&&&&&\\[-11pt]
\bf{brane} & \textbf{num.} & $X^0$ & $X^1$ & $X^2$ & $X^3$ & $X^4$ & $X^5$ & $X^6$ & $X^7$ & $X^8$ & $X^9$  \\\hline\hline
D9 & $N$ & $\bullet$ & $\bullet$ & $\bullet$ & $\bullet$ & $\bullet$ & $\bullet$ & $\bullet$ & $\bullet$ & $\bullet$ & $\bullet$\\\hline
D5 & $k$ &  &  &  &  & $\bullet$ & $\bullet$ & $\bullet$ & $\bullet$ & $\bullet$ & $\bullet$\\\hline
\end{tabular}\end{center}
${}$\\[-48pt]
\begin{align}
{}\hspace{2.9cm}\underbrace{\hspace{3.8cm}}_{\text{space-time}\,\sim\,\mathbb{R}^4}\underbrace{\hspace{1.9cm}}_{T^2}\underbrace{\hspace{3.75cm}}_{K3\,\sim\, T^4/\mathbb{Z}_2}\nonumber
\end{align}}
\end{center}
This configuration describes instantons of winding number $k$  in a gauge theory with $SU(N)$ gauge group. In order to see this, let us describe the massless spectrum of the theory. The notation used here is summarised in Appendix \ref{appendix:spinors}. More precisely, it is natural to decompose the ten-dimensional Lorentz group into 
\begin{align}
SO(10)\rightarrow SO(4)_{ST}\times SO(6)_{\text{int}}\rightarrow SO(4)_{\text{ST}}\times SO(2)_{T^2} \times SO(4)_{K3}\,,\label{DecomposeSo10}
\end{align}
reflecting the product structure of our geometry. In this way $I,J=1,\ldots,10$ denote indices of the full $SO(10)$, $\mu,\nu$ are indices of the space-time $SO(4)_{ST}$, with $\alpha,\beta$ ($\dot\alpha,\dot\beta$) the corresponding (anti-)chiral spinor indices, while $a,b$ denote internal $SO(6)_{\text{int}}$ indices.  For $SO(4)_{K3}\sim SU(2)_+\times SU(2)_-$, we introduce indices $A,B=1,2$ for fields transforming in the $(\mathbf{2},\mathbf{1})$ representation (positive chirality) and $\hat{A},\hat{B}=3,4$ for fields in the $(\mathbf{1},\mathbf{2})$ representation (negative chirality).  Following \cite{Green:2000ke,Billo:2006jm}, we associate upstairs indices of $SU(2)_+$ (resp. $SU(2)_-$) with charge $+1/2$ (resp. $-1/2$) of $SO(2)_{T^2}$ and downstairs indices with charge $-1/2$ (resp. $+1/2$) respectively. In this way, internal indices cannot be raised and lowered with the help of $\epsilon$-tensors, but we have to keep track of their position.

In this setting, there are three kinds of open string sectors which are relevant for our subsequent discussion. They can be characterised according to the location of their endpoints and are very similar to the ones studied in the Section \ref{ch:GaugeTheoryFromStringTheory}.

\begin{enumerate}
\item 9-9 sector\\
The massless excitations consist of a number of $\mathcal{N}=2$ vector multiplets, each of which containing a vector field $A^\mu$, a complex scalar $\phi$ as well as four gaugini $(\Lambda^{\alpha A}, \Lambda_{\dot\alpha A})$ in the $(\mathbf{2},\mathbf{1})$ representation of $SU(2)_{+}\times SU(2)_-$ with $SO(2)_{T^2}$ charges $(1/2,-1/2)$ respectively. The bosonic degrees of freedom stem from the NS sector, while the fermionic ones from the R sector. These fields separately realise a Yang-Mills theory living on the four-dimensional space-time. 
\item 5-5 sector\\
These states are moduli (i.e. non dynamical fields) from a string perspective, due to the instantonic nature of the corresponding D5-branes. Indeed, the states in this sector cannot carry any momentum because of the Dirichlet boundary conditions in all directions except along $K3\times T^2$. From the Neveu-Schwarz (NS) sector, we have six bosonic moduli, which we  write as a real vector $a^\mu$ and a complex scalar $\chi$. From the point of view of the SYM theory living on the world-volume of the D9-branes, $a^\mu$ corresponds to the position of gauge theory instantons. From the Ramond sector, we have eight fermionic moduli, which we denote as $M^{\alpha A}$, $\lambda_{\dot\alpha A}$.
\item 5-9 and 9-5 sectors\\
Also this sector contains moduli from a string point of view. From the NS sector, the fermionic coordinates have integer modes giving rise to two Weyl spinors of $SO(4)_{ST}$ which we call $(\omega_{\dot\alpha},\bar\omega_{\dot\alpha})$. Notice that, as in the D(-1)/D3 case, these fields all have the same chirality, which in our case is anti-chiral. From a SYM point of view, these fields control the size of the instanton. In the R sector, fields are half-integer moded giving rise to two fermions $(\mu^A, \bar\mu^A)$ transforming in the $(\mathbf{2},\mathbf{1})$ representation of $SU(2)_+$ with positive charge under $SO(2)_{T^2}$.
\end{enumerate}
For the reader's convenience, the field content is compiled in Table~\ref{Tab:FieldsBis}.
\begin{table}[htbp]
\begin{center}
\begin{tabular}{|c||c||c|c|c|c|c|}\hline
&&&&&&\\[-11pt]
\bf{sector} & \bf{field} & $SO(4)_{ST}$ & $c_{T^2}$ & $SU(2)_+\times SU(2)_-$ & \bf{statistic} & \bf{R / NS} \\\hline\hline
&&&&&&\\[-11pt]
9-9 & $A^\mu$ & $(\mathbf{1/2},\mathbf{1/2})$ & $0$ & $(\mathbf{1},\mathbf{1})$ & boson & NS \\
&&&&&&\\[-11pt]\hline
&&&&&&\\[-11pt]
 & $\Lambda^{\alpha A}$ & $(\mathbf{1/2},\mathbf{0})$ & $1/2$ & $(\mathbf{2},\mathbf{1})$ & fermion & R \\
 &&&&&&\\[-11pt]\hline
 &&&&&&\\[-11pt]
 & $\Lambda_{\dot{\alpha} A}$ & $(\mathbf{0},\mathbf{1/2})$ & $-1/2$ & $(\mathbf{2},\mathbf{1})$ & fermion & R \\
 &&&&&&\\[-11pt]\hline
 &&&&&&\\[-11pt]
 & $\phi$ & $(\mathbf{0},\mathbf{0})$ & $-1$ & $(\mathbf{1},\mathbf{1})$ & boson & NS \\ &&&&&&\\[-11pt]\hline\hline
 &&&&&&\\[-11pt]
 5-5 & $a^\mu$ & $(\mathbf{1/2},\mathbf{1/2})$ & $0$ & $(\mathbf{1},\mathbf{1})$ & boson & NS \\
&&&&&&\\[-11pt]\hline
&&&&&&\\[-11pt]
 & $\chi$ & $(\mathbf{0},\mathbf{0})$ & $-1$ & $(\mathbf{1},\mathbf{1})$ & boson & NS \\
&&&&&&\\[-11pt]\hline
&&&&&&\\[-11pt]
& $M^{\alpha A}$ & $(\mathbf{1/2},\mathbf{0})$ & $1/2$ & $(\mathbf{2},\mathbf{1})$ & fermion & R \\
&&&&&&\\[-11pt]\hline
&&&&&&\\[-11pt]
& $\lambda_{\dot{\alpha} A}$ & $(\mathbf{0},\mathbf{1/2})$ & $-1/2$ & $(\mathbf{2},\mathbf{1})$ & fermion & R \\
&&&&&&\\[-11pt]\hline\hline
&&&&&&\\[-11pt]
5-9 & $\omega_{\dot{\alpha} }$ & $(\mathbf{0},\mathbf{1/2})$ & $0$ & $(\mathbf{1},\mathbf{1})$ & boson & NS \\
&&&&&&\\[-11pt]\hline
&&&&&&\\[-11pt]
 & $\mu^A$ & $(\mathbf{0},\mathbf{0})$ & $1/2$ & $(\mathbf{2},\mathbf{1})$ & fermion & R \\[2pt]\hline
\end{tabular}
\end{center}
\caption{Overview of the massless open string spectrum relevant for the disc amplitude computation. We display the transformation properties under the groups $SO(4)_{ST}\times SU(2)_+\times SU(2)_-$, while the column $c_{T^2}$ denotes the charge under $SO(2)_{T^2}$. The last two columns denote whether the field is bosonic or fermionic and the sector it stems from.}
\label{Tab:FieldsBis}
\end{table}
A similar analysis to the one in Section \ref{ch:GaugeTheoryFromStringTheory} shows that the string tree-level effective action involving only the vector multiplets of the 9-9 sector exactly reproduces, in the field theory limit, the pure $\mathcal{N}=2$ super-Yang-Mills theory with $SU(N)$ gauge group. Inclusion of the remaining moduli fields gives rise to the ADHM action describing instantonic corrections with instanton\footnote{When taking the field theory limit, one should pay attention to the dimensionality of the various fields. In particular, a rescaling of the ADHM moduli is necessary in order for the field theory limit to be well-defined as an appropriate double scaling limit in which $g_{\textrm{YM}}$ is held fixed.} number $k$. 

Furthermore, by coupling the theory to a constant anti-self-dual graviphoton background \cite{Billo:2006jm}, the resulting effective action coincides with the ADHM action in the $\Omega$-background used in \cite{Nekrasov:2002qd} in the case where one of the deformation parameters vanishes (say, $\epsilon_+=0$). While the ADHM action is exact under a nilpotent Q-symmetry, the latter is still present after the deformation with $\epsilon_+$. Hence, one can use localisation techniques in order to compute the instanton partition function \cite{Nekrasov:2002qd}. 

From a practical perspective, this deformation is obtained by computing string disc diagrams with bulk insertions of the anti-self-dual graviphoton. Due to its anti-self-duality, the only instanton contributions come from the diagrams with insertions from the 5-5 sector and no mixed diagrams\footnote{Recall that by mixed diagrams we refer to disc diagrams whose boundary lies on both D9- and D5-branes.} contribute.
\begin{figure}[h!t]
\begin{center}
\parbox{8cm}{\epsfig{file=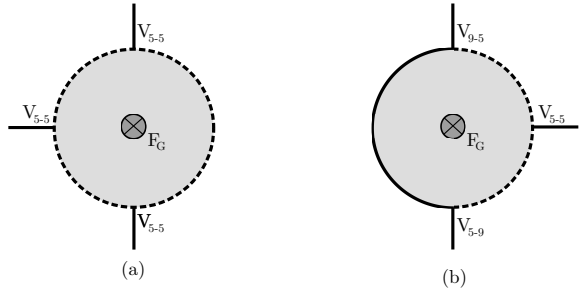,width=7.5cm}}
\end{center}
\parbox{\textwidth}{\caption{Four-point disc diagrams with graviphoton bulk-insertion. Diagram (a) involves two boundary insertions from the 5-5 sector, whereas diagram (b) two insertions from the 5-9 sector. While the whole boundary of diagram (a) lies on the D5-branes, diagram (b) lies partly on the D9- and partly on the D5-branes. The latter mixed boundary diagrams vanish due to the anti-self-duality of the graviphoton.}}
\label{fig-ee}
\end{figure}

In the following section, we generalise this construction to the case where the background includes, in addition to the anti-self-dual graviphoton field strength, the self-dual field strength of the $\bar{S}'$-vector. We show that this generalised background, discussed in Section \ref{Sec:TypeIamp} as a string theory uplift of the $\Omega$-background in Type I, reproduces the ADHM action in the presence of a general $\Omega$-background, therefore providing a non-perturbative check of the latter proposal.


\section{Refined Instanton Effective Action}\label{Sect:RefinedADHM}

\subsection{Vertex Operators}\label{Sect:DiskDiagrams}

We denote the ten-dimensional bosonic and fermionic worldsheet fields collectively as $X^I$ and $\psi^I$ respectively, with $I=1,\ldots,10$. More precisely, we use $(X^{\mu},\psi^{\mu})$, $(Z,\Psi)$ and $(Y^i,\chi^i)$ for worldsheet fields along the four-dimensional space-time, $T^2$ and $K3$ directions respectively. In the following, for simplicity, we consider an orbifold representation of  $K3$ as $T^4/\mathbb{Z}_2$. However, we expect our results to be valid also for a generic (compact) $K3$. The vertex operators relevant for the disc amplitudes involve the ADHM moduli appearing in the massless spectrum. From the 5-5 sector, we need
\begin{align}
 V_{a}(z)&= g_6\,a_{\mu}\psi^{\mu}(z)e^{-\varphi(z)}\,,\\
 V_{\chi}(z)&= \frac{\chi}{\sqrt{2}}\bar\Psi(z)e^{-\varphi(z)}\,,\\
 V_{M}(z)&= \frac{g_6}{\sqrt{2}}\,M^{\alpha A}S_{\alpha}(z)S_{A}(z)e^{-\tfrac{1}{2}\varphi(z)}\,.
\end{align}
From the 5-9 and 9-5 sectors, we use
\begin{align}
 V_{\omega}(z)=& \frac{g_6}{\sqrt{2}}\,\omega_{\dot\alpha}\Delta(z)S^{\dot\alpha}(z)e^{-\varphi(z)}\,,\\
 V_{\bar\omega}(z)=& \frac{g_6}{\sqrt{2}}\,\bar\omega_{\dot\alpha}\bar\Delta(z)S^{\dot\alpha}(z)e^{-\varphi(z)}\,.
\end{align}
Here $\Delta(z),\,\bar\Delta(z)$ are twist and anti-twist fields with conformal weight $1/4$, which act by changing the boundary conditions and $g_6$ is the D5-instanton coupling constant. Finally, we turn to the closed string background defined in Section \ref{ch:GaugeTheoryFromStringTheory}. The vertex operator of the anti-self-dual graviphoton in the $(-1)$-ghost picture at zero momentum is given by:
\begin{align}
	V^{F^G}(y,\bar y)=\frac{1}{8\pi\,\sqrt{2}}F_{\mu\nu}^{G} \bigg[\,\psi^{\mu}\psi^{\nu}(y)e^{-\varphi(\bar y)}{\bar\Psi}(\bar y)\,+e^{-\varphi(y)}\bar\Psi(y)\psi^{\mu}\psi^{\nu}(\bar y)&\nonumber\\
-\frac{i}{2}\, e^{-\tfrac{1}{2}(\varphi(y)+\varphi(\bar y))} S_{\alpha}(y) (\sigma^{\mu\nu})^{\alpha\beta} 
S_{\beta}(\bar y)\,\epsilon^{AB}\,S_{A}(y)S_{B}(\bar y)\bigg]&\,.\label{GraviphotonVertex}
\end{align}
The vertex operator for the self-dual field strength tensor of the $\bar{S}'$-vector in the $(-1)$-ghost picture and at zero momentum is given as a sum of an NS-NS part (first line) and a R-R part (second line):
\begin{align}\label{UbarVec}
V^{F^{\bar{S}'}}(y,\bar y)=\frac{1}{8\pi\,\sqrt{2}}F_{\mu\nu}^{\bar{S}'} \bigg[\,\psi^{\mu}\psi^{\nu}(y)e^{-\varphi(\bar y)}{\bar\Psi}(\bar y)\,+e^{-\varphi(y)}\bar\Psi(y)\psi^{\mu}\psi^{\nu}(\bar y)&\nonumber\\
+\frac{i}{2}\, e^{-\tfrac{1}{2}(\varphi(y)+\varphi(\bar y))} S_{\dot{\alpha}}(y) (\bar{\sigma}^{\mu\nu})^{\dot{\alpha}\dot{\beta}} 
S_{\dot{\beta}}(\bar y)\,\epsilon_{\hat A\hat B}\,S^{\hat A}(y)S^{\hat B}(\bar y)\bigg]&\,.
\end{align}
The disc diagrams involving the bulk insertions of the RR part of the anti-self-dual graviphoton \eqref{GraviphotonVertex} have already been extensively studied in the literature (see, for example, \cite{Billo:2006jm}). Following the analysis performed below, one can show that the NS-NS part also gives the same contribution leading to the $\epsilon_{-}$-dependent part of the $\Omega$-deformed ADHM action. Here, we focus on the more interesting case of the self-dual insertions of $\bar{S}'$-field strengths. The relevant disc diagrams correspond to the following correlators and involve the bosonic instanton moduli only\footnote{One can show that the insertions of fermionic moduli leads to vanishing disc amplitudes.}, and these are
\begin{align}
 \mc{D}_{aa\chi^{\dag} F^{\bar S'}}(x_1,x_2,x_3,z,\bar z)&=\left<V_{a}(x_1)\,V_a(x_2)\,V_{\chi^{\dag}}(x_3)\,V_{F^{\bar{S}'}}(z,\bar z)\right>_{\text{Disc}}\,,\label{UnmixedDiag}\\
 \mc{D}_{\omega\bar\omega\chi^{\dag}F^{\bar S'}}(x_1,x_2,x_3,z,\bar z)&=\left<V_{\omega}(x_1)\,V_{\bar\omega}(x_2)\,V_{\chi^{\dag}}(x_3)\,V_{F^{\bar{S}'}}(z,\bar z)\right>_{\text{Disc}}\,.\label{MixedDiag}
\end{align}
We emphasise that a crucial requirement for all our scattering amplitudes is that the external fields involved be physical, \emph{i.e.} that they be annihilated by the BRST operator $Q_{\text{BRST}}=\oint J_{\text{BRST}}$ \eqref{BRSTcharge}. This turns out to be important in order to determine the precise structure of the various vertex operators in different pictures. In addition, as we show below, these amplitudes take the form of contact terms $p_i.p_j/p_i.p_j$ where $p_i$ are the momenta of the various vertex operator insertions. These contact terms give non-trivial results in the limit $p_i\to 0$. To be able to compute them in a well-defined manner, we keep $p_i$ generic in all intermediate steps, which also acts as a regularisation of the worldsheet integrals, and take the limit only at the end of the calculation. However, due to the nature of our vertex insertions, we cannot switch on momenta in an arbitrary fashion: since the four-dimensional space-time corresponds to directions with Dirichlet boundary conditions for the D5-instantons, none of the ADHM moduli can carry momenta along $X^\mu$. Similarly, the $\bar{S}'$-vector insertions cannot carry momenta along $T^2$ once we impose BRST invariance (\emph{i.e.} transversality and decoupling of longitudinal modes). As a way out, we take all vertices to carry momenta along the $K3$ directions and complexify them, if necessary, to make all integrals well-defined. In fact, technically, we first replace $K3$ by $\mathbb{R}^4$ and compute an effective action term on the D-instanton world-volume $T^2 \times \mathbb{R}^4$. Since the relevant fields that appear in these couplings  survive the orbifold projection (or more generally on a smooth $K3$ manifold they give rise to zero-modes), the corresponding couplings exist also in the case where $\mathbb{R}^4$ is replaced by $K3$.

\subsection[\texorpdfstring{$D_5-D_5$ disc diagrams}{D5-D5 disc diagrams}]{\texorpdfstring{$\boldsymbol{D_5-D_5}$ disc diagrams}{D5-D5 disc diagrams}}

\begin{figure}[h!t]
\begin{center}
\parbox{4.2cm}{\epsfig{file=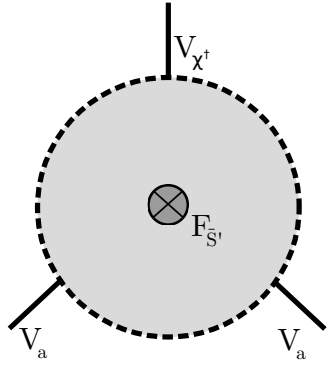,width=4.2cm}}
\end{center}
\parbox{\textwidth}{\caption{Four-point disc diagram with bulk-insertion of the $\bar{S}'$ field strength tensor and  three boundary insertions stemming from the 5-5 sector of the string setup.}\label{fig-fourpt}}
\end{figure}

We start by evaluating the amplitude \eqref{UnmixedDiag}, depicted in Fig. \ref{fig-fourpt}, corresponding to a disc diagram whose boundary lies entirely on the stack of D5-branes and with the $\bar{S}'$-modulus inserted in its bulk. Since the vertex operator for $F^{\bar{S}'}$ consists of two parts (NS and R), we can split the correlator (\ref{UnmixedDiag}) accordingly
\begin{align}
\mc{D}_{aa\chi^{\dag} F^{\bar S'}}(x_1,x_2,x_3,z,\bar z)= \mc{D}_{aa\chi^{\dag} F^{\bar S'}}^{\textrm{NS}}(x_1,x_2,x_3,z,\bar z)+\mc{D}_{aa\chi^{\dag} F^{\bar S'}}^{\textrm{R}}(x_1,x_2,x_3,z,\bar z)\,,\label{Corr11NSR}
\end{align}
with
\begin{align}
 \mathcal{D}_{aa\chi^{\dag} F^{\bar S'}}^{\textrm{NS}}&=\langle V^{(0)}_{a}(x_1) V_{a}^{(-1)}(x_2) V_{\chi^{\dag}}^{(0)}(x_3) V^{(-1,-1)}_{F^{\bar{S}'}}(z,\bar{z}) V_{\text{PCO}}(y) \rangle\label{4ptFctNS}\,,\\
 \mathcal{D}_{aa\chi^{\dag} F^{\bar S'}}^{\textrm{R}}&=\langle V^{(0)}_{a}(x_1) V_{a}^{(-1)}(x_2) V_{\chi^{\dag}}^{(0)}(x_3) V^{(-\frac{1}{2},-\frac{1}{2})}_{F^{\bar{S}'}}(z,\bar{z})\rangle\label{4ptFctR}\,.
\end{align}
Here  $V^{(-1,-1)}_{F^{\bar{S}'}}$ is the NS-NS part of the closed string $F^{\bar{S}'}$ field strength tensor introduced in (\ref{UbarVec}), $V^{(-\frac{1}{2},-\frac{1}{2})}_{F^{\bar{S}'}}$ is its R-R part and $V_{\text{PCO}}$ is the picture changing operator, recall (\ref{BosBeta},\ref{BosGamma}),
\begin{equation}
 V_{\text{PCO}}= Q_{\textrm{BRST}}\,\xi\,.
\end{equation}
In (\ref{4ptFctNS}) we have kept the PCO insertion at a fixed position $y$, even though the final result should not depend on $y$ \cite{Friedan:1985ey}. Setting $y$ to $z$, $\bar{z}$ or $x_2$, converts the ghost picture of the corresponding vertex operators to \footnote{In (\ref{UnmixedDiag}) an insertion of $\xi$ at an arbitrary position is understood, in order to soak up the $\xi$ zero mode.}
\begin{align}
\mathcal{D}_{aa\chi^{\dag} F^{\bar S'}}^{\textrm{NS}}\big|_{y=z}&=\langle V^{(0)}_{a}(x_1) V_{a}^{(-1)}(x_2) V_{\chi^{\dag}}^{(0)}(x_3) V^{(0,-1)}_{F^{\bar{S}'}}(z,\bar{z}) \rangle\,,\label{4ptyz}\\
\mathcal{D}_{aa\chi^{\dag} F^{\bar S'}}^{\textrm{NS}}\big|_{y=\bar{z}}&=\langle V^{(0)}_{a}(x_1) V_{a}^{(-1)}(x_2) V_{\chi^{\dag}}^{(0)}(x_3) V^{(-1,0)}_{F^{\bar{S}'}}(z,\bar{z})\rangle\,,\label{4ptybz}\\
\mathcal{D}_{aa\chi^{\dag} F^{\bar S'}}^{\textrm{NS}}\big|_{y=x_2}&=\langle V^{(0)}_{a}(x_1) V_{a}^{(0)}(x_2) V_{\chi^{\dag}}^{(0)}(x_3) V^{(-1,-1)}_{F^{\bar{S}'}}(z,\bar{z}) \rangle\,,\label{4ptyx}
\end{align}
respectively.
Using the doubling trick, we can convert the disc into the full plane with a $\mb Z_2$-involution, and the four-point amplitude \eqref{UnmixedDiag} becomes a five-point function with  vertices at $(x_1,x_2,x_3,z,\bar{z})$. Here we split
\begin{equation}
 V^{(-1,-1)}_{F^{\bar{S}'}}(z,\bar{z})=V_{F^{\bar{S}'}}(z)\,V_{F^{\bar{S}'}}(\bar{z})\,,
\end{equation}
where the left-right symmetrisation is implicit. $SL(2,\mb R)$ invariance implies that we can fix three real positions, which is related to the existence of three $c$-ghost zero modes on the sphere. The latter are soaked up by attaching $c$ to three dimension one vertices such that the resulting operators are BRST closed.  The dimension of these vertices becomes zero and they remain unintegrated. Since the last two terms in \eqref{BRSTcharge} annihilate any operator in the $(-1)$-picture, any physical operator with dimension one and negative ghost picture becomes BRST invariant in this manner.\footnote{Indeed, the first two terms in (\ref{BRSTcharge}) combined together annihilate $c V$ for any $V$ corresponding to a dimension one Virasoro primary operator, irrespective of the ghost picture of $V$.} However, recall from \eqref{PhysFixedVertexOperator} that for a zero-picture operator, the correct dimension zero BRST invariant combination is $c V^{(0)}+ \gamma V^{(-1)}$. Therefore, for simplicity, we choose the zero-picture vertices to be of dimension one (such that their positions are integrated),  and all the $(-1)$-picture vertices to be of dimension zero (such that their positions remain unintegrated).

Let us first consider the NS-NS contributions \eqref{4ptFctNS}, for which the vertex operators are
\begin{eqnarray}
V_{a}(x_1)&=& g_6\,a_{\mu}(\partial X^\mu-2 i  p_1\cdot\chi \,\psi^{\mu})e^{2 i p_1\cdot Y}(x_1)\,,\label{Va}\\
V_{a}(x_2)&=& g_6\,a_{\nu}\, c e^{-\varphi}\psi^{\nu}e^{2 i p_2\cdot Y}(x_2)\,,\\
V_{\chi^{\dag}}(x_3)&=&\frac{\chi^{\dag}}{\sqrt{2}}\, (\partial Z-2 i p_3\cdot\chi \,\Psi)e^{2 i p_3\cdot Y}(x_3)\label{Vchi}\,,\\
V_{F^{\bar{S}'}}(z)&=&  c e^{-\varphi}\bar{\Psi}e^{i(P_\mu X^\mu+ P\cdot Y)}(z)\,,\\
V_{F^{\bar{S}'}}(\bar{z})&=&-\frac{i\epsilon_{\lambda}}{8\pi\sqrt{2}}\,  c e^{-\varphi}\psi^{\lambda}e^{i(-P_\mu X^\mu+ P\cdot Y)}(\bar{z})\,,
\end{eqnarray}
with $F_{\mu\nu}^{\bar{S}'}\equiv \epsilon_{[\mu}P_{\nu]}$, and the only relevant terms in $V_{\text{PCO}}$ are (since the total background charge of the superghost is $-2$)
\begin{equation}
e^{\varphi}\, T_F(y)= ie^{\varphi}(\psi^\mu \partial X_\mu+\Psi \partial \bar{Z}+\bar{\Psi} \partial{Z}+\chi^i \partial Y^i)(y)\,.
\end{equation}
Here, $Y^i\in \{X^6,X^7,X^8,X^9\}$ parametrise the internal $\mathbb{R}^4$ (which we eventually replace by $K3$). The momenta $p_i$ are along these directions, while the momentum of $V_{F^{\bar{S}'}}$ is written as $(P_\mu, P)$, where $P_\mu$ is the space-time part and $P$ is along the $Y^i$ directions. Note that after using the doubling trick, the Neumann directions $( Z, \bar{Z},Y^i)$ are mapped onto themselves, whereas the Dirichlet ones pick an additional minus sign $X^\mu \rightarrow -X^{\mu}$. This is consistent with the fact that the momenta along Neumann directions are conserved $\sum_i p_i + P =0$, which follows from integrating the zero modes of $Y^i$. On the other hand, integrating over the zero modes of the Dirichlet directions $X^\mu$ does not give any conservation law for  $P_\mu$.

The three open string vertices contain Chan-Paton labels which need to be suitably ordered. For instance, if we are interested in computing the term $\textrm{Tr}(a_{\mu}\,a_{\nu}\,\chi^{\dag})$, the range of integration is the following:
\begin{equation}\label{IntRange}
\begin{cases}
 \textrm{for } x_1>x_2\,, & x_3\in ]x_2,x_1[\,,\\
 \textrm{for } x_2>x_1\,, & x_3\in]-\infty,x_1[\cup]x_2,\infty[\,.
\end{cases}
\end{equation}
For the other inequivalent ordering $\textrm{Tr}(a_{\mu}\,\chi^{\dag}\,a_{\nu})$, the range of the $x_3$-integration is opposite. It is easy to show that the sum of these two orderings vanishes so that the amplitude is of the form $\textrm{Tr}(a_{\mu}[\chi^{\dag}, a_{\nu}])$.

For definiteness, let us focus on the term $\textrm{Tr}(a_{\mu}\,a_{\nu}\,\chi^{\dag})$. The contraction of $\varphi$ and $c$ and the contraction of the exponentials in momenta yield
\begin{eqnarray}
A_0&=&\left\{-\frac{ig_6^2}{16\pi}\,\textrm{Tr}(a_{\mu}\,a_{\nu}\,\chi^{\dag})\epsilon_{\lambda}\right\}|y-z|^2(y-x_2)\nonumber\\
&\times&\prod_{1\leq i<j\leq3}(x_i-x_j)^{4 p_i\cdot p_j}\prod_{k=1}^{3}|x_k-z|^{4 p_k\cdot P}(z-\bar{z})^{- P_\mu P^\mu+ P_i P_i}\,.
\end{eqnarray}
This is a common factor that multiplies each of the remaining contractions. Now let us consider the contribution of $\partial{Z}(x_3)$ to the amplitude. This must contract with $\partial \bar{Z}(y)$ in $V_{\textrm{PCO}}$ and then $\Psi(y)$ contracts with $\bar\Psi(z)$. Then $\psi^{\lambda}(\bar{z})$ necessarily contracts with $\psi^\nu(x_2)$ and from $x_1$ only $\partial X^{\mu}(x_1)$ can contribute. The result is
\begin{equation}
A_1= \frac{i \delta^{\nu \lambda} P^\mu (z-\bar{z})}{(y-x_3)^2 (y-z)(x_2-\bar{z})|x_1-z|^2}\,.
\end{equation}
Next consider the contribution of the second term in \eqref{Vchi}. Here, there are two separate contributions.  If $p_3\cdot\chi(x_3)$ contracts with $p_1\cdot\chi(x_1)$, then $\psi^\mu(x_1)$, $\psi^{\nu}(x_2)$, $\psi^{\lambda}(\bar{z})$ and a space-time fermion $\psi^{\sigma}(y)$ from the picture changing operator must contract, leaving $\partial X^\sigma(y)$ which can only contract with the momentum parts of vertices at $z$ and $\bar{z}$ resulting in a term proportional to $P_\sigma$. Notice that the term arising from the contraction of $\psi^\mu$ with $\psi^{\nu}$ is killed by the transversality condition (a necessary condition for the operator to be in the kernel of $Q_{\textrm{BRST}}$). The total result is
\begin{equation}
A_2= \frac{4 i p_1\cdot p_3\  (z-\bar{z})}{(x_3-z)(x_3-x_1)|y-z|^2}\left[\frac{\delta^{\nu \lambda} P^\mu}{(x_2-\bar{z})(y-x_1)}-\frac{\delta^{\mu \lambda} P^\nu}{(x_1-\bar{z})(y-x_2)}\right]\,.
\end{equation}
On the other hand, if $p_3\cdot\chi(x_3)$ contracts with $\chi(y)$ in $V_{\textrm{PCO}}$, then $\partial Y(y)$ contracts with momentum dependent parts of the vertices. Thus, $\psi^{\lambda}(\bar{z})$ must contract with $\psi^\nu(x_2)$ and only $\partial X^\mu$ at $x_1$ can contribute so that one obtains
\begin{equation}
A_3=\frac{4 i\delta^{\nu \lambda} P^\mu (z-\bar{z})}{(x_3-z)(y-x_3)(x_2-\bar{z})|x_1-z|^2}\left[\frac{p_3\cdot p_1}{y-x_1}+\frac{p_3\cdot p_2}{y-x_2}+\frac{p_3\cdot P}{2(y-z)}+
\frac{p_3\cdot P}{2(y-\bar{z})}\right]\,.
\end{equation}
The total correlation function is thus 
\begin{equation}
 \mathcal{D}_{aa\chi^{\dag} F^{\bar S'}}^{\textrm{NS}}=A_0(A_1+A_2+A_3)\,,
\end{equation}
which must be integrated over $x_1$ and $x_3$. Note that all the terms in $A_1$, $A_2$ and $A_3$ come with one power of \ST momentum $P^\mu$ which is exactly what is required to obtain a coupling to the field strength of the closed string gauge field. However, both $A_2$ and $A_3$ are quadratic in the momenta along the $Y^i$ directions and  they can only contribute to the amplitude in the zero-momentum limit if the integration over $x_1$ and $x_3$ gives a pole of the form $1/(p_a\cdotp p_b)$. Clearly, $A_0\cdotp A_3$ cannot provide such a pole (we are assuming a generic value of $y$ in the complex plane i.e. $\textrm{Im}(y)\neq 0$). On the other hand, the integral over $x_3$ for  $A_0\cdotp A_2$ gives a pole of the form $1/(p_1\cdotp p_3)$. Performing the  $x_3$ integral in both the regions \eqref{IntRange} yields precisely the same result, hence, the $x_1$ integral over the entire real line reads
\begin{equation}
A_0 A_2 = \frac{g_6^2}{8\pi}\,\textrm{Tr}\left[a_{\mu}\,a_{\nu}\,\chi^{\dag}\right]\int_{-\infty}^{\infty} dx_1 \frac{(z-\bar{z})}{|x_1-z|^2}\epsilon_{\lambda}\left[\frac{(x_1-\bar{z})(y-x_2)}{(x_2-\bar{z})(y-x_1)}P^{\mu}\delta^{\nu\lambda}-P^{\nu}\delta^{\mu\lambda}\right]\,,
\label{A0A2}
\end{equation}
where we have set all the momenta along the $Y^i$ directions to zero since there are no singularities in the remaining $x_1$ integral. Notice that $A_0\cdotp A_2$ alone does not lead to a gauge invariant answer. As for the $A_0 A_1$ term the $x_3$ and $x_1$ integrals have no singularities and therefore the momenta along the $Y^i$ directions can be set to zero. The resulting $x_3$ integral for both regions \eqref{IntRange} gives precisely the same result:
\begin{equation}
A_0 A_1 = -\frac{g_6^2}{8\pi}\,\textrm{Tr}\left[a_{\mu}\,a_{\nu}\,\chi^{\dag}\right]\int_{-\infty}^{\infty} dx_1 \frac{(z-\bar{z})}{|x_1-z|^2}\epsilon_{\lambda}\frac{(y-\bar{z})(x_1-x_2)}{(x_2-\bar{z})(y-x_1)}P^{\mu}\delta^{\nu\lambda}\,.\label{A0A1}
\end{equation}
Adding the two terms \eqref{A0A2} and \eqref{A0A1}, we see that the result is gauge invariant. Performing the $x_1$ integration yields\footnote{Notice the additional factor of 2 due to the left-right symmetrisation in the closed string vertex.}:
\begin{align}
 \left<\left<V_{a}\,V_{a}\,V_{\chi^{\dag}}\,V_{F^{\bar S'}}^{\textrm{NS}}\right>\right>&=\frac{8\pi^2}{g_{\textrm{YM}}^2}\int\mathcal{D}_{aa\chi^{\dag} F^{\bar S'}}^{\textrm{NS}}\nonumber\\
 &=-2i\,\textrm{Tr}\left[a_{\mu}\,a_{\nu}\,\chi^{\dag}\right]\epsilon_{\lambda}(P^{\mu}\delta^{\nu\lambda}-P^{\nu}\delta^{\mu\lambda})\,.\label{Phys4ptNS}
\end{align}
Here, we have used $2/(2\pi\alpha' g_6)^2=8\pi^2/ g_{YM}^{2}$ and restored the dimensionality of the fields in terms of $2\pi\alpha'$.

Finally, let us consider the R-R contributions \eqref{4ptFctR}. The vertex operators are the same as above, except for the $\bar S'$-vector part which is given by
\begin{align}
V^{(-\frac{1}{2},-\frac{1}{2})}_{F^{\bar{S}'}}(z,\bar{z})=&\frac{i}{16\pi\sqrt{2}}\,F_{\bar S'}^{\rho\lambda}\,c\,e^{-\frac{\varphi}{2}}\,S_{\dot{\alpha}}\,S^{\hat A}\,e^{i(P\cdotp X+P\cdotp Y)}(z)\nonumber\\
&\times\epsilon_{\hat A\hat B}{(\bar{\sigma}_{\rho \lambda})^{\dot{\alpha}}}_{\dot{\beta}}\, 
c\,e^{-\frac{\varphi}{2}}\,S^{\dot{\beta}}\,S^{\hat B}\,e^{i(-P\cdotp X+P\cdotp Y)}(\bar{z})\,.
\end{align}
Since the total superghost charge of the vertices is $-2$, there is no need for a picture changing operator. The total charge in the torus plane implies that only $p_3\cdot\chi\,\Psi(x_3)$ in \eqref{Vchi} contributes so that only $p_1\cdot\chi\,\psi^{\mu}(x_1)$ in \eqref{Va} contributes. This term is proportional to $p_1\cdotp p_3$. Once again the integral over $x_3$ gives a pole $1/(p_1\cdotp p_3)$ in the channel $x_3 \rightarrow x_1$. Performing the integrals over $x_1$ and $x_3$ as above leads to the same result:
\begin{align}
 \left<\left<V_{a}\,V_{a}\,V_{\chi^{\dag}}\,V_{F^{\bar S'}}^{\textrm{NS}}\right>\right>&=\left<\left<V_{a}\,V_{a}\,V_{\chi^{\dag}}\,V_{F^{\bar S'}}^{\textrm{R}}\right>\right>\,.
\end{align}
Summing over the inequivalent orderings of the open vertex operators yields
\begin{equation}\label{UnmixedPhys}
 \left<\left<V_{a}\,V_{a}\,V_{\chi^{\dag}}\,V_{F^{\bar S'}}\right>\right>=-4i\,\textrm{Tr}\left[\chi^{\dag},a_{\mu}\right]a_{\nu}\,F_{\bar S'}^{\mu\nu}\,.
\end{equation}
In \cite{Morales:1996bp,Antoniadis:2010iq}, a different string background was studied, which involved the self-dual field strength of the vector partner of the dilaton~$S$. However, repeating the computation of the above disc diagrams with the insertion of the latter leads to a vanishing result. This indicates that the background \cite{Morales:1996bp,Antoniadis:2010iq} does not give rise to the deformed ADHM action on a general $\Omega$-background.


\subsection[\texorpdfstring{$D_5-D_9$ disc diagrams}{D5-D9 disc diagrams}]{\texorpdfstring{$\boldsymbol{D_5-D_9}$ disc diagrams}{D5-D9 disc diagrams}}

\begin{figure}[h!t]
\begin{center}
\parbox{4.2cm}{\epsfig{file=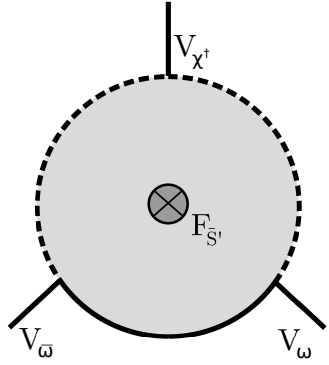,width=4.2cm}}
\end{center}
\parbox{\textwidth}{\caption{Four-point disc diagram with bulk-insertion of the $\bar{S}'$ field strength tensor and three boundary insertions stemming from the 5-9, 9-5 and 5-5 sectors of the string setup.}\label{fig-fourptBIS}}
\end{figure}

Let us consider the RR part of the physical amplitude depicted in Fig. \ref{fig-fourptBIS}
\begin{equation}
\left<V_{\omega}^{(0)}(x_1) V_{\bar{\omega}}^{(-1)}(x_2) V_{\chi^\dag}^{(0)}(x_3) V^{(-1/2,-1/2)}_{F_{\bar{S}'}}(z,\bar{z})\right> 
\end{equation}
The vertex operators for the 5-9 and 9-5 states are
\begin{eqnarray}
V_{\omega}^{(0)}(x_1)&=&\frac{g_6}{\sqrt{2}}\,\omega_{\dot\delta}\,\bigl{(}\tilde{\Delta}^\mu\,(\bar{\sigma}_\mu)^{\dot{\delta} \delta} S_\delta+i p_1\cdot\chi\,\Delta\,S^{\dot{\delta}}\big{)}  e^{i p_1\cdot Y}(x_1)\,,\nonumber\\
V_{\bar{\omega}}^{(-1)}(x_2)&=&\frac{g_6}{\sqrt{2}}\,\bar\omega_{\dot\gamma}\,ce^{-\varphi}\,\bar{\Delta}\,S^{\dot{\gamma}}\,e^{i p_2\cdot Y}(x_2)\,,
\end{eqnarray}
where $\tilde{\Delta}^\mu$ are dimension $3/4$ operators defined as
\begin{equation}
\tilde{\Delta}^\mu(x_1) =\lim_{x\rightarrow x_1} (x-x_1)^{\frac{1}{2}}\,\partial X^{\mu}(x) \Delta(x_1)
\end{equation}
Finally, we recall the vertex operator for the 5-5 state:
\begin{equation}
V_{\chi^{\dag}}(x_3)=\frac{\chi^{\dag}}{\sqrt{2}}\, (\partial Z-2 i p_3\cdot\chi \,\Psi)e^{2 i p_3\cdot Y}(x_3)\label{VchiBis}\,.
\end{equation}
Once again $\phi_3$ charge conservation implies that only $p_3\cdot \chi\,\Psi(x_3)$ term contributes from $V_\chi^{(0)}(x_3)$ and can only contract with 
$p_1\cdot\chi(x_1)$ in the vertex at $x_1$. Thus, after taking the limit of vanishing $K3$ momenta, we obtain
\begin{align}
\left<\left<V_{\omega}\,V_{\bar\omega}\,V_{\chi^\dag}\,V_{F^{\bar S'}}^{\textrm{NS}}\right>\right>&=-\frac{p_1\cdot p_3}{4\pi}\,\omega_{\dot\alpha}\,\bar\omega_{\dot\beta}\,\chi^{\dag}\,F_{\bar S'}^{\dot\beta\dot\alpha}\int dx_1 dx_3\frac{(z-\bar{z})(x_1-x_3)^{4p_1\cdot p_3-1}}{|x_1-z||x_3-z|}\Bigr|_{p_i=0}\nonumber\\
&=\frac{i}{4}\,\omega_{\dot\alpha}\,\bar\omega_{\dot\beta}\,\chi^{\dag}\,F_{\bar S'}^{\dot\beta\dot\alpha}\,,
\end{align}
where we have used the notation $F^{\dot\alpha\dot\beta}=(\bar\sigma_{\mu\nu})^{\dot\alpha\dot\beta}\,F^{\mu\nu}$.

We now turn to the NS-NS part of the bulk vertex operator which we choose to be in the $(-1,0)$-picture:
\begin{equation}
V^{\textrm{NS}}_{F^{\bar S'}}(z,\bar{z})=\frac{\epsilon_\lambda}{8\pi\sqrt{2}} e^{-\varphi}\,\bar{\Psi}\,e^{i(P\cdot X+P\cdot Y)}(z) (-\partial X^{\lambda}-i(P\cdot\psi-P\cdot\chi)\psi^{\lambda})e^{i(-P\cdot X+P\cdot Y)}(\bar z)\,.\label{VertexOpNS}
\end{equation}
Here, we are using the D5-brane map for the right-moving part of the vertex operators to the lower half plane $(\tilde{X}^\mu, \tilde{\psi}^\mu)\rightarrow 
(-X^\mu(\bar{z}),-\psi^{\mu}(\bar{z}))$. Recall that, in the present case, the boundary between $x_1$ and $x_2$ where $x_3$ sits, is on a D5-brane while the 
remaining part of the boundary is on a D9-brane. The above choice means that we are taking the branch cut (due to the twist fields at $x_1$ and $x_2$) to be on the D9 part of the boundary.

Once again $\bar\Psi(z)$ can only contract with the $p_3\cdot\chi\,\Psi(x_3)$ part of the vertex at $x_3$. However, $p_3\cdot\chi(x_3)$ can either contract with $p_1\cdot\chi(x_1)$ from vertex at $x_1$ or with $P\cdot\chi\,\psi^\lambda(\bar{z})$. In the first case, from the operator at $\bar{z}$, $\partial X^{\lambda}(\bar{z})$ cannot contribute. Indeed, since 
$\left<\Delta(x_1)\,\bar{\Delta}(x_2)\,\partial X^{\lambda}(\bar{z})\right>=0$, the only possibility is to bring down a $P\cdot X$ term from the momentum part of the vertex at $z$ or $\bar{z}$. In this case $\left<\Delta(x_1)\,\bar{\Delta}(x_2)\,\partial X^{\lambda}(\bar{z})\,P\cdot X(z,\bar z)\right> \propto P^{\lambda}$ and, hence, vanishes by the transversality condition\footnote{We only consider a single power in $P$ as required by the structure of the coupling we are evaluating.}. Consequently, only the fermionic part of the operator at $\bar{z}$ can contribute, leading to
\begin{align}
\left<\left<V_{\omega}\,V_{\bar\omega}\,V_{\chi^\dag}\,V_{F^{\bar S'}}^{\textrm{R}}\right>\right>&=\frac{p_1\cdot p_3}{4\pi}\,\omega_{\dot\alpha}\,\bar\omega_{\dot\beta}\,\chi^{\dag}\,F_{\bar S'}^{\dot\beta\dot\alpha}\int dx_1 dx_3\frac{(z-\bar{z})(x_1-x_3)^{4p_1\cdot p_3-1}}{(x_3-z)(x_1-\bar{z})}\Bigr|_{p_i=0}\nonumber\\
&=\frac{i}{4}\,\omega_{\dot\alpha}\,\bar\omega_{\dot\beta}\,\chi^{\dag}\,F_{\bar S'}^{\dot\beta\dot\alpha}\,.
\end{align}
where only the term $P\cdot\psi\,\psi^\lambda(\bar{z})$ in \eqref{VertexOpNS} contribute. Notice the additional factor of $2$ due to the left-right symmetrisation in the NS-NS vertex operator.

In the second case, the resulting term vanishes as we now show. Such a term is proportional to $\frac{p_3.P}{x_3-\bar{z}}$. The correlator of the space-time fermion $\psi^\lambda(\bar{z})$ with $S^\delta(x_1) S^{\dot{\gamma}}(x_2)$ gives a kinematic factor $(\sigma_\lambda)^{\delta \dot{\gamma}}$. Finally, as above, bringing down a single power of $P.X$ from $e^{i P\cdot X}(z)$
or $e^{-i P\cdot X}(\bar{z})$ yields the total correlation function (apart from the contraction of the momentum parts of the vertex):
\begin{equation}
p_3\cdot P\,(\bar{\sigma}_{\mu \lambda})^{\dot{\delta}\dot{\gamma}}\,\frac{z-\bar{z}}{|x_3-z|^2}\frac{(x_2-\bar{z})^{\frac{1}{2}}}{(x_1-\bar{z})^{\frac{1}{2}}}
\left[\left<\tilde{\Delta}^\mu(x_1)P\cdot X(z)\bar{\Delta}(x_2)\right>\pm\left<\tilde{\Delta}^\mu(x_1)P\cdot X(\bar{z})\bar{\Delta}(x_2)\right>\right]\,,
\end{equation}
where we have used the transversality condition, and the $\pm$ sign depends on the choice of the branch cut \footnote{The $+$ sign, which corresponds to D9 boundary conditions, 
means that the branch cut is on the D5 boundary, the $-$ sign means that the branch cut lies on the D9 one.}. In order to obtain a non-zero result in the limit of vanishing $K_3$ momenta, we need a $1/p\cdot p$ pole. Since there is no singularity in $x_3$ integral, the amplitude is non-vanishing only if $\left<\tilde{\Delta}^\mu(x_1)P\cdot X(z,\bar z) \bar{\Delta}(x_2)\right>$ has a first order pole in $(x_1-x_2)$ which can be deduced from the OPE of $\Delta^\mu(x_1)$ and $\bar{\Delta}(x_2)$. The right hand side of the OPE can only have untwisted operators, i.e. operators involving products of derivatives of $\partial X$, with $X$ being the spatial coordinates. These operators are of non-negative integer dimensions. Since the dimensions of $\tilde\Delta^\mu(x_1)$ and $\bar{\Delta}(x_2)$ are $3/4$ and $1/4$ respectively, the OPE must be
\begin{equation}
 \Delta^\mu(x_1)\bar{\Delta}(x_2)\sim \sum_{n=0}^\infty(x_1-x_2)^{n-1}\,c_{i}^{n}\,{\cal O}^i_n(x_2)
\label{defDeltamu}
\end{equation}
where $n$ denotes the dimension, $i$ counts the degeneracy of the operators $\cal O$ and $ c_{i}^{n}$ are the structure constants. Notice that $\left<\Delta^\mu(x_1)\bar{\Delta}(x_2)\right>=0$ implies\footnote{For $n=0$ there is a unique ground state in the
untwisted sector.} that the term $n=0$ is absent in \eqref{defDeltamu}. In addition, the $n=1$ term is non-vanishing and corresponds to ${\cal O}_1\equiv\partial X^\mu$. This can be more explicitly seen by considering the four-point correlation function 
\begin{eqnarray}
F(x_1,y,w,x_2)&=&\left<\Delta(x_1) \partial X^{\mu}(y) \partial X^\nu(w) \bar{\Delta}(x_2)\right>\nonumber\\
&=&-\frac{\delta^{\mu \nu}}{2}\frac{1}{(x_1-x_2)^{\frac{1}{2}}(y-w)^2}.\frac{(y-x_1)(w-x_2)+(y-x_2)(w-x_1)}{\bigl{(}(y-x_1)(y-x_2)(w-x_1)(w-x_2)\bigr{)}^{\frac{1}{2}}}\,,\nonumber\\
\end{eqnarray}
which can be easily deduced from the OPEs
\begin{equation}
\Delta(x_1) \bar{\Delta}(x_2)\sim\frac{1}{\sqrt{x_1-x_2}}\,,
\end{equation}
and
\begin{equation}
\partial X^{\mu}(y)\partial X^\nu(w) \sim-\frac{\delta^{\mu \nu}}{(y-w)^2}\,.
\end{equation}
Using the definition (\ref{defDeltamu}), we find that
\begin{equation}
\left<\tilde{\Delta}^\mu(x_1) \partial X(w) \bar{\Delta}(x_2)\right> =\frac{1}{(w-x_2)^{\frac{1}{2}}(w-x_1)^{\frac{3}{2}}}\,.
\end{equation}
Consequently, in the limit $x_1 \rightarrow x_2$, $\tilde{\Delta}^\mu(x_1) \bar{\Delta}(x_2) \rightarrow \partial X^{\mu}(x_2)$, and
\begin{equation}
\left<\tilde{\Delta}^\mu(x_1) P\cdot X(w) \bar{\Delta}(x_2)\right>\underset{x_1\rightarrow x_2}{\longrightarrow} -P^\mu \frac{1}{x_2-w}\,.
\end{equation}
Hence, the absence of poles in $(x_1-x_2)$ implies that, in the limit of vanishing $K_3$ momenta, this contribution is zero.

Summarising our result, the D5-D9 diagram in the presence of the $\bar S'$-vector leads to the following coupling in the instanton effective action:
\begin{align}\label{MixedPhys}
\left<\left<V_{\omega}\,V_{\bar\omega}\,V_{\chi^\dag}\,V_{F^{\bar S'}}\right>\right>=\frac{i}{2}\,\textrm{Tr}\Bigr\{\bar\omega_{\dot\alpha}\,\chi^{\dag}\,\omega_{\dot\beta}(\bar\sigma^{\mu\nu})^{\dot\alpha\dot\beta}F_{\mu\nu}^{\bar S'}\Bigr\}
\end{align}

These diagrams are sufficient to compute the tree-level string effective action involving D5-instanton. In the following section, we consider the field theory limit and compare the result to the ADHM action on a general $\Omega$-background.


\subsection{ADHM Action and Nekrasov Partition Function}\label{Sec:ADHM}

In this section we compare our results with the deformed ADHM action appearing in \eqref{NekPartK}. The latter describes instantons in a gauge theory with $SU(N)$ gauge group on a general $\Omega$-background.  Recall that the ADHM instanton action can be expressed as
\begin{align}
 S_{\textrm{ADHM}}=-\textrm{Tr}&\left\{[\chi^{\dag},a_{\alpha\dot\beta}]\left([\chi,a^{\dot\beta\alpha}]+\epsilon_{-}(a\tau_3)^{\dot\beta\alpha}\right)-\chi^{\dag}\,\bar\omega_{\dot\alpha}\left(\omega^{\dot\alpha}\chi-\tilde a\,\omega^{\dot\alpha}\right)-\left(\chi\bar\omega_{\dot\alpha}-\bar\omega_{\dot\alpha}\,\tilde a\right)\omega^{\dot\alpha}\,\chi^{\dag}\right.\nonumber\\
   &\left.+\epsilon_+\left[\chi^{\dag},a_{\alpha\dot{\beta}}\right](\tau_3 a)^{\dot\beta\alpha}-\epsilon_+\,\bar{\omega}_{\dot{\alpha}}\,{(\tau_3)^{\dot{\alpha}}}_{\dot{\beta}}\,\chi^{\dag}\,\omega^{\dot{\beta}}\right\}\,.
\end{align}
where we only display the part relevant for our discussion. Here, we have introduced the vev $\tilde a$ of the $\N=2$ vector multiplet that higgses the $SU(N)$ gauge group. The terms in the second line correspond to the $\epsilon_{+}$-dependent deformation of the ADHM action and which we want to compare to the effective couplings of $F^{\bar S'}$ to the ADHM moduli. To this end, we parametrise the vev of the $\bar S'$ field strength using the self-dual 't Hooft symbols (see (\ref{tHooft})):
\begin{equation}
 F_{\mu\nu}^{\bar S'}\equiv\bar\eta_{\mu\nu}^{c}\,F_c^{\bar S'}\equiv\bar\eta_{\mu\nu}^{c}\,\delta_{3c}\,\frac{\epsilon_+}{2}\,.
\end{equation}
The contribution of the diagram \eqref{UnmixedPhys} becomes
\begin{align}
 \left<\left<V_{a}\,V_{a}\,V_{\chi^{\dag}}\,V_{F^{\bar S'}}\right>\right>= -4i \,\textrm{Tr}\Bigr\{\left[\chi^{\dag},a_\mu\right]a_{\nu}\,F^{\mu\nu}_{\bar S'}\Bigr\}&=-\epsilon_+ \,\textrm{Tr}\Bigr\{\left[\chi^{\dag},a_{\alpha\dot\beta}\right]a^{\dot\gamma\alpha}\,{(\tau^{3})^{\dot\beta}}_{\dot\gamma}\Bigr\}\,.
\end{align}
Similarly, the contribution \eqref{MixedPhys} of the mixed diagrams can be recast as 
\begin{align}
 \left<\left<V_{\omega}\,V_{\bar\omega}\,V_{\chi^\dag}\,V_{F^{\bar S'}}\right>\right>=\frac{i}{2}\,\textrm{Tr}\Bigr\{\bar\omega_{\dot\alpha}\,\chi^{\dag}\,\omega_{\dot\beta}(\bar\sigma^{\mu\nu})^{\dot\alpha\dot\beta}F_{\mu\nu}^{\bar S'}\Bigr\}=\epsilon_+\,\textrm{Tr}\Bigr\{\bar\omega_{\dot\alpha}\,\chi^{\dag}\,{(\tau_{3})^{\dot\alpha}}_{\dot\beta}\,\omega^{\dot\beta}\Bigr\}\,.
\end{align}
Therefore, the D5-instanton world-volume theory in our background gives a six-dimensional field theory on $K3\times T^2$ that contains the deformed ADHM couplings. In the limit where the world-volume of the D5-instantons becomes 
small (\emph{i.e.} of order $\alpha'$ or smaller), we can reduce the six-dimensional field theory to a zero-dimensional one which exactly reproduces the deformed ADHM action for the four-dimensional gauge theory.

The ADHM action is a key ingredient to compute the non-perturbative part \eqref{NekPartK} of the Nekrasov partition function $Z^{\text{Nek}}$ of the supersymmetric gauge theory. Recall that it can be factorised in the following form:
\begin{align}
Z^{\text{Nek}}(\epsilon_+,\epsilon_-)=Z^{\text{Nek}}_{\text{pert}}(\epsilon_+,\epsilon_-)\, Z^{\text{Nek}}_{\text{n.p.}}(\epsilon_+,\epsilon_-)\,.\label{MarEq}
\end{align}
While the perturbative piece $Z^{\text{Nek}}_{\text{pert}}$ does not receive contributions beyond the one-loop order, the non-perturbative part $Z^{\text{Nek}}_{\text{n.p.}}$ is defined as a path integral over the instanton moduli space, with the integral measure given by the deformed ADHM action $S_{\text{ADHM}}$~\cite{Nekrasov:2002qd,Nekrasov:2003rj}.

\section{Channel Factorisation and Auxiliary Fields}

\begin{figure}[H]
\begin{center}
\parbox{13cm}{\epsfig{file=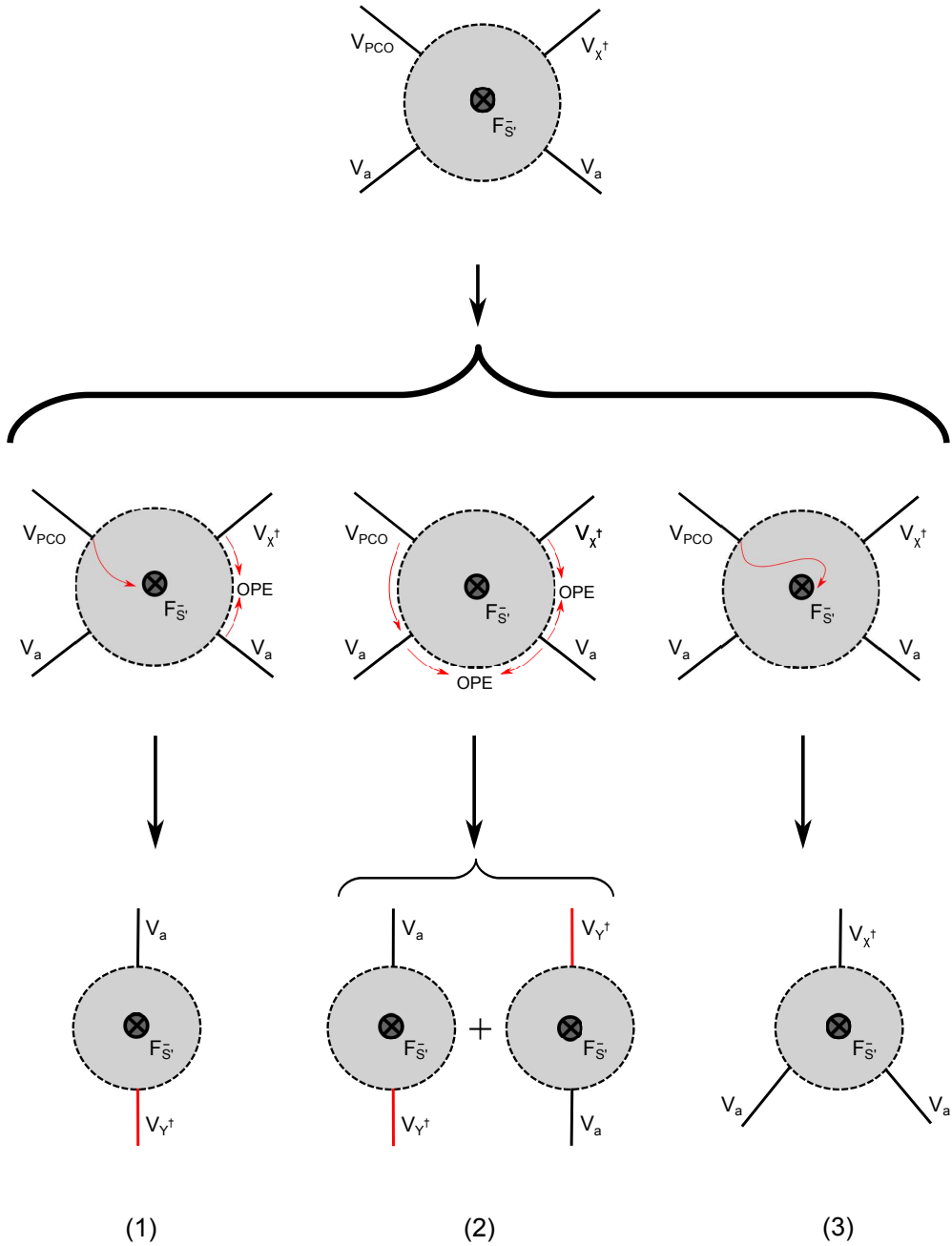,width=13cm}}
\end{center}
\parbox{\textwidth}{\caption{Factorisation channels of the disc. The diagram at the top illustrates the four-point function of physical vertices with an insertion of a PCO. The diagrams at the bottom depict various choices for the position of the latter. In case (1), the result takes the form of a contact term, leading to an effective auxiliary field vertex operator insertion on the boundary of the disc. Similarly, case (2) can be re-expressed as a sum of two contact terms, while in case (3) no such interpretation is possible.}\label{fig-auxiliary}}
\end{figure}

In this section, we show that the same results derived above can be recovered by using auxiliary fields which linearise the superalgebra. These are given by
\begin{align}
 V_{Y}(z)&=\sqrt{2}\,g_6\,Y_{\mu}\bar\Psi(z)\psi^{\mu}(z)\,,&&\,V_{Y^{\dag}}(z)=\sqrt{2}\,g_6\,Y_{\mu}^{\dag}\Psi(z)\psi^{\mu}(z)\,,\label{Auxiliary1}\\
 V_{X}(z)&=g_6\, X_{\dot\alpha}\Delta(z)S^{\dot\alpha}\bar\Psi(z)\,,&&\,V_{X^{\dag}}(z)=g_6\, X_{\dot\alpha}^{\dag}\Delta(z)S^{\dot\alpha}\Psi(z)\,,\\
 V_{\bar X}(z)&=g_6\, \bar X_{\dot\alpha}\bar\Delta(z)S^{\dot\alpha}\bar\Psi(z)\,,&&\,V_{\bar X^{\dag}}(z)=g_6\, \bar X_{\dot\alpha}^{\dag}\bar\Delta(z)S^{\dot\alpha}\Psi(z)\,.\label{Auxiliary3}
\end{align}
Of course, auxiliary fields are not BRST-closed and inserting them in a correlation function does not lead, in general, to a sensible result. For instance, as discussed in \cite{Antoniadis:2013mna}, changing the ghost pictures of the vertex operators in the presence of auxiliary fields leads to different results, hence rendering the correlation function ambiguous. However, here, we prove that in a particular factorisation limit, the result can be understood as a correlator involving an auxiliary field. This gives a rigorous justification to the use of auxiliary fields in \cite{Ito:2010vx,Ito:2011cr,Green:2000ke,Billo:2002hm,Billo:2006jm}. Similar arguments, though in a different context, have been made in \cite{Atick:1987qy}.
Consider the NS-NS contribution \eqref{4ptFctNS} and take the limit where $y$ goes to the points $\bar{z}$, $z$ and $x_2$, respectively, corresponding to using different pictures for the associated operators.  The three cases below are illustrated in Fig. \ref{fig-auxiliary}.

\begin{enumerate}
\item $y\rightarrow \bar{z}$

In this case $A_0\,A_1=0$, whereas $A_0\,A_3$ still cannot produce any pole in the momenta. On the other hand, $A_0\,A_2$ simplifies to
\begin{equation}
\frac{4i p_1\cdot p_3}{x_3-x_1}\frac{z-\bar{z}}{(x_3-z)(x_1-\bar{z})}(\delta^{\nu \lambda}P^{\mu}-\delta^{\mu \lambda}P^{\nu}) \,.
\end{equation}
Notice that the longitudinal mode manifestly decouples. Looking at the vertices (\ref{Va}) and (\ref{Vchi}), we recognise that
$ p_3\cdot p_1/(x_3-x_1)$ appears from contracting $p_1\cdot\chi(x_1)$ with $p_3\cdot\chi(x_3)$. The pole $1/p_1\cdot p_3$ appears from $x_3\rightarrow x_1$. Therefore, in this limit, the result is effectively reproduced by the OPE of the vertices at $x_1$ and $x_3$, resulting in an effective vertex $\Psi\,\psi^\mu$ at $x_1$. This is why in the $(-1,0)$-picture for the NS-NS part of $V_{F^{\bar S'}}$, the auxiliary vertex \eqref{Auxiliary1} gives the correct effective coupling calculated as the three-point function $\left<\left<V_{Y^\dag}\,V_a\,V_{F^{\bar S'}}\right>\right>$.

\item $y\rightarrow x_2$

In this case again $A_0\,A_1$ vanishes. From $A_0\,A_2$, only the kinematic structure $\delta^{\mu \lambda}P^{\nu}$ survives, with the same answer as above. However the $p_3\cdot p_2/(x_3-x_2)$ factor in $A_0 A_3$ contributes to the other kinematic structure, $\delta^{\nu \lambda}P^{\mu}$. The final result is of course the same but the total result comes from two different factorisation limits $x_3 \rightarrow x_1$ and $x_3 \rightarrow x_2$ for the two different kinematic structures.

\item $y\rightarrow z$

In this case $A_0\,A_3$ vanishes, but both the remaining terms contribute. In particular it is not clear if the $A_0\,A_1$ term can even be thought of as a contact term.
\end{enumerate}
Hence, only for $y\rightarrow\bar{z}$ can the result be understood as the factorisation in a single channel, such that it can be effectively reproduced by replacing $V_a(x_1)$ and $V_{\chi}(x_3)$ by their OPE, which is simply the auxiliary vertex \eqref{Auxiliary1}.


Finally, in the R-R contributions \eqref{4ptFctR}, the `contact term' appears only from the channel $x_3 \rightarrow x_1$ and the result can be obtained by a three-point function involving the vertices at $x_2, z, \bar{z}$ and an auxiliary vertex $\Psi\,\psi^\mu$ at $x_1$.  The analysis of the mixed D5-D9 diagrams is very similar and leads to the same conclusion, that is, in the $(-1,0)$-picture for the NS-NS part and $(-1/2,-1/2)$-picture for the RR part of $V_{F^{\bar S'}}$, the entire result comes from contact terms in a single channel and the result can be reproduced by a three-point function involving the auxiliary vertex \eqref{Auxiliary3}.


As mentioned above, even though the calculations are performed on the target space $\mathbb R^4 \times T^2\times\mathbb{R}^4$, the couplings we have obtained are non-vanishing for non-trivial momenta only along the first $\mathbb{R}^4$ (space-time). When we compactify $\mathbb R^4 \times T^2\times\mathbb{R}^4\to \mathbb R^4\times T^2\times K3$, these couplings are unchanged up to possible $\alpha'$ corrections. However, the latter are irrelevant in the field theory limit that we take in order to compare with the non-perturbative part of the $\Omega$-deformed gauge theory partition function. In addition, we have focused on the gauge theory coming from D9-branes for which the relevant D-instanton is the D5-brane wrapping the internal space. However, it is straightforward to extend our calculation to other setups by applying T-duality. For example, for a gauge theory realised by D5-branes wrapping $T^2$, the relevant D-instanton is the D1-brane wrapped on $T^2$. The corresponding couplings can be obtained from 
the above calculations by performing four T-dualities along the $Y^i$ directions.

\section{Interpretation and the Refined Topological String}\label{Sec:RefTop}

In Section \ref{Sec:Hetamps}, we have put forward a promising proposal in terms of a particular class of higher derivative terms in the string effective action.  At the component level, it involves terms of the form
\begin{align}
\mathcal{I}_{g,n} = \int d^4x\,&\mathcal{F}_{g,n}\, R_{(-)\, \mu\nu\rho\tau} 
R_{(-)}^{\mu\nu\rho\tau}\, \left[F^G_{(-)\,\lambda\sigma} F^{G\,\lambda\sigma}_{(-)}\right]^{g-1}\,\left[F_{(+)\,\rho\sigma}F^{\rho\sigma}_{(+)}\right]^n
\label{EffectiveCouplingBis}\,,
\end{align}
for $g\geq1$ and $n\geq 0$. Here, $R_{(-)}$ denotes the anti-self-dual Riemann tensor, $F^G_{(-)}$ the anti-self-dual field strength tensor of the graviphoton and $F_{(+)}$ the self-dual field strength tensor of an additional vector multiplet gauge field. In Heterotic $\mathcal{N}=2$ compactifications on $K3\times T^2$, the latter is identified with the super-partner of the K\"ahler modulus of $T^2$, while in the dual Type I setting, it is mapped to the vector partner of the $\bar{S}'$ modulus. 

For $n=0$, the $\mathcal{I}_{g,n}$ in (\ref{EffectiveCouplingBis}) are BPS-saturated and were first discussed in \cite{Antoniadis:1993ze}. The $F_g=\mathcal{F}_{g,n=0}$ are exact at the $g$-loop level in Type II string theory compactified on an elliptically fibered Calabi-Yau threefold and compute the corresponding genus $g$ partition function of topological string theory, see Section \ref{ch:TopoAmp}. In the dual heterotic theory,  $F_g$ starts receiving contributions at the one-loop level~\cite{Antoniadis:1995zn} and, in the point particle limit, is related to the perturbative part of Nekrasov's partition function for a gauge theory on an $\Omega$-background with only one non-trivial deformation parameter. The latter is then identified with the topological string coupling.

For $n>0$ the leading contribution to $\mathcal{F}_{g,n}$ is still given by a one-loop amplitude in the heterotic theory and was computed to all orders in $\alpha'$ in Section \ref{Sec:Hetamps}. The coupling functions $\mathcal{F}_{g,n}$ in (\ref{EffectiveCouplingBis}) can be compactly expressed in the form of a generating functional
\begin{align}\label{FullAmplitudeHetBis}
 \mathcal{F}\textrm(\epsilon_{-},\epsilon_{+})&=\sum_{g,n\geq0}\epsilon_-^{2g}\epsilon_+^{2n}\,\mathcal{F}_{g,n}\,.
\end{align}
In the point particle limit, the one-loop contribution to $\mathcal{F}(\epsilon_-,\epsilon_+)$ captures the perturbative part of the Nekrasov partition function (\ref{MarEq}) on a generic $\Omega$-background, whose deformation parameters are identified with the expansion parameters $\epsilon_\pm$ in (\ref{FullAmplitudeHetBis}). Thus, the $\mathcal{F}_{g,n}$ are  one-parameter extensions of the topological amplitudes $F_g$ which are (perturbatively) compatible with a refinement in the gauge theory limit.

The instanton computations performed in the previous sections are indeed evidence for this proposal. As we showed, also the non-perturbative contributions to the $\mathcal{F}_{g,n}$ in the point particle limit are compatible with the structure expected from gauge theory and capture precisely the non-perturbative part of the full Nekrasov partition function $Z^{\textrm{Nek}}_{\textrm{n.p.}}(\epsilon_{+},\epsilon_{-})$. This proves that the couplings~(\ref{EffectiveCouplingBis}) indeed provide a string theoretic realisation of the full $\Omega$-background in gauge theory. This is precisely what one would expect from a worldsheet realisation of the refined topological string.


\part{Towards a Worldsheet Definition of the Refined Topological String}
\label{ch:conc}
\chapter{Concluding Remarks}

In this work, we analysed some of the intricate connections between topological string theory, the string effective action and supersymmetric gauge theories in the $\Omega$-background.

In the first part of the manuscript, we briefly reviewed the basics of string theory and the underlying conformal field theory. In particular, we discussed some aspects of the compactification of superstring theory and its dualities.

In a second part, we recalled the construction of topological string theory through the topological twist on the superconformal algebra. This leads to a theory possessing a topological symmetry and which can be viewed as a sub-sector of string theory. For instance, the genus $g$ partition function of the topological string is calculated by a coupling $F_g$ in the string effective action involving, at genus $g$ in Type II, two anti-self-dual gravitons and $2g-2$ anti-self-dual graviphotons. The topological nature of these quantities manifests itself through the fact that they receive contributions from BPS states only. Moreover, when the Type II compactification admits a Heterotic dual, this coupling starts receiving corrections at genus one in Heterotic. Hence, this renders possible an explicit calculation of the coupling at the perturbative level. On the other hand, the field limit of the latter is given by the partition function of the $\N=2$ gauge theory in the $\Omega$-background, in the limit where one of the parameters of the $\Omega$-background, $\epsilon_+$, is set to zero. The other parameter, $\epsilon_-$, is identified with the constant background of the anti-self-dual graviphoton. Thus, the natural question is to identify, in string theory, the signification of the parameter $\epsilon_+$. Alternatively, the gauge theory in the $\Omega$-background suggests the existence of the refined topological string whose partition function reduces to the gauge theory one in the field theory limit. A \WS definition of the refined topological string using string amplitudes was a central question in this work.

The third part of the manuscript was dedicated to answering the first question. Namely, we showed that the parameter $\epsilon_+$ can be identified with the self-dual background of a particular vector multiplet. In Heterotic string theory compactified on $K3\times T^2$, it is given by the self-dual vector partner of the $\bar T$-modulus. We calculated the coupling $F_{g,n}$ of two anti-self-dual gravitons, $2g-2$ anti-self-dual graviphotons and $2n$ self-dual $\bar T$ field strengths exactly at one loop. In the field theory limit around a Wilson line enhancement point, it precisely reduces to the perturbative part of the Nekrasov partition function. On the other hand, the same conclusion was reached in the dual Type I theory in which the additional insertion corresponds to the self-dual field strength of the $\bar S'$-vector. Furthermore, by realising gauge instantons in Type I using D-brane bound states, we computed non-perturbative corrections to the couplings $F_{g,n}$. More precisely, we derived the instanton effective action in the presence of the anti-self-dual graviphotons and self-dual $\bar S'$-vectors by calculating all disc diagrams with boundary insertions of the instanton moduli. The resulting action matched the instanton measure of the path integral leading to the Nekrasov partition function. Therefore, we proved that the couplings $F_{g,n}$ agree with the latter perturbatively and non-perturbatively, and the proposed background is a good string theory uplift of the $\Omega$-background. Even though the topological nature of our background is not yet fully understood, our findings provide promising perspectives towards a \WS definition of the refined topological string.

From the \ST point of view, our A-model understanding of the refined topological string relies on a definition of its partition function in terms of an index counting BPS states. In particular, by going to M-theory, the latter are M2-branes carrying left and right spins with respect to the $SO(4)$ Lorentz group \cite{Hollowood:2003cv,Dijkgraaf:2006um}. The refined partition function can be calculated using the refined topological vertex which extends the topological vertex techniques. However, similarly to the instanton calculus in gauge theory, this formalism is applicable only in the asymptotic region of the moduli space. In this context, a \WS, B-model realisation of the refined topological string is crucial in order for it to be defined at any point in the moduli space.

On the other hand, since the refined partition function is an index, one might expect it to be calculated by some string amplitude. Our findings suggest that this is indeed possible since the amplitude studied in this manuscript reproduces the only independent explicit result, namely the partition function of the $\N=2$ gauge theory in the $\Omega$-background. Nevertheless, its BPS, topological properties remain to be unveiled. In order to achieve this, one can study the dual Type II theory compactified on an elliptically fibered Calabi-Yau, in which the refined partition function is expected to be captured by a higher genus amplitude. By identifiying the additional insertion in the dual setup, one can write down the Type II amplitude and, using the methods described in Section \ref{ch:TopoAmp}, re-express the amplitude in terms of a correlation function in the twisted, topological theory. This would provide a \WS definition of the refined topological string partition function.

It is worth mentioning that, in this work, the string theory setup was based on a compact Calabi-Yau threefold, even though it is widely believed that the refined topological string makes sense only on a non-compact Calabi-Yau manifold. More specifically, the latter is necessary in order to turn on an $SU(2)$ R-symmetry current and define a refined BPS index. However, our setup should be viewed as a convenient framework in which we successfully interpreted the $\Omega$-background in terms of physical fields. In addition, we did not introduce any R-symmetry current but rather generated it effectively in the path integral. Hence, one might take an appropriate non-compact limit of the internal space and analyse the possible topological properties that our refined couplings might acquire.

Besides, it would be interesting to analyse the holomorphicity properties of the refined couplings studied above. Indeed, the refined topological string partition function is expected to satisfy a recursion relation \cite{Krefl:2010fm,Huang:2010kf} generalising the holomorphic anomaly equation presented in Section \ref{sec:holan}. A possible agreement would then furnish additional evidence in favour of the generalised holomorphic anomaly equations and also of our proposal for the refinement. Finally, an open issue is the connection, to our approach, of the flux-trap realisation of the $\Omega$-background \cite{Hellerman:2011mv,Hellerman:2012zf} which naturally lifts the gauge theory construction to string/M-theory.

Exploring these ideas would increase our understanding of the interesting connections between topological amplitudes and supersymmetric gauge theories and, I believe, shed light on the deep structure of the refined topological string

\begin{appendices}

\chapter{Spinors, Gamma Matrices}\label{appendix:spinors}
Let us begin by discussing our conventions for various index structures. Raising and lowering of $SO(4)_{ST}$ spinor indices is achieved with the help of the epsilon-tensors $\epsilon^{12}=\epsilon_{12} = +1$, and $\epsilon^{\dot{1}\dot{2}}=\epsilon_{\dot{1}\dot{2}} = -1$, i.e.
\begin{align}
&\psi^\alpha = +\epsilon^{\alpha\beta}\psi_\beta \,,&&\psi_\alpha=-\epsilon_{\alpha\beta}\psi^\beta\,,&&\psi^{\dot\alpha} = -\epsilon^{\dot\alpha\dot\beta}\psi_{\dot\beta} \,,&&\psi_{\dot\alpha}=+\epsilon_{\dot\alpha\dot\beta}\psi^{\dot\beta}\,.
\end{align}
Furthermore we introduce the $\sigma$-matrices $(\sigma^\mu)_{\alpha\dot\alpha}$ and $(\bar\sigma^\mu)^{\dot\alpha\alpha}$ of $SO(4)_{ST}$
\begin{align}
	(\sigma^\mu)_{\alpha\dot\alpha} = (1\!\!1,-i\boldsymbol{\sigma})\,, && (\bar\sigma^\mu)^{\dot\alpha\alpha} = (1\!\!1,+i\boldsymbol{\sigma})\,,
\end{align}
which are related to one-another by raising and lowering of the spinor indices
\begin{align}
	(\sigma^\mu)^{\beta\dot\beta}\equiv\epsilon^{\beta\alpha}(\sigma^\mu)_{\alpha\dot\alpha}\epsilon^{\dot\alpha\dot\beta} = (\bar\sigma^\mu)^{\dot\beta\beta}\,, &&
	(\bar\sigma^\mu)_{\dot\beta\beta}\equiv \epsilon_{\dot\beta\dot\alpha}(\bar\sigma^\mu)^{\dot\alpha\alpha}\epsilon_{\alpha\beta}= (\sigma^\mu)_{\beta\dot\beta}\,.
\end{align}
In addition, we introduce the Lorentz generators $\sigma^{\mu\nu}, \bar\sigma^{\mu\nu}$ of $SO(4)_{ST}$
\begin{align}
{(\sigma_{\mu\nu})_\alpha}^\beta \equiv \frac{1}{2}{\bigr( \sigma_\mu\bar\sigma_\nu-\sigma_\nu\bar\sigma_\mu\bigr)_{\alpha}}^{\beta}\,,&&{(\bar\sigma_{\mu\nu})^{\dot\alpha}}_{\dot\beta} \equiv \frac{1}{2}{\bigr( \bar\sigma_\mu\sigma_\nu-\bar\sigma_\nu\sigma_\mu\bigr)^{\dot\alpha}}_{\dot\beta}\,,
\end{align}
which are symmetric in the spinor indices $(\sigma_{\mu\nu})_{\alpha\beta}=+(\sigma_{\mu\nu})_{\beta\alpha}$ and $(\bar\sigma_{\mu\nu})_{\dot\alpha\dot\beta}=+(\sigma_{\mu\nu})_{\dot\beta\dot\alpha}$. They are (anti-)self-dual in the sense
\begin{align}
(\sigma^{\mu\nu})_{\alpha\beta} = +\frac{1}{2}\epsilon^{\mu\nu\rho\sigma} (\sigma_{\rho\sigma})_{\alpha\beta}\,, &&(\bar\sigma^{\mu\nu})_{\dot\alpha\dot\beta} = -\frac{1}{2}\epsilon^{\mu\nu\rho\sigma} (\bar\sigma_{\rho\sigma})_{\dot\alpha\dot\beta}\,.
\end{align}
Therefore, we can use them to define \emph{(anti-)self-dual} tensors. In particular we write for the self-dual $F_{\mu\nu}^{(+)}$ and anti-self-dual $F_{\mu\nu}^{(-)}$ part of the field strength tensor of a given gauge field
\begin{align}
F^{(+)}_{\dot\alpha\dot\beta}\equiv (\bar\sigma^{\mu\nu})_{\dot\alpha\dot\beta}F^{(+)}_{\mu\nu}\,,&&F^{(-)}_{\alpha\beta}\equiv (\sigma^{\mu\nu})_{\alpha\beta} F^{(-)}_{\mu\nu}\,.\label{SelfDualDef}
\end{align}
since, indeed, $F^{(\pm)}_{\mu\nu} = \mp \frac{1}{2} \epsilon_{\mu\nu\rho\sigma} (F^{(\pm)})^{\rho\sigma}$. 
Also, note the following identities:
\begin{align}
	&(\sigma^{\mu\nu})_{\alpha\beta}(\sigma^{\rho\sigma})^{\alpha\beta} = 2(\delta^{\mu\rho}\delta^{\nu\sigma}-\delta^{\mu\sigma}\delta^{\nu\rho}+\epsilon^{\mu\nu\rho\sigma})\,,\\
	&(\bar\sigma^{\mu\nu})_{\dot\alpha\dot\beta}(\bar\sigma^{\rho\sigma})^{\dot\alpha\dot\beta} = 2(\delta^{\mu\rho}\delta^{\nu\sigma}-\delta^{\mu\sigma}\delta^{\nu\rho}-\epsilon^{\mu\nu\rho\sigma})\,.
\end{align}
Using the above relations, one may invert (\ref{SelfDualDef}) to obtain
\begin{align}
	& F_{\mu\nu}^{(+)} = \frac{1}{8}(\bar\sigma_{\mu\nu})^{\dot\alpha\dot\beta}F^{(+)}_{\dot\alpha\dot\beta}\,,&&
	& F_{\mu\nu}^{(-)} = \frac{1}{8}(\sigma_{\mu\nu})^{\alpha\beta}F^{(-)}_{\alpha\beta}\,.
\end{align}
Finally, we define the (anti-)self-dual 't Hooft symbols by decomposing the sigma-matrices:
\begin{align}
 {(\bar\sigma_{\mu\nu})^{\dot\alpha}}_{\dot\beta}\equiv i\,\bar\eta_{\mu\nu}^{c}\,{(\tau_c)^{\dot\alpha}}_{\dot\beta}\,,&&{(\sigma_{\mu\nu})_{\alpha}}^{\beta}\equiv i\,\eta_{\mu\nu}^{c}\,{(\tau_c)_{\alpha}}^{\beta}\,.\label{tHooft}
\end{align}
The notation we use closely follows \cite{Billo:2006jm}. Self-dual spin-fields of $SO(4)_{ST}$ are denoted $S_{\dot\alpha}$ and the anti-self-dual ones $S_{\alpha}$. In this notation, the graviphoton field $G$ is anti-self-dual and the $\bar S'$ field strength is self-dual.

Concerning the convention for the internal manifold, we denote spin fields by $S_A$, $S^A$, $S_{\hat A}$, $S^{\hat A}$. As already mentioned, while indices $(A,\hat{A})$ are indices of $SO(2)_{\pm}$ respectively, covariant and contravariant indices also reflect charges $\pm 1/2$ with respect to $SO(2)_{T^2}$ according to the decomposition (\ref{DecomposeSo10}). Thus, indices $(A,\hat{A})$ cannot be raised or lowered, but care has to be taken regarding their position. Our conventions for internal spin fields are summarised below.
\begin{align}\renewcommand{\arraystretch}{1.4}
	\begin{array}{c |c| c}
					\textrm{Spin field}  & SO(2) & SO(4) \\ \hline \hline
					S_A & - & (--),(++)\\ \hline
					S^A & + & (++),(--) \\ \hline
					S_{\hat{A}} & + & (-+),(+-) \\ \hline 
					S^{\hat{A}} & - & (+-),(-+) \\ \hline
	\end{array}\label{TabConventions}
\end{align}

\chapter{OPEs and CFT Correlators}\label{appendix:OPEs}
{\allowdisplaybreaks
We start by recalling the OPEs of fermionic fields using the notation explained in the previous section. In the general case of an $O(2N)$ current algebra, the OPEs of the fermions $\psi^M$ and the spin fields $S^A$ are
\begin{alignat}{3}
&\psi^M(z)S^A(w)\,&\sim&\,\frac{1}{\sqrt{2}}\frac{{(\Gamma^{M})^A}_B\,S^B(w)}{(z-w)^{1/2}}\,,\\
&J^{MN}(z)S^A(w)\,&\sim&\,-\frac{i}{2}\frac{{(\Gamma^{MN})^A}_B\,S^B(w)}{z-w}\,,\\
&S^A(z)\,S^B(w)\,&\sim&\,\frac{C^{AB}}{(z-w)^{N/4}}+\frac{1}{\sqrt{2}}\frac{(\Gamma^{M})^{AB}\psi_{M}(w)}{(z-w)^{N/4-1/2}}+\frac{i}{2}\frac{(\Gamma^{MN})^{AB}\psi_M\psi_N(w)}{(z-w)^{N/4-1}}\,,
\end{alignat}
where $C^{AB}$ is the charge conjugation matrix and $J^{MN}\equiv\psi^{[M}\psi^{N]}$. Applying these formulas to the ten-dimensional case of interest and decomposing the spinor indices as in the previous sections, we obtain the OPEs for the fermionic fields in the \ST directions
\begin{alignat}{5}
&S^{\dot\alpha}(z)S_{\beta}(w)\,&\sim&\,\frac{1}{\sqrt{2}}\,{(\bar\sigma^\mu)^{\dot\alpha}}_{\beta}\psi_\mu(w)\,,\,&S_{\alpha}(z)S^{\dot\beta}(w)\,&\sim\,\frac{1}{\sqrt{2}}\,{(\sigma^\mu)_{\alpha}}^{\dot\beta}\psi_\mu(w)\,,\\
&S^{\dot\alpha}(z)S^{\dot\beta}(w)\,&\sim&\,-\frac{\epsilon^{\dot\alpha\dot\beta}}{(z-w)^{1/2}}\,,\,&S_{\alpha}(z)S_{\beta}(w)\,&\sim\,\frac{\epsilon_{\alpha\beta}}{(z-w)^{1/2}}\,,\\
&\psi^{\mu}(z)S^{\dot\alpha}(w)\,&\sim&\,\frac{1}{\sqrt{2}}\frac{(\bar\sigma^\mu)^{\dot\alpha\beta}S_{\beta}(w)}{(z-w)^{1/2}}\,,\,&\psi^{\mu}(z)S_{\alpha}(w)\,&\sim\,\frac{1}{\sqrt{2}}\frac{(\sigma^\mu)_{\alpha\dot\beta}S^{\dot\beta}(w)}{(z-w)^{1/2}}\,,\\
&J^{\mu\nu}(z)S^{\dot\alpha}(w)\,&\sim&\,-\frac{1}{2}\frac{{(\bar\sigma^{\mu\nu})^{\dot\alpha}}_{\dot\beta}S^{\dot\beta}(w)}{z-w}\,,\,\,\,&J^{\mu\nu}(z)S_{\alpha}(w)\,&\sim\,-\frac{1}{2}\frac{{(\sigma^{\mu\nu})_{\alpha}}^{\beta}S_{\beta}(w)}{z-w}\,,
\end{alignat}
and in the internal ones
\begin{alignat}{5}
&S^{A}(z)S_{B}(w)\,&\sim&\,\frac{i{\delta^{A}}_{B}}{(z-w)^{3/4}}\,,\,&S_{A}(z)S^{B}(w)\,&\sim\,\frac{i{\delta_{A}}^{B}}{(z-w)^{3/4}}\\
&S^{A}(z)S^{B}(w)\,&\sim&\,\frac{i}{\sqrt{2}}\frac{(\Sigma^m)^{AB}\psi_{m}}{(z-w)^{1/4}}\,,\,&S_{A}(z)S_{B}(w)\,&\sim\,-\frac{i}{\sqrt{2}}\frac{(\Sigma^m)_{AB}\psi_{m}}{(z-w)^{1/4}}\,,\\
&\psi^{m}(z)S_{A}(w)\,&\sim&\,\frac{1}{\sqrt{2}}\frac{(\bar\Sigma^m)_{AB}S^{B}(w)}{(z-w)^{1/2}}\,,\,&\psi^{m}(z)S^{A}(w)\,&\sim\,-\frac{1}{\sqrt{2}}\frac{(\Sigma^m)^{AB}S_{B}(w)}{(z-w)^{1/2}}\,,\\
&J^{mn}(z)S^{A}(w)\,&\sim&\,\frac{1}{2}\frac{{(\bar\Sigma^{mn})^{A}}_{B}S^{B}(w)}{z-w}\,,\,\,\,&J^{mn}(z)S_{A}(w)\,&\sim\,\frac{1}{2}\frac{{(\Sigma^{mn})_{A}}^{B}S_{B}(w)}{z-w}\,.
\end{alignat}

Using the above relations and further decomposing the internal index $m$, \emph{e.g.} with respect to a $K3\times T^2$ compactification, we obtain the following disc correlation functions:

\begin{alignat}{3}
&\Bigl<\Psi(z_1)\bar\Psi(z_2)\Bigr>\, &=&\, z_{12}^{-1}\,,\nonumber\\
&\Bigl<\Psi(z_1)S^{\hat A}(z_2)S^{\hat B}(z_3)\Bigr>\,&=&\,-i\,\epsilon^{\hat A\hat B}\,z_{12}^{-1/2}z_{13}^{-1/2}z_{23}^{-1/4} \,,\nonumber\\
&\Bigl<\psi^{\mu}(z_1)\psi^{\nu}(z_2)\psi^{\rho}(z_3)\psi^{\sigma}(z_4)\Bigr>\,&=&\,\delta^{\mu\nu}\delta^{\rho\sigma}\,z_{12}^{-1}z_{34}^{-1}-\delta^{\mu\rho}\delta^{\nu\sigma}\,z_{13}^{-1}z_{24}^{-1}\nonumber\\
&&&+\delta^{\mu\sigma}\delta^{\nu\rho}\,z_{14}^{-1}z_{23}^{-1}\,,\nonumber\\
&\Bigl<S^{\dot\alpha}(z_1)S^{\dot\beta}(z_2)S^{\dot\gamma}(z_3)S^{\dot\delta}(z_4)\Bigr>\,&=&\,\frac{\epsilon^{\dot\alpha\dot\beta}\epsilon^{\dot\gamma\dot\delta}\,z_{14}z_{23}-\epsilon^{\dot\alpha\dot\delta}\epsilon^{\dot\beta\dot\gamma}\,z_{12}z_{34}}{(z_{12}z_{13}z_{14}z_{23}z_{24}z_{34})^{1/2}}\,,\nonumber\\
&\Bigl<\psi^{\mu}(z_1)\psi^{\nu}(z_2)S^{\dot\alpha}(z_3)S^{\dot\beta}(z_4)\Bigr>\,&=&\,-\frac{1}{2}\frac{z_{13}z_{24}+z_{23}z_{14}}{z_{12}(z_{13}z_{14}z_{23}z_{24}z_{34})^{1/2}}\delta^{\mu\nu}\epsilon^{\dot\alpha\dot\beta}\nonumber\\
&&&-\frac{1}{2}\frac{z_{34}^{1/2}}{(z_{13}z_{14}z_{23}z_{24})^{1/2}}(\bar\sigma^{\mu\nu})^{\dot\alpha\dot\beta}\,.
\label{Correlators}
\end{alignat}
In addition, we give the useful correlators of twist fields and superghosts:
\begin{align}
\Bigl<\bar{\Delta}(z_1)\Delta(z_2)\Bigr>&=-z_{12}^{-1/2}\,,\nonumber\\
\Bigl<e^{-\varphi(z_1)}e^{-\tfrac{1}{2}\varphi(z_2)}e^{-\tfrac{1}{2}\varphi(z_3)}\Bigr>&= z_{12}^{-1/2}z_{13}^{-1/2}z_{23}^{-1/4}\,.
\label{CorrelatorsBis}
\end{align}
}

\chapter{Modular forms}\label{appendix:ModularFunctions}
\section{Theta Functions and Prime Forms}\label{appendix:ModularFunctionsGeneral}

Consider a (closed) Riemann surface $\Sigma_g$ of genus $g$ and define the canonical basis for the homology cycles $\delta_I=a_i,b_i$:
\begin{align}
 a_i\cap a_j&=0\,,\\
 b_i\cap b_j&=0\,,\\
 a_i\cap b_j&=\delta_{ij}\,,
\end{align}
with $i=1,\ldots,g$ and $I=1,\ldots,2g$. In addition, one can choose a basis of one-forms $\omega_i$ such that
\begin{align}
 \oint_{a_i}\omega_j=\delta_{ij}\,,&&\oint_{b_i}\omega_j=\tau_{ij}\,.
\end{align}
Here, $\tau$ is a symmetric matrix with positive definite imaginary part called the \emph{period matrix} of the Riemann surface. The periods $\Pi_I\in\mb C^g$ defined as
\begin{equation}
 \Pi_I=\left(\int_{\delta_I}\omega_j\right)_{j=1,\ldots,g}
\end{equation}
are linearly independent vectors and, thus, form a lattice $\Lambda$ in $\mb C^g$. It can be used to construct the \emph{Jacobian variety} $\mc J(\Sigma_g)$ of $\Sigma_g$:
\begin{equation}
 \mc J(\Sigma_g)=\mb C^g/\Lambda=\mb C^g/(\mb Z^g+\tau\mb Z^g)\,.
\end{equation}
Coordinates on a genus $g$ Riemann surface are naturally defined as follows. Choose a base point $p_0$ on $\Sigma_g$ and cut the latter open along its homology cycles. The coordinates of a point $p$ different than $p_0$ is given by the \emph{Jacobi map}
\begin{align}
 \Sigma_g&\rightarrow\mc J(\Sigma_g)\nonumber\\
\mu\,:\quad  p\,\,&\mapsto\left(\int_{p_0}^p\omega_i\right)_{i=1,\ldots,g}\,.
\end{align}
More generally, the Jacobi map can be defined on a degree zero divisor on $\Sigma_g$ as
\begin{align}
 \textrm{Div}^{0}(\Sigma_g)&\rightarrow\mc J(\Sigma_g)\nonumber\\
\mu\,:\quad  \sum{p_i}-{q_i}&\mapsto\left(\sum_i\int_{q_i}^{p_i}\omega_j\right)_{j=1,\ldots,g}\,,
\end{align}
with $\sum\int_q^p\omega$ being the \emph{Abelian sums}. In most cases, we drop the distinction between points on a Riemann surface and its coordinates via the Jacobi map.

The Riemann (or genus $g$) theta function on $\Sigma_g$ is
\begin{equation}
 \vartheta(z,\tau)=\sum_{n\in\mb Z^g}e^{\,i\pi n^{T}\tau\,n+2i\pi n^{T}z}\,,
\end{equation}
with $z\in\mc J(\Sigma_g)$. It satisfies
\begin{align}
 \vartheta(z+\tau\,n+m,\tau)&=e^{-i\pi(n^{T}\tau\,n+2n^{T}z)}\vartheta(z,\tau)\,,\label{GenusgPeriodicity}
\end{align}
under shifts in the lattice $\Lambda$. One of the important results in the theory of Riemann theta functions is the Riemann vanishing theorem. It states that there exists a vector $\Delta\in\mb C^g$ such that for all $z\in\mb C^g$, $\vartheta(z,\tau)=0$ if and only if there exist $g-1$ points $p_1,\ldots,p_{g-1}$ on $\Sigma_g$ such that
\begin{equation}
 z=\Delta-\sum_{k=1}^{g-1}p_k\,.
\end{equation}
$\Delta$ is called the Riemann class and is an equivalence class of divisors of degree $g-1$ for $\Sigma_g$. The Riemann theta function can be generalised by including a spin structure $s=(s_1,s_2)\in(\frac{1}{2}\mb Z/\mb Z)^{2g}$:
\begin{equation}
 \vartheta_s(z,\tau)=e^{\,i\pi(s_1^{T}\tau s_1+2s_1^{T}(z+s_2))}\vartheta(z+\tau s_1+s_2,\tau)\,.\label{ThetaGenusg}
\end{equation}

Consider the function $f_{s}(z,w)\equiv\vartheta_s(z-w,\tau)$ which has a simple zero at $z=w$. Differentiating with respect to $w$ around the latter defines a one-form called $g_s(z)$ whose `square-root' is a holomorphic $\frac{1}{2}$-differential denoted $h_s(z)$. The prime form $E(z,w)$ is then defined as
\begin{equation}
 E(z,w)=\frac{f_s(z,w)}{h_s(z)h_s(w)}\,.
\end{equation}
One can show that the prime form is independent of the spin structure and that it has a simple zero at $z=w$:
\begin{equation}
 E(z,w)=z-w+O(1)\,.
\end{equation}

Most of the identities derived for theta functions are deduced from Riemann theta identities \cite{Mumford:1983}. Here, we only state the specific Riemann identity of interest:
\begin{align}
 2^{-g}\sum_s \vartheta_s(u)&\vartheta_s(v)\vartheta_s(x)\vartheta_s(y)=\nonumber\\
 &\vartheta(\frac{u+v+x+y}{2})\vartheta(\frac{u+v-x-y}{2})\vartheta(\frac{u-v+x-y}{2})\vartheta(\frac{u-v-x+y}{2})\,.\nonumber\\\label{RiemannSummation}
\end{align}

\section{The Genus One Case}\label{appendix:ModularFunctionsGenusOne}

In this section, we focus on the modular forms appearing at genus one which are relevant for the various one-loop amplitudes considered in this manuscript.

From \eqref{ThetaGenusg}, the genus one $\vartheta$-function with characteristics is given by
\begin{equation}
\vartheta[^a_b](z,\tau) = \sum\limits_{n\in\mathbb{Z}} e^{i\pi\tau (n-\frac{a}{2})^2}e^{2\pi i(z-\frac{b}{2})(n-\frac{a}{2})}\,,\label{ThetaGenusOne}
\end{equation}
with $a,b\in\mb R$ and $\tau\in\mb C^+$. Equivalently, we also use the following notation:
\begin{alignat}{3}
\vartheta_1(z,\tau)&=\,\vartheta[^1_1](z,\tau)\,,\,&\vartheta_2(z,\tau)&=\,\vartheta[^1_0](z,\tau)\,,\nonumber\\
\vartheta_3(z,\tau)&=\,\vartheta[^0_0](z,\tau)\,,\,&\vartheta_4(z,\tau)&=\,\vartheta[^0_1](z,\tau)\,.\label{ThetaOtherNotation}
\end{alignat}
From the definition \eqref{ThetaGenusOne}, it is easy to derive the $z$-periodicity property (see also \eqref{GenusgPeriodicity})
\begin{equation}
\vartheta[^a_b](z+\alpha\tau+\beta,\tau)=e^{-i\pi\tau\alpha^2-i\pi\alpha(2z-b)-2i\pi\alpha\beta}\vartheta[^{a-2\alpha}_{b-2\beta}](z,\tau)\,.\label{zPeriodicity}
\end{equation}
Moreover, for $a,b\in\mb Z$,
\begin{align}
\vartheta[^{a+2}_{\,\,\,\,b}](z;\tau)&=\,\vartheta[^a_b](z,\tau)\,,&\,\vartheta[^{\,\,\,\,a}_{b+2}](z;\tau)&=\,(-)^a\vartheta[^a_b](z,\tau)\,,\\
\vartheta[^{-a}_{-b}](z,\tau)&=\,\vartheta[^a_b](-z,\tau)\,,&\,\vartheta[^a_b](-z,\tau)&=\,(-)^{ab}\vartheta[^a_b](z,\tau)\,.
\end{align}
In order to check modular invariance, we use the behaviour of $\vartheta$ under the $S$- and $T$-modular transformations
\begin{align}
T:\,&\tau\rightarrow\tau+1\,,\\
S:\,&\tau\rightarrow-\frac{1}{\tau}\,,
\end{align}
that is,
\begin{alignat}{3}
&\vartheta[^a_b](z,\tau+1)\,&=&\,e^{-\frac{i\pi}{4}a(a-2)}\,\vartheta[^{\,\,\,\,\,\,\,a}_{a+b-1}](z,\tau)\,,\\
&\vartheta[^a_b](\tfrac{z}{\tau},-\tfrac{1}{\tau})&=&\,\sqrt{-i\tau}\,e^{\frac{i\pi}{2}(ab+2\,z^2/\tau)}\,\vartheta[^{\,\,\,b}_{-a}](z,\tau)\,.
\end{alignat}
The $z$-derivative of the $\vartheta$-function is related to another function, the Dedekind $\eta$-function
\begin{equation}\label{DedekindEta}
\eta(\tau)=q^{\frac{1}{24}}\prod_{n\geq1}(1-q^n)\,,
\end{equation}
through
\begin{equation}
\frac{\partial\,\vartheta_1(z,\tau)}{\partial z}\Bigr|_{z=0}=2\pi\,\eta^{3}(\tau)\,.
\end{equation}
Here, we have defined $q=e^{2i\pi\tau}$. The $\eta$-function transforms under $S$ and $T$ as
\begin{alignat}{3}
&\eta(-\tfrac{1}{\tau})&=&\,\sqrt{-i\tau}\,\eta(\tau)\,,\\
&\eta(\tau+1)\,&=&\,\eta(\tau)\,.
\end{alignat}
The $\vartheta$-function satisfy a useful identity known as the \emph{product formula} ($a,b\in\mb Z_2$):
\begin{align}
\vartheta[^a_b](z,\tau)=&(1+a)\cos\pi a(z-\tfrac{b}{2})\nonumber\\
&\times q^{\frac{a^2}{8}}\prod_{n\geq1}(1-q^n)(1+q^{n+\frac{a-1}{2}}e^{2i\pi(z-\frac{b}{2})})(1+q^{n+\frac{a-1}{2}}e^{-2i\pi(z-\frac{b}{2})})\,,
\end{align}
from which one can extract the asymptotic limit $\tau_2\rightarrow\infty$, known as the $\mc C$-identity:
\begin{equation}\label{CIdentity}
\vartheta[^a_b](z,\tau)\overset{\tau_2\rightarrow\infty}{\longrightarrow}(1+a)\cos\pi a(z-\tfrac{b}{2})\,q^{\frac{a^2}{8}}\,.
\end{equation}
An important Riemann theta identity satisfied by a product of four $\vartheta$-functions is the \emph{Abstrusa} identity
\begin{equation}
\frac{1}{2}\sum_{a,b\in\mb Z_2}(-)^{a+b}\prod_{i=1}^4\vartheta[^{a+h_i}_{b+g_i}](z_i,\tau)=e^{-i\pi h_1}\prod_{i=1}^{4}\vartheta[^{1+\tilde h_i}_{1+\tilde g_i}](\tilde z_i,\tau)\,,
\end{equation}
with $\sum h_i=0$. The tilded quantities are defined via the matrix
\begin{equation}\left(
\begin{array}{rrrr}
1&1&1&1\\
1&1&-1&-1\\
1&-1&1&-1\\
1&-1&-1&1
\end{array}
\right)\,,
\end{equation}
using the change of basis ($X=z,h,g$)
\begin{align}
\tilde X=\frac{1}{2}MX\,.
\end{align}

A modular form $F(\tau)$ of weight $w$ is a holomorphic function on the upper-half plane such that it is holomorphic at the cusp $z\rightarrow i\infty$ and transforms under the modular group
\begin{equation}
SL(2,\mb Z)=\left\{\left(\begin{array}{cc}a&b\\c&d\end{array}\right),\,a,b,c,d\in\mb Z\,,ad-bc=1\right\}
\end{equation}
as
\begin{equation}
F(\tfrac{a\tau+b}{c\tau+d})=(c\tau+d)^w\,F(\tau)\,.
\end{equation}
A typical example is the holomorphic \emph{Eisenstein series} which is a modular form of weight $2k$:
\begin{equation}
E_{2k}=\sum_{(m,n)\in(\mb Z^2)^\ast}(m+n\tau)^{-2k}\,,
\end{equation}
with $k>1$. For $k=1$, one defines the quasi-modular form\footnote{There are no non-trivial modular forms of weight $2$.}
\begin{equation}
 \hat E_2=E_2-\frac{3}{\pi\tau_2}\,.
\end{equation}
The real analytic Eisenstein series
\begin{align}\label{RealEisenstein}
E(\tau,s)=\frac{1}{2}\sum_{(c,d)=1}\frac{\tau_2^s}{|c\tau+d|^{2s}}
\end{align}
is invariant under modular transformations. However, it is rather a \emph{Maass form} since it is not holomorphic. It is defined for $\textrm{Re}(s)>1$ and admits an analytic continuation in $s$ to a meromorphic function on the full complex plane with a unique pole at $s=1$.

One can define modular series of weight $(w,\bar w)$, $\bar w$ being the modular weight under $\bar\tau$. For instance, the more general Eisenstein series used in this manuscript,
\begin{align}\label{Poincare}
	\Phi_{\alpha,\beta}(\tau,\bar\tau) =\sum\limits_{(m,n)\neq(0,0)}\frac{\tau_2^{\alpha}}{|m+n\tau|^{2\alpha}(m+n\tau)^{\beta-2\alpha}}\,,
\end{align}
carries weight $(\beta-2\alpha,0)$. In particular, for $\beta=2\alpha$, it reduces to the real Eisenstein series:
\begin{equation}
 \Phi_{\alpha,2\alpha}=2\zeta(2\alpha)E(\tau,\alpha)\,.
\end{equation}

Finally, many useful properties of modular series can be derived using the \emph{Poisson summation formula}
\begin{equation}
\sum_{m_i\in\mb Z}e^{-\pi\,A_{ij}m_im_j+\pi\,B_im_i}=(\det A)^{-1/2}\sum_{m_i\in\mb Z}e^{-\pi(m_i+iB_i/2)(A^{-1})_{ij}(m_j+iB_j/2)}\,,
\label{PoissonResummation}
\end{equation}
which can be obtained by applying the identity
\begin{equation}
\sum_{n\in\mb Z}f(n)=\sum_{n\in\mb Z}\mc F[f](n)
\end{equation}
relating a function $f$ to its Fourier transform $\mc F[f]$, to a Gaussian function.

\section{Application of Theta Function Identities}\label{App:ThetaFunctions}


Modular invariance puts stringent constraints on one-loop amplitudes in (closed) string theory. However, some quantities like the sign between the odd and even spin structures remain independent of this property. This is important in order to perform the sum over the spin structures. In Section \ref{Sec:TypeIamp}, in order to fix that sign, we inserted a number of fermionic vertices and then imposed monodoromy invariance of the resulting amplitude. We now present the details of this derivation using the properties of the $\vartheta$-functions.

For simplicity, we mostly suppress the $\tau$-dependence in the argument of the $\vartheta$-functions. The latter satisfy the following shift identities under $x \rightarrow x+1$ as a consequence of \eqref{zPeriodicity}:
\begin{align}
&\vartheta_1(\tfrac{x}{2}+y) \rightarrow \vartheta_2(\tfrac{x}{2}+y) \qquad,\qquad \vartheta_2(\tfrac{x}{2}+y)\rightarrow 
-\vartheta_1(\tfrac{x}{2}+y)\,,\nonumber\\
&\vartheta_3(\tfrac{x}{2}+y)\rightarrow \vartheta_4(\tfrac{x}{2}+y)\qquad,\qquad \vartheta_4(\tfrac{x}{2}+y)\rightarrow   
\vartheta_3(\tfrac{x}{2}+y)\,.
\end{align}
On the other hand, under $x \rightarrow  x+\tau$ one obtains:
\begin{alignat}{4}
&\vartheta_3(\tfrac{x}{2}+y)\,\, &\rightarrow&\,\,   +q^{-\tfrac{1}{8}}\,  e^{-i\pi (\tfrac{x}{2}+y)}\,  &\vartheta_2(\tfrac{x}{2}+y)\,,  \nonumber\\
&\vartheta_4(\tfrac{x}{2}+y)\,\, &\rightarrow&\,\,   +q^{-\tfrac{1}{8}}\,  e^{-i\pi (\tfrac{x}{2}+y-\tfrac{1}{2})}\,  &\vartheta_1(\tfrac{x}{2}+y)\,,\nonumber\\
&\vartheta_2(\tfrac{x}{2}+y)\,\, &\rightarrow&\,\,  +q^{-\tfrac{1}{8}}\,  e^{-i\pi (\tfrac{x}{2}+y)}\,  &\vartheta_3(\tfrac{x}{2}+y)\,,\nonumber\\
&\vartheta_1(\tfrac{x}{2}+y)\,\, &\rightarrow&\,\,  +q^{-\tfrac{1}{8}}\,  e^{-i\pi (\tfrac{x}{2}+y-\tfrac{1}{2})}\, &\vartheta_4(\tfrac{x}{2}+y)\,,\nonumber\\
&\vartheta_1(x-y)\,\,&\rightarrow&\,\,- q^{-\tfrac{1}{2}}\, e^{-2 i \pi( x-y)}\, &\vartheta_1(x-y)\,.
\end{alignat}
We can use these identities to explicitly perform the sum over spin structures in (\ref{SpinStructureAi}) and (\ref{SpinStructureExpressions}). The idea is to impose monodromy invariance under the shift of one of the insertion points.  For instance, if we are interested in the monodromy properties with respect to just one of the gravitini -- say, the one at position $x_1$, the relevant contribution of the prime forms with argument $x_1$ to $\mathbb{B}_{i}$ is of the form:
\begin{equation}
\frac{\prod_{i=1}^{n_1}\vartheta_1(x_1-z_i)\prod_{j=1}^{n_3}\vartheta_1(x_1-w_j)}{\vartheta_1(x_1-y_2)\prod_{k=1}^{n_2}\vartheta_1(x_1-z'_k) 
\prod_{l=1}^{n_4}\vartheta_1(x_1-\hat{w}'_l)}~.
\end{equation}
Here we have put back the indices for the positions, since we need to know how many prime forms involve $x_1$. Using the constraint (\ref{ConstraintsSum}), we see that there is one extra prime form in the denominator. Using the theta-function identities above, we can now show that the combination
\begin{equation}
G\equiv G[3]-G[4]-G[2]+G[1]~,
\end{equation}
is invariant under monodromies $x_1\rightarrow x_1+1$ and $x_1\rightarrow x_1+\tau$. This combination corresponds to taking the difference between the $SO(8)$ Vector and  Spinor conjugacy classes, with the weights $(k_1,k_2,k_3,k_4)$ of the Vector class determined by the condition 
\begin{align}
&k_i \in \mathbb{Z}~,&& \text{with} &&\sum_{i=1}^4
k_i\in\mathbb{Z}_{\text{odd}}~.
\end{align} 
Similarly, the Spinor class is defined by the condition:
\begin{align}
&k_i \in \mathbb{Z}+\frac{1}{2}\,,&&\text{with} &&\sum_{i=1}^4 k_i\in\mathbb{Z}_{\text{odd}}~.
\end{align} 
The triality map leaving the Spinor class invariant while exchanging Vector and Conjugate Spinor classes is:
\begin{equation}
(k_1,k_2,k_3,k_4)\rightarrow
(\tfrac{k_1+k_2+k_3+k_4}{2},\tfrac{k_1+k_2-k_3-k_4}{2},\tfrac{k_1-k_2+k_3-k_4}{2},\tfrac{k_1-k_2-k_3+k_4}{2}) ~.\label{RiemannCoeffs}
\end{equation}
Therefore, we can express the result of the spin structure sum in (\ref{SpinStructureExpressions}) in the following form:
\begin{align}
G_{(i)}=&\vartheta_1(x_1-y_2+z-z'+w-\hat{w}')\vartheta_1(x_2-y_1+z-z'-w'+\hat{w})\nonumber\\
&\times\vartheta_{h}(u-u'+v-\hat{v}')\vartheta_{-h}(u-u'-v'+\hat{v})\,\mathbb{B}_{(i)}\,,\\
G_{(ii)}=&\vartheta_1(x_1-y_2+z-z'+w-\hat{w}'+\hat{x}_1+\hat{x}_2-\hat{y}_1-\hat{y}_2)\nonumber\\
&\times \vartheta_1(x_2-y_1+z-z'-w'+\hat{w}+\hat{x}_1+\hat{x}_2-\hat{y}_1-\hat{y}_2)\nonumber\\
&\times\vartheta_{h}(u-u'+v-\hat{v}')\vartheta_{-h}(u-u'-v'+\hat{v})\,\mathbb{B}_{(ii)}\,.
\end{align}
Taking into account $\mathbb{B}$ as well as the bosonisation identities of \cite{Verlinde:1986kw}, these can then be re-written in terms of fermionic correlators, as in (\ref{SpinResult2}).
\chapter{On Regularisation of Functional Determinants}\label{appendix:regularization}
\section{Zeta-Function Regularisation}\label{sec:ZetaReg}
{\allowdisplaybreaks
In this section, we present the regularisation method used in order to compute some of the functional determinants. For definiteness, we focus on the case $\Delta_f\equiv\Delta_+\Delta_-$ with
\begin{align}
 \Delta_{\pm}=\prod_{\left(m,n\right)\neq\left(0,0\right)}\left(m-n\tau \right) \left(m-n\tau\mp2i\epsilon \right)\,,
\end{align}
$\tau$ being an arbitrary complex number.

Using $\zeta$-function regularisation and a Sommerfeld-Watson transformation, we rewrite $\Delta_+$ as
\begin{align}
 \log\Delta_{+} = & -\frac{\partial}{\partial s}\sum_{\left(m,n\right)\neq\left(0,0\right)}\left[\left(m-n\tau \right)                \left(m-n\tau-2i\epsilon \right)+M\right]^{-s} \Biggr|_{s,M=0}~\notag\\
                = & -\frac{\partial}{\partial s}\sum_{\left(m,n\right)\in\mathbb{Z}^2}\left[ \left(m-z_0\right)  \left(m-z_+\right)\right]^{-s} +\log M\Biggr|_{s,M=0}~\notag\\
                = & -\frac{\partial}{\partial s}\sum_{n\in\mathbb{Z}}\oint_{\mathcal C}dz I\left(z\right) \left[ \left(z-z_0\right)  \left(z-z_+\right)\right] ^{-s}+\log M\Biggr|_{s,M=0}~\label{RegProduct}
\end{align}
where we have added an infrared regulator $M$ and defined
\begin{align}
 z_+ &\equiv n\tau+2i\epsilon+\frac{iM}{2\epsilon}+O\left( M\right)\equiv n\tau_1+it_+\,, \label{RootPlus}\\
 z_0 &\equiv n\tau-\frac{iM}{2\epsilon}+O\left( M\right)\equiv n\tau_1+it_0\,.\label{RootZero}
\end{align}
Also, we choose $\epsilon$ to be real and positive and $M$ to be real and negative (and of course smaller than $\epsilon$). The contour $\mathcal C$ goes all around the real axis from $-\infty^+$ to $\infty^+$ then from $\infty^-$ to $-\infty^-$, and the interpolating function $I\left( z\right) $ is defined as
\begin{align}
 I\left( z\right)\equiv\frac{e^{i\pi z}}{e^{i\pi z}-e^{-i\pi z}}-\frac{1}{2}=-\frac{e^{-i\pi z}}{e^{-i\pi z}-e^{i\pi z}}+\frac{1}{2}.
\end{align}
For this choice, we have two different branch cut structures. First notice that
\begin{equation}
\left[ \left(z-z_0\right)\left(z-z_+\right)\right] ^{-s}
\end{equation}
has three branch points so that we need to \textsl{link} all of them (having thus two branch cuts). Moreover, we have to avoid crossing the real axis with the branch cuts in our case. Hence, for $n\geq0$, because both roots are in the upper half-plane, we choose to have a branch cut going from $z_0$ up to $z_+$ and then from $z_+$ to $i\infty$. One important thing to keep in mind is that when we go from a point above $z_+$ at the right of the branch cut to a point above $z_+$ but on the other side, we pick up a double phase $e^{4i\pi s}$ instead of only one (you can see this easily by choosing one branch cut from $z_+$ to $i\infty$ and the other one from $z_-$ to $i\infty$ by slightly going at the left of the first branch cut, and then look at the monodromies of the integrand). For $n<0$, both are in the lower half-plane and we choose similar branch cuts going to $-i\infty$. Now the standard procedure is to deform the contour all the way to complex infinity avoiding the branch cuts so that the only remaining contributions are the line integrals over the branch cuts (see Fig. \ref{fig:ContourPos}).
\begin{figure}[ht]
\begin{center}
\includegraphics[width=0.8\textwidth]{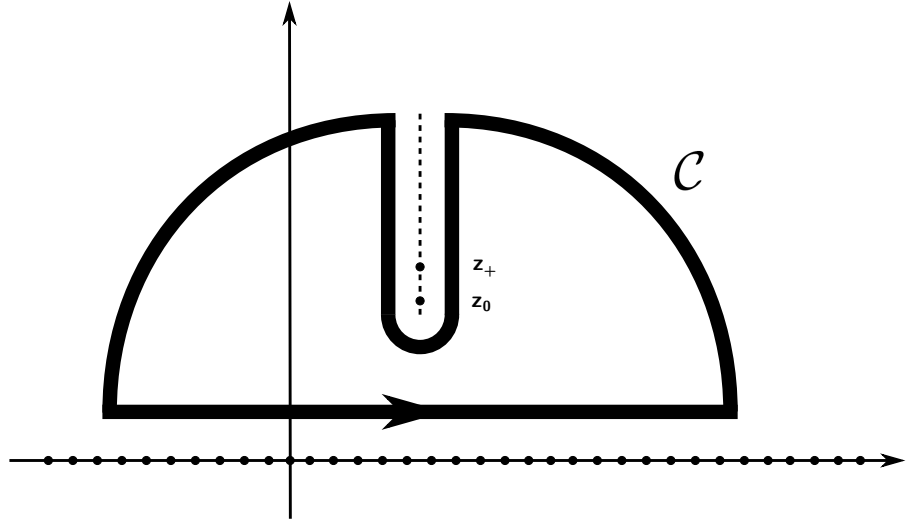}
\caption{Contour deformation for $\Delta_+$ with $n\geq0$. The dots represent poles and the dashed line is a choice of branch cut, with the dots on the latter being branch points.}
\label{fig:ContourPos}
\end{center}
\end{figure}

\noindent Taking into account this branch cut structure, \eqref{RegProduct} becomes
\begin{align}
 \log\Delta_{+} & =2\frac{\partial}{\partial s}\sin\left(\pi s \right) \sum_{n\geq0}\int_{t_0}^{t_+}dt I\left(n\tau_1+it\right) \left[ \left(t-t_0\right)  \left(t-t_+\right)\right] ^{-s}\notag\\
 & +2\frac{\partial}{\partial s}\sin\left(\pi s \right) \sum_{n\leq-1}\int_{t_+}^{t_0}dt I\left(n\tau_1+it\right) \left[ \left(t-t_0\right)  \left(t-t_+\right)\right] ^{-s}\notag\\
 & +2\frac{\partial}{\partial s}\sin\left(\pi s \right) \sum_{n\geq0}\int_{t_+}^{\infty}dt I\left(n\tau_1+it\right) \left[ \left(t-t_0\right)  \left(t-t_+\right)\right] ^{-s}\notag\\
 &+2\frac{\partial}{\partial s}\sin\left(\pi s \right) \sum_{n\leq-1}\int_{t_0}^{-\infty}dt I\left(n\tau_1+it\right) \left[ \left(t-t_0\right)  \left(t-t_+\right)\right] ^{-s}+\log M\Biggr|_{s,M=0}~.
\end{align}
For the non-constant part of the interpolating function, the result converges for $s>0$ and gives a non zero result only if the derivative hits the sine function. Thus, we rewrite the determinant as
\begin{align}
 \log\Delta_{+} & =-\frac{\partial}{\partial s}\sin\left(\pi s \right) \sum_{n\geq0}\int_{t_0}^{t_+}dt \left[ \left(t-t_0\right)  \left(t-t_+\right)\right] ^{-s}+\frac{\partial}{\partial s}\sin\left(\pi s \right) \sum_{n\leq-1}\int_{t_+}^{t_0}dt\left[ \left(t-t_0\right)  \left(t-t_+\right)\right] ^{-s}\notag\\
 & -\frac{\partial}{\partial s}\sin\left(2\pi s \right) \sum_{n\geq0}\int_{t_+}^{\infty}dt \left[ \left(t-t_0\right)  \left(t-t_+\right)\right] ^{-s}+\frac{\partial}{\partial s}\sin\left(2\pi s \right) \sum_{n\leq-1}\int_{t_0}^{-\infty}dt\left[ \left(t-t_0\right)  \left(t-t_+\right)\right] ^{-s}\notag\\
 & +2\pi\sum_{n\geq0}\int_{t_0}^{t_+}dt\frac{e^{i\pi n\tau_1-\pi t}}{e^{i\pi n\tau_1-\pi t}-e^{-i\pi n\tau_1+\pi t}} -2\pi\sum_{n\leq-1}\int_{t_+}^{t_0}dt\frac{e^{-i\pi n\tau_1+\pi t}}{e^{-i\pi n\tau_1+\pi t}-e^{i\pi n\tau_1-\pi t}}\notag\\
 & +4\pi\sum_{n\geq0}\int_{t_+}^{\infty}dt\frac{e^{i\pi n\tau_1-\pi t}}{e^{i\pi n\tau_1-\pi t}-e^{-i\pi n\tau_1+\pi t}} -4\pi\sum_{n\leq-1}\int_{t_0}^{-\infty}dt\frac{e^{-i\pi n\tau_1+\pi t}}{e^{-i\pi n\tau_1+\pi t}-e^{i\pi n\tau_1-\pi t}}\notag\\
 & +\log M\Biggr|_{s,M=0}~.
\end{align}
Performing the integrals yields
\begin{align}
 \log\Delta_{+} & =-\frac{\partial}{\partial s}\sin\left(\pi s \right)B\left(2s-1,1-s \right)\left(1-\left( -\right) ^{1-2s} \right)\left[\sum_{n\geq0}\left( t_0-t_+\right)^{1-2s}-\sum_{n\leq-1}\left( t_+-t_0\right)^{1-2s}\right]\notag\\
 & -\frac{\partial}{\partial s}\sin\left(2\pi s \right)B\left(2s-1,1-s \right)\left[\sum_{n\geq0}\left( t_0-t_+\right)^{1-2s}-\sum_{n\leq-1}\left( t_+-t_0\right)^{1-2s}\right]\notag\\
 & +\sum_{n\geq0}\log\left(1-e^{2i\pi n\tau_1-2\pi t_+}\right)\left( 1-e^{2i\pi n\tau_1-2\pi t_0}\right)\notag\\
 & +\sum_{n\leq-1}\log\left(1-e^{-2i\pi n\tau_1+2\pi t_+}\right)\left( 1-e^{-2i\pi n\tau_1+2\pi t_0}\right) \notag\\
 & +\log M\Biggr|_{s,M=0}~.\label{RegDet}
\end{align}
$B\left(x,y\right)$ denotes the Euler integral of the first kind. The $n=0$ term in the denominator of the first logarithm cancels the regulator term so that we can set it to zero in the rest of the expression. Using the definition of $t_+$ and $t_0$ in \eqref{RootPlus} and \eqref{RootZero}, \eqref{RegDet} reads:
\begin{align}
 \log\Delta_{+} & =-\frac{\partial}{\partial s}\sin\left(\pi s \right)B\left(2s-1,1-s \right)\left(1-\left( -\right) ^{1-2s} \right)\sum_{n\geq0}\left( t_0-t_+\right)^{1-2s}\Biggr|_{s,M=0}\notag\\
 & +\frac{\partial}{\partial s}\sin\left(\pi s \right)B\left(2s-1,1-s \right)\left(1-\left( -\right) ^{1-2s} \right) \sum_{n\leq-1}\left( t_+-t_0\right)^{1-2s}\Biggr|_{s,M=0}\notag\\
 & -\frac{\partial}{\partial s}\sin\left(2\pi s \right)B\left(2s-1,1-s \right)\sum_{n\geq0}\left( t_0-t_+\right)^{1-2s}\Biggr|_{s,M=0}\notag\\
 & +\frac{\partial}{\partial s}\sin\left(2\pi s \right)B\left(2s-1,1-s \right)\sum_{n\leq-1}\left( t_+-t_0\right)^{1-2s}\Biggr|_{s,M=0}\notag\\
 & +\log\frac{\left(1-e^{-4\pi\epsilon} \right)\prod_{n\geq1}\left(1-q^{n}e^{-4\pi\epsilon}\right)\left(1-q^{n}e^{4\pi\epsilon}\right)\left(1-q^{n}\right)^2 }{\pi\epsilon}~.
\end{align}
Finally, we compute the first four terms by noting that $t_0-t_+=-2\epsilon$. Both have a very similar structure. By summing of $n$ in both cases and taking directly the limit of $s$ going to zero, we show that the first term cancels against the second one and the third term against the fourth one. Thus, the only remaining piece is

\begin{align}
 \log\Delta_{+} = \log\left[ \frac{\left(1-e^{-4\pi\epsilon} \right)\prod_{n\geq1}\left(1-q^{n}e^{-4\pi\epsilon}\right)\left(1-q^{n}e^{4\pi\epsilon}\right)\left(1-q^{n}\right)^2 }{\pi\epsilon}\right] ~.
\end{align}
Exponentiating the previous result gives $\Delta_+$ which turns out to be analytic in $\epsilon$. However, to be more careful (the choice of lambda in the regularisation should be the same for all the factors), we carry out the same analysis for $\log\Delta_{-}$ rather than going through the previous argument. In this case, we have two main differences. The pole structure looks the same if we use the same regularisation except for the fact that both poles are exchanged, i.e. $z_-$ has a smaller imaginary part than $z'_0$:

\begin{align}
 z_- &\equiv n\tau-2i\epsilon-\frac{iM}{2\epsilon}+O\left( M\right)\equiv n\tau_1+it_-\,, \\
 z'_0 &\equiv n\tau+\frac{iM}{2\epsilon}+O\left( M\right)\equiv n\tau_1+it'_0\,.
\end{align}
Moreover, for the same reason, the zero mode part is now included in the negative mode sum rather than the positive one. Hence, going through the same steps, we obtain exactly the same result as for $\Delta_+$ even for the zero-mode part. Normalising the result to $1$ at $\epsilon=0$ yields
\begin{align}
\Delta_f=\left[\frac{\vartheta_1(2i\epsilon,\tau)}{4\pi i\epsilon\,\eta(\tau) ^3}\right] ^2~e^{\frac{4\pi}{\tau_2}\epsilon^2}\,.
\end{align}
}
\section{Functional Determinants}\label{sec:FuncDet}

\subsection{Heterotic Functional Determinants and Poincar\'e Series}\label{FourierExp}

In this appendix we discuss a modular-invariant regularisation of the bosonic determinant (\ref{Factorization}), 
using   properties of Poincar\'e series in order to extract the corresponding Fourier expansion. Our analysis 
closely follows \cite{Angelantonj:2011br,Angelantonj:2012gw}.
The factorisation of the modular invariant determinant (\ref{Factorization}) is defined in terms of the following functions:
\begin{align}\label{BosHol}
 G_{\textrm{ahol}}(\epsilon_{-},\epsilon_{+})\equiv&\prod\limits_{(m,n)\neq(0,0)}\Biggr[\left(\frac{2\pi}{\tau_2^2}\right)^2
\Bigr(A\bar{A}+(\tilde\epsilon_{-}-\tilde{\epsilon}_{+})A\Bigr)\Bigr(A\bar{A}+(\tilde\epsilon_{-}+\tilde{\epsilon}_{+})A
\Bigr)\Biggr]^{-1}\,,
\end{align}
\begin{align}\label{BosNhol}
 G_{\textrm{non-hol}}(\epsilon_{-},\epsilon_{+})\equiv&\prod\limits_{(m,n)\neq(0,0)}\Biggr[\Bigr(1+\frac{\tilde\epsilon_{+}A-
\check\epsilon_{+}\bar{A}}{A(\bar A+\tilde\epsilon_{-}-\tilde\epsilon_{+})}\Bigr)\Bigr(1+\frac{\tilde\epsilon_{+}A-
\check\epsilon_{+}\bar{A}}{A(\bar A-\tilde\epsilon_{-}-\tilde\epsilon_{+})}\Bigr)\Biggr]^{-1}\, ,
\end{align}
where we use the shorthand notation
\begin{align}\label{Adef}
	A \,\equiv\, m-\tau n \qquad \text{and} \qquad \bar A \,\equiv\, m-\bar\tau n \,.
\end{align}
The explicit representation of the almost holomorphic piece \eqref{BosHol} in terms of elliptic functions has already 
been calculated in \eqref{Amodel} using the method of Section \ref{sec:ZetaReg}. An alternative, more efficient way of obtaining the same result is the following. First of all, we take the logarithm of \eqref{BosHol}:
\begin{align}
 \log G_{\textrm{ahol}}&=-\lim_{s\rightarrow0}\frac{\partial}{\partial s}\sum_{(m,n)\in(\mb Z^2)^\ast}\frac{\tau_2^s}{\left[|m+n\tau|^2+(\tilde\epsilon_--\tilde\epsilon_+)(m+n\tau)\right]^s}+(\epsilon_-\rightarrow-\epsilon_-)\nonumber\\
 &=-\lim_{s\rightarrow0}\frac{\partial}{\partial s}\sum_{k>0}\binom{-s}{k}\left(\frac{\tilde\epsilon_--\tilde\epsilon_+}{\tau_2}\right)^k\,\Phi^\ast_{s,2s+k}+(\epsilon_-\rightarrow-\epsilon_-)\,.
\end{align}
Here, $\Phi$ is the modular series defined in (\ref{Poincare}) which regularises \eqref{BosHol} for large $s$. It is related to the usual Eisenstein series
\begin{equation}
 E(s,w)=\frac{1}{2}\sum_{(c,d)=1}\frac{\tau_2^{s-w/2}}{|c\tau+d|^{2s-w}}(c\tau+d)^{-w}
\end{equation}
through
\begin{equation}
 \Phi_{\alpha,\beta}=2\zeta(\beta)E(\beta/2,\beta-2\alpha)\,.
\end{equation}
Notice that the terms with odd $k$ are identically zero. Moreover, the term $k=0$ gives
\begin{equation}
 -\log\sqrt{\tau_2}\eta\bar\eta\,,
\end{equation}
which is already taken into account in the full result \eqref{FullAmplitudeHet}, since we normalise the functional determinants to one when $\epsilon_\pm$ are set to zero and simply weigh the $\epsilon_\pm$-deformed part of the amplitude with the partition function. Hence, we restrict the summation to $k\geq2$. For $k\geq 4$, $\Phi$ is a well-defined modular series even for $s=0$ and leads to the anti-holomorphic Eisenstein series $\bar E_{k}$. For $k=2$, one can show that in the limit $s\rightarrow0$, $\Phi^\ast_{s+2,2s+2}$ reduces to the quasi-modular form $\hat{\bar E}_2$. Indeed, since $\Phi^\ast_{s+2,2s+2}$ has a simple pole in $s$ and using the expansion of the binomial coefficient
\begin{equation}
 \binom{-s}{k}=(-)^k\frac{s}{k}+\mc O(s^2)\,,
\end{equation}
the pole vanishes upon acting with the derivative with respect to $s$. The constant term in the expansion of $\Phi^\ast_{s+2,2s+2}$ is $2\zeta(2)\tau_2^2\hat{\bar E}_2$. Therefore, taking the exponential yields
\begin{equation}
 G_{\textrm{ahol}}(\epsilon_{-},\epsilon_{+})=-\exp\left\{\sum_{k\geq1}\frac{\zeta(2k)}{k}(\tilde\epsilon_--\tilde\epsilon_+)^{2n}\hat{\bar E}_{2k}\right\}\exp\left\{\sum_{k\geq1}\frac{\zeta(2k)}{k}(\tilde\epsilon_-+\tilde\epsilon_+)^{2n}\hat{\bar E}_{2k}\right\}\,,
\end{equation}
with the notation $\hat E_{2k}=E_{2k}$ for $k\geq2$. Using the formula
\begin{equation}
 \frac{2\pi\lambda\eta^3}{\vartheta_1(\lambda,\tau)}e^{\frac{\pi\lambda^2}{2\tau_2}}=\exp\left\{\sum_{k\geq1}\frac{\zeta(2k)}{k}\lambda^{2n}\hat{E}_{2k}\right\}\,,
\end{equation}
we recover the result of Eq. \eqref{Amodel}.

We now focus on the non-holomorphic piece \eqref{BosNhol}. One way to see that the field theory limit of \eqref{BosNhol} trivialises at the Wilson line enhancement point (\ref{EnhancementPoint}) is to compute the $n=0$ contribution. Indeed, as can be seen by performing a Sommerfeld-Watson transformation, $n$ labels the oscillator number and thus corresponds to the mass excitation level. Consequently, the latter is exponentially suppressed except for the $n=0$ term. Now using the identity
\begin{align}
	\prod\limits_{m\neq 0}\Bigr(1+\frac{\alpha}{m+\beta}\Bigr) = \frac{\pi \beta}{\sin \pi\beta}~ \frac{\sin\pi(\alpha+\beta)}
{\pi(\alpha+\beta)}~,
\end{align}
it is straightforward to show that the $n=0$ term in \eqref{BosNhol} reads
\begin{align}\label{n=0}
	\Biggr[
\frac{\textrm{sinc}\,\pi(\tilde\epsilon_{-}-\check\epsilon_{+})}{\textrm{sinc}\, \pi(\tilde\epsilon_{-}-\tilde\epsilon_{+})}~ \frac{\textrm{sinc}\,\pi(\tilde\epsilon_{-}+\check\epsilon_{+})}
{\textrm{sinc}\, \pi(\tilde\epsilon_{-}+\tilde\epsilon_{+})}  \Biggr]^{-1}~,
\end{align}
with $\textrm{sinc}(x)\equiv \sin(x)/x$. At the enhancement point, $\tilde\epsilon_{\pm} = \check\epsilon_{\pm}$ (because $P_L=P_R$) and, hence, \eqref{n=0} trivialises. 

We  now prove this statement by regularising the full infinite product (\ref{Factorization}) at the string level in a 
modular-invariant fashion. We start by taking the logarithm:

\begin{align}
 \log[G^{\textrm{bos}}(\epsilon_-,\epsilon_+)]&=-\sum\limits_{(m,n)\neq(0,0)}\log\left(A\bar A+\tilde\epsilon_-A-\check\epsilon_+\bar A\right)+(\check\epsilon_{+}\rightarrow -\check\epsilon_{+})\nonumber\\
 &=\lim_{s\rightarrow0}\frac{\partial}{\partial s}\sum\limits_{(m,n)\neq(0,0)}\sum_{k\geq0}\binom{-s}{k}(A\bar A)^{-s}\left(\frac{\tilde\epsilon_-}{\bar A}-\frac{\check\epsilon_+}{A}\right)^k+(\check\epsilon_{+}\rightarrow -\check\epsilon_{+})\nonumber\\
 &=\lim_{s\rightarrow0}\frac{\partial}{\partial s}\sum_{\genfrac{}{}{0pt}{}{k\geq0}{k\in2\mb Z}}\sum_{\genfrac{}{}{0pt}{}{0\leq \ell\leq k}{l\in2\mb Z}}\binom{-s}{k}\binom{k}{\ell}\tilde\epsilon_-^{\,\ell}\,\check\epsilon_+^{\,k-\ell}\,\tau_2^{\,\ell-k-s}\,\Phi^\ast_{k-l+s,k+2s}\,,\label{prescription}
\end{align}
with $\Phi_{\alpha,\beta}$ being the modular series of weights $(0,\beta-2\alpha)$ defined in \eqref{Poincare}. It is absolutely convergent for $\beta>2$. Notice that in \eqref{prescription} $\beta=k+2s$ and, hence, the limit $s\rightarrow0$ is well-defined for $k\geq4$. The cases where $k=2$ are discussed separately below\footnote{The term $k=0$ is taken into account in the normalisation of the full amplitude \eqref{FullAmplitudeHet}.}.
\begin{enumerate}
	\item For $l=2$, similarly to the analysis for $G_{\textrm{ahol}}$, $\Phi_{s,2s+2}$  precisely reproduces the quasi-holomorphic Eisenstein series $2\zeta(2)\hat{E}_2$ upon taking the limit $s\rightarrow0$. 
	\item For $l=0$, using the identity
\begin{align}
	\Phi_{\alpha,\beta}(\tau,\bar\tau) = \tau_2^{2\alpha-\beta} \Bigr[\Phi_{\beta-\alpha,\beta}(\tau,\bar\tau)\Bigr]^* ~,
\end{align}
	it is easy to see that one similarly obtains $2\zeta(2)\tau_2^2\hat{\bar{E}}_2$.
\end{enumerate}
As a result, the Eisenstein series appearing in (\ref{prescription}) are well-defined in the limit where the regulator $s$ is set to zero:
\begin{align}
  \log[G^{\textrm{bos}}(\epsilon_-,\epsilon_+)]&=\zeta(2)(\tilde\epsilon_-^{\,2}\,\hat{\bar E}_2+\check\epsilon_+^{\,2}\,\hat E_2)\nonumber\\
   &+2\sum_{\genfrac{}{}{0pt}{}{k\geq4}{k\in2\mb Z}}\sum_{\genfrac{}{}{0pt}{}{0\leq \ell\leq k}{l\in2\mb Z}}\binom{k}{\ell}\frac{\zeta(k)}{k}\,\tilde\epsilon_-^{\,\ell}\,\check\epsilon_+^{\,k-\ell}\,\tau_2^{\,\ell-k}\bar E(k/2,2\ell-k)\,.\label{FullRegBos}
\end{align}

In order to derive the asymptotic behaviour of $G^{\textrm{bos}}$, we study the properties of the modular series in the expansion \eqref{prescription}. Indeed, the Fourier expansion of the modular series (\ref{Poincare}) is organised into an `asymptotic' contribution and an `oscillator' part:
\begin{align}\label{Fourier}
	\tau_2^{-\alpha}\,\Phi_{\alpha,\beta}(\tau,\bar\tau)=2\zeta(\beta)+2\tau_2^{1-\beta}\Biggr\{ C_0^{\alpha,\beta}+
\sum\limits_{n>0}\Bigr[C_n^{\alpha,\beta}(\tau_2)\,q^n + I_n^{\alpha,\beta}(\tau_2)\,\bar{q}^n \Bigr]\Biggr\}~.
\end{align}
The coefficients $C_0, C_n, I_n$ are given by:
\begin{equation}
	 \begin{dcases}
		 & C_n^{\alpha,\beta}(\tau_2) = \frac{(2\pi)^\beta (-i)^{\beta-2\alpha}}{\Gamma(\beta-\alpha)}\,(n\tau_2)^{\beta-1}\,
\sigma_{1-\beta}(n)\,(4\pi n\tau_2)^{-\frac{\beta}{2}}\,e^{2\pi n\tau_2}\,W_{\frac{\beta}{2}-\alpha,\frac{\beta-1}{2}}(4\pi n\tau_2)\\
		 & I_n^{\alpha,\beta}(\tau_2) = \frac{(2\pi)^\beta (-i)^{\beta-2\alpha}}{\Gamma(\alpha)}\,(n\tau_2)^{\beta-1}\,
\sigma_{1-\beta}(n)\,(4\pi n\tau_2)^{-\frac{\beta}{2}}\,e^{2\pi n\tau_2}\,W_{\alpha-\frac{\beta}{2},\frac{\beta-1}{2}}(4\pi n\tau_2)\\
		 & C_0^{\alpha,\beta}  =  2^{2-\beta}\pi (-i)^{\beta-2\alpha} \frac{\Gamma(\beta-1)\zeta(\beta-1)}{\Gamma(\alpha)
\Gamma(\beta-\alpha)}
	\end{dcases}  
\end{equation}
where $W_{\lambda,\mu}(z)$ is the Whittaker $W$-function and $\sigma_s(n)=\sum\limits_{d|n}{d^s}$ is the divisor function. 

Using the asymptotic properties of the Whittaker function, it is easy to show that the oscillator modes in (\ref{Fourier}), 
$\sum_n(C_n q^n + I_n\bar{q}^n)$, are exponentially suppressed in the limit, $\tau_2\rightarrow\infty$. In addition, the 
zero-frequency term in the curly brackets decays polynomially in the same limit. Hence, the dominant contribution in the 
field theory limit comes from the `asymptotic' part:
\begin{align}
	\tau_2^{-\alpha}\Phi_{\alpha,\beta} ~\overset{\tau_2\rightarrow\infty}{\longrightarrow}~ 2\zeta(\beta)~.
\end{align}
Consequently, in the limit $\tau_2\rightarrow\infty$, we obtain
\begin{align}
 \lim_{\tau_2\rightarrow\infty}\log[G^{\textrm{bos}}(\epsilon_-,\epsilon_+)]&=\zeta(2)(\tilde\epsilon_-^{\,2}+\check\epsilon_+^{\,2})\nonumber\\
 &+\sum_{k\geq2}\frac{\zeta(2k)}{k}(\tilde\epsilon_--\check\epsilon_+)^{2k}+(\check\epsilon_{+}\rightarrow -\check\epsilon_{+})\nonumber\\
 &=\sum_{k\geq1}\frac{\zeta(2k)}{k}(\tilde\epsilon_--\check\epsilon_+)^{2k}+(\check\epsilon_{+}\rightarrow -\check\epsilon_{+})\,.
\end{align}
Using the definition of the Riemann zeta function, the sum over $k$ can be performed and leads to the expected field theory limit:
\begin{align}
 \lim_{\tau_2\rightarrow\infty}\log[G^{\textrm{bos}}(\epsilon_-,\epsilon_+)]&=\sum_{n,k\geq1}\frac{1}{k}\left(\frac{\tilde\epsilon_--\check\epsilon_+}{n}\right)^{2k}+(\check\epsilon_{+}\rightarrow -\check\epsilon_{+})\nonumber\\
 &=-\sum_{n\geq1}\log\left[1-\left(\frac{\tilde\epsilon_--\check\epsilon_+}{n}\right)^{2}\right]+(\check\epsilon_{+}\rightarrow -\check\epsilon_{+})\nonumber\\
 &=-\log\left[\textrm{sinc}\,\pi(\tilde\epsilon_--\check\epsilon_+)\textrm{sinc}\,\pi(\tilde\epsilon_-+\check\epsilon_+)\right]\,.
\end{align}



\subsection{Type I Functional Determinants}\label{TypeIDetReg}

In this appendix we discuss the regularisation of the infinite products appearing in the functions 
$H_s(\tfrac{\epsilon}{2};\tfrac{g}{2};\tau)$, introduced in (\ref{Hfunction}). We start by choosing the regularisation 
parameter $s$ such that $\textrm{Re}(s)>1$ so that we are considering instead the exponential of
\begin{align}
f_g^s(\epsilon)=\sum\limits_{\genfrac{}{}{0pt}{}{m\in\mathbb{Z}}{n>0}}\log\left(1-\frac{(\epsilon/2)^2}{|m+\frac{g}{2}+\frac{\epsilon}{2}-
\frac{n\tau}{2}|^{2s}}\right) = -\sum\limits_{k=1}^\infty\frac{(\epsilon/2)^{2k}}{k}\sum\limits_{\genfrac{}{}{0pt}{}{m\in\mathbb{Z}}{n>0}}\frac{1}
{|m+\frac{g}{2}+\frac{\epsilon}{2}-\frac{n\tau}{2}|^{2sk}}~.\nonumber
\end{align}
For sufficiently large $s$, the sums are absolutely convergent. The series in $m,n$ can be viewed as a limit of a deformed 
real Eisenstein series $E(s;\tau)$. In order to study its behaviour in the the large-$\tau_2$ limit, we obtain an 
expansion in $q=e^{-\pi\tau_2}$. Using techniques similar to the ones used in extracting the Fourier expansion of  
Poincar\'e series (\emph{cf}. \cite{Angelantonj:2011br,Angelantonj:2012gw} for more details), we can obtain the analogue 
of the Chowla-Selberg formula:
 \begin{align}\label{PoincareExpansion}
f_g^s(\epsilon)= -\sum\limits_{k=1}^\infty\frac{(\epsilon/2)^{2k}}{k}\left(\frac{\tau_2}{2}\right)^{1-2sk}\sum\limits_{n>0}
\frac{1}{n^{2sk}}\sum\limits_{m\in(\mathbb{Z}/n\mathbb{Z})}\sum\limits_{c\in\mathbb{Z}}e^{2\pi i \frac{c}{n}(m+\frac{g+\epsilon}{2})}
\int\limits_{-\infty}^{\infty} dt~e^{-\pi i c t \tau_2} (t^2+1)^{-sk}~.
 \end{align}
This integral can be explicitly performed \cite{Angelantonj:2011br} as
\begin{align}
	\int\limits_{-\infty}^{\infty} dt~e^{-\pi i c t \tau_2} (t^2+1)^{-sk} = \begin{cases}
				\frac{2^{2-2sk}\pi\Gamma(2sk-1)}{[\Gamma(sk)]^2}  & ,~\textrm{for}~c=0 \\
				\frac{ (2\pi)^{2sk}(c\tau_2/2)^{2sk-1}}{[\Gamma(sk)]^2}e^{-\pi |c|\tau_2} 
\sigma(2\pi |c|\tau_2;sk) & ,~\textrm{for}~c\neq 0 \\
	\end{cases}~,
\end{align}
where $\sigma(z;s)$ is a dressed Bessel function, stripped off its asymptotic behaviour:
\begin{align}
	\sigma(z;s) = \pi^{-\frac{1}{2}}\Gamma(s) z^{\frac{1}{2}-s} e^{z/2} K_{s-\frac{1}{2}}(\tfrac{z}{2})~.
\end{align}
Indeed, for $z\rightarrow\infty$, it converges to $\sigma(z;s)\rightarrow 1$. As a result, the `mode expansion'  
(\ref{PoincareExpansion}) is exponentially suppressed in the limit $\tau_2\rightarrow \infty$ for the non-vanishing 
`frequencies', $c\neq 0$.
Special care is required in the treatment of the $c=0$ term which is potentially divergent. Ordinary (completed) 
Eisenstein series $E^\star(s;\tau)\equiv\zeta^\star(2s)E(s;\tau)$ have a meromorphic continuation to the full $s$-plane, 
except for simple poles at $s=0,1$. In our case, this problematic behaviour may arise from the $k=1$ term, as we try 
to remove the regulator, $s\rightarrow 1$. However, this naive divergence cancels out between the bosonic and fermionic 
determinants.

Indeed, let us pick the $c=0$ mode contribution in the above sum:
\begin{align}
	 -\sum\limits_{k=1}\frac{(\epsilon/2)^{2k}}{k}\left(\frac{\tau_2}{2}\right)^{1-2sk}\frac{2^{2-2sk}\pi \Gamma(2sk-1)
\zeta(2sk-1)}{[\Gamma(sk)]^2}~.
\end{align}
It is clear from the properties of the Riemann $\zeta$-function that the $k=1$ term has a simple pole at $s=1$. However, 
there is only an overall multiplicative $\epsilon$-dependence for this term. Taking the logarithm of the full ratio of 
fermionic and bosonic determinants appearing in (\ref{FullCorrelatorTypeI}), regularising each sum by introducing the 
$s$-parameter and extracting the $c=0$ term we observe that the dangerous $k=1$ terms cancel:
\begin{align}\nonumber
	 -\left(\tfrac{\tau_2}{2}\right)^{1-2s}&\frac{2^{2-2s}\pi \Gamma(2s-1)\zeta(2s-1)}{[\Gamma(s)]^2} \\
	 		&\times \Biggr[ 2\left(\frac{\epsilon_{-}}{2}\right)^2+2\left(\frac{\epsilon_{+}}{2}\right)^2-
\left(\frac{\epsilon_{-}-\epsilon_{+}}{2}\right)^2-\left(\frac{\epsilon_{-}+\epsilon_{+}}{2}\right)^2\Biggr] =0~.
\end{align}
Notice the relative factors of 2 in the first two terms in the square brackets, arising due to the fact that the fermionic 
products contain positive and negative $n\neq 0$ contributions, whereas the bosonic ones are restricted to $n>0$ only.

As a result, we can remove the regulator $s\rightarrow 1$ and obtain a well-defined expansion. In particular, in order to 
study the field theory limit, the relevant terms are those with $c=0$, $k>1$. It is easy to see that they decay power-like 
as $\tau_2^{1-2k}$. Hence, taking the exponential, the ratio of the infinite products (and, hence, the ratio of 
$H$-functions) goes to $1$ in the limit $\tau_2\rightarrow\infty$.

\end{appendices}

\backmatter


\bibliographystyle{plainnat}


\printindex

\mbox{}\clearpage
\thispagestyle{plain}
\begin{newquotation}
\noindent\textit{``Quand les mouettes suivent un chalutier,}\\
\noindent\textit{c'est qu'elles pensent qu'on va leur jeter des sardines.''}\\[0.5cm]
\begin{flushright}
 Eric Cantona
\end{flushright}
\end{newquotation}

\end{document}